\documentclass[epsfig,a4paper,11pt,titlepage]{book}
\usepackage{epsfig}
\usepackage{fancyhdr}
\usepackage{plain}

\usepackage{cite}

\usepackage{longtable}

\usepackage{multirow}

\usepackage{tocbibind}

\usepackage{array, xcolor}
\usepackage{enumitem}

\usepackage{picins} % for adding picture

\newenvironment{license}
  {\clearpage           % we want a new page
   \thispagestyle{empty}% no header and footer
   \vspace*{\stretch{3}}% some space at the top 
   \raggedright          % flush to the left margin
  }
  {\par % end the paragraph
   \vspace{\stretch{1}} % space at bottom is three times that at the top
   \clearpage           % finish off the page
  }
\newenvironment{dedication}
  {\clearpage           % we want a new page
   \thispagestyle{empty}% no header and footer
   \vspace*{\stretch{1}}% some space at the top 
   \itshape             % the text is in italics
   \raggedleft          % flush to the right margin
  }
  {\par % end the paragraph
   \vspace{\stretch{3}} % space at bottom is three times that at the top
   \clearpage           % finish off the page
  }
\makeatletter
\renewcommand\part{%
  \if@openright
    \cleardoublepage
  \else
    \clearpage
  \fi
  \thispagestyle{empty}%
  \if@twocolumn
    \onecolumn
    \@tempswatrue
  \else
    \@tempswafalse
  \fi
  \null\vfil
  \secdef\@part\@spart}
\makeatother

\newcommand{\clearemptydoublepage}{\newpage{\pagestyle{empty}\cleardoublepage}}

\makeindex
\usepackage{amssymb}
\usepackage{amsmath}
\usepackage{amsthm}

\newtheorem{definition}{Definition}
\newtheorem{theorem}{Theorem}

\usepackage{wrapfig}
\usepackage{graphicx}%,epstopdf}

\usepackage{algcompatible}
\usepackage{algorithm}
\usepackage{algorithmicx}
\usepackage{algpseudocode}

\numberwithin{algorithm}{chapter}

\algnewcommand{\algorithmicinput}{\textbf{\textit{Description:}}}
\algnewcommand\INPUT{\item[\algorithmicinput]}

\newcommand{\algofontsize}{\fontsize{11}{12}\selectfont} % {7}{8}
\usepackage{makecell}

\usepackage{subfigure}

\usepackage{AMMALanguages}

\newcommand{\Keywords}{\lstset{keywords={if,then,can,be,active,in,execute}}}

\newcommand{\footstar}[1]{%
\begingroup
\renewcommand\thefootnote{$\star$}\footnote{#1}%
\addtocounter{footnote}{-1}%
\endgroup
}

\usepackage[hyphens]{url}
\usepackage[breaklinks=true]{hyperref}
\usepackage{breakurl}

\usepackage[nonumberlist,acronym,toc]{glossaries} % makes a separate acronym list

\makeglossaries 

\newacronym{ABE}{ABE}{Attribute-Based Encryption}

\newacronym{ACL}{ACL}{Access Control List}

\newacronym{AES}{AES}{Advanced Encryption Standard}

\newacronym{BDH}{BDH}{Bilinear Diffie-Hellman}

\newacronym{BPM}{BPM}{Business Process Management}

\newacronym{CC}{CC}{Contextual Condition}

\newacronym{CCE}{CCE}{Contextual Condition Evaluation}

\newacronym{CDN}{CDN}{Content Delivery Network}

\newacronym{CPABE}{CP-ABE}{Ciphertext-Policy Attribute-Based Encryption}
\newacronym{CW}{CW}{Chinese Wall}

\newacronym{DAC}{DAC}{Discretionary Access Control}

\newacronym{DDH}{DDH}{Decisional Diffie-Hellman}

\newacronym{DSoD}{DSoD}{Dynamic Separation of Duties}

\newacronym{DTN}{DTN}{Delay Tolerant Network}

\newacronym{ECC}{ECC}{Elliptic Curve Cryptography}

\newacronym{EGRANT}{E-GRANT}{EnforcinG encRypted dynAmic security constraiNts in The cloud}

\newacronym{EHR}{EHR}{Electronic Health Record}

\newacronym{ERBAC}{ERBAC}{Encrypted Role-Based Access Control}
\newacronym{ERM}{ERM}{Enterprise Resource Management}

\newacronym{ESPOON}{ESPOON}{Enforcing Sensitive Policies in Outsourced envirOnmeNts}

\newacronym{ESPOONERBAC}{ESPOON$_{\mathit{ERBAC}}$}{Enforcing Sensitive Policies in Outsourced envirOnmeNts with Encrypted Role-Based Access Control}
\newacronym{HBDSoD}{HBDSoD}{History-Based Dynamic Separation of Duties}
\newacronym{IETF}{IETF}{Internet Engineering Task Force}

\newacronym{INDCPA}{IND-CPA}{INDistinguishable under Chosen Plaintext Attack}

\newacronym{IT}{IT}{Information Technology}

\newacronym{KB}{KB}{Kilo Byte}

\newacronym{KE}{KE}{Keyword Encryption}

\newacronym{KPABE}{KP-ABE}{Key-Policy Attribute Based Encryption}
\newacronym{MAC}{MAC}{Mandatory Access Control}

\newacronym{ms}{ms}{milliseconds}

\newacronym{MSSE}{MSSE}{Multi-user Searchable Symmetric Encryption}

\newacronym{ObDSoD}{ObDSoD}{Object-Based Dynamic Separation of Duties}

\newacronym{OpDSoD}{OpDSoD}{Operational Dynamic Separation of Duties}

\newacronym{OEM}{OEM}{Outsourced Enforcement Module}

\newacronym{PA}{PA}{Permission Assignment}

\newacronym{PBC}{PBC}{Pairing-Based Cryptography}

\newacronym{PDP}{PDP}{Policy Decision Point}

\newacronym{PEKS}{PEKS}{Public-key Encryption with Keyword Search}

\newacronym{PEP}{PEP}{Policy Enforcement Point}

\newacronym{PIDGIN}{PIDGIN}{Privacy preserving Interest anD content sharinG in opportunIstic Networks}

\newacronym{PIP}{PIP}{Policy Information Point}

\newacronym{PIR}{PIR}{Private Information Retrieval}

\newacronym{PKC}{PKC}{Public Key Cryptography}

\newacronym{PKI}{PKI}{Public Key Infrastructure}

\newacronym{PPT}{PPT}{Probabilistic Polynomial Time}

\newacronym{PRES}{PRES}{Proxy Re-Encryption with keyword Search}

\newacronym{QoS}{QoS}{Quality of Service}

\newacronym{RA}{RA}{Role Assignment}

\newacronym{RBAC}{RBAC}{Role-Based Access Control}

\newacronym{RH}{RH}{Role Hierarchy}

\newacronym{SaaS}{SaaS}{Software-as-a-Service}

\newacronym{SDE}{SDE}{Searchable Data Encryption}

\newacronym{SDSoD}{SDSoD}{Simple Dynamic Separation of Duties}

\newacronym{SP}{SP}{Search Permission}

\newacronym{SR}{SR}{Search Role}

\newacronym{SRH}{SRH}{Search Role Hierarchy}

\newacronym{TKMA}{TKMA}{Trusted Key Management Authority}

\newacronym{TTL}{TTL}{Time To Live}

\newacronym{XACML}{XACML}{eXtensible Access Control Markup Language}

\newacronym{XML}{XML}{eXtensible Markup Language}

\newenvironment{changemargin}[3]{%
\begin{list}{}{%
\setlength{\topsep}{0pt}%
\setlength{\leftmargin}{#1}%
\setlength{\rightmargin}{#2}%
\setlength{\topmargin}{#3}%
\setlength{\listparindent}{\parindent}%
\setlength{\itemindent}{\parindent}%
\setlength{\parsep}{\parskip}%
}%
\item[]}{\end{list}}

\begin{document}
\pagestyle{plain}

\mathchardef\mhyphen="2D

\newpage
\clearemptydoublepage
\thispagestyle{empty}

\begin{center}
\Large \textsc{PhD Dissertation} \\ \tiny
\hrulefill 

\vspace{7cm}

\LARGE\textbf{Privacy Preserving Enforcement of Sensitive Policies \\ in Outsourced and Distributed Environments}

\vspace{1cm}

\Large \textbf{Muhammad Rizwan Asghar} \\

\vspace{4cm}

Source: \url{http://eprints-phd.biblio.unitn.it/1124/}

\vspace{4cm}

\hrulefill

\normalsize
December $2013$

\end{center}

\newpage
\clearemptydoublepage
\thispagestyle{empty}
% left right top
\begin{changemargin}{-7mm}{-7mm}{0mm}
\begin{center}

\Large \textsc{PhD Dissertation} \\ \tiny

\hrulefill 
%\\\

\begin{figure}[h!]
\vspace{4mm}
  \centerline{\psfig{file=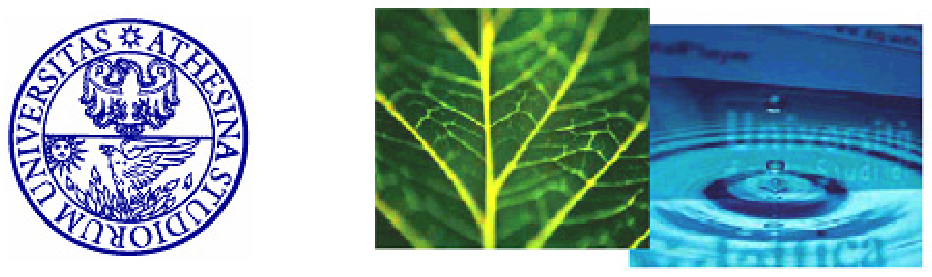,width=0.7\textwidth}}
\end{figure}

%\textbf{\large International Doctorate School in Information
%  and\\Communication Technologies}\\
\Large International Doctoral School in \\ Information and Communication Technologies (ICT) \\
\Large University of Trento, Italy \\

\vspace{7mm}

% TODO: find acronym
%\LARGE\textsc{Privacy Preserving Enforcement of Sensitive Policies in Outsourced and Distributed Environments}
\LARGE\textbf{Privacy Preserving Enforcement of Sensitive Policies \\ in Outsourced and Distributed Environments}

\vspace{2mm}

\begin{center}
\begin{tabular}{l}
%\Large \uppercase{\textbf{Muhammad Rizwan Asghar}} \\
\Large \textbf{Muhammad Rizwan Asghar} \\
\end{tabular}
\end{center}

\vspace{5mm}

{\normalsize %\scshape
\uppercase{
submitted to %\\
the Department of \\ Information Engineering and Computer Science (DISI) \\
%University of Trento %\\
in the partial fulfilment of the requirements for the degree of \\
\textbf{Doctor of Philosophy}
}
}

\vspace{13mm}

{
\large
\begin{tabular}{ll}
\textit{Advisors:} & 
Associate Prof. Dr. Bruno Crispo, University of Trento, Italy \vspace{1mm} \\
%
%\textit{Co-Advisor:} 
& Dr. Giovanni Russello, The University of Auckland, New Zealand \vspace{1mm} \\
\textit{Tutors:} & 
Prof. Dr. Imrich Chlamtac, CREATE-NET and University of Trento, Italy \vspace{1mm} \\
%
%\textit{Co-Tutor:} 
& Dr. Daniele Miorandi, CREATE-NET, Italy \vspace{1mm} \\
\textit{Examiners:}
& Associate Prof. Dr. Alessandro Armando, FBK and University of Genova, Italy \vspace{1mm} \\
& Dr. Ashish Gehani, SRI International, California, USA \vspace{1mm} \\
& Prof. Dr. Pierangela Samarati, University of Milan, Italy
\end{tabular}
}

\hrulefill

\normalsize
December $2013$
\end{center}
\end{changemargin}

\newpage
\clearemptydoublepage
\thispagestyle{empty}
\begin{license}
\textbf{$\copyright$ 2013 Muhammad Rizwan Asghar} \\
\vspace{4mm}
% left bottom right top
\includegraphics[trim=0mm 0mm 0mm 0mm,clip,width=.25\textwidth]{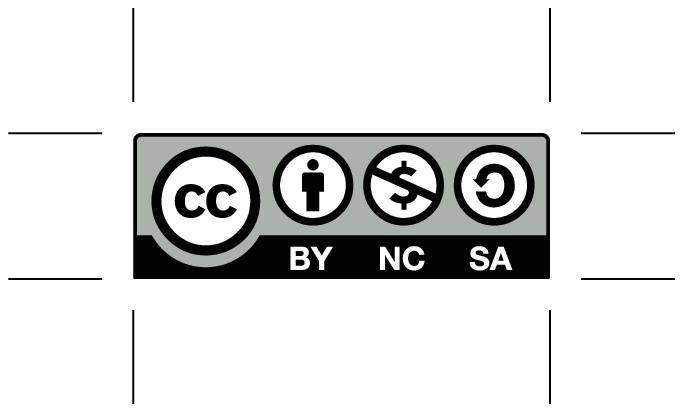} \\
\vspace{2mm}
This work is licensed under a \\ 
\vspace{2mm}
\textbf{Creative Commons \\ Attribution-NonCommercial-ShareAlike 3.0 Unported License} \\
\vspace{2mm}
To view a copy of this license, visit the following website: \\
\url{http://creativecommons.org/licenses/by-nc-sa/3.0/}
\end{license}

\newpage
\clearemptydoublepage
\thispagestyle{empty}
\begin{dedication}
To my family
%This work is dedicated to my family (including my parents, brothers, wife and son) for their utmost care, endless love and support.
%for their utmost caring, endless love and support.
\end{dedication}

\newpage
\clearemptydoublepage
\thispagestyle{empty}
\large

\pagenumbering{roman}

\chapter*{Abstract}
\addcontentsline{toc}{chapter}{Abstract}

\iffalse
\emph{
  The abstract goes here.  The abstract should be selfcontained
  and:
  1. clearly state the problem dealt with by the thesis
  2. give a synthetic description of the proposed solution
  3. highlight the sense in which the proposed solution enhances the
  state-of-the-art.
  The abstract must be limited to max. one page with no bibliographic
  references, nor external references on any kind.
}  
\fi 

% Context
The enforcement of sensitive policies in untrusted environments is still an open challenge for policy-based systems. On the one hand, taking any appropriate security decision requires access to these policies. On the other hand, if such access is allowed in an untrusted environment then confidential information might be leaked by the policies. The key challenge is how to enforce sensitive policies and protect content in untrusted environments. In the context of untrusted environments, we mainly distinguish between outsourced and distributed environments. The most attractive paradigms concerning outsourced and distributed environments are cloud computing and opportunistic networks, respectively.

% ESPOON
In this dissertation, we present the design, technical and implementation details of our proposed policy-based access control mechanisms for untrusted environments. First of all, we provide full confidentiality of access policies in outsourced environments, where service providers do not learn private information about policies during the policy deployment and evaluation phases. Our proposed architecture is such that we are able to support expressive policies and take into account contextual information before making any access decision. The system entities do not share any encryption keys and even if a user is deleted, the system is still able to perform its operations without requiring any action.
% ERBAC and EGRANT
For complex user management, we have implemented a policy-based \gls{RBAC} mechanism, where users are assigned roles, roles are assigned permissions and users execute permissions if their roles are active in the session maintained by service providers. Finally, we offer the full-fledged \gls{RBAC} policies by incorporating role hierarchies and dynamic security constraints.

% PIDGIN
In opportunistic networks, we protect content by specifying expressive access control policies. In our proposed approach, brokers match subscriptions against policies associated with content without compromising privacy of subscribers. As a result, an unauthorised broker neither gains access to content nor learns policies and authorised nodes gain access only if they satisfy fine-grained policies specified by publishers. Our proposed system provides scalable key management in which loosely-coupled publishers and subscribers communicate without any prior contact. Finally, we have developed a prototype of the system that runs on real smartphones and analysed its performance.

\vspace{8mm}
\noindent
{\bf Keywords:}
Policy Protection, Sensitive Policy Enforcement, Encrypted \acrshort{RBAC}, Secure Opportunistic Networks, Encrypted \acrshort{CPABE} Policies

%[put here a list of 3-5 keywords]

\chapter*{Acknowledgements}
\addcontentsline{toc}{chapter}{Acknowledgements}

\noindent \\

It would not have been possible to write this dissertation without the support of several individuals. It gives me great pleasure to acknowledge all the people who helped me, in different ways, during the adventurous journey of my life. \\

First and foremost, I would like to extend my sincere gratitude to my Ph.D. advisor Associate Prof. Dr. Bruno Crispo. Next, I would like to sincerely thank my second Ph.D. advisor Dr. Giovanni Russello. Both of them introduced me to scientific research and also provided constant guidance and advice throughout my research work. \\

I would like to say thanks to my Ph.D. tutor Prof. Dr. Imrich Chlamtac. As a president of Create-Net, he also offered me a research position that funded me during the course of my Ph.D. Furthermore, I am thankful to my second Ph.D. tutor Dr. Daniele Miorandi who was head of the iNSPIRE area I was part of. He was always available to guide, encourage and support me during my stay at Create-Net. \\

I am grateful to the members of my Ph.D. assessment committee comprising of Dr. Ashish Gehani, Associate Prof. Dr. Alessandro Armando and Prof. Dr. Pierangela Samarati. Moreover, I am thankful to Dr. Ashish Gehani and SRI International for providing me opportunity to visit the Computer Science Laboratory in Menlo Park, California, USA. \\

I have been fortunate in making good friends in my academic life, too many to mention one by one. I am really thankful to all of my friends who joined me during coffee breaks, online conversations and social occasions (such as dinner parties and excursions), as well as all my colleagues at the University of Trento, Create-Net and SRI International. Specifically, I am grateful to those who reviewed and provided their input for improving the quality of my research work. \\ \\

My doctoral studies were supported by EU FP7 COMPOSE (grant number 317862), EIT ICT Labs and EU FP7 ENDORSE (grant number 257063), which I gratefully acknowledge. \\

Above all, I am highly indebted to my family. My parents and brothers have given me their unequivocal support throughout my studies and they have always been by my side despite the distance. I would proudly mention my caring wife for her personal support and great patience at all times. I appreciate the spiritual support of my family. I love them very much. \\ \\ \\ \\

\hfill Muhammad Rizwan Asghar

\hfill Trento, Italy

\hfill December 2013

\tableofcontents
%\addcontentsline{toc}{chapter}{Contents}
%\clearemptydoublepage % TODO: uncomment if it not removing page number from blank page

%\addcontentsline{toc}{chapter}{\listtablename}
\listoftables

%\addcontentsline{toc}{chapter}{\listfigurename}
\listoffigures
%\clearemptydoublepage

\listofalgorithms
\addcontentsline{toc}{chapter}{\listalgorithmname}
%\addcontentsline{toc}{chapter}{List of Algorithms}
%\clearemptydoublepage

\printglossaries
%\addcontentsline{toc}{chapter}{List of Acronyms}

\chapter*{Table of Notations}
\addcontentsline{toc}{chapter}{Table of Notations}

\begin{longtable}{l l}

$\mathit{params}$ & The public parameters \\
$\mathit{msk}$ & The system wide master secret key \\ \\
$p$ and $q$ & Two primes of size $1^k$ \\ \\
$\mathbb{Z}^*_p$ and $\mathbb{Z}^*_q$  & Cyclic groups \\
$g$ & A generator \\
$\mathbb{G}$ & A unique order subgroup of $\mathbb{Z}^*_p$ \\ \\
$H$ & A collision-resistant hash function \\
$f$ & A pseudorandom function \\ \\
$K_{u_i}$ & The client side key set \\
$K_{s_i}$ & The server side key set \\ \\
$c^*_i (e)$ & The client encrypted element (by user $i$) \\
$c (e)$ & The server encrypted element \\ \\
$T^*_j (e)$ & The client generated trapdoor (by user $j$) \\
$T^ (e)$ & The server generated trapdoor \\ \\
$\mathit{CONDITION}$ & The contextual condition that is represented as a tree \\ \\
$\langle S, A, T \rangle$ & \makecell[l]{A tuple representing the subject $S$ can execute \\ the action $A$ on the target $T$} \\ \\
$KS$ & Key Store \\ \\
$AT$ & Access Time \\ \\
%
% TODO: remove it if conflicting with any other notations
$m$, $m_1$ and $m_2$ & \makecell[l]{Number of string attributes or \\ Number of string comparisons in a contextual condition} \\ \\
$n$, $n_1$ and $n_2$ & \makecell[l]{Number of numerical attributes or \\ Number of numerical comparisons in a contextual condition} \\ \\
$s$ & \makecell[l]{Size of a numerical attribute or \\ Size of a numerical comparison} \\ \\
$\mathit{ACT} = (i, R)$ & \makecell[l]{A role activation request that includes identity Requester $i$ \\ along with role $R$ to be activated} \\ \\
$\mathit{REQ} = (R, A, T)$ & \makecell[l]{An access request that includes role $R$ a Requester is active in and \\ action $A$ to be taken over target $T$} \\ \\
$L_r$ & List of roles \\
$|L_r|$ & Number of roles in the list \\ \\
$L_p$ & List of permissions \\
$|L_p|$ & Number of permissions in the list \\ \\
$G_{RH}$ & The role hierarchy graph \\
$|G_{RH}|$ & Number of roles in the role hierarchy graph \\ \\
$Y$ & Number of actions in \acrshort{HBDSoD} \\
$Z$ & Number of domains in \acrshort{CW} \\ \\
$c$ & Number of constraints \\
$r$ & Number of records \\ \\
$A$ & A list of attributes \\
$|A|$ & Number of attributes in the list \\ \\
$A_P^*$ & A list of attributes used to encrypt content \\
$|A_P^*|$ & Number of attributes used to encrypt content \\ \\
$A_S^*$ & A list of attributes used to encrypt interest \\
$|A_S^*|$ & Number of attributes used to encrypt interest \\ \\
$C$ & Content \\
$|C|$ & Content size \\ \\
$I$ & A list of keywords a subscriber is interested in \\
$|I|$ & Number of keywords a subscriber is interested in \\ \\
$T$ & A list of search tags associated with content \\
$|T|$ & Number of search tags associated with content \\ \\
\end{longtable}

\clearemptydoublepage

\pagestyle{fancy}

\pagenumbering{arabic}

\chapter{Introduction}
\label{cha:intro}

% NOTE!!!!! Short introductions signify the start of deeper problems: candidates are unaware of the research area or the theoretical framework. In the case of introductions and conclusions in doctoral theses, size does matter.

The recent advancements in technology have changed the way how electronic data is stored and retrieved. Nowadays, individuals and enterprises are increasingly utilising remote services (such as Dropbox \cite{Dropbox}, Google Cloud Storage \cite{Google:2013} and Amazon Simple Storage Service \cite{Amazon:2013}), mainly for economical benefits. These services not only enable information sharing but also ensure availability of data from anywhere at any time. However, the growing use of remote services raises serious privacy issues by putting personal data at risk, particularly when the servers offering such services are untrusted. Unfortunately, servers get direct access to the data they store and process. For protecting sensitive data from servers in untrusted environments, data could be encrypted before leaving trusted boundaries. Regardless of whether the data is encrypted or not, the server will need to decide who will gain access to it. For regulating access to the data, access control policies could be specified. These are access control policies that will describe who can gain access to the data. State-of-the-art policy-based systems can ensure enforcement of these policies. However, the matter becomes complicated when sensitive policies, which may leak private information, have to be enforced in untrusted environments.

\section{Motivation and Problem Statement}
\label{sec:intro-motivation}

% Context
The enforcement of sensitive policies in untrusted environments is still an open challenge for policy-based systems. On the one hand, taking any appropriate security decision requires access to these policies. On the other hand, if such access is allowed in an untrusted environment then confidential information might be leaked by the policies. The key challenge is how to enforce sensitive policies and protect data in untrusted environments. This challenge arises from a fundamental question, i.e., \emph{how can we establish trust in untrusted environments?} By establishing trust in untrusted environments, we will enable individuals and enterprises to leverage business models based on untrusted environments. At the same time, we would be fostering trust of end-users by ensuring privacy and security of their personal data.

According to Gartner, the cloud-based security (including access management) services market will be worth \$2.1 billion in 2013 and it will rise to \$3.1 billion in 2015 \cite{Gartner:2013}. This implies that security (access management in particular) of outsourced data is a key problem from a business analyst's point of view. It is important to know that outsourced environments are naturally untrusted. In the context of untrusted environments, we mainly distinguish two scenarios: (i) outsourced environments and (ii) distributed environments. The most attractive paradigms concerning outsourced and distributed environments are cloud computing and opportunistic networks, respectively.

% motivation - outsourcing
\subsection{Cloud Computing}
Cloud computing is an emerging paradigm offering outsourced services to enterprises for storing and processing a huge amount of data at very competitive costs. It promises higher availability, scalability and more effective quality of service than in-house solutions. In cloud computing, the outsourced piece of data is within easy reach of cloud service providers. Unfortunately, one of the strong obstacles in widespread adoption of the cloud is to preserve confidentiality of the data \cite{Armbrust:2010}. There are several techniques that can guarantee confidentiality of data stored in outsourced environments while supporting basic search capabilities \cite{Hore:2012, Kamara:2012, Bosch:2011, Cao:2011, Li:2011, Yang:2011, Zhu:2011, Li:2010, Wang:2010, Yang:2009}. However, they do not support access control policies to regulate access to a particular subset of the stored data. State-of-the-art policy based mechanisms can work only when they are deployed and operated within a trusted domain \cite{Sandhu:1996}. In an untrusted environment, access policies may reveal sensitive information about the data they aim to protect.

% healthcare scenario
To understand how access policies may reveal sensitive information in outsourced environments, let us imagine a scenario where a healthcare provider has outsourced its health record management services to a third party service provider. In this scenario, we do not trust the service provider to preserve data confidentiality. Therefore, we can encrypt health records before storing them in the outsourced environment. Furthermore, health records are associated with an access policy in order to prevent any unintended access. Let us consider the following access policy: \emph{only a Cardiologist may access the health record}, which is attached to the health record. Even if the data is encrypted, a curious service provider may still infer private information about the patient's medical conditions. In the example policy, a curious service provider may easily deduce that the patient could have heart problems. A misbehaving service provider may sell this information to banks that could deny the patient a loan given her health conditions.

% sota on policy enforcement
There are solutions that aim at providing the fine-grained access control on data stored in outsourced environments \cite{DeCapitanidiVimercati:2013, Raykova:2012, Vimercati:2007:CSAW, Vimercati:2007:VLDB}. However, those solutions are not suitable for scenarios where administrative actions are taken dynamically; this is because any administrative actions including updating access rights, adding users (or resources) and removing users (or resources) require re-distribution of new keys, as well as re-encryption of existing data with those keys. \emph{The core research issue is to develop an efficient scheme with flexible key management that can enforce expressive access control policies in outsourced environments without revealing private information to service providers.}

% opportunistic networks
\subsection{Opportunistic Networks}
Opportunistic networks are an emerging paradigm that has enabled individuals and enterprises to offer new services instantaneously. The fundamental reason behind this flexibility is that this paradigm aims at providing services without requiring any in-house \gls{IT} infrastructure \cite{Pelusi:2006}. Basically, opportunistic networks eliminate the need of any Internet connectivity.

In opportunistic networks, nodes can publish their own content and subscribe to others' content by indicating their interest. Any node can also act as a broker (also called a relay) that opportunistically receives content and interest, matches them and possibly delivers that content to other nodes. These opportunistic networks could be applied to the exchange of information in a wide range of domains from social media to military applications. Like cloud service providers, unauthorised brokers in opportunistic networks may infer private information from cleartext policies even when contents are encrypted.

Let us consider a battlefield scenario where soldiers are interested in sharing or acquiring sensitive information. We assume that there is no Internet connectivity in the battlefield. However, soldiers can exchange information via the short-range communication offered by smartphones. Soldiers can publish their content and subscribe for content of their interest. There are soldiers, known as brokers, who help to exchange content from one place to another. However, those soldiers must not be able to get access to content. For regulating access to content, a soldier, who is publishing, can encrypt content using state-of-the-art encryption techniques and specify an access policy describing which group of soldiers can get access. For instance, the policy could be \emph{either a Soldier from the Infantry unit or a Major can get access}. Although the content is encrypted, soldiers serving as brokers and attackers (enemy having access to smartphones of brokers), may infer private information from cleartext policies, i.e., who will receive this content. Furthermore, subscription information (containing interest of subscribers) might compromise privacy of subscribers.

There are schemes that preserve predicate privacy \cite{Shen:2009, Katz:2013} and assume that the predicate is evaluated at the receiver's end. Shikfa \emph{et al.} \cite{Shikfa:2010} propose a method that provides privacy and confidentiality in context-based forwarding. However, their proposed scheme disseminates information in one direction, i.e., from publishers to subscribers, without taking into account whether a subscriber is interested or not. In the context of publish-subscribe systems, there are many solutions that address privacy and security issues \cite{Choi:2010, Shang:2010, Srivatsa:2007}. However, state-of-the-art techniques are mainly based on centralised solutions that cannot be applied to opportunistic networks, where each node may serve as a publisher, a broker and a subscriber. \emph{The challenging research problem is to enable exchange of content and interest without (i) revealing content and its associated policies to unauthorised brokers and (ii) compromising the privacy of subscribers in opportunistic networks.}

\section{Research Contributions}
In this dissertation, we present the design, technical and implementation details of our proposed policy-based access control mechanisms for untrusted environments. In this section, we first discuss our research contributions in outsourced environments followed by advancements in opportunistic networks.

\subsection{Enforcement of Encrypted Policies in Outsourced Environments}
One of the main research goals is to enforce access control decisions while protecting access policies in outsourced environments. The core contributions concerning this part are as follows:

\begin{itemize}
	
	\item We provide full confidentiality of access policies such that service providers in outsourced environments do not learn private information about policies during the policy deployment and evaluation phases.
	
	\item We support expressive access control policies, consider contextual conditions and take into account contextual information before making any access decision. In particular, our proposed solution is capable of handling complex policies involving non-monotonic boolean expressions and range queries.
	
	\item The system entities do not share any encryption keys and even if a user is deleted or revoked, the system is still able to perform its operations without requiring re-encryption of data or access policies.
	
	\item For complex user management, we extend the basic policy enforcement mechanism to introduce the basic \gls{RBAC} policies, where users are assigned roles, roles are assigned permissions and users execute permissions if their roles are active in the session maintained by the service provider.
	
	\item The basic \gls{RBAC} policies are augmented with role hierarchies by enabling role inheritance.
	
	\item Finally, we integrate dynamic security constraints (including \acrlong{DSoD} and \acrlong{CW}) to provide the full-fledged \gls{RBAC} policies in an outsourced environment. The full-fledged \gls{RBAC} policies are enforced without revealing any private information to a curious service provider.

\end{itemize}

\subsection{Enforcement of Encrypted Policies in Opportunistic Networks}
The second research goal, which is even more challenging, is to propose a scheme that can enable exchange of content and interest without (i) revealing content and its associated policies to unauthorised brokers and (ii) compromising the privacy of subscribers. In the following, we describe main contributions related to the aforementioned goal:

\begin{itemize}

	\item We protect content by specifying access control policies. In opportunistic networks, brokers match subscriber's interest against policies associated with content without compromising the subscriber's privacy (say, by learning their interest or attributes).
	
	\item In our proposed solution, an unauthorised broker neither gains access to content nor learns access policies and authorised nodes gain access only if they satisfy fine-grained policies specified by the publishers.

	\item The system provides scalable key management in which loosely-coupled publishers and subscribers communicate with each other without any prior contact.
	
	\item Finally, we have developed and analysed the performance of a prototype running on real smartphones in order to show the feasibility of our approach.

\end{itemize}

\section{Organisation of the Dissertation}

This dissertation consists of the follows chapters:

\begin{description}

	\item[Chapter \ref{cha:espoon}] proposes a policy-based access control mechanism that can deploy and enforce sensitive policies in an encrypted manner. The proposed mechanism maintains a clear separation between the security policies and the actual enforcement mechanism without loss of confidentiality. Moreover, we show performance overheads of the proposed algorithms.
	
	\item[Chapter \ref{cha:erbac}] extends the proposed solution in Chapter \ref{cha:espoon} for supporting the basic \gls{RBAC} policies. In this chapter, we also explain how the basic \gls{RBAC} policies can incorporate role hierarchies. Furthermore, we provide a security analysis. Finally, we compare performance overheads incurred by access control mechanisms with and without \gls{RBAC} models.
	
	\item[Chapter \ref{cha:egrant}] explains how dynamic security policies (including \acrlong{DSoD} and \acrlong{CW}) can be enforced and integrated with \gls{RBAC} models. This chapter also shows performance overheads of the proposed algorithms.
	
	\item[Chapter \ref{cha:pidgin}] investigates how content could be encrypted and access control policies could be enforced in distributed environments, in particular in opportunistic networks. We propose a design and implement a scheme that can run on smartphones. Furthermore, we report some benchmarks of running the proposed cryptographic operations on smartphones.
	
	\item[Chapter \ref{cha:conclusion}] concludes the dissertation by summarising the chapters presented. It also points out some future research directions emerging from this work.
	
	\item[Appendix \ref{app:publications}] reports a list of publications (with the corresponding abstracts) related to the work presented in this dissertation, as well as other publications.

\end{description}

\iffalse

\chapter{Background}
\label{cha:background}
In Section~\ref{cha:intro} ... 

\section[Attribute-Based Encryption]{\acrlong{ABE}}
%Attribute-Based Encryption}

\section{Public Key Encryption with Keyword Search}
\gls{PEKS}~\cite{Boneh:2004}

\fi

%%%%%%%%%%%%%%%%%%%%%%%%% CHAPTER ESPOON %%%%%%%%%%%%%%%%%%%%%%%%%
%\footnotemark[*]
\chapter[Enforcing Policies in Outsourced Environments]{\acrshort{ESPOON}: Enforcing Encrypted Security Policies in Outsourced Environments\footstar{The preliminary version of this chapter has appeared in \cite{Asghar2011-ARES}.}}
\label{cha:espoon}

Data outsourcing is a growing business model offering services to individuals and enterprises for processing and storing a huge amount of data. It is not only economical but also promises higher availability, scalability, and more effective quality of service than in-house solutions. Despite all its benefits, data outsourcing raises serious security concerns for preserving data confidentiality. There are solutions for preserving confidentiality of data while supporting search on the data stored in outsourced environments. However, such solutions do not support access policies to regulate access to a particular subset of the stored data. 

The enforcement of sensitive policies in outsourced environments is still an open challenge for policy-based systems. On the one hand, taking the appropriate security decision requires access to the policies. However, if such access is allowed in an untrusted environment then confidential information might be leaked by the policies. Current solutions are based on cryptographic operations that embed security policies with the security mechanism. Therefore, the enforcement of such policies is performed by allowing the authorised parties to access the appropriate keys. We believe that such solutions are too rigid because they strictly intertwine authorisation policies with the enforcement mechanism. In this chapter, we address the issue of enforcing security policies in an outsourced environment while protecting the policy confidentiality. Our solution aims at providing a clear separation between security policies and the enforcement mechanism. The proposed technique does not reveal access policies and the access request.

\section{Introduction}

In recent years, data outsourcing has become a very attractive business model. It offers services to individuals and enterprises for processing and storing a huge amount of data at very low cost. It promises higher availability, scalability, and more effective quality of service than in-house solutions. Many sectors including government and healthcare, initially reluctant to data outsourcing, are now adopting it \cite{Ondo:2006}.

Despite all its benefits, data outsourcing raises serious security concerns for preserving data confidentiality. The main problem is that the data stored in outsourced environments is within easy reach of service providers that could gain unauthorised access. There are several solutions for guaranteeing confidentiality of data in outsourced environments. For instance, solutions as those proposed in \cite{Dong:2011, Kamara:2010} offer a protected data storage while supporting basic search capabilities performed on the server without revealing information about the stored data \cite{Hore:2012, Kamara:2012, Bosch:2011, Cao:2011, Li:2011, Yang:2011, Zhu:2011, Li:2010, Wang:2010, Yang:2009}. However, such solutions do not support access policies to regulate the access to a particular subset of the stored data.

\subsection{Motivation}
Solutions for providing access control mechanisms in outsourced environments have mainly focused on encryption techniques that couple access policies with a set of keys, such as the one described in \cite{Vimercati:2008, Vimercati:2010}. Only users possessing a key (or a set of hierarchy-derivable keys) are authorised to access the data. The main drawback of these solutions is that security policies are tightly coupled with the security mechanism, thus incurring high processing cost for performing any administrative change for both the users and the policies representing the access rights.

A policy-based solution, such the one described for the Ponder language in \cite{Russello:2007}, is more flexible and easy to manage because it clearly separates the security policies from the enforcement mechanism. However, policy-based access control mechanisms are not designed to operate in outsourced environments. Such solutions can work only when they are deployed and operated within a trusted domain (i.e., the computational environment managed by the organisation owning the data). If these mechanisms are outsourced to an untrusted environment, the access policies that are to be enforced on the server may leak information on the data they are protecting. As an example, let us consider a scenario where a hospital has outsourced its healthcare data management services to a third party service provider. We assume that the service provider is honest-but-curious, similar to the existing literature on data outsourcing (such as \cite{Vimercati:2007:VLDB}), i.e., it is honest to perform the required operations as described in the protocol but curious to learn information about stored or exchanged data. In other words, the service provider does not preserve data confidentiality. A patient's medical record should be associated with an access policy in order to prevent an unintended access. The data is stored with an access policy. As an example, let us consider the following access policy: \emph{only a Cardiologist may access the data}. From this policy, it is possible to infer important information about the user's medical conditions (even if the actual medical record is encrypted). This policy reveals that a patient could have heart problems. A misbehaving service provider may sell this information to banks that could deny the patient a loan given her health conditions.

\subsection{Research Contributions}
In this chapter, we present a policy-based access control mechanism for outsourced environments where we support full confidentiality of access policies. We named our solution \textbf{\gls{ESPOON}}. One of the main advantages of \gls{ESPOON} is that we maintain the clear separation between the security policies and the actual enforcement mechanism without loss of confidentiality. This can be guaranteed under the assumption that the service provider is honest-but-curious. Our approach allows us to implement the access control mechanism as an outsourced service with all the benefits associated with this business model without compromising the confidentiality of the policies. Summarising, the research contributions of our approach are threefold. First of all, the service provider does not learn private information about policies and the requester's attributes during the policy evaluation process. Second, \gls{ESPOON} is capable of handling complex policies involving non-monotonic boolean expressions and range queries. Third, the system entities do not share any encryption keys and even if a user is deleted or revoked, the system is still able to perform its operations without requiring re-encryption of the policies. As a proof-of-concept, we have implemented a prototype of our access control mechanism and analysed its performance to quantify the incurred overhead.

\subsection{Chapter Outline}
The rest of this chapter is organised as follows. In Section \ref{sec:espoon-related-work}, we review the related work. Section \ref{sec:espoon-proposed-approach} describes the proposed approach. Solution and algorithmic details are explained in Section \ref{sec:espoon-solution-details} and Section \ref{sec:espoon-algorithmic-details}, respectively. The performance overhead of the proposed solution is reported in Section \ref{sec:espoon-performance-analysis}. A discussion is provided in Section \ref{sec:espoon-discussion}. Finally, we summarise this chapter in Section \ref{sec:espoon-summary}.

\iffalse
\section{Related Work} \ref{sec:espoon-related-work}
\section{The ESPOON Approach} \ref{sec:espoon-proposed-approach}
\section{Solution Details} \ref{sec:espoon-solution-details}
\section{Algorithmic Details} \ref{sec:espoon-algorithmic-details}
\section{Performance Analysis} \ref{sec:espoon-performance-analysis}
\section{Discussion} \ref{sec:espoon-discussion}
\section{Chapter Summary} \ref{sec:espoon-summary}
\fi

\section{Related Work}
\label{sec:espoon-related-work}

Work on outsourcing data storage to a third party has been focusing on protecting the data confidentiality within the outsourced environment. Several techniques have been proposed allowing authorised users to perform efficient queries on the encrypted data while not revealing information on the data and the query \cite{Song:2000, Boneh:2004, Golle:2004, Curtmola:2006, Hwang:2007, Boneh:2007, Wang:2008, Baek:2008, Rhee:2010, Shao:2010, Dong:2011}. However, these techniques do not support the case of users having different access rights over the protected data. Their assumption is that once a user is authorised to perform search operations, there are no restrictions on the queries that can be performed and the data that can be accessed \cite{Hore:2012, Kamara:2012, Bosch:2011, Cao:2011, Li:2011, Yang:2011, Zhu:2011, Li:2010, Wang:2010, Yang:2009}.

The idea of using an access control mechanism in an outsourced environment was initially explored in \cite{Vimercati:2007:CSAW, Vimercati:2007:VLDB, Vimercati:2010}. In this approach, De Capitani di Vimercati \emph{et al.} provide a selective encryption strategy for enforcing access control policies. The idea is to have a selective encryption technique where each user has a different key capable of decrypting only the resources a user is authorised to access. In their scheme, a public token catalogue expresses key derivation relationships. However, the public catalogue contains tokens in the clear that express the key derivation structure. The tokens could leak information on access control policies and on the protected data. To circumvent the issue of information leakage, in \cite{Vimercati:2008} De Capitani di Vimercati \emph{et al.} provide an encryption layer to protect the public token catalogue. This requires each user to obtain the key for accessing a resource by traversing the key derivation structure. The key derivation structure is a graph built (using access key hierarchies \cite{Atallah:2009}) from a classical access matrix. There are several issues related to this scheme. First, the algorithm of building key derivation structure is very time consuming. Any administrative actions to update access rights require the users to obtain new access keys derived from the rebuilt key derivation structure and it consequently requires data re-encryption with new access keys. Therefore, the scheme is not very scalable and may be suitable for a static environment where users and resources do not change very often. Second, the scheme does not support complex policies where contextual information may be used for granting access rights. For instance, only specific time and location information associated with an access request may be legitimate to grant access to a user.

Another possible approach for implementing an access control mechanism is protecting the data with an encryption scheme where the keys can be generated from the user's credentials (expressing attributes associated with that user). Although these approaches are not devised particularly for outsourced environments, it is still possible to use them as access control mechanisms in outsourced settings. For instance, a recent work by Narayan \emph{et al.} \cite{Narayan:2010} employ the variant of \gls{ABE} proposed in \cite{Bethencourt:2007} (i.e., \gls{CPABE}) to construct an outsourced healthcare system where patients can securely store their \gls{EHR}. In their solution, each \gls{EHR} is associated with a secure search index to provide search capabilities while guaranteeing no information leakage. However, one of the problems associated with \gls{CPABE} is that the access structure, representing the security policy associated with the encrypted data, is not protected. Therefore, a curious storage provider might get information on the data by accessing the attributes expressed in the \gls{CPABE} policies. The problem of having the access structure expressed in cleartext affects in general all the \gls{ABE} constructions \cite{Sahai:2005, Goyal:2006, Ostrovsky:2007, Bethencourt:2007}. Therefore, this mechanism is not suitable for guaranteeing confidentiality of access control policies in outsourced environments.

Related to the issue of the confidentiality of the access structure, the hidden credentials scheme presented in \cite{Holt:2003} allows one to decrypt ciphertexts while the involved parties never reveal their policies and credentials to each other. Data can be encrypted using an access policy containing monotonic boolean expressions which must be satisfied by the receiver to get access to the data. A passive adversary may deduce the policy structure, i.e., the operators (AND, OR, m-of-n threshold encryption) used in the policy but she does not learn what credentials are required to fulfil the access policy unless she possesses them. Bradshaw \emph{et al.} \cite{Bradshaw:2004} extend the original hidden credentials scheme to limit the partial disclosure of the policy structure and speed up the decryption operations. However, in this scheme, it is not easy to support non-monotonic boolean expressions and range queries in the access policy. Furthermore, hidden credentials schemes assume that the involved parties are online all the time to run the protocol.

The homomorphic encryption schemes \cite{Gentry:2009, Dijk:2010, Brakerski:2011, Naehrig:2011, Gentry:2011, HElib:2013, Paillier:1999} allow untrusted parties to perform mathematical operations on encrypted data without compromising the encryption. There are a number of issues with these schemes. The major issue is scalability. Unfortunately, state-of-the-art schemes are not suitable in practice for processing a huge amount of data due to computational limitations. Another problem is the key management. These schemes consider a single user that can perform the decryption. Basically, we are interested in schemes that can offer encryption and decryption in a multi-user setting, where each user should have her private key (i.e., different from other users).

The data could be distributed along with the sticky policy attached to it \cite{Mont:2003, Chadwick:2008, Pearson:2011}. The data is basically encrypted with the sticky policy. For getting access to the data, the recipient needs to contact trusted authorities. The trusted authority grants access to the data by forwarding the decryption key to the recipient. Before sending the decryption key, the trusted authority verifies credentials of the recipient. Furthermore, this approach enables the trusted authority to take into account consent of the data owner before granting the access. However, approaches based on the sticky policies are not privacy preserving because both policies and credentials are in cleartext.

\gls{PIR} protocols allow users to retrieve information without revealing queries to the server \cite{Chor:1998, PIR, Yekhanin:2010, Williams:2008, Goldberg:2007, Camenisch:2009, Camenisch:2012, Olumofin:2012}. Basically, they can be deployed for fetching information from curious servers without compromising privacy of users, though they are computationally intensive. However, it is not clear how \gls{PIR} protocols can help in a situation where the policy enforcement mechanism is delegated to third parties.

\begin{figure} [htp]
\centering
% left bottom right top
\includegraphics[trim=55mm 55mm 50mm 45mm,clip,width=\textwidth]{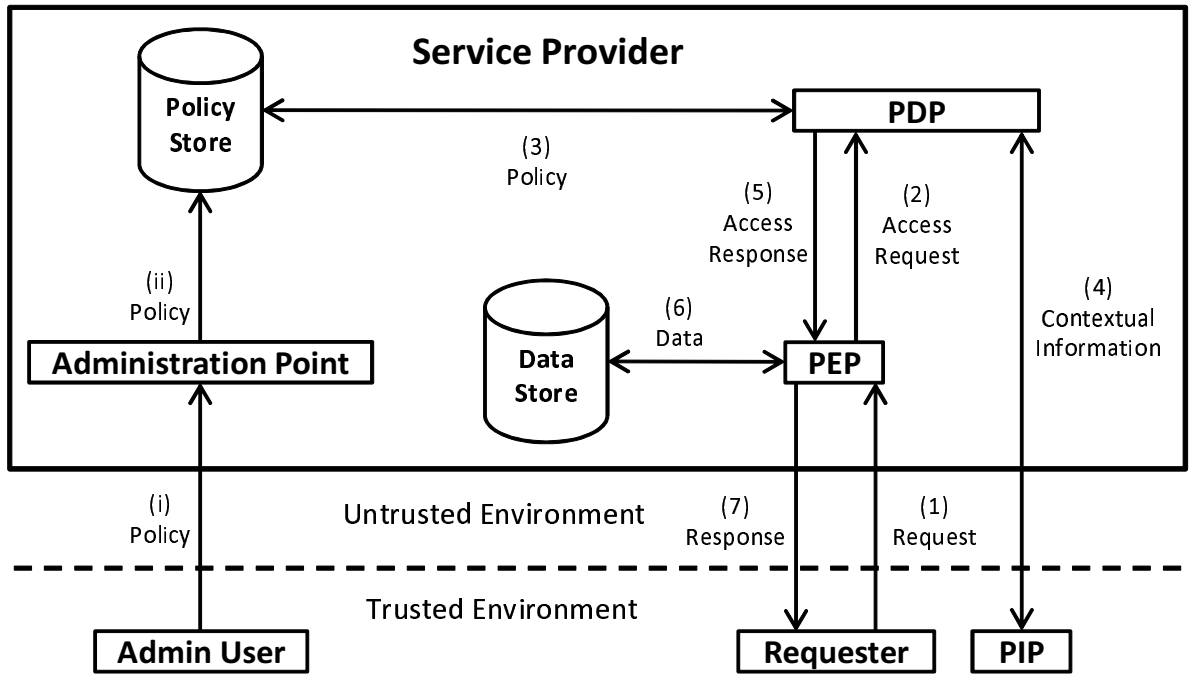}
\caption[The ESPOON architecture for enforcing policies]{The \gls{ESPOON} architecture for enforcing policies in outsourced environments}
\label{fig:espoon-abstract_picture}
\end{figure}

\section[The ESPOON Approach]{The \gls{ESPOON} Approach}
\label{sec:espoon-proposed-approach}
We propose \acrfull{ESPOON} that aims at providing a policy-based access control mechanism that can be deployed in an outsourced environment. Figure \ref{fig:espoon-abstract_picture} illustrates the proposed architecture that has similar components as the widely accepted architecture for policy-based management proposed by \gls{IETF} \cite{Yavatkar:2000}. In \gls{ESPOON}, the \textbf{Admin User} deploys (i) access policies to the \textbf{Administration Point} that stores (ii) the policies in the \textbf{Policy Store}. Whenever a \textbf{Requester}, say a doctor, needs to access the data, a request is sent to the \textbf{\gls{PEP}} (1). This request includes the Requester's identifier (subject), the requested data (target) and the action to be performed. The \gls{PEP} (2) forwards the access request to the \textbf{\gls{PDP}}. The \gls{PDP} (3) obtains the policies matching against the access request from the Policy Store and (4) retrieves the contextual information from the \textbf{\gls{PIP}}. The contextual information may include the environmental and Requester's attributes under which an access can be considered valid. For instance, a doctor should only access the data during office hours. For simplicity, we assume that the \gls{PIP} collects all required attributes including the Requester's attributes and sends all of them together in one go. Moreover, we assume that the \gls{PIP} is deployed in the trusted environment. However if attributes forgery is an issue, then the \gls{PIP} can request a trusted authority to sign the attributes before sending them to the \gls{PDP}. The \gls{PDP} evaluates the policies against the attributes provided by the \gls{PIP} checking if the contextual information satisfies any policy conditions and sends to the \gls{PEP} the access response (5). In case of \emph{permit}, the \gls{PEP} forwards the access action to the Data Store (6). Otherwise, in case of \emph{deny}, the requested action is not forwarded. Optionally, a response can be sent to the Requester (7) with either \emph{success} or \emph{failure}.

The main difference with the standard proposed by \gls{IETF} is that the \gls{ESPOON} architecture for the policy-based access control is outsourced in an untrusted environment (see Figure \ref{fig:espoon-abstract_picture}). The trusted environment comprises only a minimal \gls{IT} infrastructure that is the applications used by the Admin Users and Requesters, together with the \gls{PIP}. This reduces the cost of maintaining an \gls{IT} infrastructure. Having the reference architecture in the cloud increases its availability and provides a better load balancing compared to a centralised approach. Additionally, \gls{ESPOON} guarantees that the confidentiality of the policies is protected while their evaluation is executed in the outsourced environment. This allows a more efficient evaluation of the policies. For instance, a naive solution would see the encrypted policies stored in the cloud and the \gls{PDP} deployed in the trusted environment. At each evaluation, the encrypted policies would be sent to the \gls{PDP} that decrypts the policies for a cleartext evaluation. After that, the policies need to be encrypted and send back to the cloud. The \textbf{Service Provider} where the architecture is outsourced is honest-but-curious. This means that the provider allows the \gls{ESPOON} components to follow the specified protocols, but it may be curious to find out information about the data and the policies regulating the accesses to the data. As for the data, we assume that the confidentiality data is protected by one of the several techniques available for outsourced environments (see \cite{Dong:2008, Rhee:2010, Shao:2010, Dong:2011}). However, to the best of our knowledge, there is no solution that can address the problem of guaranteeing policy confidentiality while allowing an efficient evaluation mechanism that is clearly separated from the security policies. Most of the techniques discussed in the related work section require the security mechanism to be tightly coupled with the policies. In the following section, we can show that it is possible to maintain a generic \gls{PDP} separated from the security policies and able to take access decisions based on the evaluation of encrypted policies. In this way, the policy confidentiality can be guaranteed against a curious provider and the functionality of the access control mechanism is not restricted.

\subsection{The System Model}
\label{sec:esooon-system-model}
Before presenting the details of the scheme used in \gls{ESPOON}, it is necessary to discuss the system model. In this section, we identify the following system entities.

\begin{itemize}
\item \textbf{Admin User:} This type of user is responsible for the administration of the policies stored in the outsourced environment. An Admin User can deploy new policies or update/delete the policies already deployed.
\item \textbf{Requester:} A Requester is a user that requests an access (e.g., read, write, search, etc.) over the data residing in the outsourced environment. Before the access is permitted, the policies deployed in the outsourced environment are evaluated.

\item \textbf{Service Provider:} The Service Provider is responsible for managing the outsourced computation environment, where the \gls{ESPOON} components are deployed and to store the data, and access policies. It is assumed the Service Provider is honest-but-curious, i.e., it allows the components to follow the protocol to perform the required actions but curious to deduce information about the exchanged and stored policies.

\item \textbf{\gls{TKMA}:} The \gls{TKMA} is fully trusted and responsible for generating and revoking the keys. For each type of authorised users (both the Admin User and Requester), the \gls{TKMA} generates a key pair and securely transmits one part of the generated key pair to the user and the other to the Service Provider. The \gls{TKMA} is deployed on the trusted environment. Although requiring a \gls{TKMA} seems at odds with the needs of outsourced the \gls{IT} infrastructure, we argue that the \gls{TKMA} requires less resources and less management effort. Securing the \gls{TKMA} is much easier since a very limited amount of data needs to be protected and the \gls{TKMA} can be kept offline most of time.
\end{itemize}

It should be clarified that in our settings an Admin User is not interested in protecting the confidentiality of access policies from other Admin Users and Requesters. Here, the main goal is to protect the confidentiality of access policies from the Service Provider.

\subsection{Representation of Policies}
\label{sec:espoon-policy-representation}
In this section, we provide an informal description of the policy representation used in our approach. In this chapter, we deal with only positive authorisation policies. This means that, as default no actions are allowed unless at least one authorisation policy can be applicable to the request.

\begin{figure} [htp]
\Keywords
\begin{lstlisting}[style=AMMA,numbers=none,breaklines,mathescape,rulesepcolor=\color{black}]
if $\langle \mathit{CONDITION} \rangle$ then can $\langle S, A, T \rangle$

\end{lstlisting}
\caption[Representation of policies in ESPOON]{Representation of policies in \gls{ESPOON}}
\label{fig:espoon-policy-representation}
\end{figure}

In our approach, an authorisation policy is represented as a condition and a tuple as illustrated in Figure \ref{fig:espoon-policy-representation}. This authorisation policy is interpreted as follows: if $\mathit{CONDITION}$ is true then the subject $S$ can execute the action $A$ on the target $T$. At the time when a request is made, the information about the subject, the action that is requested and the target resource is collected by the Requester. The \gls{PIP} collects several attributes representing the context in which the request is being executed and sends them to the \gls{PDP}.

\begin{figure} [htp]
\centering
% left bottom right top
\includegraphics[trim=75mm 60mm 70mm 40mm,clip,width=.6\textwidth]{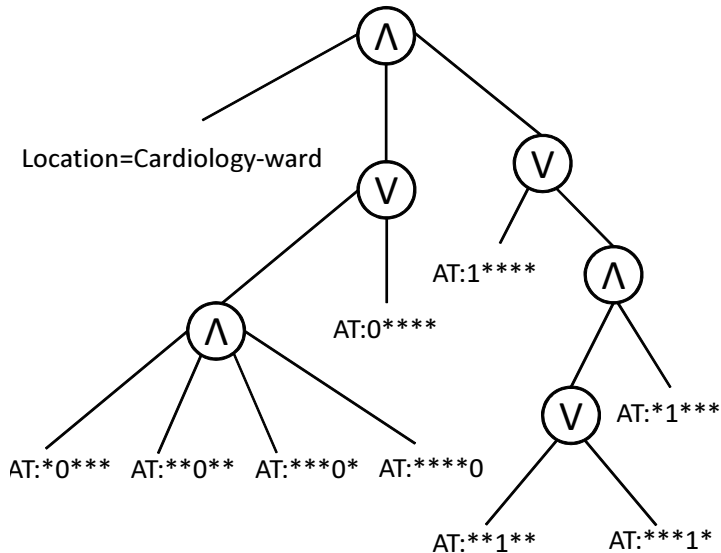}
\caption[An example of a contextual condition]{An example of a contextual condition illustrating $Location=Cardiology \mhyphen ward$ and $AT > 9\#5$ and $AT < 17\#5$}
\label{fig:erbac-cc}
\end{figure}

The \gls{PIP} collects and sends required contextual information to the \gls{PDP}. To represent contextual conditions, we use the tree structure described in \cite{Bethencourt:2007} for \gls{CPABE} policies. This tree structure allows an Admin User to express contextual conditions as conjunctions and disjunctions of equalities and inequalities. Internal nodes of the tree structure are AND, OR or threshold gates (e.g., 2 of 3) and leaf nodes are values of condition predicates either string or numerical. In the tree structure, a string comparison is represented by a single leaf node. However, the tree structure uses the \emph{bag of bits} representation to support comparisons between numerical values that could express time, date, location, age, or any numerical identifier. For instance, let us consider a contextual condition stating that the Requester location should be $Cardiology \mhyphen ward$ and that the access time should be between 9:00 and 17:00 hrs. Figure \ref{fig:erbac-cc} illustrates the tree structure representing this contextual condition, where access time is in a 5-bit representation (i.e., \#5).

% request
% TODO: on update remove $ $ that is added to fix a format issue
Let us consider $\mathit{CONDITION}$ illustrated in Figure \ref{fig:erbac-cc} requiring location of Requester and access time. We assume the Requester makes the request when she is in $\mathit{Cardiology \mhyphen}$ $\mathit{ward}$ and Access Time (AT) is 10:00 hrs. The \gls{PIP} collects and then transforms this contextual information as follows: $\mathit{Location = Cardiology \mhyphen ward}$, $\mathit{AT : 0****}$, $\mathit{AT: *1***}$, $\mathit{AT: **0**}$, $\mathit{AT: ***1*}$, $\mathit{AT: ****0}$, where AT is in a 5-bit representation (same as it is in $\mathit{CONDITION}$). After performing transformation, the \gls{PIP} sends contextual information to the \gls{PDP}. The \gls{PDP} receives contextual information and then evaluates $\mathit{CONDITION}$ by first matching attributes in contextual information against leaf-nodes in the $\mathit{CONDITION}$ tree and then evaluating internal nodes according to AND and OR gates.

In this policy representation, the $\langle S, A, T \rangle$ tuple and the leaf nodes in the condition tree are in clear text. Therefore, such information is easily accessible in the outsourced environment and may leak information about the data that the policies protect. In the following, we show how such representation can be protected while allowing the \gls{PDP} to evaluate the policies against the request.

\section[Solution Details of ESPOON]{Solution Details of \gls{ESPOON}}
\label{sec:espoon-solution-details}

The main idea of our approach is to use an encryption scheme for protecting the confidentiality of the policies while allowing the \gls{PDP} to perform the correct evaluation of the policies. We noticed that the operation performed by the \gls{PDP} for evaluating policies is similar to the search operation executed in a database. In particular, in our case the condition of a policy is the query; and the data that is matched against the query is represented by the attributes that the Requester sends in the request.

As a starting point, we consider the multiuser \gls{SDE} scheme proposed by Dong \emph{et al.} in \cite{Dong:2011}. The \gls{SDE} scheme allows an untrusted server to perform searches over encrypted data without revealing to the server information on both the data and elements used in the request. The advantage of this method is that it offers multi-user access without requiring key sharing between users. Each user in the system has a unique set of keys. The data encrypted by one user can be decrypted by any other authorised user. However, the \gls{SDE} implementation in \cite{Dong:2011} is only able to perform keyword comparison based on equalities. One of the major extensions of our implementation is that we are able to support the evaluation of contextual conditions containing complex boolean expressions such as non-conjunctive and range queries in multi-user settings.

In general, we distinguish four phases in \gls{ESPOON} for managing lifecycle of policies in outsourced environments. These phases include \emph{initialisation}, \emph{\textbf{policy deployment}}, \emph{\textbf{policy evaluation}} and \emph{user revocation}. In the following, we provide the details of the algorithms used in each phase.

\subsection{The Initialisation Phase}
Before the policy deployment and policy evaluation phases, the \gls{SDE} scheme needs to be initialised. This is required for generating the required key material. The following two algorithms that need to be run:

\begin{itemize}

\item The initialisation algorithm \textbf{Init} (Algorithm \ref{algo:erbac-init}) is run by the \gls{TKMA}. It takes as input the security parameter $1^k$ and outputs the public parameters $params$ and the master secret key set $msk$.

\item The user key sets generation algorithm \textbf{KeyGen} (Algorithm \ref{algo:erbac-keygen}) is run by the \gls{TKMA}. It takes as input the master secret key set $msk$ and the user (Admin User or Requester) identity $i$ and generates two key sets $K_{u_i}$ and $K_{s_i}$. The \gls{TKMA} sends key sets $K_{u_i}$ and $K_{s_i}$ to the user $i$ and the Key Store, respectively. Only the Administration Point, \gls{PDP} and \gls{PEP} are authorised to access the Key Store.

\end{itemize}

\subsection{The Policy Deployment Phase}
The policy deployment phase is executed when a new set of policies needs to be deployed on the Policy Store (or an existing version of policies needs to be updated). This phase is executed by the Admin User who edits the policies in a trusted environment. Before the policies leave the trusted environment, they need to be encrypted. Our policy representation consists of two parts: one for representing the condition and the other for the $\langle S, A, T \rangle$ tuple. Each part is encrypted using the following algorithms:

\begin{itemize}

\item The access policy condition encryption algorithm \textbf{ConditionEnc} (Algorithm \ref{algo:erbac-deploy-contextual-condition-client-side}) is run by the Admin User $i$. It takes as input a contextual condition and the user side key set $K_{u_i}$ corresponding to Admin User $i$ and outputs the encrypted contextual condition.

\item The access policy $\langle S, A, T \rangle$ tuple encryption algorithm \textbf{SATEnc} (Algorithm \ref{algo:espoon-deploy-sat-client-side}) is run by the Admin User $i$. It takes as input the $\langle S, A, T \rangle$ tuple and $K_{u_i}$ and outputs the client encrypted tuple $c^*_i (\langle S, A, T \rangle)$.

\end{itemize}

When the encrypted policy is sent to the outsourced environment, then another encryption round is performed. This is accomplished using the following algorithms:

\begin{itemize}

\item The access policy condition re-encryption algorithm \textbf{ConditionReEnc} (Algorithm \ref{algo:erbac-deploy-contextual-condition-server-side}) is run by the Administration Point. It takes as input the client encrypted contextual condition and the key $K_{s_i}$ corresponding to the Admin User $i$ and outputs the server encrypted contextual condition.

\item The access policy $\langle S, A, T \rangle$ tuple re-encryption algorithm \textbf{SATReEnc} (Algorithm \ref{algo:espoon-deploy-sat-server-side}) is run by the Administration Point. It takes as input the client encrypted tuple $c^*_i (\langle S, A, T \rangle)$ and the key $K_{s_i}$ corresponding to the Admin User $i$ and outputs the re-encrypted tuple $c(\langle S, A, T \rangle)$.

\end{itemize}

The access policy can be now stored in the Policy Store. The stored policies do not reveal any information about the data because they are stored as encrypted.

\subsection{The Policy Evaluation Phase}
The policy evaluation phase is executed when a Requester makes a request to access the data. Before the access permission is granted, the \gls{PDP} evaluates the matching policies in the Policy Store on the Service Provider. The request contains the $\langle S, A, T \rangle$ tuple. This information is encrypted using the following algorithm before it leaves the trusted environment:

\begin{itemize}

\item The $\langle S, A, T \rangle$ request encryption algorithm \textbf{SATRequest} (Algorithm \ref{algo:espoon-client-sat-td}) is run by Requester $j$. It takes as input the $\langle S, A, T \rangle$ tuple and $K_{u_j}$ and outputs the client encrypted tuple $T^*_j (\langle S, A, T \rangle)$.

\end{itemize}

The Requester sends the encrypted $\langle S, A, T \rangle$ tuple to the Service Provider. The policy evaluation phase on the Service Provider side starts with searching all the policies in the Policy Store matching against the Requester $\langle S, A, T \rangle$ tuple. This is accomplished by the following algorithm:

\begin{itemize}

\item The $\langle S, A, T \rangle$ tuple search algorithm \textbf{SATSearch} (Algorithm \ref{algo:espoon-search-sat}) is run by the \gls{PDP}. It takes as input the client encrypted tuple $T^*_j (\langle S, A, T \rangle)$ from Requester $j$ and all stored policies in the Policy Store $c(\langle S_i, A_i, T_i \rangle)_{1 \leq i \leq n}$ and returns the matching tuples in the Policy Store.

\end{itemize}

If any match is found in the Policy Store then the \gls{PDP} needs to match the contextual information against the access policy condition corresponding to the matched tuple. The \gls{PDP} fetches the contextual information including Requester and environmental attributes from the \gls{PIP}. The \gls{PIP} encrypts the contextual information using the following algorithm:

\begin{itemize}

\item The attributes encryption algorithm \textbf{AttributesRequest} (Algorithm \ref{algo:erbac-request-contextual-condition}) is run by the \gls{PIP} $j$. It takes as input the Requester and environmental attributes and $K_{u_j}$ and outputs the encrypted attributes.

\end{itemize}

After receiving the contextual information from the \gls{PIP}, the \gls{PDP} matches the \gls{PIP} attributes against the access policy condition. The \gls{PDP} calls the following algorithm to evaluate the access policy condition:

\begin{itemize}

\item The access policy condition evaluation algorithm \textbf{ConditionEvaluation} (Algorithm \ref{algo:erbac-match-contextual-condition}) is run by the \gls{PDP}. It takes as input a list of encrypted attributes, the key $K_{s_j}$ corresponding to the \gls{PIP} $j$ and encrypted access policy condition tree and outputs $\mathit{true}$ on successful policy evaluation and $\mathit{false}$ otherwise.

\end{itemize}

\subsection{The User Revocation Phase}
The proposed solution offers revocation of a user (an Admin User or a Requester). For this purpose, the Administration Point runs the following algorithm:

\begin{itemize}

\item A user (an Admin User or a Requester) revocation algorithm \textbf{UserRevocation} (Algorithm \ref{algo:egrant-user-revocation}) is run by the Administration Point. Given the user $i$, the Administration Point removes the corresponding server side key $K_{s_i}$ from the Key Store.

\end{itemize}

\section[Algorithmic Details of ESPOON]{Algorithmic Details of \gls{ESPOON}}
\label{sec:espoon-algorithmic-details}
In this section, we provide details of algorithms used in each phase for managing lifecycle of policies. All these algorithms constitute the proposed schema. 

% system init

\begin{algorithm} [htp]
{\algofontsize
\caption{\textbf{Init}}

\label{algo:erbac-init}

\begin{algorithmic}[1]

\INPUT \emph{It generates the system level keying material including public parameters and the master secret.}

\Require A security parameter $1^k$.

\Ensure The public parameters $params$ and the master secret key $msk$.

\medskip

\State Generate primes p and q of size $1^k$ such that $q$ $|$ $p - 1$ \label{line:erbac-primes}
\State Create a generator $g$ such that $\mathbb{G}$ is the unique order $q$ subgroup of $\mathbb{Z}^*_p$ \label{line:erbac-generator}
\State Choose a random $x \in \mathbb{Z}^*_q$ \label{line:erbac-master-x}
\State $h \leftarrow g^x$ \label{line:erbac-params-h}
\State Choose a collision-resistant hash function $H$ \label{line:erbac-params-H}
\State Choose a pseudorandom function $f$ \label{line:erbac-params-f}
\State Choose a random key $s$ for $f$ \label{line:erbac-master-s}
\State $params \leftarrow (\mathbb{G}, g, q, h, H, f)$ \label{line:erbac-params}
\State $msk \leftarrow (x, s)$ \label{line:erbac-master}

\Return $(params, msk)$

\end{algorithmic}
}
\end{algorithm}

\subsection{The Initialisation Phase}

In this phase, the system is initialised and then the \gls{TKMA} generates required keying material for entities in \gls{ESPOON}. During the system initlisation, the \gls{TKMA} takes a security parameter $k$ and outputs the public parameters $params$ and the master key set $msk$ by running \textbf{Init} illustrated in Algorithm \ref{algo:erbac-init}. The detail of \textbf{Init} is as follows: the \gls{TKMA} generates two prime numbers $p$ and $q$ of size $k$ such that $q$ divides $p-1$ (Line \ref{line:erbac-primes}). Then, it creates a cyclic group $\mathbb{G}$ with a generator $g$ such that $\mathbb{G}$ is the unique order $q$ subgroup of $\mathbb{Z}^*_p$ (Line \ref{line:erbac-generator}). Next, it randomly chooses $x \in \mathbb{Z}^*_q$ (Line \ref{line:erbac-master-x}) and compute $h$ as $g^x$ (Line \ref{line:erbac-params-h}). Next, it chooses a collision-resistant hash function $H$ (Line \ref{line:erbac-params-H}), a pseudorandom function $f$ (Line \ref{line:erbac-params-f}) and a random key $s$ for $f$ (Line \ref{line:erbac-master-s}). Finally, it publicises the public parameters $params = (\mathbb{G}, g, q, h, H, f)$ (Line \ref{line:erbac-params}) and keeps securely the master secret key $msk = (x, s)$ (Line \ref{line:erbac-master}).

% key gen

\begin{algorithm} [htp]
{\algofontsize
\caption{\textbf{KeyGen}}

\label{algo:erbac-keygen}

\begin{algorithmic}[1]

\INPUT \emph{For each user, it generates two key sets: one for the user while other for the server.}

\Require The master secret key $msk$, the user identity $i$ and the public parameters $params$.

\Ensure The client side key set $K_{u_i}$ and server side key set $K_{s_i}$.

\medskip

\State Choose a random $x_{i1} \in \mathbb{Z}^*_q$ \label{line:erbac-xi1}
\State $x_{i2} \leftarrow x - x_{i1}$ \label{line:erbac-xi2}
\State $K_{u_i} \leftarrow (x_{i1}, s)$ \label{line:erbac-ku}
\State $K_{s_i} \leftarrow (i, x_{i2})$ \label{line:erbac-ks}

\Return $(K_{u_i}, K_{s_i})$

\end{algorithmic}
}
\end{algorithm}

For each user (including an Admin User and a Requester), the \gls{TKMA} generates the keying material. For generating the keying material, the \gls{TKMA} takes the master secret key $msk$, the user identity $i$ and the public parameters $params$ and outputs two key sets: the client side key set $K_{u_i}$ and the server side key set $K_{s_i}$ by running \textbf{KeyGen} illustrated in Algorithm \ref{algo:erbac-keygen}. In \textbf{KeyGen}, \gls{TKMA} randomly chooses $x_{i1} \in \mathbb{Z}^*_q$ (Line \ref{line:erbac-xi1}) and computes $x_{i2} = x - x_{i1}$ (Line \ref{line:erbac-xi2}). It creates the client side key set $K_{u_i} = (x_{i1}, s)$ (Line \ref{line:erbac-ku}) and the server side key set $K_{s_i} = (i, x_{i2})$ (Line \ref{line:erbac-ks}).

% graphical representation of key gen

\begin{figure} [htp]
\centering
% left bottom right top
\includegraphics[trim=75mm 60mm 80mm 45mm,clip,width=.6\textwidth]{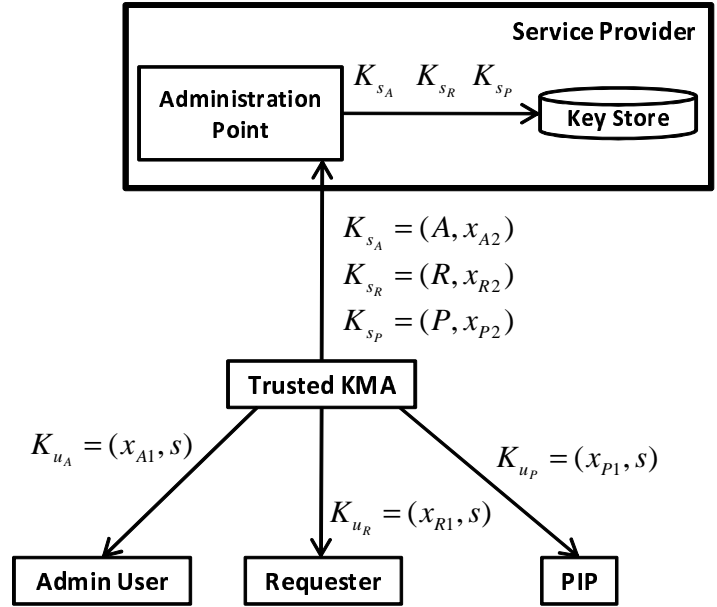} % .3
\caption[Distribution of keys in ESPOON]{Distribution of keys in \gls{ESPOON}}
\label{fig:erbac-key_generation}
\end{figure}

After running Algorithm \ref{algo:erbac-keygen}, the \gls{TKMA} sends the client side key set $K_{u_i}$ and the server side key set $K_{s_i}$ to user $i$ and the Administration Point on the Service Provider, respectively. The client side key set $K_{u_i}$ serves as a private key for user $i$. The Administration Point of the Service Provider inserts $K_{s_i}$ in the Key Store by updating it as follows: $KS = KS \cup K_{s_i}$. The Key Store is initialised as: $KS \leftarrow \phi$. Figure \ref{fig:erbac-key_generation} illustrates key distribution where Admin User $A$, Requester $R$ and \gls{PIP} $P$ receive $K_{u_A}$, $K_{u_R}$ and $K_{u_P}$, respectively. The \gls{TKMA} sends the corresponding server side key sets $K_{s_A}$, $K_{s_R}$ and $K_{s_P}$ to the Administration Point on the Service Provider. The Administration Point inserts server side key sets into the Key Store. Please note that only the Administration Point, the \gls{PDP} and the \gls{PEP} are authorised to access the Key Store.

% aux method: client enc

\begin{algorithm} [htp]
{\algofontsize
\caption{\textbf{ClientEnc}}

\label{algo:erbac-client-enc}

\begin{algorithmic}[1]

\INPUT \emph{It transforms the cleartext element into the client encrypted element.}

\Require Element $e$, the client side key set $K_{u_i}$ corresponding to Admin User $i$ and the public parameters $params$.

\Ensure The client encrypted element $c^*_i (e)$.

\medskip

\State Choose a random $r_e \in \mathbb{Z}^*_q$ \label{line:erbac-ce-r}
\State ${\sigma}_e \leftarrow f_s (e)$ \label{line:erbac-ce-sigma}
\State $\hat{c}_1 \leftarrow g^{r_e+{\sigma}_e}$ \label{line:erbac-ce-c1}
\State $\hat{c}_2 \leftarrow \hat{c}_1^{x_{i1}}$ \label{line:erbac-ce-c2}
\State $\hat{c}_3 \leftarrow H(h^{r_e})$ \label{line:erbac-ce-c3}
\State $c^*_i (e) \leftarrow (\hat{c}_1, \hat{c}_2, \hat{c}_3)$ \label{line:erbac-ce-c}

\Return $c^*_i (e)$

\end{algorithmic}
}
\end{algorithm}

% re-encryption algorithm

\begin{algorithm} [htp]
{\algofontsize
\caption{\textbf{ServerReEnc}}

\label{algo:erbac-server-re-enc}

\begin{algorithmic}[1]

\INPUT \emph{It transforms the client encrypted element into the server encrypted element.}

\Require The client encrypted element $c^*_i (e)$ and the server side key set $K_{s_i}$ corresponding to Admin User $i$.

\Ensure The server encrypted element $c(e)$.

\medskip

\State $c_1 \leftarrow (\hat{c}_1)^{x_{i2}}.\hat{c}_2 = \hat{c}_1^{x_{i1}+x_{i2}} = (g^{r_e+{\sigma}_e})^x = h^{r_e+{\sigma}_e}$ \label{line:erbac-se-c1}
\State $c_2 = \hat{c}_3 = H(h^{r_e})$ \label{line:erbac-se-c2}
\State $c(e) = (c_1, c_2)$ \label{line:erbac-se-c}

\Return $c(e)$

\end{algorithmic}
}
\end{algorithm}

% policy deployment 

\begin{figure} [htp]
\centering
% left bottom right top
\includegraphics[trim=70mm 60mm 85mm 40mm,clip,width=.6\textwidth]{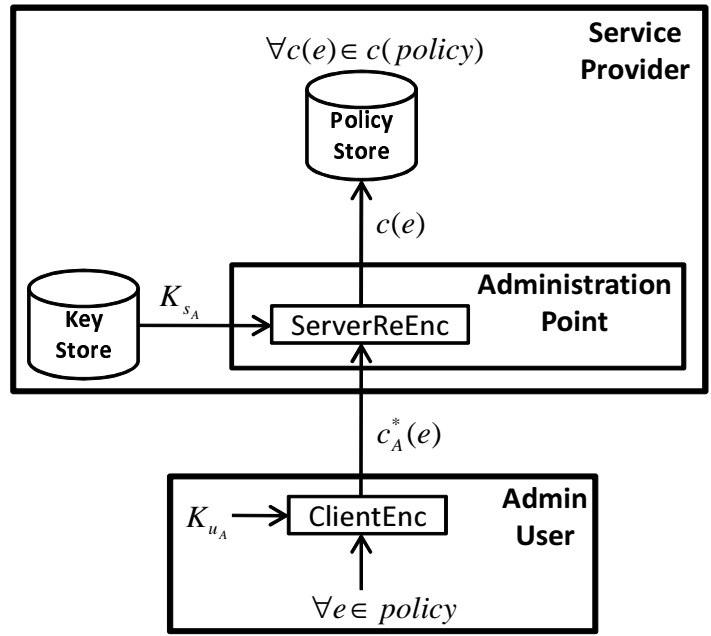} % .3
\caption{The policy deployment phase}
\label{fig:erbac-patient}
\end{figure}

\subsection{The Policy Deployment Phase}
\label{sec:espoon-policy-deployment-phase}
In the policy deployment phase, an Admin User defines and deploys policies. In general, a policy can be deployed after performing two rounds of encryptions. An Admin User performs a first round of encryption while the Administration Point on the Service Provider performs a second round of encryption. For performing a first round of encryption, an Admin User runs \textbf{ClientEnc} illustrated in Algorithm \ref{algo:erbac-client-enc}. \textbf{ClientEnc} takes as input (policy) element $e$, the client side key set $K_{u_i}$ corresponding to Admin User $i$ and the public parameters $params$ and outputs the client encrypted element $c^*_i (e)$. In \textbf{ClientEnc}, an Admin User randomly chooses $r_{e} \in \mathbb{Z}^*_q$ (Line \ref{line:erbac-ce-r}), computes ${\sigma}_e$ as $f_s (e)$ (Line \ref{line:erbac-ce-sigma}), and then computes $\hat{c}_1$, $\hat{c}_2$ and $\hat{c}_3$ as $g^{r_e+{\sigma}_e}$ (Line \ref{line:erbac-ce-c1}), $\hat{c}_1^{x_{i1}}$ (Line \ref{line:erbac-ce-c2}) and $H(h^{r_e})$ (Line \ref{line:erbac-ce-c3}), respectively. $\hat{c}_1$, $\hat{c}_2$ and $\hat{c}_3$ constitute $c^*_i (e)$ (Line \ref{line:erbac-ce-c}). An Admin User transmits to the Administration Point the client encrypted elements of a policy as shown in Figure \ref{fig:erbac-patient}.

The Administration Point retrieves the server side key set corresponding to the Admin User and performs a second round of encryption by running \textbf{ServerReEnc} illustrated in Algorithm \ref{algo:erbac-server-re-enc}. \textbf{ServerReEnc} takes as input the client encrypted element $c^*_i (e)$ and the server side key set $K_{s_i}$ corresponding to Admin User $i$ and outputs the server encrypted element $c(e)$. The Administration Point calculates $c_1$ and $c_2$ as $(\hat{c}_1)^{x_{i2}}.\hat{c}_2 = \hat{c}_1^{x_{i1}+x_{i2}} = (g^{r_e+{\sigma}_e})^x = h^{r_e+{\sigma}_e}$ (Line \ref{line:erbac-se-c1}) and $\hat{c}_3 = H(h^{r_e})$ (Line \ref{line:erbac-se-c2}), respectively. Both $c_1$ and $c_2$ form $c(e)$ (Line \ref{line:erbac-se-c}). The Administration Point stores the server encrypted policies in the Policy Store as shown in Figure \ref{fig:erbac-patient}.

% deploy contextual condition: client side

\begin{algorithm} [htp]
{\algofontsize
\caption{\textbf{ConditionEnc}}

\label{algo:erbac-deploy-contextual-condition-client-side}

\begin{algorithmic}[1]

\INPUT \emph{It transforms the cleartext condition into the client encrypted condition.}

\Require The contextual condition $T$, the client side key set $K_{u_i}$ corresponding to Admin User $i$ and the public parameters $params$.

\Ensure The client encrypted contextual condition $T_{C_i}$.

\medskip

\State $T_{C_i} \leftarrow T$ \label{line:erbac-deploy-cc-cs-copy}

\For {each leaf node $e$ in $T_{C_i}$} \label{line:erbac-deploy-cc-cs-loop}

	\State $c^*_i (e) \leftarrow$ call \textbf{ClientEnc} ($e$, $K_{u_i}$, $params$) \label{line:erbac-deploy-cc-cs-call}
	
	\State replace $e$ of $T_{C_i}$ with $c^*_i (e)$ \label{line:erbac-deploy-cc-cs-replace}

\EndFor

\Return $T_{C_i}$

\end{algorithmic}
}
\end{algorithm}

% deploy contextual condition: server side

\begin{algorithm} [htp]
{\algofontsize
\caption{\textbf{ConditionReEnc}}

\label{algo:erbac-deploy-contextual-condition-server-side}

\begin{algorithmic}[1]

\INPUT \emph{It transforms the client encrypted condition into the server encrypted condition.}

\Require The client encrypted contextual condition $T_{C_i}$ and identity of Admin User $i$.

\Ensure The server encrypted contextual condition $T_{S}$.

\medskip

\State $K_{s_i} \leftarrow KS[i]$ {\algofontsize \Comment{retrieve the server side key corresponding to Admin User $i$}} \label{line:erbac-deploy-cc-ss-ks}

\State $T_{S} \leftarrow T_{C_i}$ \label{line:erbac-deploy-cc-ss-copy}

\For {each client encrypted leaf node $c^*_i (e)$ in $T_{S}$} \label{line:erbac-deploy-cc-ss-loop}

	\State $c(e) \leftarrow$ call \textbf{ServerReEnc} ($c^*_i (e)$, $K_{s_i}$) \label{line:erbac-deploy-cc-ss-call}
	
	\State replace $c^*_i (e)$ of $T_{S}$ with $c(e)$ \label{line:erbac-deploy-cc-ss-replace}

\EndFor

\Return $T_{S}$

\end{algorithmic}
}
\end{algorithm}

\noindent \emph{\textbf{Deployment of Contextual Conditions:}}
The contextual condition can be deployed in two steps. In the first step, an Admin User performs a first round of encryption by running Algorithm \ref{algo:erbac-deploy-contextual-condition-client-side}. This algorithm takes as input the contextual condition $T$, the client side key set $K_{u_i}$ corresponding to Admin User $i$ and the public parameters $params$ and outputs the client encrypted contextual condition $T_{C_i}$. First, it copies $T$ to $T_{C_i}$ (Line \ref{line:erbac-deploy-cc-cs-copy}). For each leaf node in $T_{C_i}$ (Line \ref{line:erbac-deploy-cc-cs-loop}), it generates the client encrypted element by calling \textbf{ClientEnc} illustrated in Algorithm \ref{algo:erbac-client-enc} (Line \ref{line:erbac-deploy-cc-cs-call}) and then updates $T_{C_i}$ by replacing element $e$ with the client encrypted element $c^*_i (e)$ (Line \ref{line:erbac-deploy-cc-cs-replace}). An Admin User sends the client encrypted contextual condition to the Administration Point.
In the second step, the Administration Point performs another round of encryption by running Algorithm \ref{algo:erbac-deploy-contextual-condition-server-side}. This algorithm takes as input the client encrypted contextual condition $T_{C_i}$ and identity of Admin User $i$ and outputs the server encrypted contextual condition $T_{S}$. First, it retrieves from the Key Store the server side key $K_{s_i}$ corresponding to Admin User $i$ (Line \ref{line:erbac-deploy-cc-ss-ks}). Next, it copies $T_{C_i}$ to $T_{S}$ (Line \ref{line:erbac-deploy-cc-ss-copy}). For each each client encrypted leaf node in $T_{S}$ (Line \ref{line:erbac-deploy-cc-ss-loop}), it generates the server encrypted element by calling \textbf{ServerReEnc} illustrated in Algorithm \ref{algo:erbac-server-re-enc} (Line \ref{line:erbac-deploy-cc-ss-call}). Then, it replaces the client encrypted element $c^*_i (e)$ of $T_{S}$ with the server encrypted element $c(e)$ (Line \ref{line:erbac-deploy-cc-ss-replace}).

% deploy SAT tuple: client side

\begin{algorithm} [htp]
{\algofontsize
\caption{\textbf{SATEnc}}

\label{algo:espoon-deploy-sat-client-side}

\begin{algorithmic}[1]

\INPUT \emph{It transforms the cleartext tuple into the client encrypted tuple.}

\Require The $\langle S, A, T \rangle$ tuple, the client side key set $K_{u_i}$ corresponding to Admin User $i$ and the public parameters $params$.

\Ensure The client encrypted tuple $c^*_i (\langle S, A, T \rangle)$.

\medskip

\State $c^*_i (S) \leftarrow$ call \textbf{ClientEnc} ($S$, $K_{u_i}$, $params$) \label{line:espoon-deploy-s-client-call}

\State $c^*_i (A) \leftarrow$ call \textbf{ClientEnc} ($A$, $K_{u_i}$, $params$) \label{line:espoon-deploy-a-client-call}

\State $c^*_i (T) \leftarrow$ call \textbf{ClientEnc} ($T$, $K_{u_i}$, $params$) \label{line:espoon-deploy-t-client-call}

\State $c^*_i (\langle S, A, T \rangle) \leftarrow (c^*_i (S), c^*_i (A), c^*_i (T))$ \label{line:espoon-deploy-sat-client-assignment}

\Return $c^*_i (\langle S, A, T \rangle)$

\end{algorithmic}
}
\end{algorithm}

% deploy SAT tuple: server side

\begin{algorithm} [htp]
{\algofontsize
\caption{\textbf{SATReEnc}}

\label{algo:espoon-deploy-sat-server-side}

\begin{algorithmic}[1]

\INPUT \emph{It transforms the client encrypted tuple into the server encrypted tuple.}

\Require The client encrypted tuple $c^*_i (\langle S, A, T \rangle)$ and identity of Admin User $i$.

\Ensure The server encrypted tuple $c (\langle S, A, T \rangle)$.

\medskip

\State $K_{s_i} \leftarrow KS[i]$ {\algofontsize \Comment{retrieve the server side key corresponding to Admin User $i$}} \label{line:espoon-deploy-sat-ss-ks}

\State $c(S) \leftarrow$ call \textbf{ServerReEnc} ($c^*_i (S)$, $K_{s_i}$) \label{line:espoon-deploy-s-server-call}

\State $c(A) \leftarrow$ call \textbf{ServerReEnc} ($c^*_i (A)$, $K_{s_i}$) \label{line:espoon-deploy-a-server-call}

\State $c(T) \leftarrow$ call \textbf{ServerReEnc} ($c^*_i (T)$, $K_{s_i}$) \label{line:espoon-deploy-t-server-call}

\State $c (\langle S, A, T \rangle) \leftarrow (c (S), c (A), c (T))$ \label{line:espoon-deploy-sat-server-assignment}

\Return $c (\langle S, A, T \rangle)$

\end{algorithmic}
}
\end{algorithm}

\noindent \emph{\textbf{Deployment of a $\langle S, A, T \rangle$ Tuple:}}
For deploying any $\langle S, A, T \rangle$ tuple, an Admin User performs the first round of encryption using her private key as illustrated in Algorithm \ref{algo:espoon-deploy-sat-client-side}, where each element including $S$, $A$ and $T$ is encrypted on the client side by running \textbf{ClientEnc} (Algorithm \ref{algo:erbac-client-enc}) as shown in Line \ref{line:espoon-deploy-s-client-call}, Line \ref{line:espoon-deploy-a-client-call} and Line \ref{line:espoon-deploy-t-client-call}, respectively. The Administration Point on the server side receives the client encrypted tuple and performs the second round of encryption using the server side key corresponding to the Admin User as illustrated in Algorithm \ref{algo:espoon-deploy-sat-server-side}, where the Administration Point first retrieves the server side key corresponding to Admin User $i$ from the Key Store (see Line \ref{line:espoon-deploy-sat-ss-ks}) and then re-encrypts $c^*_i (S)$, $c^*_i (A)$ and $c^*_i (T)$ by running \textbf{ServerReEnc} (Algorithm \ref{algo:erbac-server-re-enc}) as shown in Line \ref{line:espoon-deploy-s-server-call}, Line \ref{line:espoon-deploy-a-server-call} and Line \ref{line:espoon-deploy-t-server-call}, respectively. Finally, the server encrypted tuple is stored in the Policy Store.

% aux method: client td

\begin{algorithm} [htp]
{\algofontsize
\caption{\textbf{ClientTD}}

\label{algo:erbac-client-td}

\begin{algorithmic}[1]

\INPUT \emph{It transforms the cleartext element into the client generated trapdoor.}

\Require Element $e$, the client side key set $K_{u_i}$ corresponding to user $i$ and the public parameters $params$.

\Ensure The client generated trapdoor $td^*_i (e)$.

\medskip

\State Choose a random $r_{e} \in \mathbb{Z}^*_q$ \label{line:erbac-c-td-choose}
\State ${\sigma}_{e} \leftarrow f_s (e)$ \label{line:erbac-c-td-sigma}
\State $t_1 \leftarrow g^{-r_{e}} g^{{\sigma}_{e}}$ \label{line:erbac-c-td-t1}
\State $t_2 \leftarrow h^{r_{e}} g^{-x_{i1}r_{e}} g^{x_{i1}{\sigma}_{e}} = g^{x_{i2}r_{e}} g^{x_{i1}{\sigma}_{e}}$ \label{line:erbac-c-td-t2}

\State $td^*_i (e) \leftarrow (t_1, t_2)$ \label{line:erbac-c-td-td}

\Return $td^*_i (e)$

\end{algorithmic}
}
\end{algorithm}

% aux method: server td

% server td
\begin{algorithm} [htp]
{\algofontsize
\caption{\textbf{ServerTD}}

\label{algo:erbac-server-td}

\begin{algorithmic}[1]

\INPUT \emph{It transforms the client generated trapdoor into the server generated trapdoor.}

\Require The client generated trapdoor $td^*_i (e)$ and the server side key set $K_{s_i}$ corresponding to user $i$.

\Ensure The server generated trapdoor $td(e)$.

\medskip

\State $td(e) \leftarrow t_1^{x_{i2}} . t_2 = g^{x{\sigma}_{e}}$ \label{line:erbac-s-td-calculate}

\Return $td(e)$

\end{algorithmic}
}
\end{algorithm}

% aux method: match element

\begin{algorithm} [htp]
{\algofontsize
\caption{\textbf{Match}}

\label{algo:erbac-match}

\begin{algorithmic}[1]

\INPUT \emph{It matches the serer encrypted element against the server generated trapdoor.}

\Require The server encrypted element $c(e) = (c_1, c_2)$ and the server generated trapdoor $td(e) = T$.

\Ensure $\mathit{true}$ or $\mathit{false}$.

\medskip

\If {$c_2 \stackrel{?}{=} H(c_1 . T^{-1})$} \label{line:erbac-match-condition}

	\Return $\mathit{true}$ \label{line:erbac-match-true}
	
\Else

	\Return $\mathit{false}$ \label{line:erbac-match-false}
	
\EndIf

\end{algorithmic}
}
\end{algorithm}

\begin{figure} [htp]
\centering
% left bottom right top
\includegraphics[trim=70mm 50mm 70mm 40mm,clip,width=.7\textwidth]{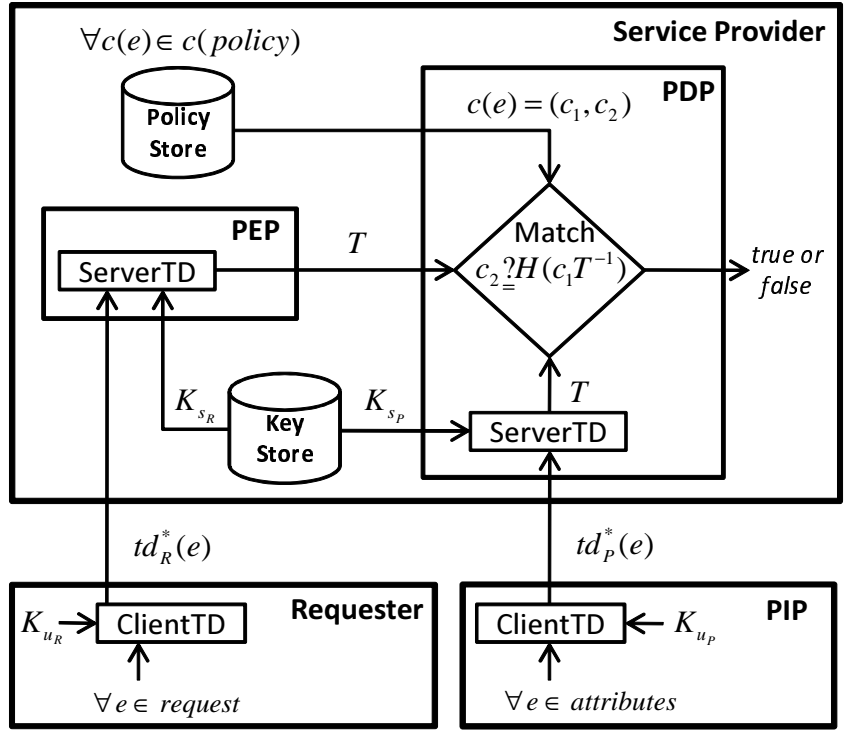} % .3
\caption{The policy evaluation phase}
\label{fig:erbac-doctor}
\end{figure}

\subsection{The Policy Evaluation Phase}

The policy evaluation phase is executed when a Requester makes a request. In this phase, a Requester sends client generated trapdoors (using Algorithm \ref{algo:erbac-client-td}) of a request to the \gls{PEP}. The \gls{PEP} converts client generated trapdoors into server generated trapdoors (using Algorithm \ref{algo:erbac-server-td}) and sends them to the \gls{PDP}. The \gls{PDP} matches server encrypted trapdoors of the request with server encrypted elements of the policy (using Algorithm \ref{algo:erbac-match}). Optionally, the \gls{PDP} may require contextual information in order to evaluate contextual conditions. The \gls{PIP} sends client generated trapdoors of contextual information to the \gls{PDP}. The \gls{PDP} converts client generated trapdoors into server generated trapdoors and then evaluates contextual conditions based on contextual information. Finally, the \gls{PDP} returns either $\mathit{true}$ or $\mathit{false}$ as shown in Figure \ref{fig:erbac-doctor}. In the following, we describe how we generate trapdoors and perform the match. 

For calculating client generated trapdoors of a request (or contextual information), a Requester (or the \gls{PIP}) runs \textbf{ClientTD} illustrated in Algorithm \ref{algo:erbac-client-td}. \textbf{ClientTD} takes as input each element $e$ of the request, the client side key set $K_{u_i}$ corresponding to user $i$ and the public parameters $params$ and outputs the client generated trapdoor $td^*_i (e)$. First, it choose randomly $r_{e} \in \mathbb{Z}^*_q$ (Line \ref{line:erbac-c-td-choose}). Next, it calculates ${\sigma}_{e}$ as $f_s (e)$ (Line \ref{line:erbac-c-td-sigma}). Then it calculates $t_1$ and $t_2$ as $g^{-r_{e}} g^{{\sigma}_{e}}$ (Line \ref{line:erbac-c-td-t1}) and $h^{r_{e}} g^{-x_{i1}r_{e}} g^{x_{i1}{\sigma}_{e}} = g^{x_{i2}r_{e}} g^{x_{i1}{\sigma}_{e}}$ (Line \ref{line:erbac-c-td-t2}), respectively. Both $t_1$ and $t_2$ form $td^*_i (e)$ (Line \ref{line:erbac-c-td-td}). A Requester sends client generated trapdoors of the request to the \gls{PEP}.
The \gls{PEP} receives client generated trapdoors and runs \textbf{ServerTD} illustrated in Algorithm \ref{algo:erbac-server-td} for calculating server generated trapdoors. \textbf{ServerTD} takes as input the client generated trapdoor $td^*_i (e)$ and the server side key set $K_{s_i}$ corresponding to user $i$ and outputs the server generated trapdoor $td(e)$. It calculates $td(e)$ as  $t_1^{x_{i2}} . t_2 = g^{x{\sigma}_{e}}$ (Line \ref{line:erbac-s-td-calculate}).

In order to match a server encrypted element of a policy with a server generated trapdoor of a request, the \gls{PDP} runs \textbf{Match} illustrated in Algorithm \ref{algo:erbac-match}. \textbf{Match} takes as input the server encrypted element $c(e) = (c_1, c_2)$ and the server generated trapdoor $td(e) = T$ and returns either $\mathit{true}$ or $\mathit{false}$. It checks the condition $c_2 \stackrel{?}{=} H(c_1 . T^{-1})$ (Line \ref{line:erbac-match-condition}). If the condition holds, it returns $\mathit{true}$ (Line \ref{line:erbac-match-true}) indicating that the match is successful. Otherwise, it returns $\mathit{false}$ (Line \ref{line:erbac-match-false}). 

In the following, we describe how to evaluate policies. For the evaluation of each policy, we follow general strategy as already described in this section and also illustrated in Figure \ref{fig:erbac-doctor}.

\begin{algorithm} [htp]
{\algofontsize
\caption{\textbf{SATRequest}}

\label{algo:espoon-client-sat-td}

\begin{algorithmic}[1]

\INPUT \emph{It transforms the cleartext tuple into the client generated trapdoor tuple.}

\Require Tuple $\langle S, A, T \rangle$, the client side key set $K_{u_i}$ corresponding to Requester $j$ and the public parameters $params$.

\Ensure The client generated trapdoor tuple $td^*_j (\langle S, A, T \rangle)$.

\medskip

\State $td^*_j (S) \leftarrow$ call \textbf{ClientTD} ($S$, $K_{u_j}$, $params$) \label{line:espoon-eval-s-client-call}
\State $td^*_j (A) \leftarrow$ call \textbf{ClientTD} ($A$, $K_{u_j}$, $params$) \label{line:espoon-eval-a-client-call}
\State $td^*_j (T) \leftarrow$ call \textbf{ClientTD} ($T$, $K_{u_j}$, $params$) \label{line:espoon-eval-t-client-call}

\State $td^*_j (\langle S, A, T \rangle) \leftarrow (td^*_j (S), td^*_j (A), td^*_j (T))$ \label{line:espoon-eval-sat-client-assignment}

\Return $td^*_j (\langle S, A, T \rangle)$

\end{algorithmic}
}
\end{algorithm}
\noindent \\
\noindent \emph{\textbf{Generating Tuples: A Client Request:}}
For making an access request, a Requester transforms the cleartext tuple into the trapdoor tuple as illustrated in Algorithm \ref{algo:espoon-client-sat-td}, which transforms each element in tuple including $S$, $A$ and $T$ into its corresponding trapdoor $td^*_j (S)$, $td^*_j (A)$ and $td^*_j (T)$ (using \textbf{ClientTD} illustrated in Algorithm \ref{algo:erbac-client-td})as shown in Line \ref{line:espoon-eval-s-client-call}, Line \ref{line:espoon-eval-a-client-call} and Line \ref{line:espoon-eval-t-client-call}, respectively. The Requester client side sends the trapdoor tuple to the \gls{PEP}.

% search SAT

\begin{algorithm} [htp]
{\algofontsize
\caption{\textbf{SATSearch}}

\label{algo:espoon-search-sat}

\begin{algorithmic}[1]

\INPUT \emph{It checks whether the access request matches with any encrypted tuple on the server side.}

\Require The client generated trapdoor tuple $td^*_j (\langle S, A, T \rangle)$, the identity of Requester $j$ and a list (of size $n$) of encrypted policies stored on the server $c(\langle S_i, A_i, T_i \rangle)_{1 \leq i \leq n})$.

\Ensure $\mathit{true}$ or $\mathit{false}$.

\medskip

\State $K_{s_j} \leftarrow KS[j]$ {\algofontsize \Comment{retrieve the server side key corresponding to Requester $j$}} \label{line:espoon-search-sat-ks}

\State $td(S) \leftarrow$ call \textbf{ServerTD} ($td^*_j (S)$, $K_{s_j}$) \label{line:espoon-search-sat-td-s}

\State $td(A) \leftarrow$ call \textbf{ServerTD} ($td^*_j (A)$, $K_{s_j}$) \label{line:espoon-search-sat-td-a}

\State $td(T) \leftarrow$ call \textbf{ServerTD} ($td^*_j (T)$, $K_{s_j}$) \label{line:espoon-search-sat-td-t}

\For {each encrypted tuple $c (\langle S, A, T \rangle)$ in $c(\langle S_i, A_i, T_i \rangle)_{1 \leq i \leq n})$} \label{line:espoon-search-sat-loop}

	\State $match_S \leftarrow$ call \textbf{Match} ($c(S)$, $td(S)$) \label{line:espoon-search-sat-match-s}
	
	\State $match_A \leftarrow$ call \textbf{Match} ($c(A)$, $td(A)$) \label{line:espoon-search-sat-match-a}
	
	\State $match_T \leftarrow$ call \textbf{Match} ($c(T)$, $td(T)$) \label{line:espoon-search-sat-match-t}

	\If {$match_S \stackrel{?}{=} true$ and $match_A \stackrel{?}{=} true$ and $match_T \stackrel{?}{=} true$} \label{line:espoon-search-sat-check-match}
		
		\Return $\mathit{true}$
		
	\EndIf

\EndFor

\Return $\mathit{false}$

\end{algorithmic}
}
\end{algorithm}

\noindent \emph{\textbf{Searching a Tuple:}}
When a Requester makes an access request, the \gls{PEP} receives the client encrypted request and then it re-encrypts the request. The Service Provider first retrieves the server side key corresponding to Requester $j$ as illustrated in Algorithm \ref{algo:espoon-search-sat} Line \ref{line:espoon-search-sat-ks}. Next, it calls \textbf{ServerTD} (Algorithm \ref{algo:erbac-server-td}) for each client encrypted element including $td^*_j (S)$, $td^*_j (A)$ and $td^*_j (T)$ and calculates $td(S)$, $td(A)$ and $td(T)$ as shown in Line \ref{line:espoon-search-sat-td-s}, Line \ref{line:espoon-search-sat-td-a} and Line \ref{line:espoon-search-sat-td-t}, respectively. Then, the Service Provider checks if any encrypted tuple in the Policy Store matches with the encrypted access request (Line \ref{line:espoon-search-sat-loop}). For performing this match, all three encrypted elements are matching using \textbf{Match} (Algorithm \ref{algo:erbac-match}) (Line \ref{line:espoon-search-sat-match-s}-\ref{line:espoon-search-sat-match-t}). If all three elements are matched (Line \ref{line:espoon-search-sat-check-match}), then this algorithm returns $\mathit{true}$. In case if no match is found, this algorithm returns $\mathit{false}$.

% contextual condition: client side: request

\begin{algorithm} [htp]
{\algofontsize
\caption{\textbf{AttributesRequest}}

\label{algo:erbac-request-contextual-condition}

\begin{algorithmic}[1]

\INPUT \emph{It transforms contextual attributes into trapdoors.}

\Require List of attributes contextual attributes $L$, the client side key set $K_{u_j}$ corresponding to \gls{PIP} $j$ and the public parameters $params$.

\Ensure The client generated list of trapdoors of contextual attributes $L_{C_j}$.

\medskip

\State $L_{C_j} \leftarrow \phi$ \label{request-cc-init}

\For {each attribute $e$ in $L$} \label{request-cc-loop}

	\State $td^*_j (e) \leftarrow$ call \textbf{ClientTD} ($r$, $K_{u_j}$, $params$) \label{request-cc-td}
	
	\State $L_{C_j} \leftarrow L_{C_j} \cup td^*_j (e)$ \label{request-cc-update}

\EndFor

\Return $T_{C_j}$

\end{algorithmic}
}
\end{algorithm}

\noindent \emph{\textbf{Generating Contextual Attributes:}} 
The \gls{PIP} runs \textbf{AttributesRequest} illustrated in Algorithm \ref{algo:erbac-request-contextual-condition} to calculate client generated trapdoors of contextual information. \textbf{AttributesRequest} takes as input a list of contextual attributes $L$, the client side key set $K_{u_j}$ corresponding to \gls{PIP} $j$ and the public parameters $params$ and outputs the client generated list of trapdoors of contextual attributes $L_{C_j}$. First, it creates and initialises new list $L_{C_j}$ (Line \ref{request-cc-init}). For each attribute $e$ in $L$ (Line \ref{request-cc-loop}), it calculates the client generated trapdoor $td^*_j (e)$ by calling Algorithm \ref{algo:erbac-client-td} (Line \ref{request-cc-td}) and adds $td^*_j (e)$ in $L_{C_j}$ (Line \ref{request-cc-update}).

% contextual condition: request and then search/matching

\begin{algorithm} [htp]
{\algofontsize
\caption{\textbf{ConditionEvaluation}}

\label{algo:erbac-match-contextual-condition}

\begin{algorithmic}[1]

\INPUT \emph{It evaluates contextual condition and returns $\mathit{true}$ on successful match and $\mathit{false}$ otherwise.}

\Require The client generated list of trapdoors of contextual attributes $L_{C_j}$, the server encrypted contextual condition $T_{S}$ and the identity of \gls{PIP} $j$.

\Ensure $\mathit{true}$ or $\mathit{false}$.

\medskip

\State $K_{s_j} \leftarrow KS[j]$ {\algofontsize \Comment{retrieve the server side key corresponding to \gls{PIP} $j$}} \label{line:erbac-cc-match-ks}

\State $L_{S} \leftarrow \phi$ \label{line:erbac-cc-match-init-list}

\For {each client generated trapdoor $td^*_j (e)$ in $L_{C_j}$} \label{line:erbac-cc-match-loop-s-td}
	
	\State $td(e) \leftarrow$ call \textbf{ServerTD} ($td^*_j (e)$, $K_{s_j}$) \label{line:erbac-cc-match-call-s-td}
	
	\State $L_{S} \leftarrow L_{S} \cup td^*_j (e)$ \label{line:erbac-cc-match-update-list}
	
\EndFor

\State $\mathit{TREE} \leftarrow T_{S}$ \label{line:erbac-cc-match-copy-tree}

\State Add decision field to each node in $\mathit{TREE}$ \label{line:erbac-cc-match-add-field}

\For {each node $n$ in $\mathit{TREE}$} \label{line:erbac-cc-match-loop-tree-init}

	\State $n.decision \leftarrow null$ \label{line:erbac-cc-match-init-field}

\EndFor

\For {each leaf node $n$ in $\mathit{TREE}$} \label{line:erbac-cc-match-loop-tree-match}
	
	\For {each server generated trapdoor $td(e)$ in $L_{S}$} \label{line:erbac-cc-match-loop-list-match}
	
		\State $n.decision \leftarrow$ call \textbf{Match} ($n.c(e)$, $td(e)$) \label{line:erbac-cc-match-decision}

		\If {$n.decision \stackrel{?}{=} true$} \label{line:erbac-cc-match-if-success}
		
			\State $\mathit{break;}$ \label{line:erbac-cc-match-stop}
			
		\EndIf  
	
	\EndFor

\EndFor

\State call \textbf{EvaluateTree} ($\mathit{TREE.root}$, $\mathit{TREE}$) {\algofontsize \Comment{see Algorithm \ref{algo:erbac-evaluate-tree}}} \label{line:erbac-cc-match-call-eval-tree}

\Return $\mathit{TREE.root.decision}$ %\label{line:erbac-cc-match-return-decision}

\end{algorithmic}
}
\end{algorithm}
\noindent \\
\noindent \emph{\textbf{Evaluating Contextual Conditions:}}
For evaluating any contextual condition, the \gls{PDP} runs \textbf{ConditionEvaluation} illustrated in Algorithm \ref{algo:erbac-match-contextual-condition}. This algorithm takes as input the client generated list of trapdoors of contextual attributes $L_{C_j}$, the server encrypted contextual condition $T_{S}$ and identity of \gls{PIP} $j$ and returns either $\mathit{true}$ or $\mathit{false}$. First, it retrieves from the Key Store the server side key $K_{s_j}$ (Line \ref{line:erbac-cc-match-ks}). Next, it creates and initialises a new list $L_{S}$ (Line \ref{line:erbac-cc-match-init-list}). For each client generated trapdoor $td^*_j (e)$ in $L_{C_j}$ (Line \ref{line:erbac-cc-match-loop-s-td}), it calculates the server generated trapdoor $td(e)$ by calling Algorithm \ref{algo:erbac-server-td} (Line \ref{line:erbac-cc-match-call-s-td}) and adds $td(e)$ in $L_{S}$ (Line \ref{line:erbac-cc-match-update-list}). Next, it copies $T_{S}$ to $\mathit{TREE}$ (Line \ref{line:erbac-cc-match-copy-tree}) and adds decision field to each node in $\mathit{TREE}$ (Line \ref{line:erbac-cc-match-add-field}). For each node $n$ in $\mathit{TREE}$ (Line \ref{line:erbac-cc-match-loop-tree-init}), it initialises $n.decision$ as $null$ (Line \ref{line:erbac-cc-match-init-field}). For each leaf node $n$ in $\mathit{TREE}$ (Line \ref{line:erbac-cc-match-loop-tree-match}), it checks if any server generated trapdoor $td(e)$ in $L_{S}$ (Line \ref{line:erbac-cc-match-loop-list-match}) matches with it by calling Algorithm \ref{algo:erbac-match} (Line \ref{line:erbac-cc-match-decision}). Next, it evaluates non-leaf nodes of $\mathit{TREE}$ by running Algorithm \ref{algo:erbac-evaluate-tree} (Line \ref{line:erbac-cc-match-call-eval-tree}). Finally, it returns either $\mathit{true}$ or $\mathit{false}$ depending upon the evaluation of $\mathit{TREE}$.% (Line \ref{line:erbac-cc-match-return-decision}). 

% aux EvaluateTree method

\begin{algorithm} [htp]

{\algofontsize
\caption{\textbf{EvaluateTree}}

\label{algo:erbac-evaluate-tree}

\begin{algorithmic}[1]

\INPUT \emph{Given a tree node, it recursively evaluates internal nodes of a policy tree and returns $\mathit{true}$ if the policy tree is satisfied and $\mathit{false}$ otherwise.}

\Require Node $n$ and tree $T$.

\Ensure $\mathit{true}$ or $\mathit{false}$.

\medskip

\If {$\mathit{n.decision} \neq null$} \label{line:erbac-evaluate-tree-if-null}

	\Return $\mathit{n.decision}$ \label{line:erbac-evaluate-tree-decision-not-null}
	
\EndIf

\For {each child $c$ of $n$ in tree $T$} \label{line:erbac-evaluate-tree-child-loop}
	
	\State call \textbf{EvaluateTree} ($c$, $T$) {\algofontsize \Comment{recursive call}} \label{line:erbac-evaluate-tree-call}
	
\EndFor

\State $t \leftarrow 0$ \label{line:erbac-evaluate-tree-init-t}

\State $m \leftarrow 0$ \label{line:erbac-evaluate-tree-init-m}

\For {each child $c$ of $n$ in tree $T$} \label{line:erbac-evaluate-tree-count-loop}

	\State $t \leftarrow t + 1$ \label{line:erbac-evaluate-tree-inc-t}
	
	\If {$\mathit{c.decision} \stackrel{?}{=} \mathit{true}$} \label{line:erbac-evaluate-tree-find-decision}
	
		\State $m \leftarrow m + 1$ \label{line:erbac-evaluate-tree-inc-m}
		
	\EndIf

\EndFor

\If {($\mathit{n.gate} \stackrel{?}{=} \mathit{AND}$ and $m \stackrel{?}{=} t$) or ($n.gate \stackrel{?}{=} \mathit{OR}$ and $m \geq 1$)} \label{line:erbac-evaluate-tree-find-gate}

	\State $\mathit{n.decision} \leftarrow \mathit{true}$ \label{line:erbac-evaluate-tree-set-decision-true}
	
\Else

	\State $\mathit{n.decision} \leftarrow \mathit{false}$ \label{line:erbac-evaluate-tree-set-decision-false}
	
\EndIf

\Return $\mathit{n.decision}$ \label{line:erbac-evaluate-tree-return-decision}

\end{algorithmic}

}

\end{algorithm}
\noindent \\
\noindent \textbf{EvaluateTree} evaluates a tree containing AND and OR gates. It takes as input root node $n$ and tree $T$ and returns either $\mathit{true}$ or $\mathit{false}$. First, it checks if the decision for $n$ is already made (Line \ref{line:erbac-evaluate-tree-if-null}). If so, it returns the decision (Line \ref{line:erbac-evaluate-tree-decision-not-null}). For each child $c$ of $n$ in tree $T$ (Line \ref{line:erbac-evaluate-tree-child-loop}), it recursively calls \textbf{EvaluateTree} (Line \ref{line:erbac-evaluate-tree-call}). Next, it creates and initialises $t$ (Line \ref{line:erbac-evaluate-tree-init-t}) and $m$ (Line \ref{line:erbac-evaluate-tree-init-m}) indicating total children of $n$ and a count of matched children, respectively. For each child $c$ of $n$ in tree $T$ (Line \ref{line:erbac-evaluate-tree-count-loop}), it counts total children (Line \ref{line:erbac-evaluate-tree-inc-t}) and matched children by checking made decisions (Line \ref{line:erbac-evaluate-tree-inc-m}). Next, it checks if non-leaf node is AND and all children are matched or non-leaf node is OR and at least one child is matched (Line \ref{line:erbac-evaluate-tree-find-gate}). If so, it is set as $\mathit{true}$ (Line \ref{line:erbac-evaluate-tree-set-decision-true}) and $\mathit{false}$ (Line \ref{line:erbac-evaluate-tree-set-decision-false}) otherwise.

% user revocation

\begin{algorithm} [htp]
{\algofontsize
\caption{\textbf{UserRevocation}}

\label{algo:egrant-user-revocation}

\begin{algorithmic}[1]

\INPUT \emph{It removes users from the system.}

\Require The user identity $i$.

\Ensure $\mathit{true}$ or $\mathit{false}$.

\medskip

\If {$exits(KS[i]) \stackrel{?}{=} false$} \label{line:egrant-user-presence}

	\Return $\mathit{false}$ \label{line:egrant-no-user-present}

\EndIf

\State $K_{s_i} \leftarrow KS[i]$ \label{line:egrant-get-user-key}
\State $KS \leftarrow KS \backslash K_{s_i}$ \label{line:egrant-remove-user-key}

\Return $\mathit{true}$ \label{line:egrant-user-removed-successfully}

\end{algorithmic}
}
\end{algorithm}

\subsection{The User Revocation Phase}
In this phase, a user (an Admin User or a Requester) can be removed from the system. This phase consists of one algorithm called \textbf{UserRevocation} illustrated in Algorithm \ref{algo:egrant-user-revocation}, which is run by the Administration Point. Given the user identity $i$, this algorithm checks whether the server side key set corresponding to user $i$ exists in the Key Store (Line \ref{line:egrant-user-presence}). If not then this algorithm returns $\mathit{false}$ (Line \ref{line:egrant-no-user-present}), indicating that no such user exists. Otherwise, the server side key set $K_{s_i}$ corresponding to user $i$ is removed from the Key Store (Line \ref{line:egrant-get-user-key}-\ref{line:egrant-remove-user-key}) and finally this algorithm returns $\mathit{true}$ (Line \ref{line:egrant-user-removed-successfully}), indicating that user $i$ has been removed from the system successfully.

\section[Performance Analysis of ESPOON]{Performance Analysis of \gls{ESPOON}}
\label{sec:espoon-performance-analysis}
In this section, we discuss a quantitative analysis of the performance of \gls{ESPOON}. It should be noticed that here we are concerned about quantifying the overhead introduced by the encryption operations performed both in the trusted and outsourced environments. In the following discussion, we do not take into account the latency introduced by the network communication.

\subsection[Implementation Details of ESPOON]{Implementation Details of \gls{ESPOON}}
We have implemented \gls{ESPOON} in Java $1.6$. We have developed all the components of the architecture required in the management lifecycle of \gls{ESPOON} policies in outsourced environments. In particular, we have implemented all the algorithms presented in Section \ref{sec:espoon-algorithmic-details}. We have tested the implementation of \gls{ESPOON} on a single node based on an Intel Core2 Duo $2.2$ GHz processor with $2$ GB of RAM, running Microsoft Windows XP Professional version $2002$ Service Pack $3$. The number of iterations performed for each of the following results is $1000$.

\subsection{Performance Analysis of the Policy Deployment Phase}
\label{sec:espoon-policy-deployment}

In this section, we analyse the performance of the policy deployment phase. In this phase, access policies are first encrypted at the Admin User side (that is a trusted domain) and then sent over to the Administration Point running in the outsourced environment. The Administration Point re-encrypts the policies and stores them in the Policy Store in the outsourced environment. The policy contains two parts (i) a contextual condition and (ii) a $\langle S, A, T \rangle$ tuple. In the following, we discuss performance overheads of deploying both parts. \\ \\
\noindent \emph{\textbf{Deploying a Contextual Condition:}}
Our policy representation consists of the tree representing the policy condition and the $\langle S, A, T \rangle$ tuple describing what action $A$ a subject $S$ can perform over the target $T$. In the tree representing contextual conditions, leaf nodes represent string comparisons (for instance, $\mathit{Location = Cardiology \mhyphen ward}$) and/or numerical comparisons (for instance, $\mathit{Access Time > 9}$). A string comparison is always represented by a single leaf node while a numerical comparison may require more than one leaf nodes. In the worst case, a single numerical comparison, represented as $s$ bits, may require $s$ separate leaf nodes. Therefore, numerical comparisons have a major impact on the encryption of a policy at deployment time.

\begin{figure} [htp]
\centering
\subfigure[]{
\includegraphics[width=.48\textwidth]{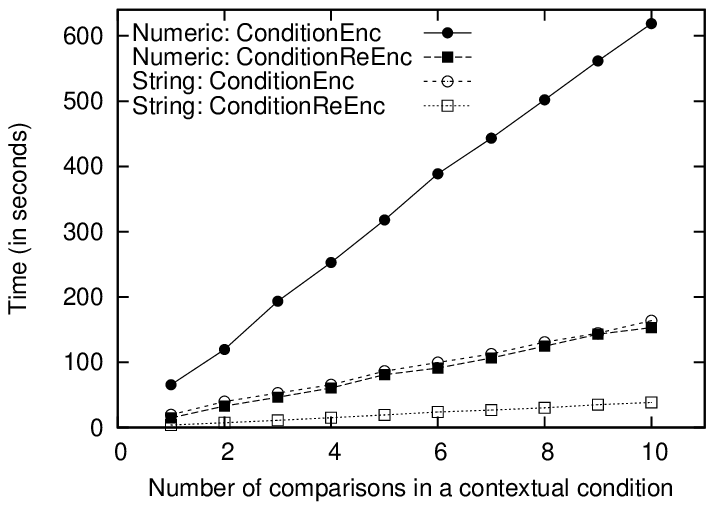}
\label{fig:erbac-deploy-context-attr}
}
\subfigure[]{
\includegraphics[width=.48\textwidth]{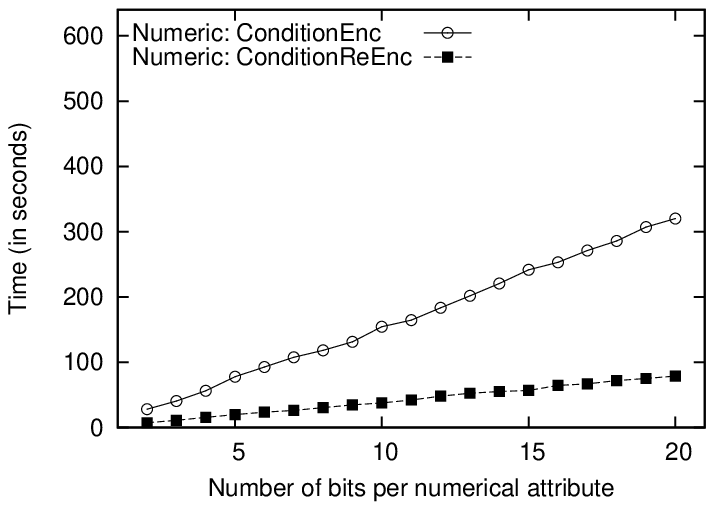}
\label{fig:erbac-deploy-context-attr-bits}
}
\caption[Performance overhead of deploying contextual conditions]{Performance overhead of deploying contextual conditions: \subref{fig:erbac-deploy-context-attr} numerical and string comparisons and \subref{fig:erbac-deploy-context-attr-bits} size of a numerical attribute}
\label{fig:erbac-policy-deployment-context}
\end{figure}

The performance overhead of deploying contextual conditions is illustrated in Figure \ref{fig:erbac-policy-deployment-context}. Figure \ref{fig:erbac-deploy-context-attr} illustrates the performance overhead of deploying numerical and string comparisons. In this graph, we increase the number of string comparisons and numerical comparisons present in the contextual condition of a policy. As the graph, the time taken by deployment functions on the client side and the server side grow linearly with the number of comparisons in the contextual condition. The numerical comparisons have a steeper line because one numerical comparison of size $s$ may be equivalent to $s$ string comparisons in the worst case. For string comparisons, we have used ``$attributeName_i$=$attributeValue_i$'', where $i$ varies from 1 to 10. For numerical comparisons, we have used ``$attributeName_i<15$\#4''.\footnote{It should be noted that using the comparison less than 15 in a 4-bit representation represents the worst case scenario requiring 4 leaf nodes.}

To check how the size of the bit representation impacts on the encryption functions during the deployment phase, we have performed the following experiment. We fixed the number of numerical comparisons in the contextual condition to only one and increased the size $s$ of the bit representation from $2$ to $20$ for the comparison ``$attributeName<2^s-1$. Figure \ref{fig:erbac-deploy-context-attr-bits} shows the performance overhead of the encryption during the policy deployment phase on the client side, as well as on the server side. We can see that the policy deployment time incurred grows linearly with the increase in the size $s$ of a numerical attribute. In general, the time complexity of the encryption of the contextual conditions during the policy deployment phase is $O (m + n \cdot s)$ where $m$ is the number of string comparisons, $n$ is the number of numerical comparisons, and $s$ represents the number of bits in each numerical comparison.

\begin{table} [htp]
\centering
\caption[Performance overhead of deploying the $\langle S, A, T \rangle$ tuple]{Performance overhead of encrypting the $\langle S, A, T \rangle$ tuple during the policy deployment}
\label{tab:espoon-sat-pol-deployment}
\begin{tabular}{ |l|c|c| }
\hline
\textbf{Algorithm Name} & \textbf{SATEnc} & \textbf{SATReEnc} \\ \hline
Time (in milliseconds) & 46.44 & 11.65 \\ \hline
\end{tabular}
\end{table}

\noindent \emph{\textbf{Deploying a $\langle S, A, T \rangle$ Tuple:}}
As for the $\langle S, A, T \rangle$ tuple, the average encryption time taken by the \textbf{SATEnc} (Algorithm \ref{algo:espoon-deploy-sat-client-side}) and \textbf{SATReEnc} (Algorithm \ref{algo:espoon-deploy-sat-server-side}) are shown in Table \ref{tab:espoon-sat-pol-deployment}. The time complexity of the encryption of the $\langle S, A, T \rangle$ tuple during the policy deployment phase is constant because it does not depend on any parameters.

During the policy deployment phase, the encryption operations performed on the Admin User side take more time to encrypt the access policy than the Service Provider side to re-encrypt the same policy (either \textbf{ConditionReEnc} or \textbf{SATReEnc}). This is because the \textbf{ConditionEnc} and \textbf{SATEnc} algorithms perform more complex cryptographic operations, such as generation of random number and hash calculations, than the respective algorithms on the Service Provider side.

\subsection{Performance Analysis of the Policy Evaluation Phase}
\label{sec:espoon-policy-evaluation}

In this section, we analyse the performance of the policy evaluation phase. In this phase, a Requester encrypts the $\langle S, A, T \rangle$ tuple before sending to the \gls{PEP} running in the outsourced environment. The \gls{PEP} re-encrypts and forwards it to the \gls{PDP}. The \gls{PDP} has to select the set of policies that are applicable to the request. Once the \gls{PDP} has found the policies then the \gls{PDP} will evaluate if the attributes in the contextual information satisfy any of the conditions of the selected policies. In the following, we discuss performance overhead of generating the encrypted $\langle S, A, T \rangle$ tuple, searching the requested $\langle S, A, T \rangle$ tuple in the policy store and evaluating contextual conditions.

\begin{table} [htp]
\centering
\caption{Performance overhead of generating the $\langle S, A, T \rangle$ request}
\label{tab:espoon-sat-pol-eval}
\begin{tabular}{ |l|c| }
\hline
\textbf{Algorithm Name} & \textbf{SATRequest} \\ \hline
Time (in milliseconds) & 47.07 \\ \hline
\end{tabular}
\end{table}

\noindent \emph{\textbf{The $\langle S, A, T \rangle$ Request Tuple:}}
To make a request, it is necessary to generate the $\langle S, A, T \rangle$ tuple representing the subject $S$ requesting to perform action $A$ on target $T$. The $\langle S, A, T \rangle$ tuple needs to be transformed into \emph{trapdoors} before it is sent over to the \gls{PEP}. The trapdoors will be used for performing the encrypted policy evaluation in the outsourced environment. The trapdoor representation does not leak information on the element of the $\langle S, A, T \rangle$ tuple. This phase takes approximately 47.07 \gls{ms} as shown in Table \ref{tab:espoon-sat-pol-eval}.

\begin{figure}[htp]
\centering
\includegraphics[angle=-90,width=.5\textwidth]{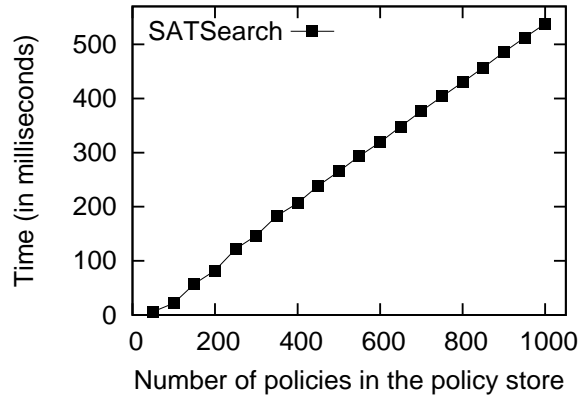}
\caption{Performance overhead of searching a $\langle S, A, T \rangle$ tuple}
\label{fig:espoon-sat-search}
\end{figure}
\noindent \\
\noindent \emph{\textbf{Searching a $\langle S, A, T \rangle$ Tuple:}}
Once the \gls{PDP} gets the request, it re-encrypts and then performs an encrypted search in the Policy Store in order to find any matching $\langle S, A, T \rangle$ tuples. Figure \ref{fig:espoon-sat-search} shows the performance overhead on the Service Provider side. In our experiment, we varied the number of encrypted policies stored in the Policy Store ranging from $50$ to $1000$. As we can observe, it takes $0.5$ \gls{ms} on average for performing an encrypted match operation between the $\langle S, A, T \rangle$ tuple of the request and the $\langle S, A, T \rangle$ tuple in the Policy Store. This means that on average it takes half a second for finding a matching policy in the Policy Store with $1000$ policies.

% context evaluation
\begin{figure} [htp]
\centering
\subfigure[]{
\includegraphics[width=.48\textwidth]{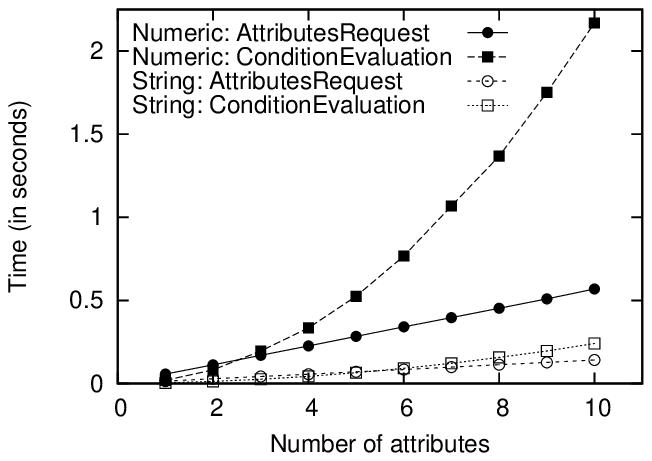}
\label{fig:erbac-request-context-attr}
}
\subfigure[]{
\includegraphics[width=.48\textwidth]{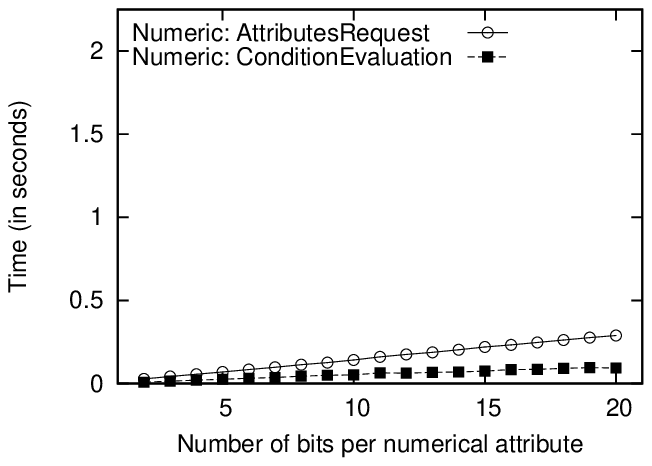}
\label{fig:erbac-request-context-attr-bits}
}
\caption[Performance overhead of evaluating contextual conditions]{Performance overhead of evaluating contextual conditions: \subref{fig:erbac-request-context-attr} numerical and string attributes and \subref{fig:erbac-request-context-attr-bits} size of a numerical attribute}
\label{fig:erbac-policy-evaluation-context}
\end{figure}

\noindent \emph{\textbf{Evaluating Contextual Conditions:}}
If any match is found in the Policy Store then the \gls{PDP} needs to fetch the contextual information from the \gls{PIP}. The \gls{PIP} is responsible to collect and send the required contextual information that includes information about the Requester (for instance, Requester's location or Requester's age) or the environment in which the request is made (for instance, time or temperature). The \gls{PIP} transforms these attributes into trapdoors before sending to the \gls{PDP} (as illustrated in Algorithm \ref{algo:erbac-request-contextual-condition}). For each single string attribute (for instance, $\mathit{Location=Cardiology \mhyphen ward}$), the \gls{PIP} generates a single trapdoor. For each numerical attribute of size s-bit (for instance, $\mathit{Access Time = 10\#5}$), the \gls{PIP} generates s trapdoors. Figure \ref{fig:erbac-policy-evaluation-context} shows the performance overhead of evaluating contextual conditions. In particular, Figure \ref{fig:erbac-request-context-attr} shows the performance overhead of generating trapdoors by the \gls{PIP} on the client side for both numerical and string attributes. In our experiment, we vary number of attributes (both string and numeric) from 1 to 10. As we can see, the graph grows linearly with the increase in number of attributes. For numerical attributes, the curve of trapdoor generation on the client side is steeper than that of the string attributes because numerical attribute is of size s bits where s is set to 4. This means that each numerical attribute requires 4 trapdoors; on the other hand, a string attribute requires only a single attribute. We observe also the behaviour of generating client trapdoors for a numerical attribute of varying size. Figure \ref{fig:erbac-request-context-attr-bits} shows behaviour of generating on the client side trapdoors of a numerical attribute of varying size ranging from 2 to 20 bits. This graph grows linearly with the increase in number of bits, representing size of a numerical attribute.

After receiving trapdoors of contextual information, the \gls{PDP} may evaluate a contextual condition. To evaluate the tree representing a contextual condition, the \gls{PDP} matches contextual information against the leaf nodes in the tree, as illustrated in Algorithm \ref{algo:erbac-match-contextual-condition}. To quantify the performance overhead of this encrypted matching, we have performed the following test. First, we have considered two cases: the first case is the one in which the \gls{PIP} provides only string attributes and the contextual condition contains only string comparisons; in the second, the \gls{PIP} provides only numerical attributes and the contextual condition consists only of numerical comparisons. For both cases, the number of attributes varies together with the number of comparisons in the tree. %In particular, if the \gls{PIP} provides $n$ different attributes then the contextual condition will contain $n$ different comparisons.

Figure \ref{fig:erbac-request-context-attr} shows also the performance overhead of evaluating string and numerical comparisons on the server side. As we can see, the condition evaluation for numerical attributes has a steeper curve. This can be explained as follows. For the first case, for each string attribute only a single trapdoor is generated. A string comparison is represented as a single leaf node in the tree representing a contextual condition. This means that $m_1$ trapdoors in a request are matched against $m_2$ leaf nodes in the tree resulting in a $O(m_1 \cdot m_2)$ complexity (however, in our experiments the number of attributes and the number of comparisons are always the same). For the case of the numerical attributes, we have also to take in to consideration the bit representation. In particular, for a give numerical attribute represented as $s$ bits, we need to generate $s$ different trapdoors. This means that $n$ numerical attributes in a request will be converted in to $n \cdot s$ different trapdoors. These trapdoors then need to be matched against the leaf nodes representing the numerical comparisons. Figure \ref{fig:erbac-request-context-attr-bits} shows the performance overhead of evaluating a numerical comparison where the size of a numerical attribute varies from 2 to 20. As we have discussed for the policy deployment phase, in the worst case scenario, a numerical comparison for a $s$-bit numerical attribute requires $s$ different leaf nodes. If there are $n_1$ numerical attributes in the request and $n_2$ different numerical comparisons (where each numerical attribute or numerical comparison is of size $s$), the complexity of evaluating numerical conditions will be $O(n_1 \cdot n_2 \cdot s^2)$ in the worst case. In general, the complexities of generating trapdoors for conditions and evaluating contextual conditions are $O (m + n \cdot s)$ and $O(m_1 \cdot m_2 + n_1 \cdot n_2 \cdot s^2)$, respectively.

\begin{table} [htp]
\centering
\caption[Time complexity of each phase in the lifecycle of ESPOON]{Summary of time complexity of each phase in the lifecycle of \gls{ESPOON}}
\label{tab:espoon-complexity-summary}
\begin{tabular}{ |l|c| } 

\hline

\textbf{Phase Name} & \textbf{Complexity in the Worst Case} \\ \hline

Deployment of contextual condition & $O (m + n \cdot s)$ \\ \hline

Attributes request & $O (m + n \cdot s)$ \\ \hline

Evaluation of contextual condition & $O (m_1 \cdot m_2 + n_1 \cdot n_2 \cdot s^2)$ \\ \hline

\end{tabular}

\end{table}

Table \ref{tab:espoon-complexity-summary} provides a summary of time complexity of each phase in the lifecycle of \gls{ESPOON}.

\section{Discussion}
\label{sec:espoon-discussion}
%This section provides the discussion about the security and privacy aspects of \gls{ESPOON}.

\subsection{Data Protection}
In this chapter, we have focused on how to enforce sensitive security policies in outsourced environments. For the data protection, we may employ existing encryption techniques, such as the proxy encryption scheme \cite{Dong:2011} or schemes based on \gls{ABE} \cite{Bethencourt:2007, Goyal:2006}. In \cite{Asghar2013:CCSW}, we have discussed how to protect data using the proxy encryption scheme. In this dissertation, we have covered the topic of data protection using \gls{CPABE} \cite{Bethencourt:2007} in Chapter \ref{cha:pidgin}.

\subsection{Revealing Policy Structure}
The access policy structure reveals information about the operators, such as AND and OR, and the number of operands used in the access policy condition. To overcome this problem, dummy attributes may be inserted in the tree structure of the access policy. Similarly, the \gls{PIP} can send dummy attributes to the \gls{PDP} at the time of policy evaluation to obfuscate the number of attributes required in a request.

\subsection{Collusion Attack}
In \gls{ESPOON}, we assume that multiple users can collude; however, they cannot gain more than what each user can access individually because each one has her own private key and combination of those keys do not reveal any further information. On the other hand, a user and the Service Provider can collude together to gain unauthorised access to the data by combining their keys, where they can recover the master secret. For withstanding against this kind of collusion, one possibility is to assume multiple instances of the Service Provider and split the server side key such that each instance gets one share. The main drawback of this approach is that it cannot work if all instances of the Service Provider are compromised. Another approach is to provide protection with an extra layer of encryption say by employing \gls{KPABE} \cite{Goyal:2006}, which is collusion-resistant.

\subsection[Impossibility of Cryptography Alone for Preserving Privacy]{On the Impossibility of Cryptography Alone for Privacy-Preserving Cloud Computing}
Van Dijk and Juels argue in \cite{VanDijk:2010} that cryptography alone is not sufficient for preserving the privacy in the cloud environment. They prove that in multi-client settings it is impossible to control how information is released to clients with different access rights. Basically, in their threat model clients do not mutually trust each other. In our settings, users are mutually trusted: our main contribution is to protect the confidentiality of access policies (and therefore of the data) from the Service Provider.

\section{Chapter Summary}
\label{sec:espoon-summary}
In this chapter, we have presented the \gls{ESPOON} architecture to support a policy-based access control mechanism for outsourced environments. Our approach separates the security policies from the actual enforcement mechanism while guaranteeing the confidentiality of the policies when given assumptions hold (i.e., the Service Provider is honest-but-curious). The main advantage of our approach is that policies are encrypted but it still allows the \gls{PDP} to perform the policy evaluation without knowing the policies. Second, \gls{ESPOON} is capable of handling complex policies involving non-monotonic boolean expressions and range queries. Finally, the authorised users do not share any encryption keys making the process of key management very scalable. Even if a user key is deleted or revoked, the other entities are still able to perform their operations without requiring re-encryption of the policies.

From performance and management perspectives, \gls{ESPOON} might be suitable for handling access policies of small to medium enterprises. However, both performance and management will be cumbersome if \gls{ESPOON} has to be deployed for handling access policies of large enterprises having a large number of users, thus requiring complex user management. In the next chapter, we propose architecture that can enforce sensitive policies of large enterprises having a large number of users.

%%%%%%%%%%%%%%%%%%%%%%%%% CHAPTER ERBAC %%%%%%%%%%%%%%%%%%%%%%%%%

\chapter[Enforcing Encrypted RBAC Policies]{\acrshort{ESPOONERBAC}: Enforcing Encrypted \acrshort{RBAC} Policies in Outsourced Environments\footstar{The preliminary version of this chapter has appeared in \cite{Asghar2013-COSE, Asghar2011-CCS}.}}
\label{cha:erbac}

For complex user management, large enterprises employ \gls{RBAC} models for making access decisions based on the role in which a user is active in. However, \gls{RBAC} models cannot be deployed in outsourced environments as they rely on trusted infrastructure in order to regulate access to the data. The deployment of \gls{RBAC} models may reveal private information about sensitive data they aim to protect. In this chapter, we aim at filling this gap by proposing \textbf{\gls{ESPOONERBAC}} for enforcing \gls{RBAC} policies in outsourced environments. \gls{ESPOONERBAC} is based on \gls{ESPOON} (discussed in Chapter \ref{cha:espoon}). Basically, \gls{ESPOONERBAC} extends \gls{ESPOON} in order to enforce \gls{RBAC} policies in an encrypted manner, where a curious service provider do not learn private information about sensitive \gls{RBAC} policies. We have implemented \gls{ESPOONERBAC} and provided its performance evaluation showing a limited overhead, thus confirming viability of our approach.

\section{Introduction}

According to \cite{Connor:2010}, \gls{RBAC} is the most widely used security model. \gls{RBAC} \cite{Sandhu:1996} makes decisions based on roles a user is active in. However, it cannot be deployed in outsourced environments because it assumes a trusted infrastructure in order to regulate access on data. In \gls{RBAC} models, \gls{RBAC} policies may leak information about the data they aim to protect. In \cite{Asghar2011-ARES}, we propose \gls{ESPOON} that aims at enforcing authorisation policies in outsourced environments. In \cite{Asghar2011-CCS}, we extend \gls{ESPOON} to support \gls{RBAC} policies and role hierarchies but our solution does not outsource all operations because we assume presence of the Company \gls{RBAC} Manager in trusted environments for the role assignment.

\subsection{Research Contributions}
In this chapter, we present an \gls{RBAC} mechanism for outsourced environments where we support full confidentiality of \gls{RBAC} policies. We named our solution \textbf{\acrfull{ESPOONERBAC}}. \gls{ESPOONERBAC} is based on \gls{ESPOON}. Like \gls{ESPOON}, \gls{ESPOONERBAC} can enforce \gls{RBAC} policies without revealing private information to the service provider that is assumed honest-but-curious. Summarising, the research contributions in this chapter are threefold. 

\begin{enumerate}

	\item The service provider does not learn private information about \gls{RBAC} policies and the requester's attributes during the policy deployment or evaluation processes.
	
	\item We extend the basic \gls{RBAC} policies to support role hierarchies. The curious service provider enforces role hierarchy without revealing information about roles in the role hierarchy graph.
	
	\item The system entities do not share any encryption keys and even if a user is deleted or revoked, the system is still able to perform its operations without requiring re-encryption of \gls{RBAC} policies.

\end{enumerate}

As a proof-of-concept, we have implemented a prototype of our \gls{ESPOONERBAC} mechanism and analysed its performance to quantify the overhead incurred by cryptographic operations used in the proposed scheme.

\subsection{Chapter Outline}
The rest of this chapter is structured as follows. Section \ref{sec:erbac-related-work} reviews the related work. In Section \ref{sec:erbac-approach}, we present the proposed architecture of \gls{ESPOONERBAC}. Section \ref{sec:erbac-solution-details} and Section \ref{sec:erbac-algorithmic-details} focus on solution details and algorithmic details, respectively. Security analysis of \gls{ESPOONERBAC} is provided in Section \ref{sec:erbac-security-analysis}. In Section \ref{sec:erbac-performance-analysis}, we analyse the performance overhead of \gls{ESPOONERBAC}. Finally, Section \ref{sec:erbac-summary} summarises this chapter.

\section{Related Work}
\label{sec:erbac-related-work}

\gls{RBAC} \cite{Sandhu:1996} is an access control model that logically maps well to the job-function specified within an organisation. In the basic \gls{RBAC} model, a system administrator or a security officer assigns permissions to roles and then roles are assigned to users. A user can make an access request to execute permissions corresponding to a role only if he or she is active in that role. A user can be active in a subset of roles assigned to him/her by making a role activation request. In \gls{RBAC}, a session keeps mapping of users to roles that are active. In \cite{Sandhu:1996}, Sandhu \emph{et al.} extend the basic \gls{RBAC} model with role hierarchies for structuring roles within an organisation. The concept of role hierarchy introduces the role inheritance. In the role inheritance, a derived role can inherit all permissions from the base role. The role inheritance incurs extra processing overhead as requested permissions might be assigned to the base role of one in which the user might be active. 

The \gls{RBAC} model may activate a role or grant permissions while taking into account the context under which the user makes the access request or the role activation request \cite{Joshi:2008, Kim:2007, Joshi:2005, Strembeck:2004, Neumann:2003, Bertino:2001, Lupu:1997}. The \gls{RBAC} model captures this context by defining contextual conditions. A contextual condition requires certain attributes about the environment or the user making the request. These attributes are contextual information, which may include access time, access date and location of the user who is making the request. The \gls{RBAC} model grants the request if the contextual information satisfy the contextual conditions. In \cite{Crampton:2008}, Crampton and Khambhammettu discuss delegation in \gls{RBAC}. Unfortunately, existing solutions \cite{Crampton:2008, Joshi:2008, Kim:2007, Joshi:2005, Strembeck:2004, Neumann:2003, Bertino:2001, Lupu:1997, Sandhu:1996} assume a trusted infrastructure to regulate access on data and they cannot be applied to outsourced environments, where a curious service provider might leak sensitive policies.

\gls{MAC} is a strict model of access control that takes a hierarchical approach to control access to resources \cite{Vimercati:2011:MAC}. In \gls{MAC}, access to resources is controlled by the system administrator. \gls{MAC} assigns security labels to resources. \gls{DAC} is a type of access control in which resource owners control access to their resources \cite{Vimercati:2011:DAC}. In \gls{DAC}, each resource object has an \gls{ACL} that contains a list of users or groups who can gain access to the resource object. Like traditional \gls{RBAC}, both \gls{MAC} and \gls{DAC} assume a trusted infrastructure in order to regulate access to the resources.

\gls{XACML} is a standard that defines an access control policy language and a processing model specifying how to evaluate access requests against deployed access control policies \cite{Godik:2002, Yavatkar:2000}. The \gls{XACML} policy language is based on \gls{XML}. For making any access decision, \gls{XACML} considers that access control policies and access requests are in cleartext. Unfortunately, cleartext policies and access requests may reveal private information.

In \cite{Asghar2011-ARES}, we propose \gls{ESPOON} that aims at enforcing authorisation policies in outsourced environments. In \gls{ESPOON}, a data owner (or someone on the behalf of data owners) may attach an authorisation policy with the data while storing it on the outsourced server. Any authorised requester may get access to the data if she satisfies the authorisation policy associated with that data. However, \gls{ESPOON} lacks to provide support for \gls{RBAC} policies. In \cite{Asghar2011-CCS}, we extended \gls{ESPOON} to support \gls{RBAC} policies and role hierarchies. However, in \cite{Asghar2011-CCS} the role assignment is performed by the Company \gls{RBAC} Manager, which is run in the trusted environment. On the other hand, in our current architecture, the role assignment is performed by the service provider running in the outsourced environment. In other words, we have eliminated the need of an additional online-trusted-server i.e., the Company \gls{RBAC} Manager.

\begin{figure} [htp]
\centering
% left bottom right top
\includegraphics[trim=35mm 45mm 30mm 45mm,clip,width=\textwidth]{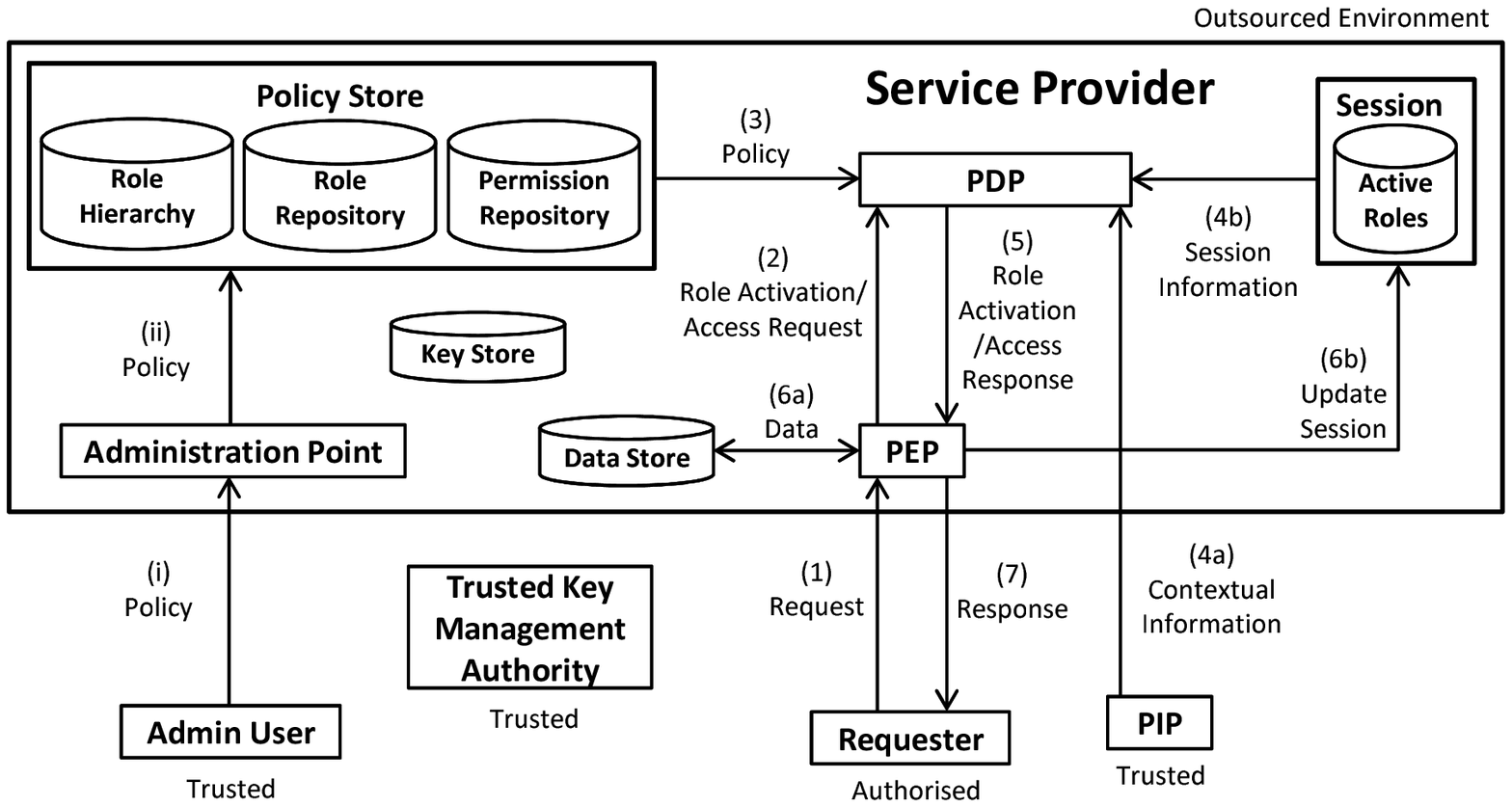}
\caption[The proposed architecture for enforcing RBAC policies]{The \gls{ESPOONERBAC} architecture for enforcing RBAC policies in outsourced environments}
\label{fig:erbac-abstract_picture}
\end{figure}

\section[The ESPOON$_{\mathit{ERBAC}}$ Approach]{The \acrshort{ESPOONERBAC} Approach}
\label{sec:erbac-approach}

\gls{ESPOONERBAC} aims at providing \gls{RBAC} mechanism that can be deployed in an outsourced environment. Figure \ref{fig:erbac-abstract_picture} illustrates the proposed architecture that has similar components to the widely accepted architecture for the policy-based management proposed by \gls{IETF} \cite{Yavatkar:2000}. In \gls{ESPOONERBAC}, an \textbf{Admin User} deploys (i) \gls{RBAC} policies and sends them to the \textbf{Administration Point} that stores (ii) \gls{RBAC} policies\footnote{In the rest of this chapter, by term \emph{policies} we mean \emph{\gls{RBAC} policies}.} in the \textbf{Policy Store}. These policies may include permissions assigned to roles, roles assigned to users and the role hierarchy graph that are stored in the Permission Repository, the Role Repository and the Role Hierarchy repository, respectively. 

% role activation
A \textbf{Requester} may send (1) the role activation request to the \gls{PEP}. This request includes the Requester's identifier and the requested role. The \gls{PEP} forwards (2) the role activation request to the \gls{PDP}. The \gls{PDP} retrieves (3) the policy corresponding to the Requester from the Role Repository of the Policy Store and fetches (4) the contextual information from the \gls{PIP}. The contextual information may include the environmental and Requester's attributes under which the requested role can be activated. For instance, consider a contextual condition where a role doctor can only be activated during the duty hours. For simplicity, we assume that the \gls{PIP} collects all required attributes and sends all of them together in one go. Moreover, we assume that the \gls{PIP} is deployed in the trusted environment. However, if attributes forgery is an issue, the \gls{PIP} can request a trusted authority to sign the attributes before sending them to the \gls{PDP}. The \gls{PDP} evaluates role assignment policies against the attributes provided by the \gls{PIP} checking if the contextual information satisfies contextual conditions and sends to the \gls{PEP} (5) the role activation response. In case of \emph{permit}, the \gls{PEP} activates the requested role by updating the \textbf{Session} containing the Active Roles repository (6a). Otherwise, in case of \emph{deny}, the requested role is not activated. Optionally, a response can be sent to the Requester (7) with either \emph{success} or \emph{failure}.

% access request
After getting active in a role, a Requester can make the access request that is sent to the \gls{PEP} (1). This request includes the Requester's identifier, the requested data (target) and the action to be performed. The \gls{PEP} forwards (2) the access request to the \gls{PDP}. After receiving the access request, the \gls{PDP} first retrieves from the Session information about the Requester if she is already active in any role (3a). If so, the \gls{PDP} evaluates if the Requester's (active) role is permitted to execute the requested action on the requested data. For this purpose, the \gls{PDP} retrieves (3) the permission assignment policy corresponding to the active role from the Permission Repository of the Policy Store and fetches (4) the contextual information from the \gls{PIP} required for evaluating contextual conditions in the permission assignment policy. For instance, consider the example where a \emph{Cardiologist} can access the cardiology report during office hours. The \gls{PDP} evaluates the permission assignment policies against the attributes provided by the \gls{PIP} checking if the contextual information satisfies any contextual conditions and sends to the \gls{PEP} (5) the access response. In case of \emph{permit}, the \gls{PEP} forwards the access action to the \textbf{Data Store} (6b). In case if no contextual condition is satisfied, the \gls{PDP} retrieves the role hierarchy from the Role Hierarchy repository of the Policy Store and then traverses this role hierarchy graph in order to find if any base role, the Requester's role might be derived from, has permission to execute the requested action on the requested data. If so, the \gls{PEP} forwards the access action to the Data Store (6b). Otherwise, in case of \emph{deny}, the requested action is not forwarded. Optionally, a response can be sent to the Requester (7) with either \emph{success} or \emph{failure}.

Since \gls{ESPOONERBAC} is based on \gls{ESPOON}, we use the same system model as already considered in \gls{ESPOON} (see Section \ref{sec:esooon-system-model}).

\begin{figure} [htp]
\Keywords
% \begin{lstlisting}[style=AMMA,breaklines,mathescape,rulesepcolor=\color{black}]
\begin{lstlisting}[style=AMMA,numbers=none,breaklines,mathescape,rulesepcolor=\color{black}]
if $\langle \mathit{CONDITION} \rangle$ then $\langle USER \rangle$ can be active in $\langle \{ R_1, R_2, \ldots, R_n \} \rangle$

\end{lstlisting}
\caption[RBAC Policy: Role assignment]{\gls{RBAC} Policy: Role assignment}
\label{fig:erbac-policy-role-assignment}
\end{figure}

\begin{figure} [htp]
\Keywords
\begin{lstlisting}[style=AMMA,numbers=none,breaklines,mathescape,rulesepcolor=\color{black}]
if $\langle \mathit{CONDITION} \rangle$ then $\langle R \rangle$ can execute $\langle \{ (A_1, T_1), (A_2, T_2), \ldots, (A_n, T_n) \} \rangle$

\end{lstlisting}
\caption[RBAC Policy: Permission assignment]{\gls{RBAC} Policy: Permission assignment}
\label{fig:erbac-policy-permission-assignment}
\end{figure}

\subsection[Representation of RBAC Policies and Requests]{Representation of \gls{RBAC} Policies and Requests}
\label{sec:representation}
In this section, we provide details about how to represent policies and requests used in our approach. An \gls{RBAC} policy contains a role assignment policy, a permission policy and a role hierarchy graph. In the following, we discuss each of them. Figure \ref{fig:erbac-policy-role-assignment} illustrates how we represent role assignment policies in \gls{ESPOONERBAC}. The meaning of role assignment policy is as follows: if contextual condition, $\mathit{CONDITION}$, is $\mathit{true}$ then $USER$ can be active in any role(s) out of role set $\{ R_1, R_2, \ldots, R_n \}$. Figure \ref{fig:erbac-policy-permission-assignment} illustrates how we represent permission assignment policies in \gls{ESPOONERBAC}. The meaning of permission assignment policy is as follows: if contextual condition, $\mathit{CONDITION}$, is $\mathit{true}$ then role $R$ can execute any permission(s) out of permission set $\{ (A_1, T_1), (A_2, T_2), \ldots, (A_n, T_n) \}$.

% CONDITION
% TODO: define CP-ABE if not done so
The \gls{PDP} evaluates contextual conditions of both role assignment and permission assignment policies before granting the access. In order to evaluate a contextual condition, the \gls{PDP} requires contextual information. The contextual information captures the context in which a Requester makes access or role activation requests.

% requests
A Requester can make a role activation request $\mathit{ACT}$ or an access request $\mathit{REQ}$. In $\mathit{ACT} = (i, R)$, a Requester includes her identity $i$ along with role $R$ to be activated. After a Requester is active in $R$, she can execute permissions assigned to $R$. For executing any permission, a Requester sends $\mathit{REQ} = (R, A, T)$ that includes $R$ she is active in, action $A$ to be taken over target $T$. A Requester sends $\mathit{ACT}$ or $\mathit{REQ}$ requests to the \gls{PEP}. 

% \gls{PIP}
The \gls{PEP} receives and forwards requests $\mathit{ACT}$ or $\mathit{REQ}$ to the \gls{PDP}. The \gls{PDP} fetches policies corresponding to requests from the Policy Store. The \gls{PDP} may require contextual information in order to evaluate contextual conditions to grant $\mathit{ACT}$ or $\mathit{REQ}$ (as already explained in Section \ref{sec:espoon-policy-representation}).

\begin{figure} [htp]
\Keywords
\begin{lstlisting}[style=AMMA,numbers=none,breaklines,mathescape,rulesepcolor=\color{black}]
$R_1$ extends $\langle \{ R_{i}, R_{ii}, \ldots, R_{k_1} \} \rangle$
$R_2$ extends $\langle \{ R_{i}, R_{ii}, \ldots, R_{k_2} \} \rangle$
$\vdots$
$R_n$ extends $\langle \{ R_{i}, R_{ii}, \ldots, R_{k_n} \} \rangle$

\end{lstlisting}
\caption[RBAC Policy: Role hierarchy]{\gls{RBAC} Policy: Role hierarchy}
\label{fig:erbac-policy-role-hierarchy}
\end{figure}

\begin{figure} [htp]
\centering
% left bottom right top
\includegraphics[trim=60mm 40mm 70mm 40mm,clip,width=.5\textwidth]{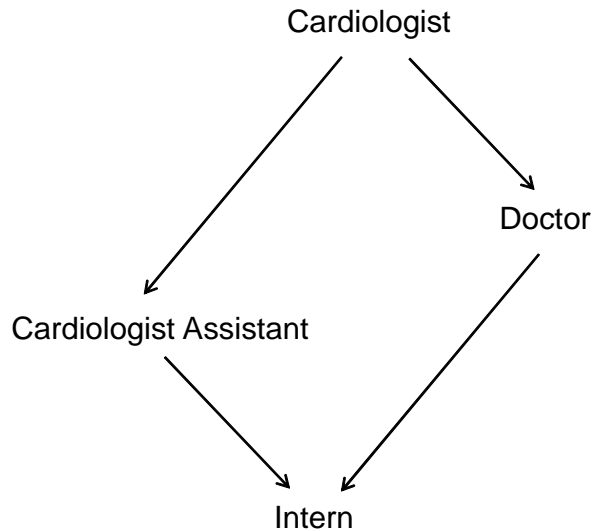}
\caption[An example of a role hierarchy graph]{An example of a role hierarchy graph illustrating that Cardiologist and Doctor roles are derived from Intern while the Cardiologist role is derived from Cardiologist and Doctor roles}
\label{fig:erbac-role_hierarchy_graph}
\end{figure}

% role hierarchy graph
The \gls{ESPOONERBAC} architecture supports role inheritance. In role inheritance, a derived role can execute all permissions from its base role. Before denying $\mathit{REQ}$, the \gls{PDP} may need to check if base role of one in $\mathit{REQ}$ can execute requested permissions. In order to find base roles, we store a role hierarchy graph on the Service Provider. In \gls{ESPOONERBAC}, the \gls{PDP} traverses in the role hierarchy graph to find base roles. Figure \ref{fig:erbac-policy-role-hierarchy} illustrates how we represent a role hierarchy graph. In Figure \ref{fig:erbac-policy-role-hierarchy}, each line represents a role that may extend a set of roles. All these inheritance rules may form a role hierarchy graph. For instance, consider an example from healthcare domain where a \emph{Cardiologist Assistant} extends \emph{Intern}, a \emph{Doctor} extends \emph{Intern} and finally a \emph{Cardiologist} extends both \emph{Cardiologist Assistant} and \emph{Doctor}. If we combine all these inheritance rules then it can form a graph as shown in Figure \ref{fig:erbac-role_hierarchy_graph}.

% why we need encryption of policies/requests
In this representation, leaf-nodes in $\mathit{CONDITION}$, $R$, $A$, $T$ of both $\mathit{ACT}$ and $\mathit{REQ}$, roles in the role hierarchy graph, and attributes in contextual information are in cleartext. Therefore, such information is easily accessible in the outsourced environment and may leak information about the data that policies protect. In the following, we show how we protect such representation while allowing the \gls{PDP} to evaluate policies against requests and contextual information.

\section[Solution Details of ESPOON$_{\mathit{ERBAC}}$]{Solution Details of \gls{ESPOONERBAC}}
\label{sec:erbac-solution-details}

\gls{ESPOONERBAC} aims at enforcing policies in outsourced environments. The main idea of our approach is to use an encryption scheme for preserving confidentiality of policies while allowing the \gls{PDP} to perform the correct evaluation. In \gls{ESPOONERBAC}, we can notice that the operation performed by the \gls{PDP} for evaluating policies (against attributes in the request and contextual information) is similar to the search operation executed in a database. In particular, in our case the policy is a query; while, attributes in the request ($\mathit{ACT}$ or $\mathit{REQ}$) and contextual information represent the data. For \gls{ESPOONERBAC}, we extend \gls{ESPOON}. In the following, we describe core phases in \gls{ESPOONERBAC}.

\subsection{The Policy Deployment Phase}

For deploying (or updating existing) policies, an Admin User performs a first round of encryption using her client side key set. An Admin User encrypts elements of policies. In role assignment policies, an Admin User encrypts all roles assigned to a user. In permission assignment policies, an Admin User encrypts both action and target parts of each permission and also encrypts the role to which these permissions are assigned. As we know that a tree represents condition conditions of both role assignment and permission assignment policies (as shown in Figure \ref{fig:erbac-cc}), an Admin User encrypts each leaf node of the tree while non-leaf (internal) nodes representing AND, OR or threshold gates are in cleartext. In a role hierarchy graph (as shown in Figure \ref{fig:erbac-role_hierarchy_graph}), an Admin User encrypts each of its node representing a role. After completing the first round of encryption on policies, an Admin User sends client encrypted policies to the Administration Point on the Service Provider. These client encrypted policies are protected but cannot be enforced as these are not in common format. To convert client encrypted policies to common format, the Administration Point performs a second round of encryption using server side key set corresponding to the Admin User. The second round of encryption serves as a proxy re-encryption. In the second round of encryption, the Administration Point encrypts all elements that are encrypted in the first round of encryption. Finally, the Administration Point stores server encrypted policies in the Policy Store.

\subsection{The Policy Evaluation Phase}
A Requester can make a role activation request $\mathit{ACT}$. Before sending $\mathit{ACT}$ to the Service Provider, a Requester generates a client trapdoor of the role in $\mathit{ACT}$. A Requester generates client trapdoor using her client side key set. The trapdoor representation does not leak information on elements of requests. Similarly, a Requester can make an access request $\mathit{REQ}$ after getting active in a role. A Requester generates a client trapdoor for each element in $\mathit{REQ}$ including the role, the action and the target. A Requester sends requests containing client generated trapdoors to the \gls{PEP} on the Service Provider. The \gls{PEP} performs another round of trapdoor generation for converting all trapdoors into a common format. After performing a second round of trapdoor generation on the server side, the \gls{PEP} forwards server generated trapdoors to the \gls{PDP}. The \gls{PDP} fetches policies from the Policy Store and then performs encrypted matching of trapdoors in request against encrypted elements in policies. The encrypted matching in outsourced environments does not leak information about elements of requests or policies. 

The \gls{PDP} may require contextual information in order to evaluate the contextual conditions of policies. The \gls{PIP} collects contextual information and generates client trapdoors for elements of contextual information using her client side key set. The \gls{PIP} sends client generated trapdoors of contextual information to the \gls{PDP}. The \gls{PDP} performs another round of trapdoor generation using server side key set corresponding to the \gls{PIP}. Finally, the \gls{PDP} evaluates the contextual condition by matching trapdoors of contextual information against encrypted leaf nodes of the tree representing the contextual condition (as shown in Figure \ref{fig:erbac-cc}). After evaluating leaf nodes, the \gls{PDP} evaluates non-leaf nodes of the tree based on AND, OR and threshold gates. The \gls{PDP} grants the access request if (the root node of) the tree evaluates to $\mathit{true}$.

The \gls{PDP} may need to find base roles corresponding to the role in $\mathit{REQ}$ considering the fact that a derived role has all permissions from its base role. In order to find base roles, the \gls{PDP} fetches the role hierarchy graph from the Policy Store. The \gls{PDP} matches trapdoor of role in $\mathit{REQ}$ against server encrypted roles in the role hierarchy graph. While deploying the role hierarchy graph, we store also server generated trapdoor of the role along with each server encrypted of role because the \gls{PDP} needs a trapdoor of each base role so that it can match this trapdoor against encrypted roles in the Permission Repository. After traversing in the role hierarchy graph, the \gls{PDP} extracts server generated trapdoors of all base roles of one that matches with trapdoor of role in $\mathit{REQ}$. The \gls{PDP} verifies if any base role has requested permissions. If so, the \gls{PDP} grants the request.

% TODO: switch to discussion section?
% deactivate role by sending or a server may deactivate after a certain time.

\section[Algorithmic Details of ESPOON$_{\mathit{ERBAC}}$]{Algorithmic Details of \gls{ESPOONERBAC}}
\label{sec:erbac-algorithmic-details}
In this section, we provide details of algorithms used in core phases (including the policy deployment phase and the policy evaluation phase) for managing lifecycle of policies. The following algorithms (along with \gls{ESPOON} algorithms described in Chapter \ref{cha:espoon}, Section \ref{sec:erbac-algorithmic-details}) constitute the proposed schema. 

\subsection{The Policy Deployment Phase}
In the following, we describe how to deploy different (parts of) policies including role assignment, permission assignment, contextual conditions and role hierarchy graph. For the deployment of each (part of) policy, we follow general strategy as already described in Section \ref{sec:espoon-policy-deployment-phase} and also illustrated in Figure \ref{fig:erbac-patient}.

% deploy: role assignment: client side

\begin{algorithm} [htp]
{\algofontsize
\caption{\textbf{RoleAssignment:ClientEnc}}

\label{algo:erbac-deploy-role-assignment-client-side}

\begin{algorithmic}[1]

\INPUT \emph{It transforms the cleartext role assignment list into the client encrypted role assignment list.}

\Require List of roles $L$ to be assigned to Requester $j$, the client side key set $K_{u_i}$ corresponding to Admin User $i$ and the public parameters $params$.

\Ensure The client encrypted role assignment list $L_{C_i}$.

\medskip

\State $L_{C_i} \leftarrow \phi$ \label{line:erbac-deploy-ra-cs-init}

\For {each role $r$ in list $L$} \label{line:erbac-deploy-ra-cs-loop}

	\State $c^*_i (r) \leftarrow$ call \textbf{ClientEnc} ($r$, $K_{u_i}$, $params$) {\algofontsize \Comment{see Algorithm \ref{algo:erbac-client-enc}}} \label{line:erbac-deploy-ra-cs-call-enc}
	
	\State $L_{C_i} \leftarrow L_{C_i} \cup c^*_i (r)$ \label{line:erbac-deploy-ra-cs-update}

\EndFor

\Return ($j$, $L_{C_i}$)

\end{algorithmic}
}
\end{algorithm}

% deploy: role assignment: server side

\begin{algorithm} [htp]
{\algofontsize
\caption{\textbf{RoleAssignment:ServerReEnc}}

\label{algo:erbac-deploy-role-assignment-server-side}

\begin{algorithmic}[1]

\INPUT \emph{It re-encrypts the client encrypted role assignment list and generates the server encrypted role assignment list.}

\Require The client encrypted role assignment list $L_{C_i}$ for Requester $j$ and identity $i$ of Admin User.

\Ensure The server encrypted role assignment list $L_{S}$.

\medskip

\State $K_{s_i} \leftarrow KS[i]$ {\algofontsize \Comment{retrieve the server side key corresponding to Admin User $i$}} \label{line:erbac-deploy-ra-ss-ks}

\State $L_{S} \leftarrow \phi$ \label{line:erbac-deploy-ra-ss-init}

\For {each client encrypted role $c^*_i (r)$ in list $L_{C_i}$} \label{line:erbac-deploy-ra-ss-loop}

	\State $c(r) \leftarrow$ call \textbf{ServerReEnc} ($c^*_i (r)$, $K_{s_i}$) {\algofontsize \Comment{see Algorithm \ref{algo:erbac-server-re-enc}}} \label{line:erbac-deploy-ra-ss-call}
	
	\State $L_{S} \leftarrow L_{S} \cup c(r)$ \label{line:erbac-deploy-ra-ss-update}

\EndFor

\Return ($j$, $L_{S}$)

\end{algorithmic}
}
\end{algorithm}

% TODO: on update remove \noindent \\ that is added to fix a format issue
\noindent \\
\noindent \emph{\textbf{Deployment of Role Assignment Policies:}} 
In order to assign roles to a Requester, an Admin User can deploy role assignment policies. For this purpose, an Admin User runs \textbf{RoleAssignment:ClientEnc} illustrated in Algorithm \ref{algo:erbac-deploy-role-assignment-client-side}. This algorithm takes as input a list of roles $L$ to be assigned to Requester $j$, the client side key set $K_{u_i}$ corresponding to Admin User $i$ and the public parameters $params$ and outputs the client encrypted role assignment list $L_{C_i}$. First, it creates and then initialises a list $L_{C_i}$ (Line \ref{line:erbac-deploy-ra-cs-init}). For each role in $L$ (Line \ref{line:erbac-deploy-ra-cs-loop}), it generates client encrypted role by calling \textbf{ClientEnc} illustrated in Algorithm \ref{algo:erbac-client-enc} (Line \ref{line:erbac-deploy-ra-cs-call-enc}) and then it updates $L_{C_i}$ by adding client encrypted role (Line \ref{line:erbac-deploy-ra-cs-update}). An Admin User sends the client encrypted role assignment list to the Administration Point. 
During the second round of encryption, the Administration Point runs \textbf{RoleAssignment:ServerReEnc} illustrated in Algorithm \ref{algo:erbac-deploy-role-assignment-server-side}. This algorithm takes as input the client encrypted role assignment list $L_{C_i}$ for Requester $j$ and identity $i$ of Admin User and ouputs the server encrypted role assignment list $L_{S}$. While running \textbf{RoleAssignment:ServerReEnc}, the Administration Point first retrieves the server side key $K_{s_i}$ corresponding to Admin User $i$ (Line \ref{line:erbac-deploy-ra-ss-ks}). It creates and initialises a list $L_{S}$ (Line \ref{line:erbac-deploy-ra-ss-init}). For each role in $L_{C_i}$ (Line \ref{line:erbac-deploy-ra-ss-loop}), it generates server encrypted role by calling \textbf{ServerReEnc} illustrated in Algorithm \ref{algo:erbac-server-re-enc} (Line \ref{line:erbac-deploy-ra-ss-call}) and updates $L_{S}$ by adding the server encrypted role (Line \ref{line:erbac-deploy-ra-ss-update}).

% deploy: permission assignment: client side

\begin{algorithm} [htp]
{\algofontsize
\caption{\textbf{PermissionAssignment:ClientEnc}}

\label{algo:erbac-deploy-permission-assignment-client-side}

\begin{algorithmic}[1]

\INPUT \emph{It transforms the cleartext permission assignment list into the client encrypted permission assignment list.}

\Require List of permissions $L$ to be assigned to role $r$, the client side key set $K_{u_i}$ corresponding to Admin User $i$ and the public parameters $params$.

\Ensure The client encrypted permission assignment list $L_{C_i}$ assigned to the client generated role $c^*_i (r)$.

\medskip

\State $c^*_i (r) \leftarrow$ call \textbf{ClientEnc} ($r$, $K_{u_i}$, $params$) \label{line:erbac-deploy-pa-cs-role}

\State $L_{C_i} \leftarrow \phi$ \label{line:erbac-deploy-pa-cs-init}

\For {each permission $(action, target)$ in $L$} \label{line:erbac-deploy-pa-cs-loop}

	\State $c^*_i (action) \leftarrow$ call \textbf{ClientEnc} ($action$, $K_{u_i}$, $params$) \label{line:erbac-deploy-pa-cs-action}
	
	\State $c^*_i (target) \leftarrow$ call \textbf{ClientEnc} ($target$, $K_{u_i}$, $params$) \label{line:erbac-deploy-pa-cs-target}
	
	\State $L_{C_i} \leftarrow L_{C_i} \cup (c^*_i (action), c^*_i (target))$ \label{line:erbac-deploy-pa-cs-update}

\EndFor

\Return ($c^*_i (r)$, $L_{C_i}$)

\end{algorithmic}
}
\end{algorithm}

% deploy: permission assignment: server side

\begin{algorithm} [htp]
{\algofontsize
\caption{\textbf{PermissionAssignment:ServerReEnc}}

\label{algo:erbac-deploy-permission-assignment-server-side}

\begin{algorithmic}[1]

\INPUT \emph{It re-encrypts the client encrypted permission assignment list.}

\Require The client encrypted permission assignment list $L_{C_i}$ for client generated role $c^*_i (r)$ and identity $i$ of Admin User.

\Ensure The server encrypted permission assignment list $L_{S}$ and the server generated role $c(r)$.

\medskip

\State $K_{s_i} \leftarrow KS[i]$ {\algofontsize \Comment{retrieve the server side key corresponding to Admin User $i$}} \label{line:erbac-deploy-pa-ss-ks}

\State $c(r) \leftarrow$ call \textbf{ServerReEnc} ($c^*_i (r)$, $K_{s_i}$) \label{line:erbac-deploy-pa-ss-role}

\State $L_{S} \leftarrow \phi$ \label{line:erbac-deploy-pa-ss-init}

\For {each client encrypted permission $(c^*_i (action), c^*_i (target))$ in list $L_{C_i}$} \label{line:erbac-deploy-pa-ss-loop}

	\State $c(action) \leftarrow$ call \textbf{ServerReEnc} ($c^*_i (action)$, $K_{s_i}$) \label{line:erbac-deploy-pa-ss-action}
	
	\State $c(target) \leftarrow$ call \textbf{ServerReEnc} ($c^*_i (target)$, $K_{s_i}$) \label{line:erbac-deploy-pa-ss-target}
	
	\State $L_{S} \leftarrow L_{S} \cup (c(action), c(target))$ \label{line:erbac-deploy-pa-ss-update}

\EndFor

\Return ($c(r)$, $L_{S}$)

\end{algorithmic}
}
\end{algorithm}

\noindent \\
\noindent \emph{\textbf{Deployment of Permission Assignment Policies:}} 
An Admin User can assign permissions to a role. In order to deploy policies regarding permissions assignment to roles, an Admin User runs Algorithm \ref{algo:erbac-deploy-permission-assignment-client-side}. This algorithm takes as input a list of permissions $L$ to be assigned to role $r$, the client side key set $K_{u_i}$ corresponding to Admin User $i$ and the public parameters $params$ and outputs the client encrypted permission assignment list $L_{C_i}$ assigned to client generated role $c^*_i (r)$. First, it generates client encrypted role $c^*_i (r)$ by calling \textbf{ClientEnc} illustrated in Algorithm \ref{algo:erbac-client-enc} (Line \ref{line:erbac-deploy-pa-cs-role}). Next, it creates and initialises new list $L_{C_i}$ (Line \ref{line:erbac-deploy-pa-cs-init}). For each permission in $L$ (Line \ref{line:erbac-deploy-pa-cs-loop}), it generates the client encrypted action $c^*_i (action)$ (Line \ref{line:erbac-deploy-pa-cs-action}) and the client encrypted target $c^*_i (target)$ (Line \ref{line:erbac-deploy-pa-cs-target}) and updates $L_{C_i}$ by adding the client encrypted permission (Line \ref{line:erbac-deploy-pa-cs-update}). An Admin User sends the client encrypted permission list along with the client encrypted role to the Administration Point. 
The Administration Point runs another round of encryption by running Algorithm \ref{algo:erbac-deploy-permission-assignment-server-side}. This algorithm takes as input the client encrypted permission assignment list $L_{C_i}$ for client generated role $c^*_i (r)$ and identity $i$ of Admin User and outputs the server encrypted permission assignment list $L_{S}$ and the server generated role $c(r)$. First, it retrieves from the Key Store the server side key set $K_{s_i}$ corresponding to Admin User $i$ (Line \ref{line:erbac-deploy-pa-ss-ks}). Next, it generates the server encrypted role by calling \textbf{ServerReEnc} illustrated in Algorithm \ref{algo:erbac-server-re-enc} (Line \ref{line:erbac-deploy-pa-ss-role}). Then, it creates and initialises new list $L_{S}$ (Line \ref{line:erbac-deploy-pa-ss-init}). For each client encrypted role in $L_{C_i}$ (Line \ref{line:erbac-deploy-pa-ss-loop}), it generates the server encrypted action (Line \ref{line:erbac-deploy-pa-ss-action}) and the server encrypted target (Line \ref{line:erbac-deploy-pa-ss-target}) and updates $L_{S}$ by adding the server encryption permission (Line \ref{line:erbac-deploy-pa-ss-update}).

% deploy role hierarchy: client side

\begin{algorithm} [htp]
{\algofontsize
\caption{\textbf{RoleHierarchy:ClientEnc}}

\label{algo:erbac-deploy-role-hierarchy-client-side}

\begin{algorithmic}[1]

\INPUT \emph{It encrypts the role hierarchy graph.}

\Require The role hierarchy graph $G$, the client side key set $K_{u_i}$ corresponding to Admin User $i$ and the public parameters $params$.

\Ensure The client generated role hierarchy graph $G_{C_i}$.

\medskip

\State $G_{C_i} \leftarrow G$ \label{line:erbac-deploy-rh-cs-copy}

\For {each node $r$ in $G_{C_i}$} \label{line:erbac-deploy-rh-cs-loop}

	\State $c^*_i (r) \leftarrow$ call \textbf{ClientEnc} ($r$, $K_{u_i}$, $params$) \label{line:erbac-deploy-rh-cs-enc}
	\State $td^*_i (r) \leftarrow$ call \textbf{ClientTD} ($r$, $K_{u_i}$, $params$) {\algofontsize \Comment{see Algorithm \ref{algo:erbac-client-td}}} \label{line:erbac-deploy-rh-cs-td}
	\State replace $r$ of $G_{C_i}$ with $(c^*_i (r), td^*_i (r))$ \label{line:erbac-deploy-rh-cs-replace}

\EndFor

\Return $G_{C_i}$

\end{algorithmic}
}
\end{algorithm}

% deploy role hierarchy: server side

\begin{algorithm} [htp]
{\algofontsize
\caption{\textbf{RoleHierarchy:ServerReEnc}}

\label{algo:erbac-deploy-role-hierarchy-server-side}

\begin{algorithmic}[1]

\INPUT \emph{It re-encrypts the client generated role hierarchy graph.}

\Require The client generated role hierarchy graph $G_{C_i}$ and identity of Admin User $i$.

\Ensure The server generated role hierarchy graph $G_{S}$.

\medskip

\State $K_{s_i} \leftarrow KS[i]$ {\algofontsize \Comment{retrieve the server side key corresponding to Admin User $i$}} \label{line:erbac-deploy-rh-ss-ks}

\State $G_{S} \leftarrow G_{C_i}$ \label{line:erbac-deploy-rh-ss-copy}

\For {each client generated node $(c^*_i (r), td^*_i (r))$ in $G_{S}$} \label{line:erbac-deploy-rh-ss-loop}

	\State $c(r) \leftarrow$ call \textbf{ServerReEnc} ($c^*_i (r)$, $K_{s_i}$) \label{line:erbac-deploy-rh-ss-enc}
	
	\State $td(r) \leftarrow$ call \textbf{ServerTD} ($td^*_i (r)$, $K_{s_i}$) {\algofontsize \Comment{see Algorithm \ref{algo:erbac-server-td}}} \label{line:erbac-deploy-rh-ss-td}
	
	\State replace $(c^*_i (r), td^*_i (r))$ of $G_{S}$ with $(c(r), td(r))$ \label{line:erbac-deploy-rh-ss-replace}

\EndFor

\Return $G_{S}$

\end{algorithmic}
}
\end{algorithm}

\noindent \emph{\textbf{Deployment of a Role Hierarchy Graph:}} 
We know that a derived role inherits all permissions from its base role. In case if requested permissions are not assigned to the Requester's role, the \gls{PDP} may need to traverse in the role hierarchy graph to find base roles corresponding to the Requester's role and then \gls{PDP} verifies if any base role can fulfil requested permissions. For this purpose, the \gls{PDP} needs a trapdoor of each base role so that it can match this trapdoor against encrypted roles in the Permission Repository. Therefore, a role hierarchy graph stores a role trapdoor along with each encrypted role. The deployment of role hierarchy graph takes place in two steps. In the first step, an Admin User runs Algorithm \ref{algo:erbac-deploy-role-hierarchy-client-side}. This algorithm takes as input the role hierarchy graph $G$, the client side key set $K_{u_i}$ corresponding to Admin User $i$ and the public parameters $params$ and outputs the client generated role hierarchy graph $G_{C_i}$. First, it copies $G$ to $G_{C_i}$  (Line \ref{line:erbac-deploy-rh-cs-copy}). For each node $r$ in $G_{C_i}$ (Line \ref{line:erbac-deploy-rh-cs-loop}), it generates the client encrypted role by calling \textbf{ClientEnc} illustrated in Algorithm \ref{algo:erbac-client-enc} (Line \ref{line:erbac-deploy-rh-cs-enc}) and the client trapdoor by calling \textbf{ClientTD} (Line \ref{line:erbac-deploy-rh-cs-td}) illustrated in Algorithm \ref{algo:erbac-client-td} that is explained later in this section. Next, it replaces $r$ of $G_{C_i}$ with the client encrypted role and the client generated trapdoor (Line \ref{line:erbac-deploy-rh-cs-replace}). An Admin User sends the client generated role hierarchy graph to the Administration Point. 
In the second step, the Administration Point runs Algorithm \ref{algo:erbac-deploy-role-hierarchy-server-side}. This algorithm takes as input the client generated role hierarchy graph $G_{C_i}$ and identity of Admin User $i$ and outputs the server generated role hierarchy graph $G_{S}$. First, it retrieves from the Key Store the server side key $K_{s_i}$ corresponding to Admin User $i$ (Line \ref{line:erbac-deploy-rh-ss-ks}). Next, it copies $G_{C_i}$ to $G_{S}$ (Line \ref{line:erbac-deploy-rh-ss-copy}). For each client generated node (Line \ref{line:erbac-deploy-rh-ss-loop}), it generates the server encrypted role by calling \textbf{ServerReEnc} illustrated in Algorithm \ref{algo:erbac-server-re-enc} (Line \ref{line:erbac-deploy-rh-ss-enc}) and the server trapdoor by calling \textbf{ServerTD} (Line \ref{line:erbac-deploy-rh-ss-td}) illustrated in Algorithm \ref{algo:erbac-server-td} that is explained later in this section and then updates $G_{S}$ by replacing the client generated node with the server generated node (Line \ref{line:erbac-deploy-rh-ss-replace}).

\subsection{The Policy Evaluation Phase}

The policy evaluation phase is executed when a Requester makes a request either $\mathit{ACT}$ or $\mathit{REQ}$. In the following, we describe how to evaluate (parts of) policies including role assignment, permission assignment, contextual conditions and role hierarchy graph. For the evaluation of each (part of) policy, we follow general strategy as already described in this section and also illustrated in Figure \ref{fig:erbac-doctor}.

% search role in repository or session

\begin{algorithm} [htp]
{\algofontsize
\caption{\textbf{SearchRole}}

\label{algo:erbac-search-role}

\begin{algorithmic}[1]

\INPUT \emph{It checks whether the requested role is in the role assignment list of the Requester.}

\Require The client generated trapdoor of role $td^*_i (r)$ and the server encrypted role assignment list (or list of active roles in session) $L_{S}$ for Requester $i$.

\Ensure $\mathit{true}$ or $\mathit{false}$.

\medskip

\State $K_{s_i} \leftarrow KS[i]$ {\algofontsize \Comment{retrieve the server side key corresponding to Requester $i$}} \label{line:erbac-search-role-ks}

\State $td(r) \leftarrow$ call \textbf{ServerTD} ($td^*_i (r)$, $K_{s_i}$) \label{line:erbac-search-role-td}

\For {each server encrypted role $c(r)$ in $L_{S}$} \label{line:erbac-search-role-loop}

	\State $match \leftarrow$ call \textbf{Match} ($c(r)$, $td(r)$) {\algofontsize \Comment{see Algorithm \ref{algo:erbac-match}}} \label{line:erbac-search-role-call}

	\If {$match \stackrel{?}{=} true$} \label{line:erbac-search-role-match}
	
		\Return $\mathit{true}$ \label{line:erbac-search-role-true}
		
	\EndIf

\EndFor

\Return $\mathit{false}$ \label{line:erbac-search-role-false}

\end{algorithmic}
}
\end{algorithm}

\noindent \\
\noindent \emph{\textbf{Searching a Role:}} 
A Requester can make a role activation request $\mathit{ACT}$ and sends it to the Service Provider. In order to grant $\mathit{ACT}$, the Service Provider runs \textbf{SearchRole} illustrated in Algorithm \ref{algo:erbac-search-role}. This algorithm takes as input the client generated trapdoor of role $td^*_i (r)$ and the server encrypted role assignment list $L_{S}$ for Requester $i$. First, it retrieves from the Key Store the server side key $K_{s_i}$ corresponding to Requester $i$ (Line \ref{line:erbac-search-role-ks}). Next, it calculates the server generated trapdoor $td(r)$ by calling Algorithm \ref{algo:erbac-server-td} (Line \ref{line:erbac-search-role-td}). For each server encrypted role $c(r)$ in $L_{S}$ (Line \ref{line:erbac-search-role-loop}), it performs matching against $td(r)$ by calling Algorithm \ref{algo:erbac-match} (Line \ref{line:erbac-search-role-call}). If any match is successful (Line \ref{line:erbac-search-role-match}), it returns $\mathit{true}$ (Line \ref{line:erbac-search-role-true}), meaning that $\mathit{ACT}$ is granted. Otherwise, it returns $\mathit{false}$ (Line \ref{line:erbac-search-role-false}).

After $\mathit{ACT}$ is granted, the \gls{PEP} updates Session by adding in the Active Roles repository the server generated trapdoor of role. Once a Requester is active in a role, she can make an access request $\mathit{REQ}$. Before granting $\mathit{REQ}$, the Service Provider checks if the Requester is already in the role in $\mathit{REQ}$. For this purpose, the Service Provider runs Algorithm \ref{algo:erbac-search-role}, where $L_{S}$ shows a list of active roles in the session. Furthermore, the \gls{PDP} also runs Algorithm \ref{algo:erbac-search-role} for searching the role in $\mathit{REQ}$ in the Permission Repository with a slight modification of ignoring the server trapdoor generation (in Line \ref{line:erbac-search-role-td}) as it is already generated when the role of $\mathit{REQ}$ is searched in the session.

% search permissions

\begin{algorithm} [htp]
{\algofontsize
\caption{\textbf{SearchPermission}}

\label{algo:erbac-search-permission}

\begin{algorithmic}[1]

\INPUT \emph{It checks whether the requested permission is present in the list of permissions assigned to the Requester.}

\Require The client generated trapdoor of permission ($td^*_i (action)$, $td^*_i (target)$ and the server encrypted permission assignment list $L_{S}$ for Requester $i$.

\Ensure $\mathit{true}$ or $\mathit{false}$.

\medskip

\State $K_{s_i} \leftarrow KS[i]$ {\algofontsize \Comment{retrieve the server side key corresponding to Requester $i$}} \label{line:erbac-search-permission-ks}

\State $td(action) \leftarrow$ call \textbf{ServerTD} ($td^*_i (action)$, $K_{s_i}$) \label{line:erbac-search-permission-td-a}

\State $td(target) \leftarrow$ call \textbf{ServerTD} ($td^*_i (target)$, $K_{s_i}$) \label{line:erbac-search-permission-td-t}

\For {each server encrypted permission $(c(action), c(target))$ in $L_{S}$} \label{line:erbac-search-permission-loop}

	\State $match_{action} \leftarrow$ call \textbf{Match} ($c(action)$, $td(action)$) \label{line:erbac-search-permission-match-a}
	
	\State $match_{target} \leftarrow$ call \textbf{Match} ($c(target)$, $td(target)$) \label{line:erbac-search-permission-match-t}

	\If {$match_{action} \stackrel{?}{=} true$ and $match_{target} \stackrel{?}{=} true$} \label{line:erbac-search-permission-check-match}
	
		\Return $\mathit{true}$ \label{line:erbac-search-permission-true}
		
	\EndIf

\EndFor

\Return $\mathit{false}$ \label{line:erbac-search-permission-false}

\end{algorithmic}
}
\end{algorithm}

\noindent \\
\noindent \emph{\textbf{Searching a Permission:}} A Requester can send $\mathit{REQ}$ for executing certain permissions. The \gls{PEP} on the Service Provider checks if the Requester is active in the role indicated in $\mathit{REQ}$ and then the searches that role in the Permission Repository by running Algorithm \ref{algo:erbac-search-role}. After a role is matched in the Permission Repository, the \gls{PEP} searches the permission in $\mathit{REQ}$ by running Algorithm \ref{algo:erbac-search-permission}. This algorithm takes as input the client generated trapdoor of permission ($td^*_i (action)$, $td^*_i (target)$ and the server encrypted permission assignment list $L_{S}$ for Requester $i$ and returns either $\mathit{true}$ or $\mathit{false}$. First, it retrieves from the Key Store from the Key Store the server side key $K_{s_i}$ corresponding to Requester $i$ (Line \ref{line:erbac-search-permission-ks}). Next, it calculates server generated trapdoors of both action (Line \ref{line:erbac-search-permission-td-a}) and target (Line \ref{line:erbac-search-permission-td-t}) by calling Algorithm \ref{algo:erbac-server-td}. For each server encrypted permission $(c(action), c(target))$ in $L_{S}$ (Line \ref{line:erbac-search-permission-loop}), it matches the server encrypted action with the server generated action (Line \ref{line:erbac-search-permission-match-a}) and the server encrypted target with the server generated taret (Line \ref{line:erbac-search-permission-match-t}), respectively, by calling Algorithm \ref{algo:erbac-match}. If both matches are successful (Line \ref{line:erbac-search-permission-check-match}) for any permission $(c(action), c(target))$ in $L_{S}$, it returns $\mathit{true}$ (Line \ref{line:erbac-search-permission-true}). Otherwise, it returns $\mathit{false}$ (Line \ref{line:erbac-search-permission-false}).

% search role hierarchy

\begin{algorithm} [htp]
{\algofontsize
\caption{\textbf{SearchRoleHierarchyGraph}}

\label{algo:erbac-search-role-hierarchy-graph}

\begin{algorithmic}[1]

\INPUT \emph{It checks whether the Requester's role is inherited from any base role in the role hierarchy graph.}

\Require The server generated trapdoor of role $td(r)$ and the server generated role hierarchy graph $G_{S}$.

\Ensure $\mathit{true}$ or $\mathit{false}$.

\medskip

\For {each server encrypted role $c(r)$ in $G_{S}$} \label{line:erbac-search-rh-loop}

	\State $match \leftarrow$ call \textbf{Match} ($c(r)$, $td(r)$) \label{line:erbac-search-rh-call}

	\If {$match \stackrel{?}{=} true$} \label{line:erbac-search-rh-match}
	
		\Return $\mathit{true}$ \label{line:erbac-search-rh-true}
		
	\EndIf

\EndFor

\Return $\mathit{false}$ \label{line:erbac-search-rh-false}

\end{algorithmic}
}
\end{algorithm}

% TODO: on update remove \noindent \\ that is added to fix a format issue
\noindent \\
\noindent \emph{\textbf{Searching Roles in Role Hierarchy Graph:}}
The \gls{PDP} may need to search base roles of one in $\mathit{REQ}$ since a derived role inherits all permissions from its base role. The \gls{PDP} runs \textbf{SearchRoleHierarchyGraph} illustrated in Algorithm \ref{algo:erbac-search-role-hierarchy-graph} to find base roles from the encrypted role hierarchy graph. This algorithm takes as input the server generated trapdoor of role $td(r)$ and the server generated role hierarchy graph $G_{S}$ and returns $\mathit{true}$ if any base role is found and $\mathit{false}$ otherwise. For each server encrypted role $c(r)$ in $G_{S}$ (Line \ref{line:erbac-search-rh-loop}), it checks if $td(r)$ matches with any $c(r)$ by calling Algorithm \ref{algo:erbac-match} (Line \ref{line:erbac-search-rh-call}). If any match is found (Line \ref{line:erbac-search-rh-match}), it returns $\mathit{true}$ (Line \ref{line:erbac-search-rh-true}). Otherwise, it returns $\mathit{false}$ (Line \ref{line:erbac-search-rh-false}).

\section{Security Analysis}
\label{sec:erbac-security-analysis} 
In this section, we provide a combined security analysis of \gls{ESPOONERBAC} and \gls{ESPOON} because \gls{ESPOONERBAC} is built on the top of \gls{ESPOON}. In other words, \gls{ESPOONERBAC} uses algorithms presented in Chapter \ref{cha:espoon}. Therefore, we have not provided any security analysis of \gls{ESPOON} in Chapter \ref{cha:espoon}. In this section, we analyse the security of the policy deployment phase that includes the \gls{RA} encryption (Algorithms \ref{algo:erbac-deploy-role-assignment-client-side} and \ref{algo:erbac-deploy-role-assignment-server-side}), the \gls{PA} encryption (Algorithms \ref{algo:erbac-deploy-permission-assignment-client-side} and \ref{algo:erbac-deploy-permission-assignment-server-side}), the \gls{CC} encryption (Algorithms \ref{algo:erbac-deploy-contextual-condition-client-side} and \ref{algo:erbac-deploy-contextual-condition-server-side}), and the \gls{RH} encryption (Algorithms \ref{algo:erbac-deploy-role-hierarchy-client-side} and \ref{algo:erbac-deploy-role-hierarchy-server-side}). We then analyse the security of the policy evaluation phase that include \gls{SR} (Algorithms \ref{algo:erbac-client-td} and \ref{algo:erbac-search-role}), \gls{SP} (Algorithms \ref{algo:erbac-client-td} and \ref{algo:erbac-search-permission}), \gls{CCE} (Algorithms \ref{algo:erbac-request-contextual-condition} and \ref{algo:erbac-match-contextual-condition}) and \gls{SRH} (Algorithms \ref{algo:erbac-client-td}, \ref{algo:erbac-server-td} and \ref{algo:erbac-search-role-hierarchy-graph}). 

We first define some basic concepts on which we build our security proofs.

\subsection{Preliminaries}
In general, a scheme is considered secure if no adversary can break the scheme with probability significantly greater than random guessing. The adversary's advantage in breaking the scheme should be a negligible function of the security parameter.

\begin{definition}[Negligible Function]
A function $f$ is negligible if for each polynomial $p(.)$ there exists $N$ such that for all integers $n > N$ it holds that 
$f(n) < \frac{1}{p(n)}$.
\end{definition}

We consider a realistic adversary that is computationally bounded and show that our scheme is secure against such an adversary. We model the adversary as a randomised algorithm that runs in polynomial time and show that the success probability of any such adversary is negligible. An algorithm that is randomised and runs in polynomial time is called a \gls{PPT} algorithm.

Our scheme relies on the existence of a pseudorandom function $f$. Intuitively, the output a pseudorandom function cannot be distinguished by a realistic adversary from that of a truly random function. Formally, a pseudorandom function is defined as:

\begin{definition}[Pseudorandom Function]
A function $f:\{0,1\}^* \times \{0,1\}^* \rightarrow \{0,1\}^*$ is pseudorandom if for all \gls{PPT} adversaries $\mathcal{A}$, there exists a negligible function $negl$ such that:
\begin{center}
$|Pr[\mathcal{A}^{f_k(\cdot)} = 1] - Pr[\mathcal{A}^{F(\cdot)} = 1]| < negl(n)$
\end{center}
where $k \rightarrow \{0,1\}^n$ is chosen uniformly randomly and $F$ is a function chosen uniformly randomly from the set of function mapping n-bit strings to n-bit strings.
\end{definition}

Our proof relies on the assumption that the \gls{DDH} is hard in a group $\mathbb{G}$, i.e., it is hard for an adversary to distinguish between group elements $g^{\alpha \beta}$ and  $g^{\gamma}$ given  $g^{\alpha}$ and  $g^{\beta}$.

% TODO: on update remove $ $ that is added to fix a format issue
\begin{definition}[\gls{DDH} Assumption]
The \gls{DDH} problem is hard regarding a group $\mathbb{G}$ if for all \gls{PPT} adversaries $\mathcal{A}$, there
exists a negligible function $negl$ such that
$|Pr[\mathcal{A}(\mathbb{G}, q, g, g^\alpha, g^\beta,$ $g^{\alpha \beta}) = 1] - Pr[\mathcal{A}(\mathbb{G}, q, g, g^\alpha, g^\beta, g^\gamma) = 1]| < negl(k)$
where $\mathbb{G}$ is a cyclic group of order $q$ $(|q| = k)$ and $g$ is a generator of $\mathbb{G}$, and $\alpha, \beta, \gamma \in \mathbb{Z}_q$ are uniformly randomly chosen.
\end{definition}

Encryption algorithms in the policy deployment phase are based on \textbf{ClientEnc} (Algorithm \ref{algo:erbac-client-enc}) and \textbf{ServerReEnc} (Algorithm \ref{algo:erbac-server-re-enc}). It is equivalent to encrypting a single keyword in the \gls{SDE} scheme \cite{Dong:2011}. Dong \emph{et al.} \cite{Dong:2011} show that the single \gls{KE} scheme is \gls{INDCPA}. A cryptosystem is considered \gls{INDCPA} secure if no \gls{PPT} adversary, given an encryption of a message randomly chosen from two plaintext messages chosen by the adversary, can identify the message choice with non-negligible probability. Dong \emph{et al.} \cite{Dong:2011} prove the following theorem about the single \gls{KE} scheme:

\begin{theorem}
If the \gls{DDH} problem is hard relative to $\mathbb{G}$, then the single keyword encryption scheme \gls{KE} is \gls{INDCPA} secure against the server $\mathit{S}$, i.e., for all \gls{PPT} adversaries $\mathcal{A}$ there exists a negligible function $negl$ such that:
\begin{equation}
\begin{array}{l}
Succ_{KE, S}^{\mathcal{A}}(k) = Pr \left[ b'=b \left|
\begin{matrix}
(params, msk) \leftarrow Init(1^k)\\
(K_u,K_s) \leftarrow KeyGen(msk,U)\\
w_0,w_1 \leftarrow \mathcal{A}^{ClientEnc(K_u, \cdot)}(K_s)\\
b \xleftarrow{R} \{0,1\}\\
c_i^*(w_b) = ClientEnc(x_{i1},w_b) \\
b' \leftarrow \mathcal{A}^{ClientEnc(K_u, \cdot)}(K_s,c^*_i(w_b))
\end{matrix}
\right]\right. \\
<\frac{1}{2} + negl(k)
\end{array}
\end{equation}
\end{theorem}

Proof. See Theorem 1 in \cite{Dong:2011}.

\subsection{Security of Encryption Algorithms in the Policy Deployment Phase}
Using the fact that the \gls{KE} scheme is \gls{INDCPA} secure, we show that the four encryption schemes: \gls{RA}, \gls{PA}, \gls{CC} and \gls{RH} are also \gls{INDCPA} against the server. We give the proof details for the Roles Assignment encryption scheme \gls{RA}. We will show that the following theorem holds:

\begin{theorem}
If the single keyword encryption \gls{KE} scheme is \gls{INDCPA} secure against the server, then the \gls{RA} encryption scheme \gls{RA} is also \gls{INDCPA}, i.e., for all \gls{PPT} adversaries $\mathcal{A}$, there exists a negligible function $negl$ such that
$Succ_{RA,S}^{\mathcal{A}}(k) < \frac{1}{2} + negl(k)$.
\end{theorem}

Proof. We prove the theorem by showing that breaking the \gls{RA} encryption reduces to breaking the \gls{KE} encryption. We define the following game in which the adversary $\mathcal{A}$ challenges the game with two lists of roles $L_0$ and $L_1$ having the same number of roles $t$. We construct the following vector containing the encryption of roles from both lists: $\vec{C}^{(i)} = C(r_0^1), \ldots, C(r_0^i), C(r_1^{i+1}), \ldots, C(r_1^t)$. The success probability of the adversary in distinguishing the encryption of the two lists of roles is defined as:

\begin{equation}
Succ_{\mathcal{A}}(k) =\frac{1}{2}
Pr[A(\vec{C}^0) = 0] + \frac{1}{2}
Pr[A(\vec{C}^t) = 1]
\end{equation}

In the following, we show that breaking the \gls{RA} scheme reduces to breaking the \gls{KE} game. In the \gls{KE} game from \cite{Dong:2011}, the adversary challenges the game with two keywords $w_0$ and $w_1$ and tries to distinguish between their encryptions. Let us consider a \gls{PPT} adversary $\mathcal{A}'$ who attempts to challenge the single keyword encryption scheme \gls{KE} using the corresponding \gls{RA} adversary $\mathcal{A}$ as a sub-routine The game is the following:
\begin{itemize}
\item $\mathcal{A}'$ is given the parameters $(\mathbb{G},q,g,h,H,f)$ as input and for each user $i$ is given $(i,x_{i2})$.
\item $\mathcal{A}'$ passes these parameters to $\mathcal{A}$.
\item $\mathcal{A}$ generates two lists of roles $L_0$ and $L_1$ having the same number of roles  $t$ and gives them to $\mathcal{A}'$.
\item $\mathcal{A}'$ chooses $i \xleftarrow{r} [1, t]$. It then uses $r^i_0, r^i_1$ to challenge the single keyword encryption \gls{KE} game. The adversary gets back $c^i_b$ as the result, where $c^i_b$ is the encryption of either $r^i_0$ or $r^i_1$. $\mathcal{A}'$ uses this result to construct a hybrid vector $(c^1_0,\ldots, c^{i-1}_0, c_b^i, c_1^{i+1},\ldots,c^t_1)$ and sends it to $\mathcal{A}$.
\item $\mathcal{A}'$ outputs $b'$, the bit output by $\mathcal{A}$.
\end{itemize}

$\mathcal{A}$ is required to distinguish $\vec{C}^{(i)}$ and $\vec{C}^{(i-1)}$ and the probability of $\mathcal{A}$'s success in distinguishing correctly is:
\begin{equation}
Succ_{\mathcal{A}}^i(k) =\frac{1}{2}
Pr[A(\vec{C}^{(i)}) = 0] + \frac{1}{2}
Pr[A(\vec{C}^{(i-1)}) = 1]
\end{equation}

Since $i$ is randomly chosen, it holds that:
\noindent
\begin{equation}
\begin{array}{lll}
Succ_{\mathcal{A}'}(k) & = &\sum_{i=1}^tSucc_{\mathcal{A}}^i(t) \cdot \frac{1}{t} \\
& = &\frac{1}{2t}Pr[A(\vec{C}^0) = 0] + \sum_{i=1}^{t-1}(Pr[A(\vec{C}^i) = 0] \\
& & + Pr[A(\vec{C}^i) = 1]) + \frac{1}{2} Pr[A(\vec{C}^t) = 1] \\
& = & \frac{1}{t} (\frac{1}{2} Pr[A(\vec{C}^0) = 0] + \frac{1}{2} Pr[A(\vec{C}^t)=1]) + \frac{t-1}{2t} \\
& = & \frac{1}{t}Succ_{\mathcal{A}}(k) + \frac{t-1}{2t}
\end{array}
\end{equation}

Because the success probability of $\mathcal{A}'$ to break the single keyword encryption scheme \gls{KE} is $Succ_{\mathcal{A}'}(k) < \frac{1}{2} + negl(k)$, it follows that $Succ_{\mathcal{A}}(k) < \frac{1}{2} + negl(k)$.

The proof for the other encryption schemes is similar and for lack of space we do not show all the details.

\subsection{Security of Algorithms in the Policy Evaluation Phase}
We now analyse the security of \gls{SR}, \gls{SP}, \gls{CCE} and \gls{SRH}. These algorithms require the Service Provider to take some client input (i.e., trapdoors computed using Algorithm \ref{algo:erbac-client-td}), process it (i.e., re-encrypt it using Algorithm \ref{algo:erbac-server-td}), and test whether it matches some information stored on the server. Though a single operation has been proved secure, we are interested in what these algorithms leak to the Service Provider. We follow the concept of non-adaptive indistinguishability security introduced for encrypted databases by \cite{Curtmola:2006} and adapted by \cite{Dong:2011} in a multi-user setting. We show that given two non-adaptively generated histories with the same length and outcome, no \gls{PPT} adversary can distinguish the histories based on what it can observe from the interaction. A history contains all the interactions between clients and the Service Provider. Non-adaptive history means that the adversary cannot choose sequences of client inputs based on previous inputs and matching outcomes.

In the following, we show the details for the \gls{SR} scheme. In this scheme, a history is defined as follows:

% TODO: on update remove $ $ that is added to fix a format issue
\begin{definition}[\gls{SR} History]
An \gls{SR} history $\mathcal{H}_i$ is an interaction between a Service Provider and all clients that connect to it, over $i$ role activation requests. $\mathcal{H}_i=(L_s^{u_1}, \ldots, L_s^{u_i}, r_{1}^{u_1},$ $\ldots, r_{i}^{u_i})$, where $u_i$ represents an identifier of the client making the requests, $L_s^{u_i}$ represents the lists of roles for client $u_i$, and $r_{i}^{u_i}$ represents the request made by the client.
\end{definition}

We formalise the information leaked to a Service Provider as a \textit{trace}. We define two kinds of traces: the trace of a single request and the trace of a history. The trace of a request leaks to the Service Provider which role in $L_s^{i}$ matches the request and can be formally defined as: $tr(r)=\{td*_i(role), L_s^{i}, idx\}$, where $idx$ is the index of the matched role, if any, in $L_s^{i}$.

We define the role matching pattern $\mathcal{P}$ over a history $\mathcal{H}_i$ to be a set of  binary matrices (one for each client) with columns corresponding to encrypted roles in the list of the client, and rows corresponding to requests. $\mathcal{P}[j,k]=1$ if request $j$ matched the $k$'s role and $\mathcal{P}[j,k]=0$ otherwise.

The trace of a history includes the encrypted role assignment lists of all clients $L_s^{u_i}$ stored by the Service Provider and which can change as new roles are added and clients leave of join the system, the trace of each request, and the role matching pattern $\mathcal{P}_i$ for each client.

During an interaction, the adversary cannot see directly the plaintext of the request, instead it sees the ciphertext. The view of a request is defined as:

\begin{definition}[View of a Request]
We define the view of a request $q_{1}^{u_1}$ under a key set $K_{u_i}$ as: 
$V_{K_{u_i}}(q^{u_i})= tr(q^{u_i})$
\end{definition}

\begin{definition}[View of a History]
We define the view of a history with $i$ interactions $\mathcal{H}_i$ as 
$V_{K_u}(H_i) = (L_s^{u_1}, \ldots, L_s^{u_i}, V_{K_{u_i}}(q_1^{u_i}), \ldots, V_{K_{u_i}}(q_i^{u_i})$.
\end{definition}

The security definition is based on the idea that the scheme is secure if nothing is leaked to the adversary beyond what the adversary can learn from traces.

We define the following game in which an adversary $\mathcal{A}$ generates two histories $\mathcal{H}_{i0}$ and $\mathcal{H}_{i1}$ with the same trace over $i$ requests. Then the adversary is challenged to distinguish the views of the two histories. If the adversary succeeds with negligible probability, the scheme is secure.

\begin{definition}[Non-adaptive indistinguishability against a curious Service Provider]
The \gls{SR} scheme is secure in the sense of non-adaptive indistinguishability against a curious Service Provider if for all $i \in \mathbb{N}$ and for all \gls{PPT} adversaries $\mathcal{A}$ there exists a negligible function $negl$ such that:
\begin{equation}
Pr \left[ b'=b \left|
\begin{array}{lll}
(params, msk) \leftarrow Init(1^k)\\
(K_u, K_s) \leftarrow KeyGen(msk, U)\\
\mathcal{H}_{io},\mathcal{H}_{i1} \leftarrow \mathcal{A}(K_s)\\
b \xleftarrow{R} \{0,1\}\\
b' \leftarrow \mathcal{A}(K_s, V_{K_u}(\mathcal{H}_{ib}))
\end{array}
\right] < \frac{1}{2} + negl(k)
\right.
\end{equation}
where $U$ is a set of user IDs, $K_u$ is the user side key sets, $K_s$ are the server side key sets, $\mathcal{H}_{i1}$ and $\mathcal{H}_{i0}$ are two histories over $i$ requests such that $Tr(\mathcal{H}_{i0}) = Tr(\mathcal{H}_{i1})$.
\end{definition}

\begin{theorem}\label{thm:3}
If the \gls{DDH} problem in hard relative to $\mathbb{G}$, then the \gls{SR} scheme is a non-adaptive indistinguishable secure scheme. The success probability of a \gls{PPT} adversary $\mathcal{A}$ in breaking the \gls{SR} scheme is defined as:
\begin{equation}
\begin{array}{l}
Succ^{\mathcal{A}}(k) = \frac{1}{2} Pr[\mathcal{A}(RA(\vec{L}_0), TD(\vec{r}_0)) = 0] + \\
\hspace{50pt} \frac{1}{2} Pr[\mathcal{A}(RA(\vec{L}_1), TD(\vec{r}_1)) = 1] \\
\hspace{41pt} < \frac{1}{2} + negl(k)
\end{array}
\end{equation}
where $RA(\vec{L}_i)$ is the role encryption of the vector of lists of $H_i$, and $TD(\vec{r}_i)$ is the \textbf{ClientTD} of the roles in the requests of $H_i$.
\end{theorem}

Proof. We consider an adversary $\mathcal{A}'$ that challenges the RE \gls{INDCPA} game using $\mathcal{A}$ as a sub-routine. $\mathcal{A}'$ does the following:
\begin{itemize}
\item $\mathcal{A}'$ receives public parameters $params$ and the server side $(i, x_{i2})$ keys. 

\item To generate a view of a history $\mathcal{H}_i=(L_1^{u_1}, \ldots, L_i^{u_i}, q_1^{u_1}, \ldots, q_i^{u_i})$. $\mathcal{A}'$ performs the following steps:
\begin{itemize}
\item For each role assignment list $L_j^{u_j}$, run Algorithm \ref{algo:erbac-deploy-role-assignment-client-side} to encrypt it as $RA(L_j^{u_j})$. 
\item For each Search Role request $q_j^{u_j}$, run \textbf{ClientTD} (Algorithm \ref{algo:erbac-client-td}) to generate the trapdoor $TD(r)$ for the role.
\end{itemize}

\item $\mathcal{A}$ outputs $\mathcal{H}_{i0},\mathcal{H}_{i1}$. $\mathcal{A}'$ encrypts $\mathcal{H}_{i1}$ by itself and challenges the \gls{RA} \gls{INDCPA} game with $\vec{L}_0$ and $\vec{L}_1$, the vectors of all roles lists in the two histories. It gets the result $RA(\vec{L}_b)$, where $b \xleftarrow{R} \{0,1\}$ and forms a view of a history $(RA(\vec{L}_b), TD(\vec{r_1}))$. It sends the view to $\mathcal{A}$.

\item $\mathcal{A}$ tries to determine which vector was encrypted and outputs $b' \in \{0,1\}$.

\item $\mathcal{A'}$ outputs $b'$.

\end{itemize}

Because the \gls{RA} scheme is \gls{INDCPA}, it follows that:

\begin{equation}
\begin{array}{l}
\frac{1}{2} + negl(k)  >  Succ_{RA}^{\mathcal{A}'}(k) \\
\hspace{47pt} = \frac{1}{2} Pr[\mathcal{A}((RA(\vec{L}_0), TD(\vec{r}_1))) = 0] + \\
\hspace{56pt} \frac{1}{2} Pr[\mathcal{A}((RA(\vec{L}_1), TD(\vec{r}_1))) = 1]
\end{array}
\end{equation}

Now let us consider another adversary $\mathcal{A}''$ who wants to distinguish the pseudorandom function $f$ using $\mathcal{A}$ as a sub-routine. The adversary does the following:
\begin{itemize}
\item It generates $(\mathbb{G}, q, g, h, H)$ as public parameters, and sends them to $\mathcal{A}$ along with $f$. For each user $i$, it chooses randomly $x_{i1}$, $x_{i2}$ such that $x_{i1} + x_{i2} = x$. It sends all $(i, x_{i2})$ to $\mathcal{A}$ and keeps all $(i, x_{i1}, x_{i2})$.
\item $\mathcal{A}$ outputs $\mathcal{H}_{i0},\mathcal{H}_{i1}$. $\mathcal{A}''$ encrypts all the roles lists in $\mathcal{H}_{i0}$ as $RA(\vec{L}_0)$. It chooses $b \xleftarrow{R} \{0,1\}$ and asks the oracle to encrypt all roles in $\mathcal{H}_{ib}$. It combines the results to form a view $(RA(\vec{L}_0),TD(\vec{r}_b))$ and returns it to $\mathcal{A}$.
\item $\mathcal{A}$ outputs $b'$. $\mathcal{A}''$ outputs $1$ if $b'=b$ and $0$ otherwise.
\end{itemize}

There are two cases to consider:
Case 1: the oracle in $\mathcal{A}''$s game is the pseudorandom function $f$, then:

\begin{equation}
\begin{array}{l}
Pr[\mathcal{A}''^{f_{s}(.)}(1^k) = 1] = \\
\hspace{50pt} \frac{1}{2} Pr[\mathcal{A}(RA(\vec{L}_0), TD(\vec{r}_0)) = 0] + \\
\hspace{50pt} \frac{1}{2} Pr[\mathcal{A}(RA(\vec{L}_0), TD(\vec{r}_1)) = 1]
\end{array}
\end{equation}

Case 2: the oracle in $\mathcal{A}''$s game is a random function $f$, then for each distinct
role $r$, $\sigma_r$ is completely random to $\mathcal{A}$. Moreover, we know the traces are identical, so $RA(\vec{L}_b)$ and $TD(\vec{r}_b)$ are completely random to $\mathcal{A}$. In this case:
\begin{equation}
Pr[\mathcal{A}''^{f_{s}(.)}(1^k) = 1] = \frac{1}{2}
\end{equation}

Because $f$ is a pseudorandom function, by definition it holds that:
\begin{equation}
\begin{array}{l}
|Pr[\mathcal{A}''^{f_{s}(.)}(1^k) = 1] - Pr[\mathcal{A}'^{f_{s}(.)}(1^k) = 1]| < negl(k)\\
\hspace{88pt} Pr[\mathcal{A}''^{f_{s}(.)}(1^k) = 1] < \frac{1}{2} + negl(k)
\end{array}
\end{equation}

Sum up $Succ_{RE}^{\mathcal{A}'}(k)$ and $Pr[\mathcal{A}''^{f_{s}(.)}(1^k) = 1]$:

\begin{equation}
\begin{array}{l}
1 + negl(k) >  \frac{1}{2} Pr[\mathcal{A}(RA(\vec{L}_0), TD(\vec{r}_0)) = 0] + \\
\hspace{56pt}\frac{1}{2} Pr[\mathcal{A}(RA(\vec{L}_0), TD(\vec{r}_1)) = 1] + \\
\hspace{56pt}\frac{1}{2} Pr[\mathcal{A}(RA(\vec{L}_0), TD(\vec{r}_1)) = 0]+ \\
\hspace{56pt}\frac{1}{2} Pr[\mathcal{A}(RA(\vec{L}_1), TD(\vec{r}_1)) = 1] \\
\hspace{47pt} = \frac{1}{2} Pr[\mathcal{A}(RA(\vec{L}_0), TD(\vec{r}_0)) = 0] + \\
\hspace{56pt}\frac{1}{2} + \\
\hspace{56pt}\frac{1}{2} Pr[\mathcal{A}(RA(\vec{L}_1), TD(\vec{r}_1)) = 1] + \\
\hspace{47pt} = \frac{1}{2} + Succ^{\mathcal{A}}(k)
\end{array}
\end{equation}

Therefore, $Succ^{\mathcal{A}}(k)< \frac{1}{2} + negl(k)$.

\section[Performance Analysis of ESPOON$_{\mathit{ERBAC}}$]{Performance Analysis of \gls{ESPOONERBAC}}
\label{sec:erbac-performance-analysis}
In this section, we discuss a quantitative analysis of the performance of \gls{ESPOONERBAC}. In particular, we focus on performance of the modules that have been modified as compared to the \gls{ESPOON} architecture presented in Chapter \ref{cha:espoon}. It should be noticed that here we are concerned about quantifying the overhead introduced by the encryption operations performed both at the trusted environment and the outsourced environment. In the following discussion, we do not take into account the latency introduced by the network communication.

\subsection[Implementation Details of ESPOON$_{\mathit{ERBAC}}$]{Implementation Details of \gls{ESPOONERBAC}}

We have implemented \gls{ESPOONERBAC} in Java $1.6$. We have developed all the components of the architecture required for performing the policy deployment and policy evaluation phases. For the cryptographic operations, we have implemented all the functions presented in Section \ref{sec:erbac-algorithmic-details}. We have tested the implementation of \gls{ESPOONERBAC} on a single node based on an Intel Core2 Duo $2.2$ GHz processor with $2$ GB of RAM, running Microsoft Windows XP Professional version $2002$ Service Pack $3$. The number of iterations performed for each of the following results is $1000$.

\begin{figure} [htp]
\centering
\subfigure[]{
\includegraphics[width=.48\textwidth]{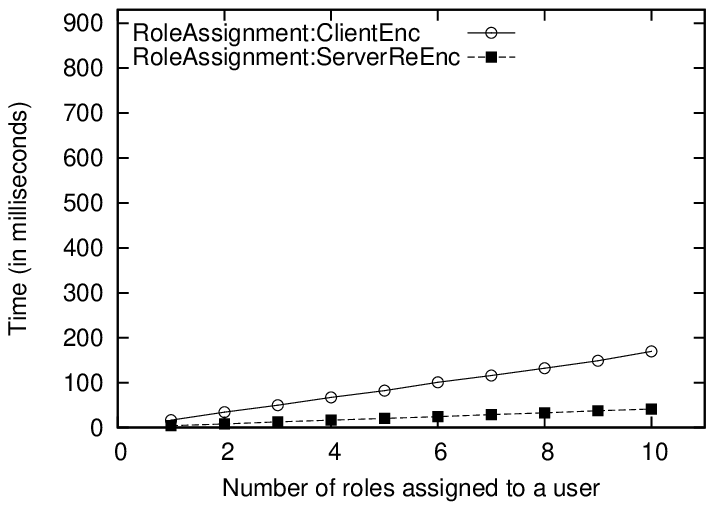} % .31
\label{fig:erbac-deploy-ura}
}
\subfigure[]{
\includegraphics[width=.48\textwidth]{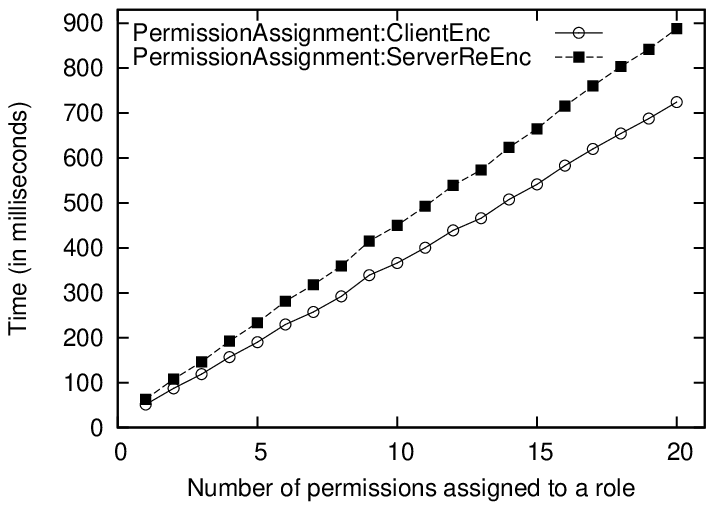} % .31
\label{fig:erbac-deploy-rpa}
}
\subfigure[]{
\includegraphics[width=.48\textwidth]{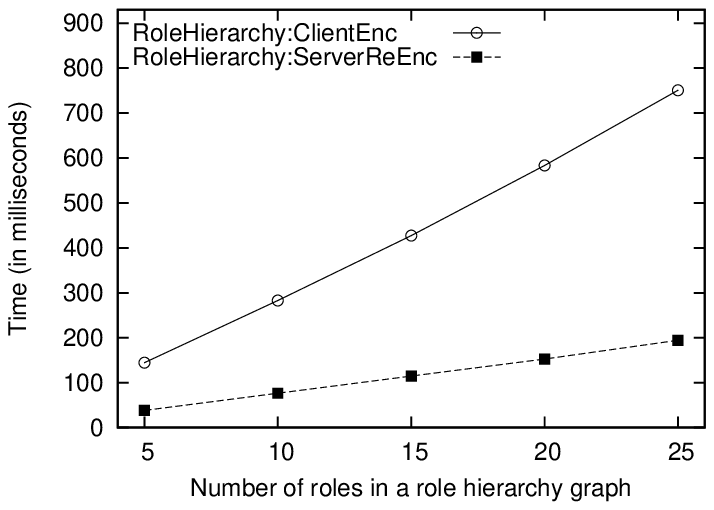} % .31
\label{fig:erbac-deploy-role-hierarchy}
}
\caption[Performance overhead of deploying RBAC policies]{Performance overhead of deploying \gls{RBAC} policies: \subref{fig:erbac-deploy-ura} a list of roles assigned to a user, \subref{fig:erbac-deploy-rpa} a list of permissions to a role and \subref{fig:erbac-deploy-role-hierarchy} a role hierarchy graph}
\label{fig:erbac-policy-deployment-rbac-policy}
\end{figure}

\subsection{Performance Analysis of the Policy Deployment Phase}
In this section, we analyse the performance of the policy deployment phase. In this phase, an Admin User encrypts policies and sends those encrypted policies to the Administration Point running in the outsourced environment. The Administration Point re-encrypts policies and stores them in the Policy Store in the outsourced environment. In the following, we analyse the performance of deploying (part of) policies including the role assignment list, the permission assignment and the role hierarchy graph (as shown in Figure \ref{fig:erbac-policy-deployment-rbac-policy}. \\ \\
\noindent \emph{\textbf{The Role Assignment List:}} 
In order to deploy a role assignment list, an Admin User performs a first round of encryption on the client side (see Algorithm \ref{algo:erbac-deploy-role-assignment-client-side}) and sends the client encrypted role assignment list to the Administration Point. The Administration Point performs another round of encryption on the server side (see Algorithm \ref{algo:erbac-deploy-role-assignment-server-side}) before storing the role assignment list in the Policy Store. Figure \ref{fig:erbac-deploy-ura} shows performance overhead on the client side, as well as on the server side in order to deploy a role assignment list. In this graph, we observe the performance by increasing number of roles in a role assignment list. As we can expect, the performance overhead increases linearly with the linear increase in the number of roles in a role assignment list. As we can notice, the graph grows linearly with the linear increase in the number of roles in the role assignment list $L_r$. Asymptotically, the complexity of this phase is ${\Theta}(| L_r |)$.

During the policy deployment phase, the encryption algorithm on the client side (Algorithm \ref{algo:erbac-client-enc}) takes more time that of the server side (Algorithm \ref{algo:erbac-server-re-enc}) as shown in Figure \ref{fig:erbac-policy-deployment-rbac-policy}. The encryption algorithm on the client side takes more time because it performs more complex cryptographic operations such as random number generation and hash calculation as illustrated in Algorithm \ref{algo:erbac-client-enc}. However, any policy is deployed very rarely; whereas, it may be evaluated quite frequently. Therefore, the performance overhead of the policy evaluation phase (discussed in Section \ref{sec:erbac-policy-evaluation}) is of great importance. \\ \\
\noindent \emph{\textbf{The Permission Assignment List:}} 
For deploying permissions assigned to a role, an Admin User performs a first round of encryption on the client side (see Algorithm \ref{algo:erbac-deploy-permission-assignment-client-side}) and sends both the client encrypted role and client encrypted permissions to the Administration Point, where each permission contains both an action and a target. The Administration Point generates the server encrypted role and server encrypted permissions after performing a second round of encryption on the server side (see Algorithm \ref{algo:erbac-deploy-permission-assignment-server-side}). Figure \ref{fig:erbac-deploy-rpa} shows the performance overhead of deploying a permission assignment list. This graph illustrates the performance of deploying a permission assignment list for a role with a number of permissions ranging from 1 to 20. As we can expect, the performance overhead increases linearly with the linear increase in the number of permissions in the permission assignment list $L_p$. Asymptotically, the complexity of this phase is ${\Theta}(| L_p |)$. \\ \\
\noindent \emph{\textbf{Contextual Conditions:}} 
Both the role assignment and the permission assignment lists include a contextual condition as we can see in Figure \ref{fig:erbac-policy-role-assignment} and Figure \ref{fig:erbac-policy-permission-assignment}, respectively. The performance of contextual condition is already analysed in Chapter \ref{cha:espoon}, Section \ref{sec:espoon-policy-deployment} (see Figure \ref{fig:erbac-policy-deployment-context}). \\ \\
\noindent \emph{\textbf{The Role Hierarchy Graph:}} 
The \gls{PDP} may search for a base role of the one in the access request $\mathit{REQ}$ since a derived role inherits all permissions from its base role. For supporting this search, we deploy a role hierarchy graph. For deploying a role hierarchy graph, an Admin User performs the first round in order to generate the client encrypted trapdoor, as well as to calculate the client generated trapdoor of each role in the graph (see Algorithm \ref{algo:erbac-deploy-role-hierarchy-client-side}). The Admin User sends the client generated role hierarchy graph to the Administration Point. The Administration Point performs the second round to generate the server encrypted trapdoor, as well as to calculate the server generated trapdoor of each role in the graph (see Algorithm \ref{algo:erbac-deploy-role-hierarchy-server-side}). The \gls{PDP} matches the trapdoor of role in $\mathit{REQ}$ with the server encrypted role and if this match is successful, it finds trapdoors of the base roles. The trapdoors of base roles are required in order to perform search in the list of server encrypted roles in the Permission Repository.

In our experiment, we consider a role hierarchy graph in which each role $R_i$ extends role $R_{i+1}$ for all values of $i$ from 0 to $n - 1$ where $n$ indicates the total number of nodes and varies from 5 to 25. Figure \ref{fig:erbac-deploy-role-hierarchy} shows the performance overhead of encrypting a role hierarchy graph both on the client side and the server side. The graph grows linearly with the number of roles in a role hierarchy graph $G_{RH}$. Asymptotically, the complexity of this phase is ${\Theta}(|G_{RH}|)$.

\begin{table} [htp]
\centering
\caption[Performance overhead of encrypting requests]{Performance overhead of encrypting requests during the policy evaluation phase}
\label{tab:erbac-request}
\begin{tabular}{ |l|c| }
\hline
\textbf{Request Type} & \textbf{Time (in milliseconds)} \\ \hline
$\mathit{ACT}$ & 16.353 \\ \hline
$\mathit{REQ}$ & 47.069 \\ \hline
\end{tabular}
\end{table}

% access request
\begin{figure} [htp]
\centering
\subfigure[]{
\includegraphics[width=.48\textwidth]{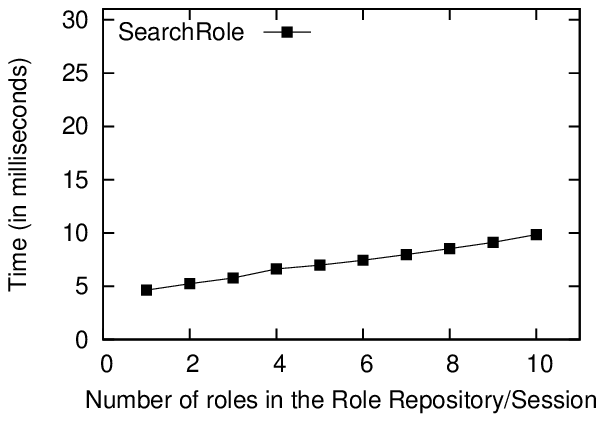} 
\label{fig:erbac-search-role}
}
\subfigure[]{
\includegraphics[width=.48\textwidth]{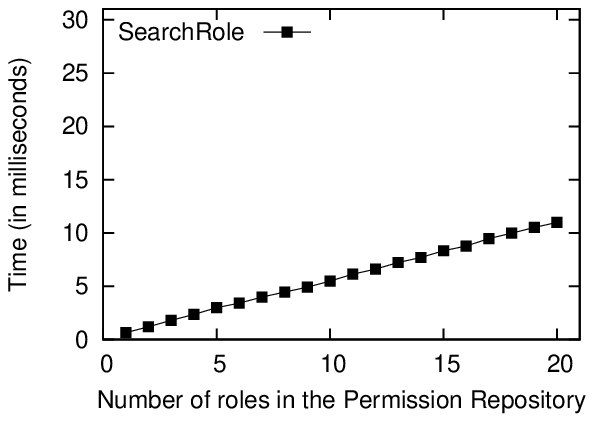}
\label{fig:erbac-request-search-req-role-in-perms}
}
\subfigure[]{
\includegraphics[width=.48\textwidth]{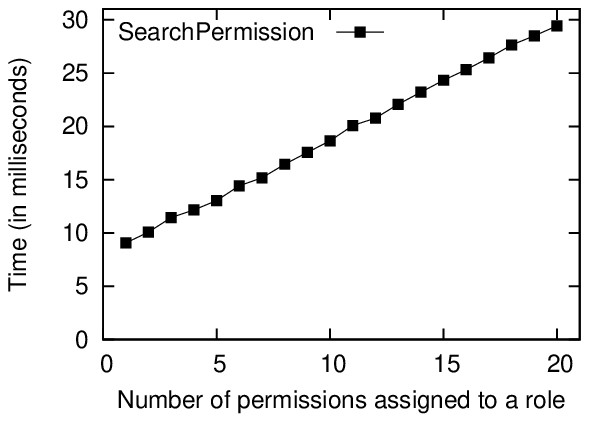}
\label{fig:erbac-request-search-req-perms}
}
\subfigure[]{
\includegraphics[width=.48\textwidth]{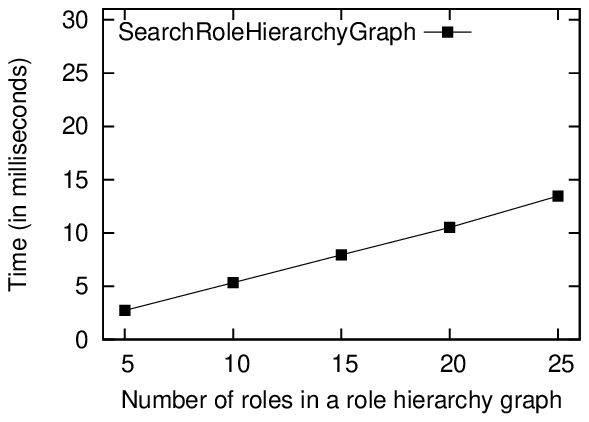}
\label{fig:erbac-search-role-hierarchy}
}
\caption[Performance overhead of evaluating RBAC policies]{Performance overhead of evaluating \gls{RBAC} policies: \subref{fig:erbac-search-role} searching roles in the Role Repository/Session, \subref{fig:erbac-request-search-req-role-in-perms} searching a role in the Permission Repository, \subref{fig:erbac-request-search-req-perms} checking the list of permissions assigned to a role and \subref{fig:erbac-search-role-hierarchy} searching a role in the role hierarchy graph}
\label{fig:erbac-policy-evaluation-rbac-policy}
\end{figure}

\subsection{Performance Analysis of the Policy Evaluation Phase}
\label{sec:erbac-policy-evaluation}

In this section, we analyse the performance of the policy evaluation phase. In this phase, a Requester sends the encrypted request to the \gls{PEP} running in the outsourced environment. The \gls{PEP} forwards the encrypted request to the \gls{PDP}. The \gls{PDP} has to select the set of policies that are applicable to the request. The \gls{PDP} may require contextual information in order to evaluate the selected policies. In the following, we calculate the performance overhead of generating requests, search a role (in the Role Repository, in the Active Roles repository or in the Permission Repository), searching a permission and searching a role in a role hierarchy graph. \\ \\
\noindent \emph{\textbf{Generating Requests:}}
A Requester may send the role activation request $\mathit{ACT}$. In order to generate $\mathit{ACT}$, a Requester calculates the client generated role (see Algorithm \ref{algo:erbac-client-td}). This trapdoor generation of role takes 16.353 \gls{ms} as illustrated in Table \ref{tab:erbac-request}. After a Requester is active in a role, she may make an access request $\mathit{REQ}$ . A Requester has to calculate trapdoor for each element (including role, action and target) in $\mathit{REQ}$. The $\mathit{REQ}$ generation takes 47.069 \gls{ms} as illustrated in Table \ref{tab:erbac-request}. We can see that $\mathit{REQ}$ generation takes 3 times of $\mathit{ACT}$ generation because $\mathit{REQ}$ has to calculate 3 trapdoors while $\mathit{ACT}$ has to generate only a single trapdoor. The request generation does not depend on any parameters and can be considered constant. \\ \\
\noindent \emph{\textbf{Searching a Role in the Role Repository/Session:}}
In order to grant $\mathit{ACT}$, the \gls{PDP} needs to search roles in the Role Repository. For searching a role, the \gls{PDP} first calculates the server generated trapdoor of role in $\mathit{ACT}$ and then matches this server encrypted trapdoor with server encrypted roles in the role assignment list as illustrated in Algorithm \ref{algo:erbac-search-role}. Figure \ref{fig:erbac-search-role} shows the performance overhead (in the worst case) of performing this search. In this graph, we can observe that it grows linearly with increase in number of roles. As the graph indicates, the search function takes initial approximately $4$ \gls{ms} to generate the server encrypted trapdoor of role in $\mathit{ACT}$ while it takes approximately $0.6$ \gls{ms} to perform encrypted match.

The \gls{PDP} grants $\mathit{ACT}$ by adding the server encrypted role of the Requester in the Active Roles repository of the Session. This implies that the Session maintains a list of active roles. Once a Requester makes an access request $\mathit{REQ}$, the \gls{PDP} has to search in the Session if she is already active in role indicated in $\mathit{REQ}$. The performance overhead of searching a role in session is same as it incurs for searching a role in the Role Repository (shown in Figure \ref{fig:erbac-search-role}). Asymptotically, the complexity of this phase is $O(| L_r |)$. \\ \\
\noindent \emph{\textbf{Searching a Role in the Permission Repository:}}
After finding the role of $\mathit{REQ}$ in the list of active roles, the \gls{PDP} has to search if the same role has the requested permission. For this purpose, the \gls{PDP} has first to search the role of $\mathit{REQ}$ in the Permission Repository and if any match is found, it has to search the requested permission in the list of permissions assigned to the found role. Figure \ref{fig:erbac-request-search-req-role-in-perms} shows the performance overhead (in the worst case) of searching a role in the Permission Repository. The graph grows linearly with the increase in the number of roles in the Permission Repository. The \gls{PDP} runs Algorithm \ref{algo:erbac-search-role} but with a slight modification of ignoring the server trapdoor generation (in Line \ref{line:erbac-search-role-td}) as it is already generated when the role of $\mathit{REQ}$ is searched in the session. This is why, searching a role in the Permission Repository (as illustrated in Figure \ref{fig:erbac-request-search-req-role-in-perms}) takes less time than searching a role in the Role Repository or Session (as illustrated in Figure \ref{fig:erbac-search-role}). Asymptotically, the complexity of this phase is $O(| L_r |)$. \\ \\
\noindent \emph{\textbf{Searching a Permission:}}
After a role is found in the Permission Repository, the \gls{PDP} searches the requested permission in the list of permissions assigned to the found role (see Algorithm \ref{algo:erbac-search-permission}). Before searching the list of permissions, the \gls{PDP} has to calculate server generated trapdoors of both the action and the target present in $\mathit{REQ}$. As we explained earlier, a single trapdoor generation on the server side takes approximately 4 \gls{ms}. The trapdoor generation of the requested permission, containing an action and a target, takes 8 \gls{ms}. Next, the \gls{PDP} match (server generated trapdoors of) this requested permission with the list of (sever encrypted) permissions assigned to the found role. Figure \ref{fig:erbac-request-search-req-perms} shows the performance overhead (in the worst case) of searching server generated trapdoor of permission with a list of server encrypted permissions. The graph grows linearly with the increase in the number of permissions in the list. For each permission match, the \gls{PDP} performs (at most) two encrypted matches each incurring approximately 0.6 \gls{ms}. Asymptotically, the complexity of this phase is $O(| L_p |)$. \\ \\
\noindent \emph{\textbf{Evaluating Contextual Conditions:}}
For evaluating the role assignment (illustrated in Figure \ref{fig:erbac-policy-role-assignment}) or the permission assignment (illustrated in Figure \ref{fig:erbac-policy-permission-assignment}) policies, the \gls{PDP} may need to evaluate contextual conditions. This part has already been discussed in Chapter \ref{cha:espoon}, Section \ref{sec:espoon-policy-evaluation} (see Figure \ref{fig:erbac-policy-evaluation-context}). \\ \\
\noindent \emph{\textbf{Searching in a Role Hierarchy Graph:}}
The \gls{PDP} may search a role in the role hierarchy graph. For performing this search, we consider a role hierarchy graph in which each role $R_i$ extends role $R_{i+1}$ for all values of $i$ from 0 to $n - 1$ where $n$ indicates the total number of nodes and varies from 5 to 25. Figure \ref{fig:erbac-search-role-hierarchy} shows the performance overhead of searching a role in the role hierarchy graph deployed on the server side. As we can expect, the graph grows linearly with the number of roles in a role hierarchy graph $G_{RH}$. Asymptotically, the complexity of this phase is $O(|G_{RH}|)$. \\ \\

\begin{table} [htp]
\centering
\caption[Time complexity of each phase in the lifecycle of ESPOON$_{\mathit{ERBAC}}$]{Summary of time complexity of each phase in the lifecycle of \gls{ESPOONERBAC}}
\label{tab:erbac-complexity-summary}

\begin{tabular}{ |l|c| } 

\hline

\textbf{Phase Name} & \textbf{Complexity in the Worst Case} \\ \hline

Deployment of the role assignment list & ${\Theta}(| L_r |)$ \\ \hline

Deployment of the permission assignment list & ${\Theta}(| L_p |)$ \\ \hline

Deployment of the role hierarchy graph & ${\Theta}(| G_{RH} |)$ \\ \hline

Searching a role & $O(| L_r |)$ \\ \hline

Searching a permission & $O(| L_p |)$ \\ \hline

Searching in the role hierarchy graph & $O(| G_{RH} |)$ \\ \hline

\end{tabular}

\end{table}

Table \ref{tab:erbac-complexity-summary} provides a summary of time complexities of different phases in the lifecycle of \gls{ESPOONERBAC}. \\ \\
\noindent \emph{\textbf{Comparing \gls{ESPOONERBAC} with \gls{ESPOON}:}}
We compare the performance overheads of the policy evaluation of \gls{ESPOONERBAC} with that of \gls{ESPOON} \cite{Asghar2011-ARES}. Before we show the comparison, we see how policies are expressed in both \gls{ESPOONERBAC} and \gls{ESPOON}. The \gls{ESPOONERBAC} policies are explained in Section \ref{sec:representation}. The \gls{ESPOON} policy is expressed as a $\langle S, A, T \rangle$ tuple with a $\mathit{CONDITION}$, meaning if $\mathit{CONDITION}$ holds then subject $S$ can take action $A$ over target $T$. For comparing the performance overheads, we consider \gls{ESPOON} policies with 50 unique subjects and each subject has 10 unique actions and targets where each $\langle S, A, T \rangle$ tuple's condition is the conjunction (AND) of the contextual condition illustrated in Figure \ref{fig:erbac-cc} and \emph{RequesterName$=<$NAME$>$}. That is, a subject can execute action over the target provided subject's name is equal to one specified in the condition, subject's location is cardiology-ward and time is between 9 AM and 5 PM. Similarly, we consider \gls{ESPOONERBAC} policies with 50 unique roles and each role has 10 unique permissions, where each user can get active in 5 roles. The introduction of \gls{RBAC} simplifies the roles and permission management because we can enforce possible conditions at role activation time instead of enforcing them at the permission grant time. For instance, we can enforce location and time checks (i.e., the condition illustrated in Figure \ref{fig:erbac-cc}) at the role activation time while the condition \emph{RequesterName$=<$NAME$>$} can be enforced at the permission grant time.

\begin{figure} [htp]
\centering
% left bottom right top
\includegraphics[trim=0mm -7mm 0mm 0mm,clip,width=.7\textwidth]{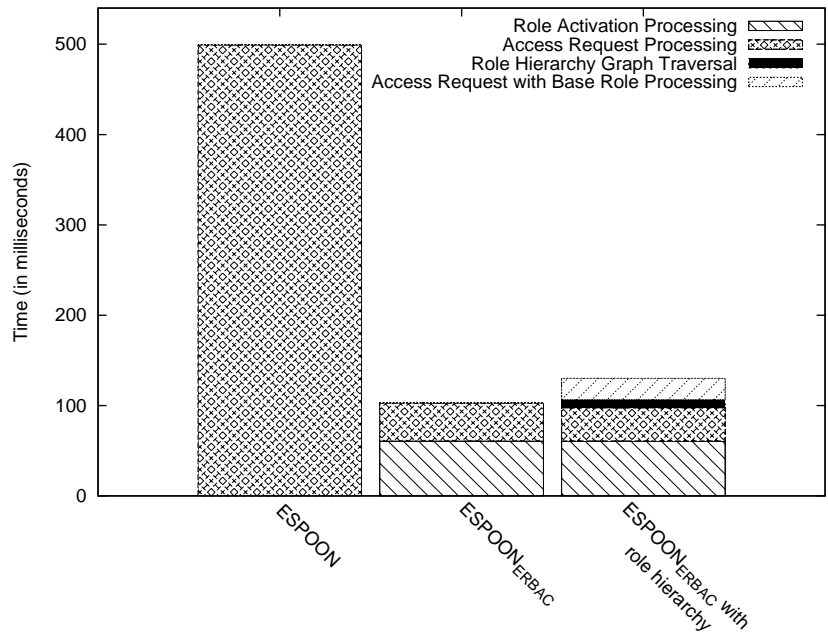}
\caption[Performance comparison of ESPOON and ESPOON$_{\mathit{ERBAC}}$]{Performance comparison of \gls{ESPOON} and \gls{ESPOONERBAC}}
\label{fig:erbac-espoon-vs-erbac}
\end{figure}

% TODO: on update remove \\ that is added to fix a format issue

Figure \ref{fig:erbac-espoon-vs-erbac} shows the performance overheads of evaluating \gls{ESPOON} and \\ \gls{ESPOONERBAC} policies. In \gls{ESPOON}, a requester's subject is matched with one in the repository of 500 entries (i.e., 50 subjects each with 10 actions and targets). If there is any match, requester's action and target are matched and then condition is evaluated. In the worst case, in \gls{ESPOON}, the access request processing can take approximately up to 500 \gls{ms}. On the other hand, in \gls{ESPOONERBAC}, a requester first gets active in a role provided condition holds. The role activation can take approximately up to 60 \gls{ms} for a user that can get active in 5 roles. After the role activation, a requester can be granted permissions assigned to its role. However, first the active role is searched in the session and then the permission can be granted if the condition associated with that permission holds. As we can see in Figure \ref{fig:erbac-espoon-vs-erbac}, grating the permission takes up to 42 \gls{ms}. The reason why \gls{ESPOONERBAC} performance is better than that of \gls{ESPOON} because (i) all possible conditions are enforced at the role activation time and (ii) introduction of roles simplified the roles and permissions management.

We also consider the effect of role hierarchies on the \gls{ESPOONERBAC} performance. In a role hierarchy, we assume that a role can inherit all permissions from its base role. This simplifies the role management and permission assignment to roles. In our experimentation, we consider 50 roles where each role has 5 permissions. Furthermore, there is a role hierarchy graph containing 25 roles, which is necessary for finding inheritance relationship between roles. Figure \ref{fig:erbac-espoon-vs-erbac} shows a very slight performance gain to evaluate the access request in case of role hierarchy in \gls{ESPOONERBAC}. Since the permission can be associated with base role, we need to traverse in the role hierarchy graph to find base roles. The performance of traversing in the role hierarchy graph is shown in Figure \ref{fig:erbac-espoon-vs-erbac}. Finally, the requested permission is granted if associated even with any base roles. The role hierarchy may improve performance but in the worst case it incurs higher overhead. However, the performance of \gls{ESPOONERBAC} with role hierarchy is still better than that of \gls{ESPOON}.

\section{Chapter Summary}
\label{sec:erbac-summary}
In this chapter, we have presented the \gls{ESPOONERBAC} architecture that can enforce sensitive \gls{RBAC} policies in an encrypted manner, where users are assigned roles and users can execute permissions if they are active in a session that manages lists of roles different users are active in. For structuring sensitive roles within an organisation, \gls{ESPOONERBAC} also supports the role hierarchies in \gls{RBAC}. The \gls{RBAC} policy is enforced such that it does not reveal information about roles and permissions managed in the outsourced environment.

In order to cope with the real-world business requirements, Sandhu \emph{et al.} \cite{Sandhu:1996} propose an \gls{RBAC} constraint model that includes both static and dynamic security constraints. The static security constraints can easily be enforced by existing \gls{ESPOONERBAC} and \gls{ESPOON} architectures. However, the challenging issue is to support dynamic security constraints in outsourced environments, where the access histories are managed by the curious Service Provider. In the next chapter, we investigate how to manage the access histories in order to enforce dynamic security constraints without leaking private information to the curious Service Provider.

%%%%%%%%%%%%%%%%%%%%%%%%% CHAPTER EGRANT %%%%%%%%%%%%%%%%%%%%%%%%%

\chapter[Enforcing Dynamic Security Constraints in RBAC]{\acrshort{EGRANT}: Dynamic Security Constraints in \acrshort{RBAC}\footstar{The final version of this chapter will appear in \cite{Asghar2013:IJIS:EGRANT}.}}
\label{cha:egrant}
Cloud computing is an emerging paradigm offering outsourced services to enterprises for storing and processing huge amount of data at very competitive costs. For leveraging the cloud to its fullest potential, organisations require security mechanisms to regulate access on data, particularly at runtime. One of the strong obstacles in widespread adoption of the cloud is to preserve confidentiality of the data. In fact, confidentiality of the data can be guaranteed by employing existing encryption schemes; however, access control mechanisms might leak information about the data they aim to protect. State of the art access control mechanisms can statically enforce constraints such as static separation of duties. The major research challenge is to enforce constraints at runtime, i.e., enforcement of dynamic security constraints (including \acrlong{DSoD} and \acrlong{CW}) in the cloud. The main challenge lies in the fact that dynamic security constraints require notion of sessions for managing access histories that might leak information about the sensitive data if they are available as cleartext in the cloud. In this chapter, we present \acrshort{EGRANT}: an architecture able to enforce dynamic security constraints without relying on a trusted infrastructure, which can be deployed as \gls{SaaS}. In \gls{EGRANT}, sessions' access histories are encrypted in such a way that enforcement of constraints is still possible. As a proof-of-concept, we have implemented a prototype and provided a preliminary performance analysis showing a limited overhead.

\section{Introduction}
With its cost-effective model, cloud-based services are very attractive for enterprises and government sectors. Initially developed as a cheap storage solution (monthly \$0.085/GB and \$0.095/GB, as of October 2013, offered by Google \cite{Google:2013} and Amazon \cite{Amazon:2013}, respectively), the cloud paradigm today is able to offer affordable software solutions. The term \acrfull{SaaS} is used to indicate software products offered as a service through the cloud. Several vendors have adopted this model to offer their products at a more affordable price. Classes of software products available as \gls{SaaS} range from document management tools (such as Google Drive \cite{Google:2012}) to image processing tools (such as Adobe Photoshop \cite{Pehrson:2011}). Recently, even \gls{BPM} solutions have become available as \gls{SaaS} from major players in this field, such SAP with its Business ByDesign \cite{SAP:2013}. \gls{BPM} solutions are at the core of modern organisations to coordinate the activities within their departments and streamline customers' requests. As empirical studies have demonstrated \cite{Kohlbacher:2009}, the use of \gls{BPM} solutions increases the productivity of the organisation and customer satisfaction.

One of the crucial aspects of \gls{BPM} systems is the enforcement of access control decisions for assigning human resources to execute tasks within a business process. If this control is too restrictive then it could hamper the productivity of the overall business process. On the other hand, a very lax approach might undermine the confidentiality of sensitive data (when accessed by unauthorised users), resulting in serious consequences for the organisation. In a \gls{BPM} system, the access control mechanism has to take into account business-related notions such as conflict-of-interests. Typical examples are that of an employee able to execute two tasks that might lead to fraudulent actions and that of an employee executing the same task over two different sets of data that could be in conflict with each other. Over the years, a huge amount of research effort has been put on this topic. The results have culminated in identifying and enforcing dynamic security constraints \cite{Kong:2007, Sandhu:1996, Nash:1990, Brewer:1989}.

If dynamic security constraints are to be correctly enforced, the system needs to maintain history of all actions executed by the entities that it controls, as well as contextual information of the requester (e.g., time and location). When the system receives a new request, it checks whether allowing the current request violates any constraints in view of the earlier actions performed by the same (group of) requesters. State of the art enforcement techniques \cite{Crampton:2009, Joshi:2005, Ahn:2000, Gligor:1998} rely on a trusted infrastructure, which expects information to be in cleartext. That is, the history of actions, contextual information, and constraints are all stored in cleartext to be readily accessible. 

With the move towards outsourced solutions, the trust assumptions in the management of the infrastructure do not hold any longer. The cloud providers that have control over the hardware, where data and security constraints are deployed (and enforced), could easily have access to them. The data can be protected using encryption techniques; however, state-of-the-art enforcement techniques \cite{Crampton:2009, Joshi:2005, Ahn:2000, Gligor:1998} cannot preserve confidentiality of dynamic security constraints because they expect all information in cleartext at both deployment and enforcement times. The problem here is that learning about the security constraints might leak information about the data itself. There are some cryptographic techniques that can enforce static security constraints in outsourced environments \cite{Asghar2013-COSE, Asghar2011-CCS, Asghar2011-ARES, Vimercati:2010, Vimercati:2007:VLDB}. Unfortunately, there is no cryptographic solution that can enforce dynamic security constraints in the cloud.

\subsection{Research Contributions}
In this chapter, we want to fill this gap and propose an enforcement mechanism for dynamic security constraints that can be offered either as a stand-alone \gls{SaaS} solution or integrated with other \gls{SaaS} products that require the enforcement of these constraints. The main idea is to outsource the enforcement of constraints without revealing sensitive information to the untrusted infrastructure. To the best of our knowledge, we are first to address the problem of enforcing dynamic security constraints in outsourced environments. We named our solution \gls{EGRANT}. \gls{EGRANT} can enforce constraints while taking into account contextual information (such as time and location of the user) without revealing any information to cloud providers. In our mechanism, an administrator can specify constraints with contextual conditions including non-monotonic boolean expressions and range queries. In \gls{EGRANT}, constraints as well as session information are encrypted. The encryption scheme we use is such that it does not require users to share any encryption keys. In case a user leaves the organisation, the system is still able to perform its operations without requiring re-encryption of constraints or access histories managed by the session. Finally, we have implemented a prototype of \gls{EGRANT} and analysed its performance to quantify the incurred overhead.

\subsection{Chapter Outline}
The rest of this chapter is organised as follows. 
Section \ref{sec:egrant-related_work} reviews the related work. 
Section \ref{sec:egrant-dynamic-security-constraints} provides an overview of the dynamic security constraints supported in \gls{EGRANT}. 
Section \ref{sec:egrant-architecture} describes the \gls{EGRANT} architecture. 
Section \ref{sec:egrant-solution-details} and Section \ref{sec:egrant-algorithmic-details} focus on solution and algorithmic details of the \gls{EGRANT} architecture, respectively. 
In Section \ref{sec:egrant-discussion}, we provide details about information disclosure in \gls{EGRANT} and the type of collusion attack that our solution is subjected to.
Section \ref{sec:egrant-performance_evaluation} describes implementation details and analyses the performance overhead of the \gls{EGRANT} prototype. 
%Section \ref{sec:egrant-related_work} reviews the related work. 
Finally, Section \ref{sec:egrant-summary} concludes this chapter.

\section{Related Work}
\label{sec:egrant-related_work}

There is a significant amount of research on enforcing dynamic security constraints including \gls{DSoD} \cite{vanTilborg:2011:DSoD, Kong:2007, Ahn:1999, Sandhu:1996, Sandhu:1990, Nash:1990} and \gls{CW} \cite{Vimercati:2011:CW, Brewer:1989}. State of the art solutions including \emph{RCL 2000} \cite{Ahn:2000}, \emph{GTRBAC} \cite{Joshi:2005}, \emph{MFOTL} \cite{Basin:2010} and \cite{Armando:2012, Crampton:2009, Crampton:2003, Gligor:1998} mainly focus on formally specifying the constraints. They assume a trusted infrastructure in order to enforce the constraints. There are some approaches that extend the enforcement mechanisms for taking into account contextual information such as time and location while making the access decision \cite{Joshi:2008, Kim:2007, Joshi:2005, Strembeck:2004}. However, none of the existing approaches are applicable when the enforcement mechanism is delegated to a third party that is not trusted. These approaches operate on the constraints that are stored in cleartext. Unfortunately, these constraints may leak information about the internal policies of an organisation and can result in serious implications if not adequately protected.

There are some approaches for enforcing static security constraints in outsourced environments \cite{Asghar2013-COSE, Asghar2011-CCS, Asghar2011-ARES, Vimercati:2010, Vimercati:2007:VLDB}. The idea of delegating the access control mechanism to an outsourced environment has initially been explored by De Capitani di Vimercati \emph{et al.} in \cite{Vimercati:2007:VLDB} and extended it in \cite{Vimercati:2010}. Their proposed solution is based on the key derivation method \cite{Atallah:2009}, where each user has a key capable of decrypting resources she is authorised to access. The main drawback of this type of approaches is that they tightly couple security policies with the enforcement mechanism; therefore, any changes in the security policies require to generate new keys and to redistribute them to the users. 

In \cite{Asghar2011-ARES}, we propose \gls{ESPOON} that aims at providing a clear separation between security policies and the enforcement mechanism. \gls{ESPOON} enforces authorisation policies in outsourced environments. In \gls{ESPOON}, a data owner may attach an authorisation policy with her data while storing it on the server running in the outsourced environment. A data consumer may request for the data and get access if the authorisation policy corresponding to the requested data is satisfied, where the evaluation is performed also by the server running in the outsourced environment. \gls{ESPOON} does not consider concept of roles at all. In \cite{Asghar2013-COSE, Asghar2011-CCS}, we extend \gls{ESPOON} for supporting an encrypted version of the \gls{RBAC} model and propose \gls{ESPOONERBAC}. Users can be associated to roles and get access rights based on the role hierarchies that are managed by the server. In \gls{ESPOONERBAC}, it is possible to enforce static security constraints, such as static separation of duties; however, it is not possible to delegate the enforcement of dynamic security constraints, such as \gls{HBDSoD} and \gls{CW}. The main issue is that the proposed architecture in \cite{Asghar2013-COSE, Asghar2011-CCS} lacks to manage encrypted session management, necessary for enforcing dynamic security constraints in outsourced environments.

The security policy enforcement is mainly based on encrypted matching schemes in untrusted environments. There are number of schemes that address encrypted matching in outsourced environments \cite{Shao:2010, Bethencourt:2007, Curtmola:2006, Goyal:2006, Boneh:2004, Song:2000}. Song \emph{et al.} \cite{Song:2000} are the first to propose an encrypted matching scheme, where documents and requests are encrypted using symmetric keys. The main drawback of this scheme is that it is a single-user scheme. \gls{MSSE} \cite{Curtmola:2006} is the first scheme to support encrypted matching in multi-user settings. In the \gls{MSSE} scheme, a data owner controls the search access by granting and revoking the search privileges to the users within her group by employing the symmetric encryption. The issue with scheme is that it requires redistribution of secret to all users once a user is revoked. Boneh \emph{et al.} \cite{Boneh:2004} are the first to propose the encrypted matching scheme in the public settings; however, it is not a multi-user scheme. Shao \emph{et al.} \cite{Shao:2010} introduce \gls{PRES} scheme that is a combination of proxy re-encryption and \gls{PEKS}. In \gls{PEKS}, a delegation key is generated for the target user. The target user re-encrypts the ciphertext with the delegated key. The re-encryption algorithm outputs another ciphertext corresponding to the public key of the target user. That is why, this scheme high performance overhead for re-encrypting ciphertext. 

There are schemes based on \gls{ABE} including \gls{CPABE} \cite{Bethencourt:2007} and \gls{KPABE} \cite{Goyal:2006}. In \gls{CPABE}, policies are attached with ciphertext; while, in \gls{KPABE}, attributes are attached with ciphertext. The main issue is that both schemes leave policies and attributes in cleartext, respectively. Unfortunately, policies and attributes in cleartext may reveal private information about the encrypted data.

\section[Dynamic Security Constraints in E-GRANT]{Dynamic Security Constraints in \acrshort{EGRANT}}
\label{sec:egrant-dynamic-security-constraints}
\gls{EGRANT} focuses mainly on enforcing dynamic security constraints. There are two variants of dynamic security constraints: (i) \gls{DSoD} \cite{Kong:2007, Sandhu:1996, Nash:1990} and (ii) \gls{CW} \cite{Brewer:1989}. Both \gls{DSoD} and \gls{CW} can be implemented by maintaining access history for each entity active in the system \cite{Hu:2006}. At each new request, the system has to check that none of the defined constraints are violated by granting the received request with respect to the earlier actions performed by the same (group of) requesters. With each variant of constraints, it is possible to specify contextual conditions i.e., enforcing constraints while taking into account contextual information, such as time and location of the requester. In the following, first we briefly explain both variants and then we describe contextual conditions.

\subsection{Dynamic Separation of Duties}
\gls{DSoD} constraints \cite{Kong:2007, Sandhu:1996, Nash:1990} aim at providing multi-user control over the resources when there is any conflict-of-interest for completing a business process. In the following, we provide a brief description of each category of \gls{DSoD} varying from coarse-grained to fine-grained levels, as discussed in \cite{Schaad:2005}.

\begin{description}

	\item[\gls{SDSoD}] In \gls{SDSoD}, a user may be a member of two mutually exclusive roles but must not be active in both roles simultaneously.
	
	\item[\gls{ObDSoD}] In \gls{ObDSoD}, a user may be active in mutually exclusive roles simultaneously, but must not act in both roles upon a single object.
	
	\item[\gls{OpDSoD}] In \gls{OpDSoD}, a user may be active in mutually exclusive roles simultaneously, but must not get authorised to execute all actions of a business process.
	
	\item[\gls{HBDSoD}] In \gls{HBDSoD}, a user may be active in mutually exclusive roles simultaneously, but the user must not get authorised to execute all actions of a business process involving the same object. For example, a user active in both clerk and manager roles can either issue or approve a particular instance of the \emph{purchase order}. \gls{HBDSoD} combines ideas behind \gls{ObDSoD} and \gls{OpDSoD}, requiring a detailed access history on each object. Thus, it is the most fine-grained category of \gls{DSoD}.

\end{description}

\subsection[Chinese Wall]{\acrlong{CW}}
A \gls{CW} constraint \cite{Brewer:1989} prevents users to access an object belonging to a domain which is in conflict-of-interest with other domain whose object is previously accessed by the same (group of) users. In other words, a \gls{CW} constraint aims at providing confidentiality by preventing illegitimate information flow between domains that are in conflict-of-interest. For instance, let us consider the consultant organisation that provides services to companies that are in conflict-of-interest, say Google and Microsoft. The \gls{CW} constraint will help the consultant organisation to enforce the policy that an employee can work at either Google or Microsoft but cannot work at both companies.

\subsection{Contextual Conditions}
In \gls{EGRANT}, both \gls{DSoD} and \gls{CW} constraints can be enforced under a certain context \cite{Asghar2013-COSE, Asghar2011-ARES, Joshi:2008, Kim:2007, Joshi:2005, Strembeck:2004}. The context can be specified as contextual conditions, which are evaluated at runtime by collecting contextual information. Usually, contextual information includes, but not limited to, the requester's location and the access time. As an example of a \gls{HBDSoD} constraint with contextual conditions, we can consider the case where a user active in two mutually exclusive roles. For instance, two roles clerk and manager cannot issue and approve the same instance of the purchase order \emph{on the same day from the same sub-office}.

\begin{figure} [htp]
\centering
% left bottom right top
\includegraphics[trim=35mm 40mm 35mm 35mm,clip,width=\textwidth]{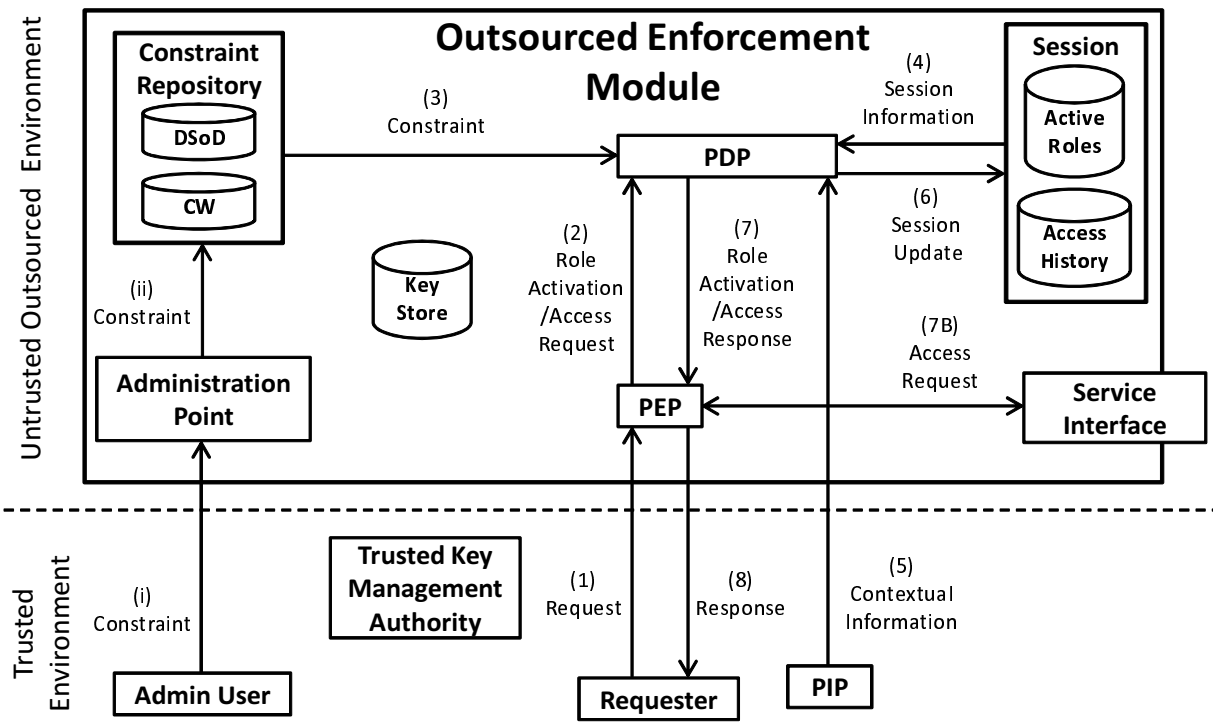}
\caption[The E-GRANT architecture for enforcing dynamic security constraints]{The \gls{EGRANT} architecture for enforcing dynamic security constraints in outsourced environments}
\label{fig:egrant-abstract_picture}
\end{figure}

\section[The E-GRANT Architecture]{The \acrshort{EGRANT} Architecture}
\label{sec:egrant-architecture}
The \gls{EGRANT} architecture aims at enforcing dynamic security constraints in outsourced environments in such a way that contents of constraints, contextual conditions, session information for maintaining access histories and contents of the request are not revealed to cloud providers because they are encrypted. Therefore, the enforcement mechanism can be deployed in the cloud without the need of fully trusting administrators of cloud providers. Our main goal here is to protect the confidentiality of information used by the enforcement mechanism for taking its access control decisions. The rationale behind this is that even if the data is protected (e.g., encrypted) a curious administrator might learn information about the data by inspecting the constraints and access histories that are typically deployed in cleartext. Figure \ref{fig:egrant-abstract_picture} illustrates the \gls{EGRANT} architecture containing the following entities:

\begin{description}

	\item[Admin User] An Admin User is responsible for deploying, updating and deleting dynamic security constraints.
	
	\item[Requester] A Requester is a user that can make requests to access resources and execute actions in the system.

	\item[\gls{OEM}] It is responsible for storing and enforcing dynamic security constraints. In \gls{EGRANT}, the \gls{OEM} is deployed as \gls{SaaS} in the outsourced environment, managed by the cloud provider. We assume that the cloud provider is \emph{honest-but-curious} (as assumed in \cite{Vimercati:2010, Vimercati:2007:VLDB}): that is, it allows the components to follow the protocol for performing requested actions but curious to deduce information about contents of constraints, access histories and requests.
	
	\item[\acrfull{TKMA}] The \gls{TKMA} is a trusted authority responsible for generating keys used for protecting data stored on the \gls{OEM}. For each user (be it an Admin User or a Requester), the \gls{TKMA} generates the client key set and the server key set that are sent to the user and the \gls{OEM}, respectively. The \gls{OEM} stores all server side key sets in the Key Store and is responsible for revoking users. The \gls{TKMA} is only the minimal infrastructure that is run within a trusted environment. However, the \gls{TKMA} can be kept offline because it generates the key only once when any user gets registered with the system.	
	
\end{description}

In \gls{EGRANT}, an Admin User can deploy new constraints and update (or delete) existing constraints. For deploying new constraints, an Admin User sends the (i) Constraint to the \gls{OEM} as shown in Figure \ref{fig:egrant-abstract_picture}. The Administration Point is a component of the \gls{OEM} that receives (i) and then stores it in the Constraint Repository (ii), which is managed by the \gls{OEM}.

A Requester can send a (1) Request to the \gls{OEM} as illustrated in Figure \ref{fig:egrant-abstract_picture}. The \gls{PEP} of the \gls{OEM} receives (1) and then identifies whether (1) is a role activation request or an access request. The \gls{PEP} forwards the (2) Role Activation/Access Request to the \gls{PDP} of the \gls{OEM}. The \gls{PDP} is the core component that can grant the request after evaluating the deployed constraints. For evaluating constraints, the \gls{PDP} fetches the (3) Constraint from the Constraint Repository and the (4) Session Information from the Session component of the \gls{OEM}. The Session component maintains two repositories including Active Roles and the Access History. Active Roles is a repository that keeps record of roles that have been activated for a Requester while the Access History is a repository that maintains what information has been accessed by a Requester. The Session Information can include information about active roles or the access history; thus, it plays a vital role in evaluating the constraints.

The constraints could be enforced under some contextual conditions. A \gls{PDP} evaluates contextual conditions after collecting contextual information, such as time and information about the Requester, e.g., her location. The Policy Information Point (\gls{PIP}) is a trusted entity that provides (5) Contextual Information to the \gls{PDP}. The contextual information must satisfy contextual conditions for the successful enforcement of constraints.

After the evaluation, the \gls{PDP} sends the (7) Role Activation/Access Response to the \gls{PEP}. The response in (7) is either \emph{allow} or \emph{deny} depending on the \gls{PDP} evaluation as explained in Section \ref{sec:egrant-solution-details}. In case of \emph{allow}, the \gls{PDP} updates the session with the role activation or access information by sending the (6) Session Update message to the Session. The \gls{PDP} forwards its decision to the \gls{PEP}. If the decision is \emph{allow}, the \gls{PEP} forwards (7B) Access Request to the Service Interface. Finally, the \gls{PEP} may send the (8) Response to the Requester.

\begin{figure} [htp]
\centering
% left bottom right top
\includegraphics[trim=50mm 80mm 30mm 65mm,clip,width=.9\textwidth]{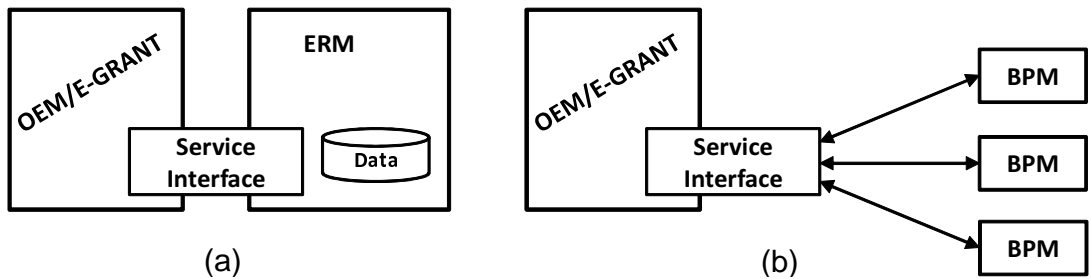}
\caption[Integration of E-GRANT with other services]{Integration of E-GRANT with other services by (a) directly importing the Service Interface (b) remotely invoking the Service Interface}
\label{fig:egrant-integration}
\end{figure}

The Service Interface is a programmable interface that can be used for integrating \gls{EGRANT} with other services. The Service Interface can be used as an entry point for forwarding access requests to the \gls{PEP} for other services. Figure \ref{fig:egrant-integration} shows two possible configurations. In Figure \ref{fig:egrant-integration}(a), the \gls{EGRANT} \gls{OEM} is integrated with an \gls{ERM} \gls{SaaS}. In this scenario, the \gls{OEM} can be used for receiving users' requests, enforcing security constraints and forwarding the granted requests to the \gls{ERM}. Another option is shown in Figure \ref{fig:egrant-integration}(b), where several \gls{BPM} \gls{SaaS} instances remotely invoke the Service Interface of the \gls{OEM} for making access control requests. It should be noted that the mechanisms used by other services to protect their data is out of the scope of \gls{EGRANT}. \gls{EGRANT} is solely responsible for the enforcement of encrypted security constraints. In the following section, we will provide a detailed description on how encrypted security constraints are deployed and enforced by the \gls{OEM}.

\section[Solution Details of E-GRANT]{Solution Details of \gls{EGRANT}}
\label{sec:egrant-solution-details}

\gls{EGRANT} aims at enforcing dynamic security constraints in outsourced environments. The main idea behind \gls{EGRANT} is to employ the encryption scheme for protecting constraints and the sessions while delegating the enforcement mechanism to the \gls{OEM}. The encryption scheme is based on the proxy re-encryption proposed by Dong \emph{et al.} \cite{Dong:2011}. Due to lack of space, we omit details of some operations (including enforcement of S\gls{DSoD}, \gls{ObDSoD} and \gls{OpDSoD}) and cover the most complex operations offered by \gls{EGRANT} including enforcement of \gls{HBDSoD} and \gls{CW}. In the following, we describe how constraints, as well as requests are represented and then we provide technical details for enforcing constraints in an encrypted way.

\begin{figure} [htp]
\centering
% left bottom right top
\includegraphics[trim=65mm 80mm 60mm 70mm,clip,width=.7\textwidth]{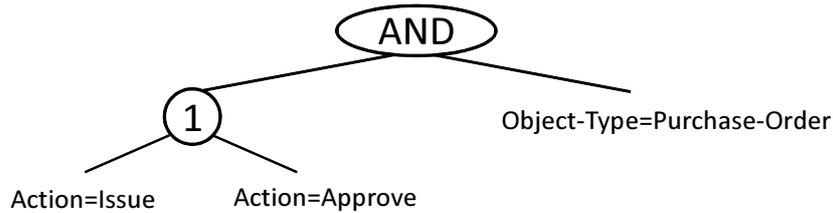}
\caption[An example of History-Based Dynamic Separation of Duties]{An example of \gls{HBDSoD} where a Requester's \emph{action can be 1-of-(Issue,Approve) AND Object-Type is Purchase-Order}}
\label{fig:egrant-hbdsod_example}
\end{figure}

\begin{figure} [htp]
\centering
% left bottom right top
\includegraphics[trim=65mm 95mm 65mm 70mm,clip,width=.7\textwidth]{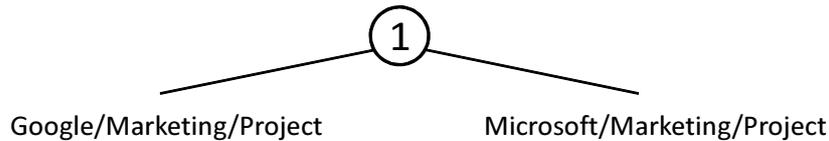}
\caption[An example of Chinese Wall]{An example of \gls{CW} illustrating two domains that are in conflict-of-interest}
%\caption{An example of CW where a Requester can work on \emph{1-of-(Google/Marketing/Project,Microsoft/Marketing/Project)}}
\label{fig:egrant-cw_example}
\end{figure}

\subsection{Representation of Constraints}
For representing both \gls{DSoD} and \gls{CW} constraints, we use the tree structure proposed by Bethencourt \emph{et al.} in \cite{Bethencourt:2007}, which they used for representing \gls{CPABE} policies. Internal nodes of the tree represent AND, OR or threshold gates (e.g., 2 out of 3) while leaf nodes represent values of the condition predicates of a constraint. Figure \ref{fig:egrant-hbdsod_example} illustrates an example of the \gls{HBDSoD} constraint, where a Requester can execute either \emph{issue} or \emph{approve} but not both actions on the same instance of the \emph{purchase order}. Similarly, we can express the \gls{CW} constraint. Figure \ref{fig:egrant-cw_example} illustrates an example of the \gls{CW} constraint, where a Requester can work exclusively on instance of either Google's marketing project or Microsoft's marketing project.

\subsection{Representation of a Request}
The access request can be represented as a tuple $\mathit{REQ} = \langle R, A, O, I \rangle$, where $R$ is role of the Requester, $A$ indicates the action to be taken, $O$ and $I$ describe type of the object being accessed and its instance identifier, respectively. For instance, consider a Requester, active in a role \emph{manager}, takes the \emph{approve} action over the instance of a \emph{purchase order}. The object type $O$ may be a fully qualified name that may include the domain hierarchy an object type may belong to. For example, consider a \gls{CW} constraint, where a Requester (employed by a consultant organisation) cannot work on instances belonging to both Google's marketing project and Microsoft's marketing project. Here, the object type $O$ is \emph{Project} while the domain hierarchy is: \emph{Google/Marketing} and \emph{Microsoft/Marketing}. In case, if it is the role activation request then a Requester just needs to send her role. Thus, the access request is more complex than the role activation request; therefore, we will focus more on the access request in rest of the chapter.

\begin{figure} [htp]
\centering
% left bottom right top
\includegraphics[trim=55mm 50mm 50mm 40mm,clip,width=.85\textwidth]{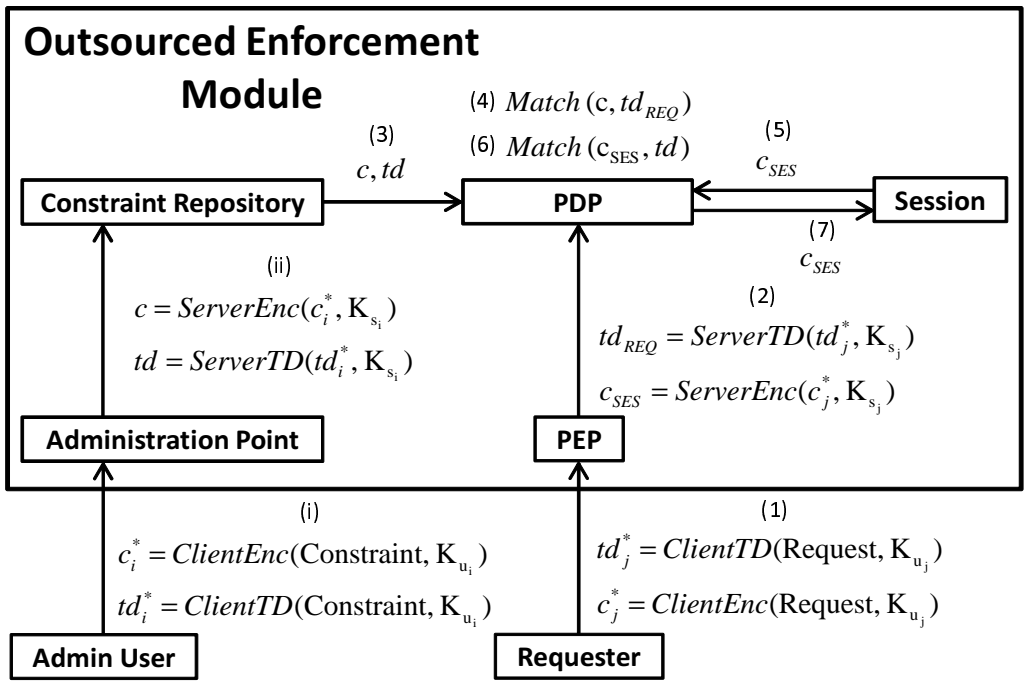}
\caption[The detailed E-GRANT architecture]{The detailed \gls{EGRANT} architecture}
\label{fig:egrant-solution_detail}
\end{figure}

\subsection[Technical Details of E-GRANT]{Technical Details of \gls{EGRANT}}
In this section, we provide technical details of the \gls{EGRANT} architecture as illustrated in Figure \ref{fig:egrant-solution_detail}. The detail of the algorithms in Figure \ref{fig:egrant-solution_detail} can be found in Chapter \ref{cha:espoon} (Section \ref{sec:espoon-algorithmic-details}) while the detail of each phase in the enforcement lifecycle of dynamic security constraints can be found in Section \ref{sec:egrant-algorithmic-details}. \\ \\
\noindent \textbf{Initialisation:} \gls{EGRANT} is based on the proxy re-encryption scheme proposed by Dong \emph{et al.} \cite{Dong:2011}, where each user (including an Admin User and a Requester) gets a client side key set from the \gls{TKMA} while the \gls{OEM} as a proxy server also receives a server side key set corresponding to that user. The \gls{OEM} maintains all these key sets in a Key Store, which can be accessed by different components of the \gls{OEM} including the Administration Point, the \gls{PDP} and the \gls{PEP}. \\ \\
\noindent \textbf{Constraint Deployment:} For deploying a constraint, an Admin User performs the first round of encryption using the client side key set. In this round of encryption, each leaf node of the constraint tree is encrypted while non-leaf nodes representing AND, OR or threshold gates are in cleartext. Next, an Admin User sends the user encrypted tree to the Administration Point of the \gls{OEM} as shown in Figure \ref{fig:egrant-solution_detail} Step (i). After the first round of encryption, constraints are protected but they cannot be enforced yet as they are not in common format. To convert constraints into a common format, the Administration Point of the \gls{OEM} performs the second round of encryption using the server side key set corresponding to the same Admin User who performed the first round of encryption as shown in Figure \ref{fig:egrant-solution_detail} Step (ii). In fact, the second round of encryption by the Administration Point serves as a proxy re-encryption. The common format implies that the constraints get encrypted with the master secret key, which is known neither to any users nor to the \gls{OEM}. Like the first round of encryption, each leaf node of the tree representing the security constraint is re-encrypted. Finally, the re-encrypted constraints are stored by the Constraint Repository.

% why trapdoors while making request
If an encrypted request satisfies any encrypted deployed constraint (i.e., Figure \ref{fig:egrant-solution_detail} Step (4)), then the session information is required to be matched against elements of the constraint (i.e., Figure \ref{fig:egrant-solution_detail} Step (6)). That is, the session information is matched with those elements of the constraint that are not present in the request. For example, let us consider the S\gls{DSoD} constraint, where a user may be a member of two mutually exclusive roles clerk and manager but must not be active in both roles simultaneously. Let us assume that the requester's role is clerk. Since the requester's role is matched against the same role in the constraint, the \gls{OEM} will consult the session to check if the same user is active in manager's role. For performing such a check, \gls{OEM} requires trapdoors of the constraint because only trapdoors could be matched with the encrypted information. That is why, trapdoors are stored along with the encrypted constraint at deployment time. For calculating these trapdoors, an Admin User performs the first round of trapdoor generation using the client side key set for each leaf node in the request (as shown in Figure \ref{fig:egrant-solution_detail} Step (i)) while the \gls{OEM} performs the second round of trapdoor generation using the server side key set corresponding to that Admin User (as shown in Figure \ref{fig:egrant-solution_detail} Step (ii)). The trapdoor representation does not leak any information.  \\ \\
%
% request - user
\noindent \textbf{Making a Request:} For making a request, a Requester generates $\mathit{REQ}$ and transforms it into trapdoors using the client side key set for each element in the request. That is, there is a trapdoor for each element in $\mathit{REQ}$. Finally, $\mathit{REQ}$ is sent over to the \gls{PEP} of the \gls{OEM} as shown in Figure \ref{fig:egrant-solution_detail} Step (1). \\ \\
%
% request - server: second round
\noindent \textbf{Constraint Evaluation:} The deployed constraints are checked when the \gls{OEM} receives a request from any Requester. The request is not in the common format yet and requires another round of the trapdoor generation. In the second round of trapdoor generation, the \gls{PEP} generates the server side trapdoors for each element in $\mathit{REQ}$ (i.e., Figure \ref{fig:egrant-solution_detail} Step (2)). After completing the second round of trapdoor generation, the \gls{PEP} forwards the request to the \gls{PDP}. The \gls{PDP} fetches encrypted constraints from the Constraint Repository (i.e., Figure \ref{fig:egrant-solution_detail} Step (3)) and matches it against the encrypted request (i.e., Figure \ref{fig:egrant-solution_detail} Step (4)). If the constraint is satisfied, then certain elements of the constraint (i.e., all elements except one that is present in the request) are required to be matched against the session information. \\ \\
%
% contextual information
\noindent \textbf{Contextual Conditions:} Optionally, constraints may include contextual conditions (already discussed in detail in Chapter \ref{cha:espoon}). For the evaluation of contextual conditions, the \gls{PDP} might require contextual information, which is fetched from the \gls{PIP}. The \gls{PIP} performs the first round of trapdoor generation using the client side key set\footnote{The \gls{PIP} is considered as a user and gets the client side key set in the same way as a normal user (an Admin User or a Requester) does.}. Let us consider that the required contextual information is current office hour and location of the Requester. We represent each string attribute as a single element. The numerical attributes are represented as a bag of bits, where each numerical attribute of size s-bit is represented by s elements (in the worst case). For the simplicity, we assume that there are total 8 office hours (from 9:00 AM to 5:00 PM) that can be represented with three bits. For instance, the first office hour can be represented as: $\mathit{t:**0}$, $\mathit{t:*0*}$ and $\mathit{t:0**}$; and the last (8th) office hour can be represented as: $\mathit{t:**1}$, $\mathit{t:*1*}$ and $\mathit{t:1**}$. Similarly, the location of the Requester can be represented as: $\mathit{location:office}$. While performing the first round of trapdoor generation, a trapdoor is generated for each element of contextual information. For instance, in the example where contextual information includes office hours (say the first hour) and location of the Requester (say office), a trapdoor is generated for each element including $\mathit{t:**0}$, $\mathit{t:*0*}$, $\mathit{t:0**}$ and $\mathit{location:office}$. After performing the first round of trapdoor generation, the \gls{PIP} sends contextual information to the \gls{PDP}. The \gls{PDP} performs the second round of trapdoor generation for each element of contextual information so that a match can be performed.

% contextual information matching with history
While performing the encrypted match between the encrypted session information and the encrypted constraint/request, the \gls{OEM} does not reveal contents. If contextual information is required to be matched, it is matched in the same way as other elements of the constraint/request are matched against the session information. After checking the session information (i.e., Figure \ref{fig:egrant-solution_detail} Step (6)), if the constraint is not satisfied, the access is permitted and the role activation (or the access) response is sent from the \gls{PDP} to the \gls{PEP} as \emph{allow}. Otherwise, the access is denied and the role activation (or the access) response is sent from the \gls{PDP} to the \gls{PEP} as \emph{deny}. \\ \\
%
% session updation
\noindent \textbf{Updating the Session:} If the evaluation is successful, the \gls{PDP} updates the session to maintain the access history, as well as active roles. For updating the session, the \gls{PDP} requires the request (and contextual information). The Requester may send encrypted request along with the trapdoors of the request as shown in Figure \ref{fig:egrant-solution_detail} Step (1). Alternatively, the \gls{PDP}/\gls{PEP} can collect this information after the \gls{PDP} evaluation is succeeded. In both cases, the \gls{OEM} performs the second round of encryption and finally updates the Session with the encrypted request as shown in Figure \ref{fig:egrant-solution_detail} Step (7). If the requested action is the access request, the \gls{PEP} additionally forwards it to the Service Interface. Finally, the \gls{PEP} may send a response to the Requester. \\ \\
%
% user revocation
\noindent \textbf{User Revocation:} In \gls{EGRANT}, users (both Admin Users and Requesters) do not share any keys and even if a compromised user is removed, there is no need to re-encrypt deployed constraints or re-distribute keys. For removing a user from the system, the Administration Point of the \gls{OEM} takes the user identifier and then removes the server side key corresponding to that user from the Key Store.

\section[Algorithmic Details of E-GRANT]{Algorithmic Details of \gls{EGRANT}}
\label{sec:egrant-algorithmic-details}
In this section, we identify all phases describing the enforcement lifecycle of dynamic security constraints in outsourced environments. For each of these phases, we list all of its algorithms in detail. In fact, these algorithms constitute the proposed schema that is based on \cite{Dong:2011}.

% TODO remove $ $: added to fix a format issue
\subsection{The Initialisation Phase}
In this phase, the system is initialised by the \gls{TKMA}. During the system initialisation, the system level master key and public parameters are generated. This phase consists of only one algorithm called \textbf{Init} illustrated in Algorithm \ref{algo:erbac-init}. After running this algorithm, the \gls{TKMA} publicises the public parameters $params = (\mathbb{G}, g, q, h,$ $H, f)$ and keeps securely the master secret key $msk = (x, s)$.

\subsection{The Key Generation Phase}
During the key generation phase, the keying material is generated for each user including an Admin User and a Requester by the \gls{TKMA}. This phase consists of only one algorithm called \textbf{KeyGen} and is illustrated in Algorithm \ref{algo:erbac-keygen}. After running the \textbf{KeyGen} algorithm, the \gls{TKMA} generates two key sets: $K_{u_i}$ and $K_{s_i}$ corresponding to user $i$. The \gls{TKMA} securely transmits $K_{u_i}$ and $K_{s_i}$ to the user $i$ and the \gls{OEM}, respectively. Each user $i$ receives the user side key set $K_{u_i}$ and stores it securely as it serves as the private key for her. The Administration Point of the \gls{OEM} receives the server side key set $K_{s_i}$ corresponding to user $i$ and inserts it in the Key Store, where the Key Store is updated as: $KS \leftarrow KS \cup K_{s_i}$. The Key Store of the \gls{OEM} is initialised as: $KS \leftarrow \Phi$.

% client generated constraint

\begin{algorithm} [htp]
{\algofontsize
\caption{\textbf{ClientGeneratedConstraint}}

\label{algo:egrant-client-generated-constraint}

\begin{algorithmic}[1]

\INPUT \emph{It transforms cleartext constraints into the (encrypted) client generated constraints, which are sent to the Administration Point as shown in Figure \ref{fig:egrant-abstract_picture} Step (i).}

\Require The constraint tree $SCT$, the client side key set $K_{u_i}$ corresponding to Admin User $i$ and the public parameters $params$.

\Ensure The client generated constraint tree $SCT_{C_i}$.

\medskip

\State $SCT_{C_i} \leftarrow SCT$

\For {each leaf-node element $e$ in tree $SCT_{C_i}$}

	\State $c^*_i (e) \leftarrow$ call \textbf{ClientEnc} ($e$, $K_{u_i}$, $params$)
	\State $td^*_i (e) \leftarrow$ call \textbf{ClientTD} ($e$, $K_{u_i}$, $params$)
	\State $ug(e) \leftarrow (c^*_i (e), td^*_i (e))$
	\State replace $e$ of $SCT_{C_i}$ with $ug(e)$

\EndFor

\Return $SCT_{C_i}$

\end{algorithmic}
}
\end{algorithm}

\subsection{The Constraint Deployment Phase}
During this phase, a constraint is deployed by an Admin User. Each constraint is deployed in two phases; therefore, this phase consists of two algorithms: Algorithm \ref{algo:egrant-client-generated-constraint} and Algorithm \ref{algo:egrant-server-generated-constraint} called \textbf{ClientGeneratedConstraint} and \textbf{ServerGeneratedConstraint}, respectively. The constraint is first transformed into a tree structure as already explained in Section \ref{sec:egrant-solution-details}. After performing transformation, each leaf node of this tree $SCT$ is encrypted (by running \textbf{ClientEnc} described in Chapter \ref{cha:espoon} as Algorithm \ref{algo:erbac-client-enc}) and client generated trapdoors (by running \textbf{ClientTD} described in Chapter \ref{cha:espoon} as Algorithm \ref{algo:erbac-client-td}) are also calculated using the client side key set $K_{u_i}$ corresponding to Admin User $i$ as shown in Algorithm \ref{algo:egrant-client-generated-constraint}, which is run by the Admin User. Finally, the client generated constraint $SCT_{C_i}$ is sent over to the Administration Point of the \gls{OEM} as illustrated in Figure \ref{fig:egrant-abstract_picture} Step (i).

% server generated constraint

\begin{algorithm} [htp]
{\algofontsize
\caption{\textbf{ServerGeneratedConstraint}}

\label{algo:egrant-server-generated-constraint}

\begin{algorithmic}[1]

\INPUT \emph{It re-encrypts the client generated constraints into the server generated constraints, which are finally deployed by the Administration Point as shown in Figure \ref{fig:egrant-abstract_picture} Step (ii).}

\Require The client generated constraint tree $SCT_{C_i}$ and Admin User $i$.

\Ensure The server generated constraint tree $SCT_{S}$.

\medskip

\State $K_{s_i} \leftarrow KS[i]$ {\algofontsize \Comment{retrieve the server side key corresponding to Admin User $i$}}

\State $SCT_{S} \leftarrow SCT_{C_i}$

\For {each leaf-node client generated element $ug(e) = (c^*_i (e), td^*_i (e))$ in tree $SCT_{S}$}

	\State $c(e) \leftarrow$ call \textbf{ServerReEnc} ($c^*_i (e)$, $K_{s_i}$)
	
	\State $td(e) \leftarrow$ call \textbf{ServerTD} ($td^*_i (e)$, $K_{s_i}$)
	
	\State $sg(e) \leftarrow (c(e), td(e))$

	\State replace $ug(e)$ of $SCT_{S}$ with $sg(e)$

\EndFor

\Return $SCT_{S}$

\end{algorithmic}
}
\end{algorithm}

The Administration Point of the \gls{OEM} receives the client encrypted constraint $SCT_{C_i}$ and performs another round of encryption (by running \textbf{ServerReEnc} described in Chapter \ref{cha:espoon} as Algorithm \ref{algo:erbac-server-re-enc}) and the trapdoor generation (by running \textbf{ServerTD} described in Chapter \ref{cha:espoon} as Algorithm \ref{algo:erbac-server-td}) using the server side key set $K_{s_i}$ corresponding to Admin User $i$ as shown in Algorithm \ref{algo:egrant-server-generated-constraint}. After running Algorithm \ref{algo:egrant-server-generated-constraint}, the Administration Point stores the server generated constraints in the Constraint Repository on the \gls{OEM} as illustrated in Figure \ref{fig:egrant-abstract_picture} Step (ii).

% client generated request

\begin{algorithm} [htp]
{\algofontsize
\caption{\textbf{ClientGeneratedRequest}}

\label{algo:egrant-client-request}

\begin{algorithmic}[1]

\INPUT \emph{It transforms the cleartext request into the client generated request, which is sent to the \gls{PEP} as shown in Figure \ref{fig:egrant-abstract_picture} Step (1).}

\Require The request $\mathit{REQ}$ containing list of elements, the client side key set $K_{u_i}$ corresponding to Requester $i$ and the public parameters $params$.

\Ensure The client generated request ${\mathit{REQ}}_{C_i}$.

\medskip

\State ${\mathit{REQ}}_{C_i} \leftarrow REQ$

\For {each element $e$ in list ${\mathit{REQ}}_{C_i}$}

	\State $td^*_i (e) \leftarrow$ call \textbf{ClientTD} ($e$, $K_{u_i}$, $params$)
	
	\State $c^*_i (e) \leftarrow$ call \textbf{ClientEnc} ($e$, $K_{u_i}$, $params$)
	
	\State $\mathit{req}^*_i(e) \leftarrow (td^*_i (e), c^*_i (e))$
	
	\State replace $e$ of ${\mathit{REQ}}_{C_i}$ with $\mathit{req}^*_i(e)$

\EndFor

\Return ${\mathit{REQ}}_{C_i}$

\end{algorithmic}
}
\end{algorithm}

\subsection{The Request Phase}
In this phase, Requester $i$ makes a request $\mathit{REQ}$, which is enciphered using her private key set $K_{u_i}$. This phase consists of one algorithm called \textbf{ClientGeneratedRequest} illustrated in Algorithm \ref{algo:egrant-client-request} in which each element in $\mathit{REQ}$ (assuming $\mathit{REQ}$ also includes contextual information) is transformed into a trapdoor (by running \textbf{ClientTD} described in Chapter \ref{cha:espoon} as Algorithm \ref{algo:erbac-client-td}). Furthermore, each element in $\mathit{REQ}$ is encrypted (by running \textbf{ClientEnc} described in Chapter \ref{cha:espoon} as Algorithm \ref{algo:erbac-client-enc}) because it is required to be stored in the session provided it is granted. Finally, the client request ${\mathit{REQ}}_{C_i}$ is sent over to the \gls{OEM}.

% server eval and session up

\begin{algorithm} [htp]
{\algofontsize
\caption{\textbf{ConstraintEval-SessionUp}}

\label{algo:egrant-evaluate-constraint-session-up}

\begin{algorithmic}[1]

% \INPUT \emph{It fetches the server generated constraints (see Figure \ref{fig:egrant-abstract_picture} Step (3)), transforms the client generated request into the server generated request, then matches the constraints against the request and finally sends either \emph{yes} or \emph{no} to the \gls{PEP} as shown in Step (7).}

\INPUT \emph{It fetches the encrypted constraints (see Figure \ref{fig:egrant-abstract_picture} Step (3)), transforms the client request into the server generated request, then matches constraints with the request.}

\Require The server generated constraint tree $SCT_{S}$, the list of client generated trapdoor ${\mathit{REQ}}_{C_i}$, Requester $i$ and session $S$.

\Ensure $\mathit{true}$ or $\mathit{false}$.

\medskip

\State $K_{s_i} \leftarrow KS[i]$ {\algofontsize \Comment{retrieve the server side key corresponding to Requester $i$}} \label{line:egrant-get-server-side-key}

\State ${\mathit{REQ}}_{S} \leftarrow REQ_{C_i}$ \label{line:egrant-re-enc-s-td-start}

\For {each client generated request element $\mathit{req}^*_i(e).td^*_i (e)$ in list ${\mathit{REQ}}_{S}$}
%\For {each client generated request element $\mathit{req}^*_i(e). = (td^*_i (e), c^*_i (e))$ in list ${\mathit{REQ}}_{S}$}
	\State $td(e) \leftarrow$ call \textbf{ServerTD} ($td^*_i (e)$, $K_{s_i}$)
%	\State $c(e) \leftarrow$ call \textbf{ServerReEnc} ($c^*_i (e)$, $K_{s_i}$)
	
%	\State $\mathit{req}(e) \leftarrow (td(e), c(e))$
	
	\State replace $\mathit{req}^*_i(e).td^*_i (e)$ of ${\mathit{REQ}}_{S}$ with $td(e)$
	
\EndFor \label{line:egrant-re-enc-s-td-end}

\State $EncryptedTree \leftarrow SCT_{S}$ \label{line:egrant-tree-match-start}

\State Add field $decision$ to each node of $\mathit{EncryptedTree}$

\For {each node $n$ in tree $\mathit{EncryptedTree}$}
	\State $n.decision \leftarrow null$ {\algofontsize \Comment{initialise decision field with $null$}}
%	\State $n.satisfiable \leftarrow false$
\EndFor

\State call \textbf{CheckTreeSatisfiability} ($EncryptedTree.root$, $\mathit{EncryptedTree}$, ${\mathit{REQ}}_{S}$) \label{line:egrant-tree-match-end}

\If {$EncryptedTree.root.decision \stackrel{?}{=} true$} \label{line:egrant-tree-matched}

	\State $TrapdoorList \leftarrow$ extract trapdoors from $\mathit{EncryptedTree}$ that needs to be searched in session $s$ \label{line:egrant-extract-tds}
	
	\State $record \mhyphen found \leftarrow false$ \label{line:egrant-session-search-start}
	
	\For {each record $r$ in session $S$}
	
%		\State $match \leftarrow false$
		
		\For {each server encrypted element $c(e)$ in $r$ to be matched with $td(e)$ in $TrapdoorList$}
			
			\State $match \leftarrow$ call \textbf{Match} ($child.c(e)$, ${\mathit{REQ}}_{S}.td(e)$)
			
			\If {$match \stackrel{?}{=} false$}
				\State break;
			\EndIf
			
		\EndFor
		
		\If {$match \stackrel{?}{=} true$}
			\State $record \mhyphen found \leftarrow true$
			\State break;
		\EndIf
		
	\EndFor \label{line:egrant-session-search-end}
	
	\If {$record \mhyphen found \stackrel{?}{=} true$} \label{line:egrant-session-match-result}
		\Return $\mathit{false}$ \label{line:egrant-no-action}
	\EndIf
	
\EndIf

\medskip {\algofontsize \Comment{steps for updating session}}

%\State $r \leftarrow$ get all corresponding $c^*_i (e)$ from ${\mathit{REQ}}_{S}$ \label{line:egrant-session-up-start}
\State $r \leftarrow \phi$ \label{line:egrant-session-up-start}

\For {each client encrypted request element $\mathit{req}^*_i(e).c^*_i (e)$ in list ${\mathit{REQ}}_{S}$}
	\State $c(e) \leftarrow$ call \textbf{ServerReEnc} ($c^*_i (e)$, $K_{s_i}$)
	\State $r \leftarrow r \cup c(e)$
\EndFor

\State $S \leftarrow S \cup r$  \label{line:egrant-session-up-end} {\algofontsize \Comment{session updation}}
	
\Return $\mathit{true}$ \label{line:egrant-session-up-action}

\end{algorithmic}
}
\end{algorithm}

% check tree satisfiability

\begin{algorithm} [htp]%{tbp}{loa}[H] %{t}{lop} % [h]
{\algofontsize
\caption{\textbf{CheckTreeSatisfiability}}

\label{algo:egrant-check-satisfiability}

\begin{algorithmic}[1]

\INPUT \emph{It checks whether the encrypted constraint satisfies the encrypted request.}

\Require The root node $n$ of encrypted constraint tree $\mathit{EncryptedTree}$ and the list of server generated trapdoors of request ${\mathit{REQ}}_{S}$.

\Ensure $\mathit{true}$ or $\mathit{false}$.

\medskip

\If {$n \stackrel{?}{=} null$} 

	\Return $\mathit{true}$ {\algofontsize \Comment{if $null$ constraint then it trivially satisfies the request}}
	
\EndIf

\If {$n.decision \neq null$}

	\Return $n.decision$ {\algofontsize \Comment{if decision is already made then return it}}
	
\EndIf

\If {$isLeaf(n) \stackrel{?}{=} true$}

	\State $n.decision \leftarrow$ call \textbf{Match} ($n.c(e)$, ${\mathit{REQ}}_{S}.td(e)$)
	
	\Return $n.decision$ {\algofontsize \Comment{if it is leaf node then perform matching and return its decision}}
	
\EndIf

\State $k' \leftarrow 0$

\For {each $child$ of $n$ in $\mathit{EncryptedTree}$} {\algofontsize \Comment{if it is non-leaf node then call this function recursively for each of its child}}
	\If {call \textbf{CheckTreeSatisfiability} ($child$, $\mathit{EncryptedTree}$, ${\mathit{REQ}}_{S}$) $\stackrel{?}{=} true$}
		\State $k' \leftarrow k' + 1$
	\EndIf
\EndFor

\If {($n.gate \stackrel{?}{=} OR$ and $k' \geq 1$) or $n.k \stackrel{?}{=} k'$}
	\State $n.decision \leftarrow true$ {\algofontsize \Comment{set decision as $true$ if (a) node's gate is $OR$ and one of its child is satisfied or (b) the number of children $n$ has is equal to number of satisfied elements, i.e., the case of both $AND$ and $threshold$ gates}}
\Else
	\State $n.decision \leftarrow false$
\EndIf

%\State $node.mark \leftarrow true$

\Return $n.decision$

\end{algorithmic}
}
\end{algorithm}

\subsection{The Constraint Evaluation and Session Update Phase}
This is the core phase in which constraints are evaluated and the session is updated with the information within the request, provided the request is granted. This phase consists of one algorithm called \textbf{ConstraintEval-SessionUp} illustrated in Algorithm \ref{algo:egrant-evaluate-constraint-session-up}, which is run by the \gls{PEP} of the \gls{OEM}. After receiving the client request ${\mathit{REQ}}_{C_i}$, the \gls{PEP} first retrieves the server side key $K_{s_i}$ corresponding to Requester $i$ (Line \ref{line:egrant-get-server-side-key}). The \gls{PEP} then performs the second round of trapdoor generation (by running \textbf{ServerTD} described in Chapter \ref{cha:espoon} as Algorithm \ref{algo:erbac-server-td}) for each element in ${\mathit{REQ}}_{C_i}$ (Line \ref{line:egrant-re-enc-s-td-start}-\ref{line:egrant-re-enc-s-td-end}). After performing the second round of trapdoor generation, the server generated request ${\mathit{REQ}}_S$ is matched against the deployed constraint $SCT_S$ (Line \ref{line:egrant-tree-match-start}-\ref{line:egrant-tree-match-end}), where it is mainly checked if the encrypted tree $\mathit{EncryptedTree}$ of the deployed constraint $SCT_S$ is satisfied by the encrypted request ${\mathit{REQ}}_S$ (Line \ref{line:egrant-tree-match-end}). The detail how $\mathit{EncryptedTree}$ is matched against ${\mathit{REQ}}_S$ is provided in Algorithm \ref{algo:egrant-check-satisfiability}.

If $SCT_S$ is matched against ${\mathit{REQ}}_S$ (Line \ref{line:egrant-tree-matched}), then the certain trapdoors of the deployed constraint are extracted (Line \ref{line:egrant-extract-tds}) and then matched against records in the Active Roles repository (in case of role activation request) or the Access History repository (in case of access request) of the session (Line \ref{line:egrant-session-search-start}-\ref{line:egrant-session-search-end}). If the match (in Line \ref{line:egrant-session-match-result}) is successful (assuming the constraint with 1-out-of-n condition for roles or actions), no action is taken and $\mathit{false}$ is returned (Line \ref{line:egrant-no-action}), indicating that the session is not updated; otherwise, each element of ${\mathit{REQ}}_S$ is re-encrypted (by running \textbf{ServerReEnc} described in Chapter \ref{cha:espoon} as Algorithm \ref{algo:erbac-server-re-enc}) and then the session is updated with the encrypted information of active roles or the access history (Line \ref{line:egrant-session-up-start}-\ref{line:egrant-session-up-end}) and $\mathit{true}$ is returned (Line \ref{line:egrant-session-up-action}), indicating that the session is updated by running Algorithm \ref{algo:egrant-evaluate-constraint-session-up}.

\section{Discussion}
\label{sec:egrant-discussion}
%\subsection{Revealing Structure of \gls{RBAC} Constraints}

This section provides the discussion about security aspects of \gls{EGRANT} including information disclosure and the collusion attack.

\subsection{Information Disclosure}
In \gls{EGRANT}, a curious \gls{OEM} may deduce the structure of security constraints. That is, a curious \gls{OEM} may learn what gates (AND, OR and k-of-n) are used in security constraints. However, the most important information is actually contents of security constraints that are not revealed to the \gls{OEM}. To partially resolve the problem of revealing structure, we may include some dummy elements in the constraint. Furthermore, a curious \gls{OEM} may also deduce how many elements are present (but does not learn about contents of elements) in the request or contextual information; once again, the Requester or the \gls{PIP} can include some dummy elements in order to obfuscate the number of elements present in the request or contextual information, respectively.

\subsection{Collusion Attack}
In \gls{EGRANT}, a single compromised user (either an Admin User or a Requester) may recover the master secret key by colluding with the \gls{OEM}. One way to withstand the collusion attack is to split the client side key set into two parts; where, one part is given to the user while the other part is managed by the organisation gateway to access the \gls{OEM}. In this case, the organisation gateway is assumed trusted. The other way to withstand the collusion attack is to consider the trusted hardware for storing the client side key set.

\section[Performance Analysis of E-GRANT]{Performance Analysis of \gls{EGRANT}}
\label{sec:egrant-performance_evaluation}

In this section, we show the effectiveness of \gls{EGRANT} for enforcing dynamic security constraints by quantifying the performance overhead incurred by the cryptographic operations performed at both the client and the server sides. During this performance evaluation, we are not taking into account the latency introduced by the network. In the following, we first describe implementation details of the prototype we have developed. Next, we show the performance evaluation of: (i) deploying dynamic security constraints, (ii) making a request, (iii) evaluating dynamic security constraints and (iv) finally updating session with the information within the request.

\subsection[Implementation Details of E-GRANT]{Implementation Details of \gls{EGRANT}}
We have developed a prototype of \gls{EGRANT} for enforcing dynamic security constraints. The prototype is implemented in Java $1.6$. For this prototype, we have designed all the components of the architecture required for deploying and evaluating constraints. In short, we have implemented all algorithms presented in Section \ref{sec:egrant-algorithmic-details}.

We have tested our \gls{EGRANT} prototype on a single node based on an Intel Core2 Duo $2.2$ GHz processor with $2$ GB of RAM, running Microsoft Windows XP Professional version $2002$ Service Pack $3$. The values of the execution time shown in the following graphs are averaged over $1000$ iterations.

\subsection{Performance Analysis of Deploying Dynamic Security Constraints}
In this section, we analyse the performance of deploying dynamic security constraints. In order to deploy a constraint, an Admin User performs on the client side the first round of encryption and the trapdoor generation for each element in the constraint as explained in Section \ref{sec:egrant-solution-details} (see Algorithm \ref{algo:egrant-client-generated-constraint}) and sends the client generated constraint to the \gls{OEM}. The Administration Point of the \gls{OEM} receives the client generated constraint and performs the second round of encryption and the trapdoor generation for each element in the client generated constraint (see Algorithm \ref{algo:egrant-server-generated-constraint}). Finally, the server generated constraint is sent to the Constraint Repository of the \gls{OEM}.

We measure the performance of deploying both types of security constraints including \gls{HBDSoD} and \gls{CW}. The simplest \gls{HBDSoD} constraint is defined with two actions at least, meaning a user cannot execute both actions. For increasing complexity of the \gls{HBDSoD} constraint, we can consider more than two actions using the following notation: $\gls{HBDSoD}(Ya)$, where $Y$ ($\geq$ 2) denotes the number of actions in the constraint. Similarly, the simplest \gls{CW} constraint is defined at the object level, meaning a user cannot access an instance of an object whose instance has already been accessed. In order to increase the complexity of the \gls{CW} constraint, we can include the domain hierarchy. Generally, the \gls{CW} constraint can be represented as: $CW(Zd/o)$, where $Z$ ($\geq$ 0) denotes the number of domains that may be present in the domain hierarchy. If the constraint is at the object level, the value of $Z$ will be 0 and constraint would become \gls{CW}(o). However, if the constraint includes any domains, then the value of $Z$ will be more than 0. For instance, if there is one domain then the constraint would be represented as \gls{CW}(d/o). Similarly, if there are two domains (i.e., one domain and one subdomain) in the domain hierarchy of an object then the constraint would be represented as \gls{CW}(2d/o) and so on. Asymptotically, the complexities of deploying \gls{HBDSoD} and \gls{CW} constraints are ${\Theta}( Y )$ and ${\Theta}( Z )$, respectively.

\begin{figure} [htp]
\centering
% left bottom=-5 right top
\includegraphics[trim=0mm -5mm 0mm 0mm,clip,width=.5\textwidth]{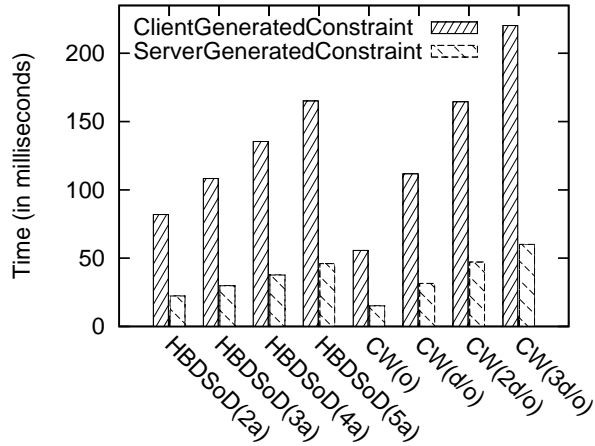}
\caption{Performance overhead of deploying dynamic security constraints}
\label{fig:egrant-constraints-deployment}
\end{figure}

Figure \ref{fig:egrant-constraints-deployment} indicates the performance overhead incurred by deploying constraints on both the client and the server sides. During the performance evaluation, we consider both \gls{HBDSoD} and \gls{CW} constraints, each with varying level of complexity, where number of actions in the \gls{HBDSoD} constraint are varied from 2 to 5 (with step size 1) and number of domains in the \gls{CW} constraint are varied from 0 to 3 (with step size 1), respectively. As we can expect, the performance overhead of each type of constraint grows linearly if we gradually increase its complexity. Furthermore, we can observe that algorithms on the client side take more time as compared to that of the server side for deploying any type of constraints. This is mainly due to the fact the client side performs more complex cryptographic operations such as random number generations and hash calculations (as shown in Algorithm \ref{algo:erbac-client-enc} and Algorithm \ref{algo:erbac-client-td} in Chapter \ref{cha:espoon}) than the respective algorithms on the server side (as shown in Algorithm \ref{algo:erbac-server-re-enc} and Algorithm \ref{algo:erbac-server-td} in Chapter \ref{cha:espoon}). However, these operations are executed only when the Admin User has to deploy a new constraint or update existing ones. On the other hand, constraints are evaluated every time a request is made. Thus, the performance of generating requests and evaluating constraints, which are measured in the following sections, is of great importance, considering the fact that it will impact the latency for providing access to the data.

\subsection{Performance Analysis of Generating Requests}

\begin{figure} [htp]
\centering
% left bottom right top
\includegraphics[width=.5\textwidth]{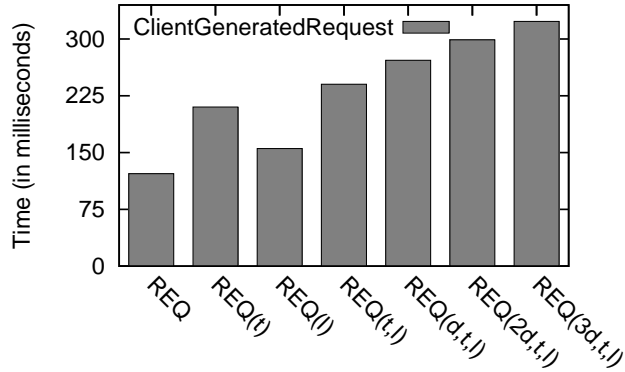}
\caption[Performance overhead of generating access requests]{Performance overhead of generating access requests on the Requester's client side}
\label{fig:egrant-integrated-request}
\end{figure}

In this section, we analyse the performance of generating access requests on the Requester's client side. To make the access request, a Requester has to generate the $\mathit{REQ} = \langle R, A, O, I \rangle$ tuple representing that role $R$ is requesting to perform action $A$ on instance $I$ of object type $O$. Each element of $\mathit{REQ}$ is transformed into trapdoors, necessary for performing the match against encrypted \gls{HBDSoD} or \gls{CW} constraints deployed on the \gls{OEM}. The trapdoor representation does not leak information on elements of $\mathit{REQ}$. Furthermore, each element of $\mathit{REQ}$ is also encrypted, necessary for storing the $\mathit{REQ}$ tuple as encrypted in the session after $\mathit{REQ}$ is granted. The time required to generate such a tuple (by running Algorithm \ref{algo:egrant-client-request}) is around 120 \gls{ms} as shown in the graph of Figure \ref{fig:egrant-integrated-request}.

The \gls{PDP} might need contextual information to make the decision whether the requested action is permitted based on deployed constraints. One way to provide such information is to send the required contextual information together with the $\mathit{REQ}$ tuple. In this case, the client side of the Requester takes the responsibility to generate the trapdoors of contextual information. The other option is to let the \gls{PDP} requests contextual information to the \gls{PIP} (running in the trusted environment) when such information is needed. The former option requires fewer interactions because the \gls{PDP} has already all required information. However, this comes at the price for the Requester's client side of generating extra encrypted data (the trapdoor representation for contextual information). The latter option requires more interaction since the \gls{PDP} has to contact the \gls{PIP}. However, this happens only if contextual information is really required by the \gls{PDP}.

In our experiments, we considered case in which the contextual information is included with every $\mathit{REQ}$ tuple. We selected two types of contextual information: the time and the location of the Requester. As we explained in Section \ref{sec:egrant-solution-details}, the time $t$ is represented as three elements indicating the office hour while the location $l$ is represented as a single string element.

The graph in Figure \ref{fig:egrant-integrated-request} shows the performance overhead incurred at the Requester's client when the $\mathit{REQ}$ tuple contains the value of time $t$ ($\mathit{REQ}(t)$ in the graph) and location $l$ ($\mathit{REQ}(l)$ in the graph). As can be seen in the graph, when the value of time is added to the $\mathit{REQ}$ tuple, there is more performance overhead to be incurred as compared to that of the location because the time value $t$ is represented as three elements, requiring generation of three trapdoors. On the other hand, the value $l$ of the location is represented by just a single element, requiring generation of only a single trapdoor. We also measured the case in which both time and location trapdoors are generated with the $\mathit{REQ}$ tuple and the overhead is combination of two previous cases ($\mathit{REQ}(t,l)$ in the graph).

When \gls{CW} constraints are enforced, it might be needed to include additional information about the target resource within the $\mathit{REQ}$ tuple. This additional information is the domain hierarchy an object type may belong to. In the domain hierarchy, there may be multiple levels of domains. The trapdoors representing this information need also to be generated by the Requester's client. We performed experiments where together with the time and location, also domain information have been added to the $\mathit{REQ}$ tuple. Moreover, we also varied the depth of the domain hierarchy from one domain level (represented as $\mathit{REQ}(t,l,d)$) to three levels (represented as $\mathit{REQ}(t,l,3d)$). The last three values in Figure \ref{fig:egrant-integrated-request} provide the measurements for these cases. As it is quite obvious, the performance overhead of generating these requests increases linearly with the increase in domains levels. However, it should be noticed that even in the worst case (where time, location and three domain levels are inserted in the $\mathit{REQ}$ tuple), the average time for generating a request is still below 325 \gls{ms}. In the worst case, the request generation phase takes ${\Theta}( Z )$.

\subsection{Performance Analysis of Evaluating Dynamic Security Constraints}

In this section, we analyse performance of evaluating security constraints on the \gls{OEM}. For evaluating constraints, the request coming from the Requester is first transformed into the common format by performing the second round of trapdoor generation (see Algorithm \ref{algo:egrant-evaluate-constraint-session-up}). During the trapdoor generation, each client generated trapdoor is transformed into the server generated trapdoor as illustrated in Algorithm \ref{algo:erbac-server-td} of Chapter \ref{cha:espoon}. This second round of encryption is necessary to perform the matching between the trapdoors of the request and the encrypted constraints. In the following, we analyse the performance overheads of evaluating both \gls{HBDSoD} and \gls{CW} constraints. \\ \\
\noindent \textbf{Evaluating \gls{HBDSoD} Constraints:} First of all, let us make a concrete example of the enforcement of \gls{HBDSoD} constraints to understand what operations are executed at the \gls{OEM}. Let us assume a Requester makes a request $\mathit{REQ}$ for executing the action \emph{approve} on the object type \emph{purchase order}. As an example of a \gls{HBDSoD} constraint, let us consider one that limits a Requester to execute only one action out of the two actions \emph{issue} and \emph{approve} that can be executed on a particular instance of a \emph{purchase order}. First, the \gls{PDP} matches the object type in $\mathit{REQ}$ with the object type of the deployed constraints in the Constraint Repository. If the match is successful, the \gls{PDP} will match the action in $\mathit{REQ}$ with one of the action specified in the \gls{HBDSoD} constraint. On the second successful match, the \gls{PDP} has to check that the Requester has not executed the \emph{issue} action on this specific instance of \emph{purchase order} in the past. To perform this check, the \gls{PDP} searches in the Access History to find all records where the object type and instance match with that of $\mathit{REQ}$ tuple. If such a record is found then the \gls{PDP} checks if the action value in the records matches the k-out-of-n condition of the \gls{HBDSoD} constraint. In particular, in our example it means the \gls{PDP} searches in the Access History to find any records containing action \emph{approve}. If this is the case, the constraint is violated and the \gls{PDP} will not grant the action. Otherwise, the Requester can \emph{issue} the \emph{purchase order}.

From the above example, it is clear that the performance of enforcing a constraint depends on three main factors. The first factor is the number of constraints deployed in the Constraint Repository. When a request arrives, the \gls{PDP} has to find in the repository a matching constraint. Finding a matching constraint clearly depends on the number of constraints in the repository. The second factor is the number of elements specified in the constraint. These elements can include two or more actions that could be executed only once by a Requester on a given instance of an object. Moreover, also contextual information can be taken into account. Finally, the other major factor is the number of records in the Access History that the \gls{PDP} has to search to check whether a given constraint is violated or not. Asymptotically, the enforcement of \gls{HBDSoD} constraints takes $O(Y \cdot c \cdot r)$, where $C$ is the number of constraints deployed in the repository and $r$ is the number of records in the Access History.

% server evaluation
\begin{figure}[htp]
\centering
\subfigure[]{
\includegraphics[width=.7\textwidth]{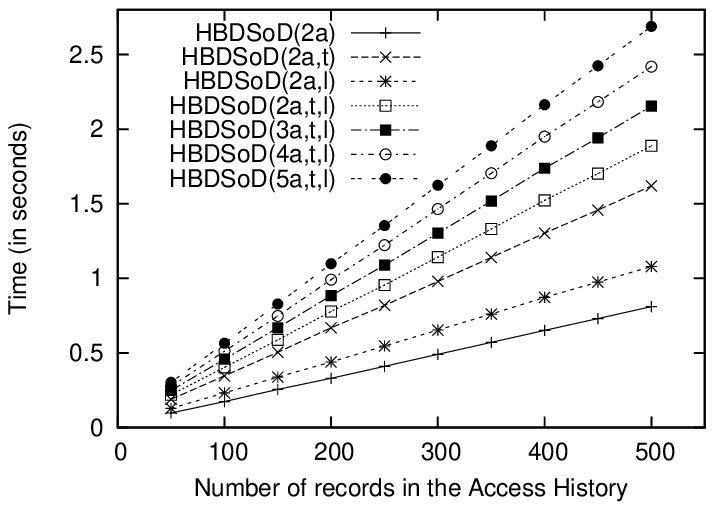} % .48
\label{fig:egrant-constraints-evaluation-server-hbdsod}
}
\subfigure[]{
\includegraphics[width=.7\textwidth]{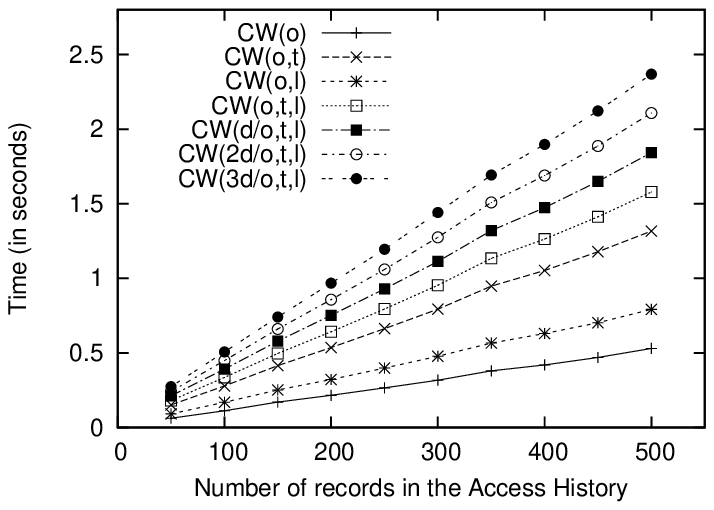} % .48
\label{fig:egrant-constraints-evaluation-server-cw}
}
\caption[Performance overhead of evaluating dynamic security constraints]{Performance overhead of evaluating dynamic security constraints \subref{fig:egrant-constraints-evaluation-server-hbdsod} \gls{HBDSoD} and \subref{fig:egrant-constraints-evaluation-server-cw} \gls{CW} on the \gls{OEM}}
\label{fig:egrant-constraint-evaluation}
\end{figure}

To measure the performance overhead, we performed the following experiments. We deployed 100 different \gls{HBDSoD} constraints in the repository such that the one that matches the incoming request is the last one. This, of course, represents the worst case scenario. We also believe that 100 different constraints is way beyond the typical needs of an enterprise. To study how the complexity of the constraint specification and number of records in the Access History affect the performance of the constraint evaluation, we execute several runs of our experiments varying the constraint complexity and number of records. Figure \ref{fig:egrant-constraints-evaluation-server-hbdsod} shows the evaluation time in seconds in different settings. As we can observe in Figure \ref{fig:egrant-constraints-evaluation-server-hbdsod}, the evaluation time increases with the increase in the number of actions in the constraint (from 2 actions up to 5) and when contextual information such as time $t$ and/or location $l$ of the Requester are also considered. Similarly, the evaluation time increases with the increase in the number of records in the Access History. \\ \\
\noindent \textbf{Evaluating \gls{CW} Constraints:} A \gls{CW} constraint enforces that a Requester cannot gain access to two mutually exclusive objects. When a request $\mathit{REQ}$ tuple is received, the \gls{PDP} has to search the \gls{CW} constraints relevant to the object type specified in the request tuple. Basically, the object type in the request tuple has to match one of the object types specified in a \gls{CW} constraint. If a match is found, the \gls{PDP} has to search in the Access History for a record containing the object type specified in the constraint that is not matched with that of the $\mathit{REQ}$ tuple (and that is relevant to the Requester). If such a record is found, it means the constraint is violated; that is, the Requester has accessed in the past a object type that is in conflict with the one specified in the current request. In this case, the action in the request will not be permitted. The \gls{CW} constraints can be specified at the level of object types. However, a fine-grained specification may be achieved if the domain hierarchy, objects may belong to, is also taken into account. In this case, we assume that $\mathit{REQ}$ and records in the Access History repository have the domain information at the same level (where level indicates number of domains) as is present in the constraint, where each element of the domain information in $\mathit{REQ}$ will be matched with the corresponding element in the constraint.

As for the \gls{HBDSoD} constraints, the time for evaluating the \gls{CW} constraints depends on the number of deployed constraints in the repository, the complexity of the constraint specification and the number of records in the Access History. Thus, the asymptotic complexity can be calculated as $O(Z \cdot c \cdot r)$. To measure the actual overhead, we performed a similar set of experiments as conducted for \gls{HBDSoD} constraints. We deployed 100 different \gls{CW} constraints and considered the worst case scenario. We then changed the number of elements in the constraint and the number of records in the Access History. The results are shown in Figure \ref{fig:egrant-constraints-evaluation-server-cw}.

\begin{table} [htp]
\centering
\caption[Time complexity of each phase in the lifecycle of E-GRANT]{Summary of time complexity of each phase in the lifecycle of \gls{EGRANT}}
\label{tab:egrant-complexity-summary}

\begin{tabular}{ |l|c| } 

\hline

\textbf{Phase Name} & \textbf{Complexity in the Worst Case} \\ \hline

Deployment of \gls{HBDSoD} constraints & ${\Theta}( Y )$ \\ \hline

Deployment of \gls{CW} constraints & ${\Theta}( Z )$ \\ \hline

Generation of requests & ${\Theta}( Z )$ \\ \hline

Evaluation of \gls{HBDSoD} constraints & $O ( Y \cdot c \cdot r )$ \\ \hline

Evaluation of \gls{CW} constraints & $O ( Z \cdot c \cdot r )$ \\ \hline

\end{tabular}

\end{table}

The above results clearly show that there is a penalty to be paid for the enforcement of encrypted constraints in outsourced environments. The execution time varies from 100 \gls{ms} to 2.5 seconds as number of records in the Access History increase from 100 to 500. To be fair, our experiments have been executed with very basic hardware. We expect that our solution would be able to perform better with more dedicated resources, such as servers deployed in a cloud infrastructure. Moreover, all the executions have been performed as a centralised solution. Clearly, having in these settings a single \gls{PEP} and a single \gls{PDP} to process all the incoming requests represent a bottleneck. To solve this problem, we are planning to develop a distributed version of our proposed architecture that can be deployed on multiple nodes and adapted to the actual request demand.

Table \ref{tab:egrant-complexity-summary} provides a summary of time complexities of different phases in the lifecycle of \gls{EGRANT}.

% session up

\begin{figure} [htp]
\centering
% left bottom right top
\includegraphics[width=.5\textwidth]{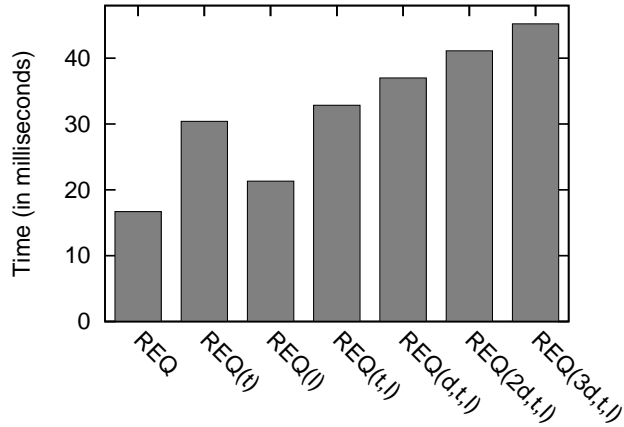}
\caption{Performance overhead of updating the Session with the request data}
\label{fig:egrant-session-up-integrated}
\end{figure}

\subsection{Performance Analysis of Session Update}
After the \gls{PDP} checks that the current request is not violating any deployed constraints and the request is granted, the Access History in the Session needs to be updated with the information in the executed request. The session update is managed by the \gls{PEP} that executes the second round of encryption before storing the encrypted data in the Session (see Algorithm \ref{algo:egrant-evaluate-constraint-session-up}). Figure \ref{fig:egrant-session-up-integrated} shows the performance overhead of encrypting the request for storing it in the Session. The graph shows the execution time of different formats of the $\mathit{REQ}$ tuple: that is, from the basic format containing only subject, action and target information to more complex ones having time, location and a domain hierarchy of objects up to three levels.

\section{Chapter Summary}
\label{sec:egrant-summary}

In this chapter, we have proposed \gls{EGRANT}, an architecture for enforcing dynamic security constraints as an outsourced service running in the cloud. The main contribution of \gls{EGRANT} is that it supports the enforcement of encrypted security constraints while maintaining the encrypted session in the cloud. In this way, cloud providers learn neither about the information stored by the session nor about the content of security constraints being enforced. The proposed approach provides a scalable key management, where users do not share any encryption keys. If users leave the organisation or keys get compromised, they can be revoked without requiring re-distribution of keys and re-encryption of deployed constraints.

The combination of Chapter \ref{cha:espoon}, Chapter \ref{cha:erbac} and Chapter \ref{cha:egrant} offers the full-fledged \gls{RBAC} model that can support role hierarchies and the constraint model. This full-fledged \gls{RBAC} model can be outsourced such that the Service Provider cannot learn private information about sensitive policies being enforced. The data (or policy) outsourcing follows the traditional client-server model, where there are two main roles, a client and a server. Our proposed solutions assume that both client and server roles run in different spaces. The issue of enforcement of sensitive policies becomes quite challenging if we consider a distributed model, where each peer can play multiple roles simultaneously. Unfortunately, our existing proposals do not work because the underlying assumption becomes invalid, i.e., both a client and a server run in the same space in distributed settings. In the next chapter, we investigate privacy and security issues in enforcing sensitive security policies in distributed environments.
%

%%%%%%%%%%%%%%%%%%%%%%%%% CHAPTER PIDGIN %%%%%%%%%%%%%%%%%%%%%%%%%

\chapter[Enforcing Policies in Distributed Environments]{\acrshort{PIDGIN}: Enforcing Security Policies in Distributed Environments\footstar{The final version of this chapter will appear in \cite{Asghar2013:IJIS:EGRANT}.}}
\label{cha:pidgin}

Opportunistic networks have recently received considerable attention from both industry and researchers. These networks can be used for many applications without the need for a dedicated \gls{IT} infrastructure. In the context of opportunistic networks, the application to content sharing in particular has attracted specific attention. To support content sharing, opportunistic networks may implement a publish-subscribe system in which users may publish their own content and indicate interest in others' content through subscription. Using a smartphone, any user can act as a broker by opportunistically forwarding both published content and interest within the network. Unfortunately, despite their provision of this great flexibility, opportunistic networks raise serious privacy and security issues. Untrusted brokers can not only compromise the privacy of subscribers by learning their interest but also can gain unauthorised access to the disseminated content. 

There are solutions that can regulate access to content by specifying access policies. However, access policies may reveal information about content they aim to protect. This chapter addresses the research challenges inherent to the exchange of content and interest without: (i) revealing content and its associated policies to unauthorised brokers and (ii) compromising the privacy of subscribers. Specifically, this chapter presents an interest and content sharing solution that addresses these security challenges and preserves privacy in opportunistic networks. We demonstrated the feasibility and efficiency of this solution by implementing a prototype and analysing its performance on real smart phones.

\section{Introduction}

In the last few years, the usage of smartphones has grown dramatically and is predicted to increase even more in coming years \cite{Emarketer:2013}. Considering the pervasive nature of smartphones, mobile opportunistic networks could be leveraged to share information. Several of the concepts behind opportunistic networks originate from \glspl{DTN} that offer flexible content sharing without requiring a dedicated \gls{IT} infrastructure \cite{ Pelusi:2006}. Haggle \cite{Haggle:2010}, an example of such a network architecture, allows smartphones to opportunistically share content via short-range communication \cite{ Nordstrom:2009}. To share content, opportunistic networks such as Haggle implement a publish-subscribe system in which nodes can publish their own content and subscribe to others' content by indicating their interest. Any node can also act as a broker (also called a relay) that opportunistically receives content and interest, matches them, and possibly delivers that content to other nodes.

The opportunistic networks could be applied to the exchange of information in a wide range of domains from social media to military applications. However, such networks also present serious privacy and security issues, particularly the need for an approach to the exchange of content and interest that neither (i) reveals content and its associated policies to unauthorised brokers nor (ii) compromises the privacy of subscribers.

For the regulation of access to content, cryptographic approaches such as \gls{ABE} which include \gls{CPABE} \cite{Bethencourt:2007} and \gls{KPABE} \cite{Goyal:2006} offer fine-grained control over content but leak information about the policies and attributes that protect that content, respectively. To protect these policies, state-of-the-art solutions exist to enforce sensitive policies in outsourced environments \cite{Asghar2013-COSE, Asghar2011-ARES, Kapadia:2007}. However, such solutions assume that the outsourced server does not collude with any client. Thus, these solutions cannot be applied in opportunistic network settings in which nodes communicate in a peer-to-peer fashion, i.e., serving as both a client and a server.

\subsection{Research Contributions}

This chapter presents \textbf{\gls{PIDGIN}}, an interest and content sharing scheme that preserves privacy. In \gls{PIDGIN},

\begin{itemize}

	\item brokers match subscriber's interest against policies associated with content without compromising the subscriber's privacy (say, by learning attributes or interest).
	
	\item an unauthorised broker neither gains access to content nor learns access policies, and authorised nodes gain access only if they satisfy fine-grained policies specified by the publishers.

	\item the system provides scalable key management in which loosely-coupled publishers and subscribers communicate with each other without any prior contact.

\end{itemize}

As a proof-of-concept, we have developed and analysed the performance of a prototype running on real smartphones in order to show the feasibility of our approach.

\subsection{Chapter Outline}
The rest of this chapter is organised into the following sections.
Section \ref{sec:pidgin-overview-scenario-challenges} provides a brief overview of opportunistic networks, describes the motivating scenario, and lists some of the major research challenges for interest and content sharing in opportunistic networks with guaranteed preservation of privacy.
In Section \ref{sec:pidgin-system-model}, we draw the system model. 
Next, we describe the proposed scheme in Section \ref{sec:pidgin-proposed-scheme}.
Section \ref{sec:pidgin-details} elaborates \gls{PIDGIN}'s details. 
In Section \ref{sec:pidgin-construction}, we provide the concrete construction.
Section \ref{sec:pidgin-security-analysis} analyses \gls{PIDGIN} from a security perspective.
In Section \ref{sec:pidgin-analysis}, we report the outcomes of the performance analysis.
Section \ref{sec:pidgin-discussion} is dedicated for discussion.
Section \ref{sec:pidgin-related-work} reviews the related work.
Finally, we conclude in Section \ref{sec:pidgin-summary}.

\section{Opportunistic Networks and Research Challenges}
\label{sec:pidgin-overview-scenario-challenges}

In this section, we provide a brief overview of opportunistic networks, a motivating scenario, and the major research challenges in opportunistic networks that we address.

\subsection{Overview of Opportunistic Networks}
\label{sec:pidgin-overview}
Conceptually, opportunistic networks originate from \glspl{DTN} that enable content exchange between nodes in a publish-subscribe fashion, generally via short-range communication. In a typical opportunistic network, such as Haggle, a subscriber node subscribes interest while a publisher node publishes content to its neighbouring nodes \cite{Nordstrom:2009}. These neighbouring nodes are intermediate nodes, known as brokers, that epidemically disseminate interest and content within the network. A resolution takes place when a broker node finds a match between the interest of a subscriber and the tags associated with published content. 
As a result of resolution, a broker forwards content to the subscriber.
In the following section, we consider a motivating scenario that can further help to understand opportunistic networks and research challenges concerning privacy and confidentiality. \\ \\
\noindent \textbf{Curiosity - A Military Mission:} Let us consider a battlefield scenario for a mission called \emph{Curiosity} in which soldiers are equipped with smartphones. During the mission, a scout collects some sensitive information about the enemy (for instance, an image of the enemy's position) using her smartphone camera. After acquiring this sensitive information, a scout desires to share it with other soldiers. For this reason, she may tag the image with the mission name, i.e., \emph{Curiosity}. Unfortunately, there is no Internet connectivity on the battlefield and the only way to share is to use the short-range communication offered by smartphones. Therefore, the scout would like to share the image with other soldiers using their smartphones. We assume that the soldiers are interested in getting information about the mission and subscribe using their smartphones.

\begin{figure} [htp]
\centering
% left bottom right top
\includegraphics[trim=15mm 80mm 15mm 65mm,clip,width=.9\textwidth]{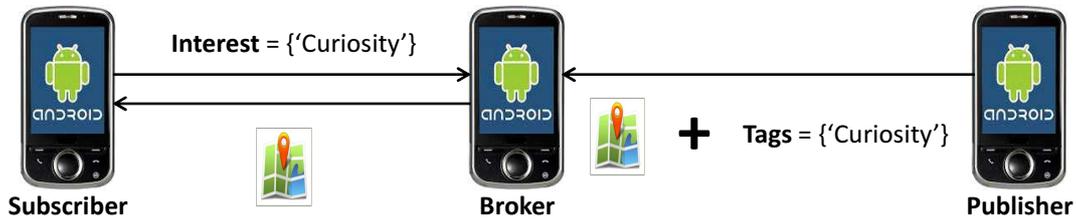}
\caption{An example of content sharing in an opportunistic network}
\label{fig:pidgin-example}
\end{figure}

\subsection{Motivating Scenario}
\label{sec:pidgin-scenario}

\noindent \textbf{Haggle: A Possible Solution:} To exchange information in such scenarios, we can leverage opportunistic networks, such as Haggle. Using Haggle, the scout publishes the image with \emph{Curiosity} as a tag. Any solider can show interest in \emph{Curiosity} by subscribing, as illustrated in Figure \ref{fig:pidgin-example}. Here, we assume that someone as a broker receives both interest and image along with the tag. Whenever that happens, the broker checks whether the interest of a subscriber matches any tag associated with the image. If so, the broker forwards the image to the subscriber(s). \\

\noindent \textbf{Privacy and Confidentiality Issues:} First of all, to preserve confidentiality, the information about the \emph{Curiosity} mission should be shared only within a particular group of soldiers. Each content item is associated with an access policy that indicates who should have access to it. For example, information about the \emph{Curiosity} mission might have a policy (P) that content is shared with either a \emph{Major} or a \emph{Soldier} from the \emph{Infantry} unit. Even if the content (i.e., image) is encrypted, the policy itself could reveal sensitive information. That is, an enemy may infer useful information from the fact that some contents are sent to a \emph{Major} or a \emph{Soldier} from the \emph{Infantry} unit. Outsiders (i.e., enemies) and insiders (i.e., soldiers) serving as brokers may gain unauthorised access to contents. Furthermore, the interest of subscribers and the tags associated with content may also reveal sensitive information. Therefore, in addition to the content itself, its associated tags, policies, and subscription information (i.e., interests) should also be protected.

This scenario motivates the need to tackle the security and privacy issues that we generally face in opportunistic networks. In the following section, we list some major research challenges inherent to these issues that we address in this chapter.

\subsection{Research Challenges}
\label{sec:pidgin-challenges}
To guarantee the preservation of privacy for interest and content sharing in opportunistic networks, the following major research challenges related to both (i) privacy and confidentiality (i.e., \textit{C1-C3}) and (ii) functionality (i.e., \textit{C4-C5}) need to be addressed:

\begin{description}

  \item[C1] In the presence of unauthorised brokers, how do we regulate access to disseminated content and preserve confidentiality of content and associated policies?
  
  \item[C2] In the presence of curious brokers, how does the network exchange content without compromising the privacy of its subscribers?
  
  \item[C3] How can a subscriber subscribe to content without exposing her interest to untrusted brokers?
	
	\item[C4] In order to minimise the flood of unnecessary traffic on the communication network, how do we ensure that a subscriber receives content if and only if authorised to decrypt?
	
	\item[C5] Assuming the loosely-coupled nature of the publish-subscribe model, how do we address the challenges above (i.e., \textit{C1-C4}) without sharing any keys between publishers and subscribers?
	
\end{description}

\section{The System Model}
\label{sec:pidgin-system-model}
Before presenting our threat model and assumptions, we identify the entities involved in the system:

\begin{description}

	\item [A Publisher] is a node that can publish the content.
	
	\item [A Subscriber] is a node that can subscribe interest.
	
	\item [A Broker] is a node that may receive and disseminate both content and interest. It evaluates whether any content matches known interest. On successful evaluation, it forwards content to the subscribers.
	
	\item [\acrfull{TKMA}] is an offline trusted entity that distributes keying material (including private keys and/or public parameters) to all nodes out of the band (usually once in the lifetime of a node, typically when the node is initialised).
	
\end{description}

\noindent \textbf{The Threat Model.} We assume that brokers are \textit{honest-but-curious}, i.e., they honestly follow the protocol, but remain curious to learn about content and interest. Also, we assume that brokers may collude. Furthermore, we consider that the \gls{TKMA} is fully trusted and plays a role at the time of system initialisation. Last but not least, we assume only passive adversaries and do not consider active adversaries that can manipulate the exchanged information.

\section{The Proposed Idea}
\label{sec:pidgin-proposed-scheme}
In this section, we describe the proposed scheme for preserving privacy during interest and content sharing in opportunistic networks. As a starting point, we consider some basic schemes that partially address research challenges listed in Section \ref{sec:pidgin-challenges}. Next, we gradually address all research challenges and finally describe the proposed scheme.

\begin{figure} [htp]
\centering
% left bottom right top
\includegraphics[trim=15mm 80mm 15mm 65mm,clip,width=.9\textwidth]{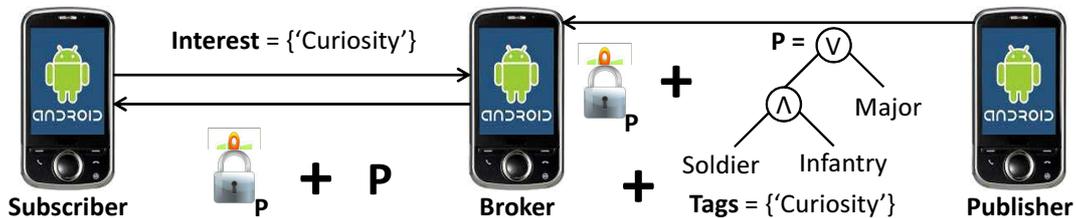}
\caption[Regulating access to content using CPABE policies]{Regulating access to content using \gls{CPABE} policies}
\label{fig:pidgin-cpabe}
\end{figure}

\subsection{Scheme I: Regulate Access on Content}
To preserve the confidentiality of content, a publisher might specify who can gain access. A possible approach for the publisher could be to regulate access on content by employing \gls{ABE}, such as \gls{CPABE} \cite{Bethencourt:2007} or \gls{KPABE} \cite{Goyal:2006}. \gls{ABE} offers fine-grained policies for content access. In this scheme, we consider \gls{CPABE} because it enables a publisher to exert control over access to content, as described in the use case scenario. In contrast, in \gls{KPABE}, a key generation authority exerts control over who can access content. Figure \ref{fig:pidgin-cpabe} illustrates this scheme in which the image is encrypted according to the policy: \emph{either a Major or a Solider from the Infantry unit can get access}. The policy is expressed as a tree whose leaf nodes represent the attributes; non-leaf nodes denote the AND, OR and threshold gates. In this scheme, a broker forwards content to the subscribers if a subscriber's interest matches with any tag associated with the content.

This approach preserves the confidentiality of disseminated contents without providing access to unauthorised brokers. This scheme, however, has a drawback. A broker might send content to subscribers who might not be able to decrypt it. In fact, a broker's role is merely to match the interest of subscribers against tags associated with content without checking whether a subscriber has access authorisation. For instance, consider a subscriber who is a soldier but neither a \emph{Major} nor a member of the \emph{Infantry} unit.

In summary, this scheme resolves the access control problem \textit{(C1)} while raising the problem of a communication network flooded with unnecessary traffic \textit{(C4)}.

\subsection{Scheme II: Perform an Authorisation Check}
This scheme extends Scheme I and resolves the flooding problem \textit{C4}. In this scheme, a subscriber may send attributes and interest to brokers so that a broker can perform an authorisation check prior to forwarding the contents. To perform the authorisation check, a broker matches leaf nodes in the policy tree with the subscriber's attributes. If there is a match, a leaf node will be marked as satisfied. After evaluating leaf nodes, a broker evaluates intermediate nodes (including AND, OR and threshold) in the policy. A broker will forward encrypted content to subscriber if and only if (i) the root node of the policy is marked as satisfied and (ii) the interest matches to the tags.

This scheme targets both the access control problem \textit{(C1)} and the flooding problem \textit{(C4)}. However, it still raises some privacy issues. First, both the cleartext attributes of subscribers and the cleartext \gls{CPABE} policies can compromise the privacy of subscribers, i.e., \textit{C2}. For example, the enemy may learn from policies that there is some information intended for a \emph{Major}. Second, the cleartext interest of a subscriber may also leak information, i.e., \textit{C3}. For instance, the enemy may learn that this content or interest concerns the \emph{Curiosity} mission.

\begin{figure} [htp]
\centering
% left bottom right top
\includegraphics[trim=15mm 80mm 15mm 65mm,clip,width=.9\textwidth]{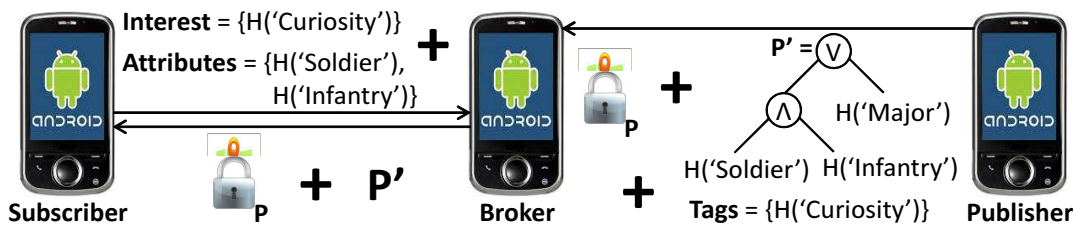}
\caption[Hiding private information using hash functions]{Private information is hidden through replacement of leaf nodes in the \gls{CPABE} policy, tags, attributes and interest items with their corresponding hashes}
\label{fig:pidgin-hash}
\end{figure}

\subsection{Scheme III: Hide Private Information Using a Hash}
In order to partially overcome the issue of subscriber privacy \textit{(C2)}, a subscriber and a publisher may hash both attributes and leaf nodes in the policy tree, respectively. Similarly, a subscriber's interest could be protected by calculating the hash values of interest items and tags associated with contents. In this scheme, a broker forwards encrypted content to subscribers if and only if (i) the hash value of the interest matches the hash value of the tag (i.e., h(`Curiosity')) and (ii) hash values of attributes (i.e., \{h(`Soldier'), h(`Infantry')\}) satisfy the policy $P'$ whose leaf nodes are also hashed, as shown in Figure \ref{fig:pidgin-hash}.

Unfortunately, this scheme is vulnerable to a pre-computed dictionary attack. That is, the enemy may pre-calculate a list of hashes for possible attributes (and leaf nodes in the policy tree) and a list of hashes for potential interest items (and tags). The pre-calculated list of hashes may easily reveal the original attributes (and leaf nodes in the policy tree) and interest (and tags).

\begin{figure} [htp]
\centering
% left bottom right top
\includegraphics[trim=15mm 80mm 15mm 65mm,clip,width=.9\textwidth]{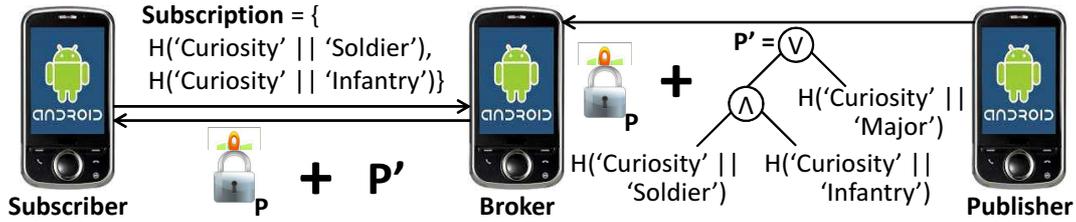}
\caption[Hardening against a pre-computed dictionary attack]{Hardening against a pre-computed dictionary attack through concatenation a pair of (i) a leaf node in the \gls{CPABE} policy and a tag (ii) an attribute and an interest item, then calculation of the hash on the final string}
\label{fig:pidgin-extended-hash}
\end{figure}

\subsection{Scheme IV: Hardening Against a Pre-Computed Dictionary Attack}
To harden against the pre-computed dictionary attack, a publisher may replace each leaf node in the policy with a hash of a concatenated pair of a tag and an attribute. Similarly, a subscriber may subscribe using the hash of a concatenated pair of an interest item and an attribute (i.e., \{H(`Curiosity' $||$ `Soldier'), H(`Curiosity' $||$ `Infantry')\}) as illustrated in Figure \ref{fig:pidgin-extended-hash}. In this scheme, a broker just needs to check whether the items in a subscription satisfy the hashed policy $P'$. Upon successful evaluation, the broker will forward the content to subscribers. The advantage of this scheme is that it not only hardens against pre-computed dictionary attacks but also decreases the number of comparisons performed at the broker's end as compared to Scheme III. This is because a broker performs integrated checks that cover both authorisation and interest matching simultaneously in contrast to Scheme III in which a broker performs two different checks: one to check the authorisation and one to match the interest. Though it enlarges the key space (which could be computationally extensive), this scheme is still vulnerable to a pre-computed dictionary attack.

\begin{figure} [htp]
\centering
% left bottom right top
\includegraphics[trim=15mm 80mm 15mm 65mm,clip,width=.9\textwidth]{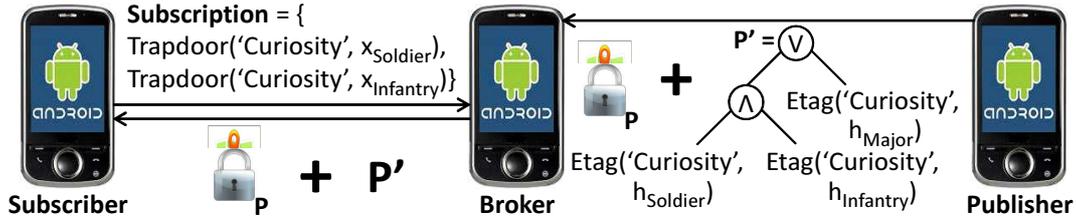}
\caption[The PIDGIN scheme protecting the content and policies]{The \gls{PIDGIN} scheme protecting the content, the policy, the tags associated with content, and the subscriber's interest and attributes}
\label{fig:pidgin-pidgin}
\end{figure}

\subsection[PIDGIN: The Proposed Scheme]{\gls{PIDGIN}: The Proposed Scheme}
Our proposed scheme, \gls{PIDGIN}, aims at addressing all research challenges (i.e., \textit{C1-C5}) listed in Section \ref{sec:pidgin-challenges}. The main idea behind \gls{PIDGIN} is regulation of access to content using \gls{CPABE} and extension of cleartext \gls{CPABE} policies with the \gls{PEKS} scheme \cite{Boneh:2004} to protect attributes, interest, tags and leaf nodes in the policy tree. The \gls{PEKS} scheme consists of four basic functions including \textbf{Keygen}, \textbf{Etag}\footnote{The Etag function is called \gls{PEKS} in \cite{Boneh:2004}.}, \textbf{Trapdoor} and \textbf{Test}. For each attribute, we run \textbf{Keygen} to calculate a key pair consisting of both public (i.e., $h_{Soldier}$) and private (i.e., $x_{Soldier}$) keys corresponding to a given attribute (i.e., \emph{Soldier}). To protect policies and tags, a publisher can replace each leaf node in the policy tree with the \textbf{Etag} function of the \gls{PEKS} scheme, which takes as input a tag (i.e., \emph{Curiosity}) and the public key of the attribute as shown in Figure \ref{fig:pidgin-pidgin}. A subscriber protects attributes and interest by replacing each interest item in the subscription list with a \textbf{Trapdoor} function which takes as input an interest item (i.e., \emph{Curiosity}) and the private key (i.e., generated by the \gls{PEKS} scheme) corresponding to the attribute. 

A broker performs encrypted matching between encrypted policies and encrypted subscriptions. It runs the \textbf{Test} function, a building block that matches a trapdoor to an encrypted tag. If an encrypted tag in the policy tree $P'$ matches with any encrypted trapdoor in the subscription list, the tree node is marked as satisfied. The broker evaluates all nodes in the policy tree starting from leaf nodes to root. If the root is satisfied, the broker will forward content along with the encrypted policy to the subscribers.

\section[Technical Details of PIDGIN]{Technical Details of \acrshort{PIDGIN}}
\label{sec:pidgin-details}

\subsection{Initialisation and Key Generation Phases}
During the initialisation phase, the system is set up to initialise both \gls{CPABE} and \gls{PEKS} schemes. In \gls{PIDGIN}, the \gls{TKMA} generates and distributes keys during the key generation phase. The \gls{TKMA} generates a private set of attributes (i.e., \gls{CPABE} private key) and sends it securely to the subscriber out of the band. The \gls{TKMA} publishes the public part of attributes (i.e., \gls{CPABE} public key) to all publishers. Since the attributes are protected using the \gls{PEKS} scheme, the \gls{TKMA} also generates a pair of keys corresponding to each attribute. Similar to the \gls{CPABE} key distribution, the \gls{TKMA} sends the private and public parts of the \gls{PEKS} key pair to the subscriber and publishers, respectively. The major difference between the \gls{CPABE} private key set and the \gls{PEKS} private key set is that the former is unique for each user, while the latter is not.

\subsection{The Publisher's Encryption Phase}
To protect the content and preserve the privacy of subscribers, a publisher encrypts content with \gls{CPABE} policies and protects those policies as well. The contents could be encrypted with a symmetric key, such as \gls{AES}, which is further encrypted with the \gls{CPABE} policy. Since the \gls{CPABE} policy may compromise the privacy of subscribers, the \gls{CPABE} policies are encrypted using \gls{PEKS}. While encrypting \gls{CPABE} policies using \gls{PEKS}, \gls{PIDGIN} also incorporates tags that are associated with content.

\begin{figure} [htp]
\centering
% left bottom right top
\includegraphics[trim=24mm 70mm 29mm 63mm,clip,width=.95\textwidth]{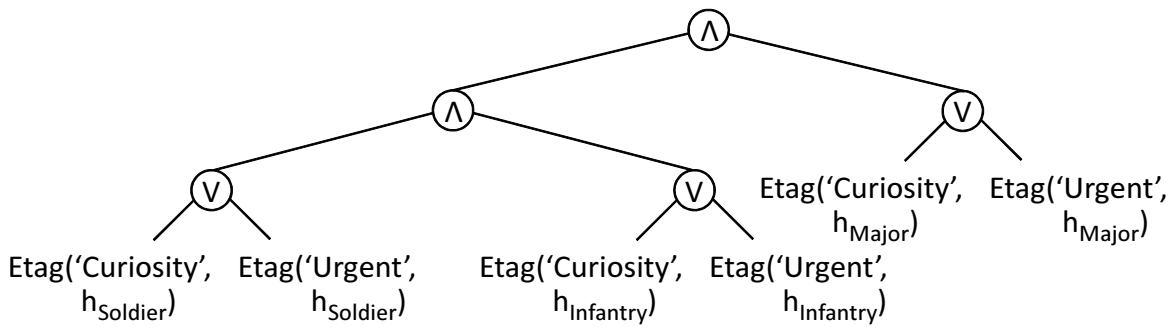}
\caption[The extended CPABE policy with multiple tags]{The extended \gls{CPABE} policy with two tags, i.e., `Curiosity' and `Urgent'.}
\label{fig:pidgin-complex-policy}
\end{figure}

To extend \gls{CPABE} policies for \gls{PEKS}, a publisher considers each leaf node in the policy tree as well as number of tags that are associated with contents. If there is only a single tag then a publisher replaces the leaf node with the \textbf{Etag} function as already illustrated in Figure \ref{fig:pidgin-pidgin}. The \textbf{Etag} function takes a tag keyword to be encrypted and the public key corresponding to the leaf node under consideration. After running the \textbf{Etag} function, a publisher gets an encrypted tag. The \textbf{Etag} function does not leak information about the tags or leaf nodes in the policy tree. In the case that there is more than one tag then a publisher runs the \textbf{Etag} function for each tag item and encrypts it with the public key corresponding to the leaf node under consideration. Finally, the leaf node attribute is replaced with the subtree where all newly generated Etags corresponding to tags are disjuncted using OR. Figure \ref{fig:pidgin-complex-policy} illustrates an example of the policy involving two tags, i.e., `Curiosity' and `Urgent'.
%TODO: optional revise further? 

\subsection{The Subscriber's Encryption Phase}
In order to protect the interest of a subscriber and its attributes, a subscriber encrypts each interest item using the private key (i.e., generated by the \gls{PEKS} scheme) corresponding to the attribute. \gls{PIDGIN} considers that a subscriber might have multiple attributes and interest items. Generally, each interest item is encrypted with each private key (i.e., generated by the \gls{PEKS} scheme) that corresponds to the attribute. Figure \ref{fig:pidgin-pidgin} describes the case in which a subscriber holds two attributes and subscribes with a single interest item. Let us assume that a subscriber has two interest items, say `Curiosity' and `Urgent', while holding attributes Solider and Infantry. The subscription list would contain four items including \textbf{Trapdoor}(`Curiosity', $x_{Soldier}$), \textbf{Trapdoor}(`Curiosity', $x_{Infantry}$), \textbf{Trapdoor}(`Urgent', $x_{Soldier}$) and \textbf{Trapdoor}(`Urgent', $x_{Infantry}$). The trapdoor representation does not leak information about the interest item and the attribute.

\subsection{The Broker's Matching Phase}

A broker opportunistically exchanges both content and subscriptions. Once a broker receives both the encrypted subscription and the encrypted content along with the encrypted policies, it evaluates whether the encrypted subscription satisfies the encrypted policy. For this evaluation, the broker runs a matching function that recursively evaluates the encrypted policy tree. The \textbf{Test} function matches each encrypted leaf node in the policy against the encrypted interest item in the subscription.

The \textbf{Test} function returns either \emph{TRUE} or \emph{FALSE}, indicating whether the encrypted tag is matched with the trapdoor or not, respectively. By running the \textbf{Test} function, a broker does not learn about the tag or the interest item because both are encrypted and they are matched in an encrypted manner. If an encrypted tag in the policy tree matches with any trapdoor in the subscription list, that node is marked as satisfied. After evaluating leaf nodes, a broker can evaluate intermediate AND, OR and threshold nodes in the policy tree to finally identify whether the root node of the policy tree is satisfied or not. If the root node is satisfied, the broker will forward content along with the encrypted policy to the subscriber.

\subsection{The Subscriber's Decryption Phase}
Once a subscriber receives the encrypted content along with the encrypted policy, it first recovers the original \gls{CPABE} policy. 
For this recovery, either leaf node (if a single tag, see Figure \ref{fig:pidgin-cpabe}) or a subtree of tags (if more than one tag, see Figure \ref{fig:pidgin-complex-policy}) is replaced with their corresponding attribute. Before sharing the encrypted interest, a subscriber builds the subscription history as a lookup table containing an attribute and its corresponding trapdoor. If the trapdoor is matched with any encrypted tag in the leaf node of the policy, the subscription history will be looked up to find the attribute corresponding to the matched trapdoor. Next, a leaf node (if a single tag) or a subtree of tags (if more than one tag) will be replaced with the found attribute. If no match is found, then a dummy attribute will be placed. This recovers the original \gls{CPABE} policy (i.e., one shown in Figure \ref{fig:pidgin-cpabe}) that can finally be used by the \gls{CPABE} decryption function to get the symmetric key that is required for decryption of the contents.

\section[Concrete Constructions of PIDGIN]{Concrete Constructions of \gls{PIDGIN}}
\label{sec:pidgin-construction}

% TODO: provide definition of bilinear maps (computable, bilinear and non-degenerate) in the journal version

%In this section, we briefly describe the policy structure, review some definitions and then provide details of core functions used in different phases of the \gls{PIDGIN} lifecycle. \\ \\

In this section, we provide some definitions and details of core functions used in different phases of the \gls{PIDGIN} lifecycle.

%TODO: in journal: delegation % f = g^{1/{\beta}}

\subsection{Definitions}

\noindent \textbf{The Policy Structure.} We assume a policy tree $P$ that represents an access structure. Each non-leaf node represents an AND, an OR or a threshold gate. Let us consider that $num_x$ denotes number of children of a node $x$ and $k_x$ represents the threshold value. For OR and AND gates, $k_x$ is $1$ and $num_x$, respectively. For the threshold gate, the value of $k_x$ is: $0 < k_x \leq num_x$. Let us consider that \textbf{parent}($x$) represents the parent of a node $x$, \textbf{att}($x$) denotes the attributes associated with leaf node $x$, and \textbf{index}($x$) returns the number associated with a node $x$, with nodes numbered from $1$ to $num$. \\ \\
\textbf{Bilinear Maps.} Let $\mathbb{G}_1$ and $\mathbb{G}_2$ be two multiplicative cyclic groups of prime order $p$. Let $g$ be a generator of $\mathbb{G}_1$ and $e : \mathbb{G}_1 \times \mathbb{G}_1 \rightarrow \mathbb{G}_2$ be a bilinear map. The bilinear map $e$ satisfies the following properties:

\begin{itemize}
	\item Computability: given $g, h \in \mathbb{G}_1$, there is a polynomial time algorithm to compute $e(g, h) \in \mathbb{G}_2$.
	\item Bilinearity: $\forall u, v \in \mathbb{G}_1$ and $a, b \in \mathbb{Z}_p$, we have $e(u^a, v^b) = e(u, v)^{ab}$.
	\item Non-degeneracy: if $g$ is a generator of $\mathbb{G}_1$ then $e(g, g)$ is a generator of $\mathbb{G}_2$, where $e(g, g) \neq 1$.
\end{itemize}
Notice that the bilinear map $e$ is symmetric since $e(g^a, g^b) = e(g, g)^{ab} = e(g^b, g^a)$. \\ \\
\textbf{Hash Functions.} We consider the hash functions: 

\begin{align*}
H_1 &:\{0,1\}^{*} \rightarrow \mathbb{G}_1 \\ 
H_2 &: \mathbb{G}_2 \rightarrow \{0,1\}^{\log{p}}
\end{align*}

\noindent \textbf{Lagrange Coefficient.} We define the Lagrange coefficient $\Delta_{i,A}$ for $i \in \mathbb{Z}_p$ and a set A of elements in $\mathbb{Z}_p$: 

\begin{align*}
\Delta_{i,A}(x)=\prod_{j \in A, j \neq i}{\frac{x-j}{i-j}}
\end{align*}

\subsection[Construction Details of PIDGIN]{Construction Details of \gls{PIDGIN}}
%\subsection{The Initialisation Phase} 
\noindent \textbf{Init($1^K$).} The init algorithm takes as input the security parameter $k$ that determines the size of $p$. 
It randomly picks two exponents $\alpha, \beta \in \mathbb{Z}_p$ and outputs the public key $PK = (\mathbb{G}_1, g, h = g^{\beta}, e(g, g)^{\alpha})$ and the master key $MK = (\beta, g^{\alpha})$. 
The public key $PK$ is published while the master key $MK$ is kept securely by the \gls{TKMA}. 
Moreover, two stores, the Search Key Secret Store ($SKSS$) and the Search Key Public Store ($SKPS$), which are managed by the \gls{TKMA}, are initialised as:
\begin{align*}
SKSS \leftarrow \phi \\ 
SKPS \leftarrow \phi
\end{align*}
%\subsection{The Key Generation Phase}
\textbf{KeyGen($MK, A$).} The key generation algorithm is run by the \gls{TKMA}. 
It takes as input a list of attributes $A$ and outputs a \gls{CPABE} decryption key and a set of search key pairs. 
To generate the decryption key, it first chooses a random $r \in \mathbb{Z}_p$ and then a random $r_j \in \mathbb{Z}_p$ for each attribute $j \in A$. Next, it computes the decryption key as: 
\begin{align*}
DK &= (D = g^{(\alpha + r)/\beta}, \\
&\qquad {} \forall \in A: D_j = g^r \cdot H_1(j)^{r_j}, D'_j = g^{r_j})
\end{align*}

Before the generation of a search key pair for an attribute $j \in A$, a search key store (either $SKSS$ or $SKPS$) can be looked up. If the search key pair already exists, then the public and private keys will be collected from $SKPS$ or $SKSS$, respectively. Otherwise, the algorithm chooses a random $x_j \in \mathbb{Z}_p^*$, calculates $h_j = g^{x_j}$, and updates both private and public key stores as:
\begin{align*}
SKSS \leftarrow SKSS \cup (j, x_j) \\ 
SKPS \leftarrow SKPS \cup (j, h_j)
\end{align*}
%, respectively. 
Next, it computes the search key secret as: $SKS = (\forall \in A: x_j)$. Finally, the $SKPS$ is publicised while the decryption key $DK$ and the search key secret $SKS$ are securely transmitted to the subscriber. \\ \\
%\subsection{The Publisher's Encryption Phase}
%
\textbf{Etag($PK, h_i, t$).} The \textbf{Etag} algorithm encrypts a given tag $t$ with $h_i$. It chooses a random $r \in \mathbb{Z}_p^*$ and computes $z = e(H_1 (t), h^r)$. 
Next, it computes $A = g^r$ and $B = H_2 (z)$ and outputs the encrypted tag as: $ET=(A, B)$. \\ \\
\textbf{Pub-Enc($PK, SKPS, C, P, T$).} The publisher encryption algorithm encrypts content $C$ under the access policy $P$ with a list of tags $T$. It also encrypts $P$. In reality, it randomly generates a symmetric key $K$ and encrypts $C$ as $\{C\}_K$ and then encrypts $K$ under $P$. To encrypt $K$ under $P$, it chooses a polynomial $q_x$ for each node $x$ in a top-down manner, starting from the root $R$, such that it sets degree $d_x$ one less than the threshold value $k_x$, i.e., $d_x = k_x - 1$. Starting from the root $R$, it chooses a random $s \in \mathbb{Z}_p$, sets $q_R (0) = s$ and chooses other $d_R$ points randomly. For any other non-root node $x$, it sets $q_R (0) = q_{parent(x)} (index(x))$ and chooses other $d_x$ points randomly. Let $Y$ be the set of leaf nodes in $P$. 
The ciphertext is computed as: 
\begin{align*}
CT &= (\tilde{E} = Ke(g,g)^{\alpha s}, E = h^s, \\
&\qquad {} \forall y \in Y: E_y = g^{q_y (0)}, E'_y = H_1 (att(y))^{q_y (0)})
\end{align*}
Next, the policy $P$ is encrypted as follows. For each leaf node $i$, it looks up the corresponding private secret key $h_i$ from the $SKPS$. Then, it runs \textbf{Etag}($h_i, t$) for each tag $t \in T$ and combines all encrypted tags corresponding to an attribute to form an OR subtree. The original leaf node attribute is replaced with this OR subtree. If only one tag exists in $T$, the original attribute is replaced with the output of the \textbf{Etag} function. This basically generates the encrypted policy $P'$. Finally, this algorithm returns $PE = (P', CT, \{C\}_K)$. \\ \\
\textbf{Trapdoor($x_i, t$).} The \textbf{Trapdoor} algorithm encrypts interest item $t$ using $x_i$. It returns the encrypted interest item $TD = H_1 (t)^{x_i}$. \\ \\
%\subsection{The Subscriber's Encryption Phase}
\textbf{Sub-Enc($I, SKS$).} The subscriber encryption algorithm encrypts interest $I$ using the attributes $SKS$. For each interest item $t \in I$, it runs \textbf{Trapdoor}($x_i, t$) using search key secret $x_i$ corresponding to each attribute $i \in SKS$. A subscriber also maintains a history of subscription $HS$ to keep track of all trapdoors belonging to a subscription. $HS$ is initialised as $HS \leftarrow \phi$ and updated as: 
\begin{align*}
\forall i \in SKS: HS \leftarrow HS \cup (i, TD_i)
\end{align*}
$HS$ maintains each trapdoor with its corresponding attribute. %JS check this, I was unsure how to read this sentence. 
Finally, this algorithm publicises $SE = (TD_1, TD_2, \ldots, TD_{|I| . |SKS|})$ and keeps $HS$ securely. \\ \\
\textbf{Test($ET, TD$).} The \textbf{Test} algorithm takes the encrypted tag and trapdoor and returns \emph{TRUE} if $H_2 (e(TD, A) \stackrel{?}{=} B$ is \emph{TRUE} and \emph{FALSE} otherwise. \\ \\
%\subsection{The Broker's Matching Phase}
\textbf{Bro-Match($P', SE$).} This algorithm takes the publisher encrypted policy $P'$ and the subscriber encrypted interest $SE$ and returns \emph{TRUE} if they match and \emph{FALSE} otherwise. To perform the match, a broker runs \textbf{Test}($ET_i, TD_j$) for each leaf node $i$ in $P'$ and trapdoor $TD_j \in SE$. If an encrypted leaf node matches with any trapdoor, it is marked as satisfied (i.e., \emph{TRUE}). After evaluating leaf nodes, the algorithm evaluates intermediate nodes (AND, OR and threshold). After this evaluation, if the root node of the encrypted policy $P'$ is satisfied, that is, \emph{TRUE}, then this algorithm returns \emph{TRUE} and \emph{FALSE} otherwise. \\ \\
%\subsection{The Subscriber's Decryption Phase}
\textbf{Sub-Dec($PE, HS, DK$)} This algorithm decrypts the policy $P'$ and then decrypts the encrypted contents $PE$. First, it matches encrypted leaf nodes with a trapdoor in $HS$ by running \textbf{Test}. If a match is found, the corresponding attribute is selected from $HS$. The leaf node (if a single tag) or a subtree of encrypted tags conjuncted with OR (if more tags) will be replaced with the selected attribute. If no match is found, then a dummy attribute will be placed. This recovers the original policy, which will be used to decrypt the symmetric key:
 if node $x$ is a leaf node then we assume $i = att(x)$ and run the following function if $i \in A$:
\begin{align*}
DecryptNode(CT, DK, x) &= \frac{e(D_i, E_x)}{e(D'_i, E'_x)} \\
 &= \frac{e(g^r . H(i)^{r_i}, g^{q_x (0)})}{e(g^{r_i}, H(i)^{q_x (0)})} \\
 &= e(g, g)^{r q_x (0)}
\end{align*}
If $i \not\in A$ then $DecryptNode(CT, DK, x) = \bot$. 
For a non-leaf node $x$, the algorithm runs $DecryptNode(CT, DK, z)$ for each child $z$ of $x$ and stores output as $F_z$. Let $A_x$ be an arbitrary $k_x$-sized set of child nodes $z$ such that $F_z \neq \bot$. If no such set exists then the node was not satisfied and the function returns $\bot$. Otherwise, it computes:
\begin{align*}
F_x &= \prod_{z \in A_{x}} F_z^{\Delta_{i,A'_x (0)}} \\ 
\intertext{(where $i = index(z)$ and $A'_x = index(z) : z \in A_x$)}
&= \prod_{z \in A_x} (e(g, g)^{r . q_z(0)})^{\Delta_{i, S'_x}(0)} \\
&= \prod_{z \in A_x} (e(g, g)^{r . q_{parent(z)}(index(z))})^{\Delta_{i, A'_x}(0)} \\
\intertext{(by construction)}
&= \prod_{z \in A_x} (e(g, g)^{r . q_x (0)})^{\Delta_{i, A'_x} (0)} \\
&= (e(g, g)^{r . q_x (0)} \\ \intertext{(using polynomial interpolation)}
\end{align*}
If the tree is satisfied by $A$, we set
\begin{align*}
G &= DecryptNode(CT, DK, R) \\
&= e(g, g)^{r q_R (0)} \\
&= e(g, g)^{rs} \\ 
\end{align*}
The symmetric key is decrypted by computing: \\
$\tilde{E} / (e(E, D)/G) = \tilde{E} / (e(h^s, g^{(\alpha + r) / \beta}) / e(g, g)^{rs}) = K$. \\ \\
Finally, $K$ is used to decrypt $\{C\}_K$ in order to access contents $C$.

\section[Security Analysis of PIDGIN]{Security Analysis of \gls{PIDGIN}}
\label{sec:pidgin-security-analysis}

In \gls{PIDGIN}, the contents are encrypted using a symmetric key, which is encrypted with the \gls{CPABE} policy. The leaf nodes in the policy tree are further encrypted using \textbf{Etag} as proposed in \gls{PEKS} by Boneh \emph{et al.} \cite{Boneh:2004}. The \gls{PEKS} is semantically secure against a chosen keyword attack in the random oracle model, assuming that the \gls{BDH} problem is hard (for proof, see Theorem 3.1 in \cite{Boneh:2004}). However, the \gls{CPABE} policy structure is not protected and leaks information about number of attributes or tags used. This leak could partially be tackled by inclusion of some dummy attributes at the cost of an increase in complexity. In \gls{PIDGIN}, brokers may collude but they cannot gain access to contents, policies or subscriptions. If a broker colludes with a subscriber, they together learn no more information than is already available to the subscriber alone. In the case that two subscribers collude to receive content that each of them alone cannot get otherwise, our scheme prevents such collusion attacks because each subscriber's (\gls{CPABE}) decryption key includes a randomness value that will prevent access to the content.

\section[Performance Analysis of PIDGIN]{Performance Analysis of \gls{PIDGIN}}
\label{sec:pidgin-analysis}

As a proof-of-concept, we have developed a prototype of \gls{PIDGIN}. The prototype is based on an extension of the open source libfenc library \cite{libfenc} written in the C language, a library of functional encryption that includes \gls{CPABE}. Since we proposed to extend \gls{CPABE} with \gls{PEKS}, we have implemented \gls{PEKS} in C using the \gls{PBC} library \cite{Lynn}, which is an underlying library also required by the libfenc library. \gls{PBC} is based on \gls{ECC}. The curve we use in our experimentation is of type A. After extending the \gls{CPABE} with \gls{PEKS} (on the x86 architecture), we cross-compiled it for the ARM architecture to test our prototype on a Samsung Galaxy SIII smartphone (Android version 4.1.2, kernel version 3.0.31, 1 GB RAM, and 1.4 GHz processor). For the deployment of this prototype, we cross-compiled both GMP \cite{GMP} (the GNU Multiple Precision arithmetic library required by \gls{PBC}) and \gls{PBC} libraries for the ARM architecture and installed both on the smartphone. The presented results are averaged over 20 runs.

In our analysis, we have not considered battery consumption because the prototype of \gls{PIDGIN} we have developed so far requires some optimisations that we have suggested in Section \ref{sec:pidgin-discussion}. However, our future plan is to analyse battery consumption after implementing possible optimisations.

\iffalse
Model number GT-I9300
Android version 4.1.2
Baseband version GT-I9300BUELL1
Kernel version 3.0.31-742798
Build number JZO54K.I9300XXELLA
Processor ARMv7
API level 16
CPU ABI: armeabi-v7a
Processor speed is 1.4 GHz
Frequency stats
Max: 1.4 GHz 1.14\%
Min: 200 MHz 87.39\%
Memory RAM 1 GB
\fi

\begin{figure} [htp]
\centering
% left bottom right top
\includegraphics[angle=-90,width=.5\textwidth]{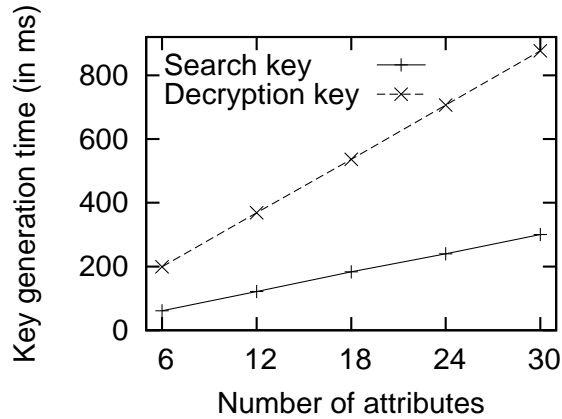} % .3 - decreased for space
\caption{Effect of attributes on the key generation time}
\label{gra:keygen}
\end{figure}

\subsection{Initialisation and Key Generation Phases}
During the initialisation phase, the system-level keying material is generated. During the key generation phase, both search and decryption keys are generated for a given set of attributes. Both phases could be run on a PC because keys are distributed out of the band. However, we consider running both phases on a smartphone (with specifications already described above). The initialisation phase takes 108.5 \gls{ms}. The generation time of search keys grows linearly with increase in number of attributes as illustrated in Figure \ref{gra:keygen}, where 30 search keys take 300 \gls{ms} (i.e., an average of 10 \gls{ms} per attribute). Similarly, the key generation time of decryption keys also grows linearly with increase in number of attributes, where 30 decryption keys take approximately 877 \gls{ms} (i.e., an average of 29.25 \gls{ms} per attribute). Asymptotically, the complexity of the key generation is ${\Theta}(|A|)$, where $|A|$ indicates number of attributes in list $A$.

\begin{figure} [htp]
\centering
% left bottom right top
%\includegraphics[trim=65mm 80mm 60mm 70mm,clip,width=.45\textwidth]{...}
\includegraphics[angle=-90,width=.5\textwidth]{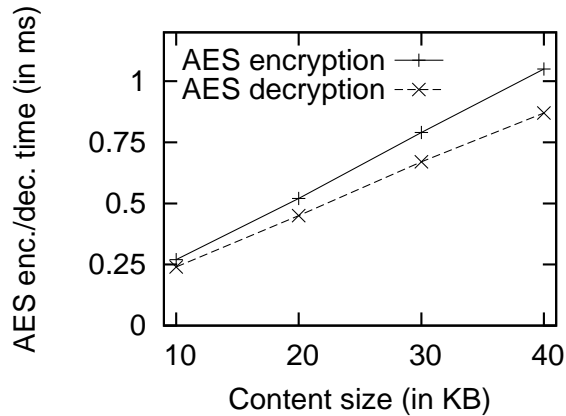} % .3 - descreased for space
\caption[Effect of content size on the AES encryption/decryption time]{Effect of content size on the \gls{AES} encryption/decryption time}
\label{gra:contentsize}
\end{figure}

\begin{figure}
\centering

\subfigure[]{
\includegraphics[angle=-90,width=.48\textwidth]{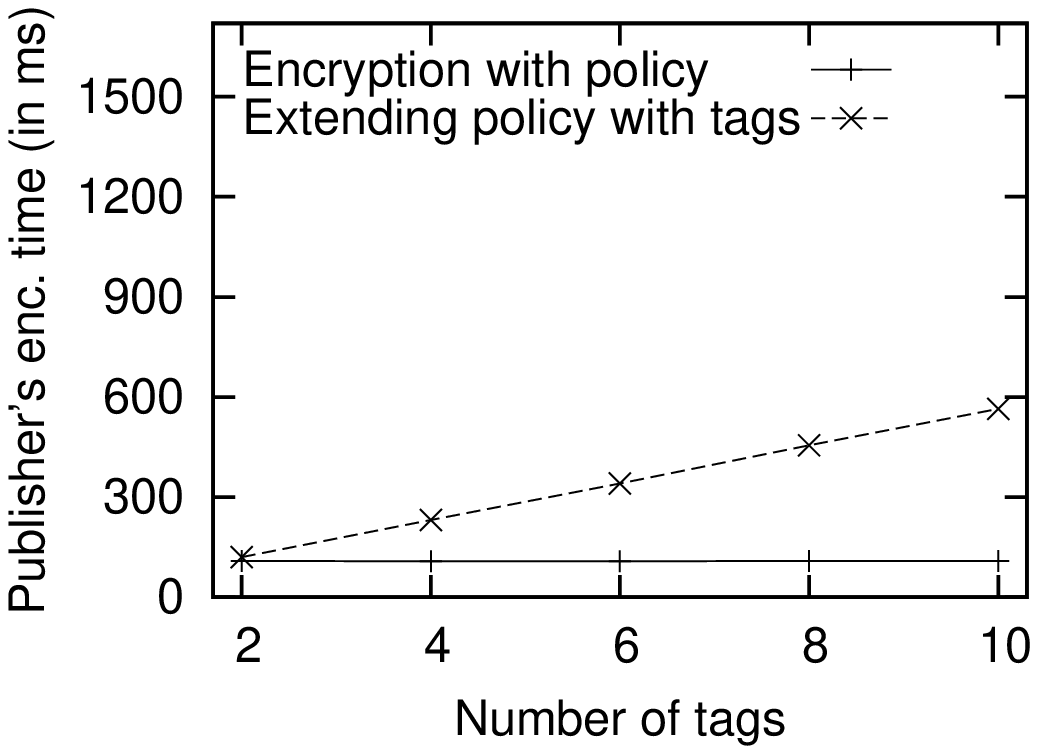} % .31
\label{gra:enc-tags}
}
\subfigure[]{
\includegraphics[angle=-90,width=.48\textwidth]{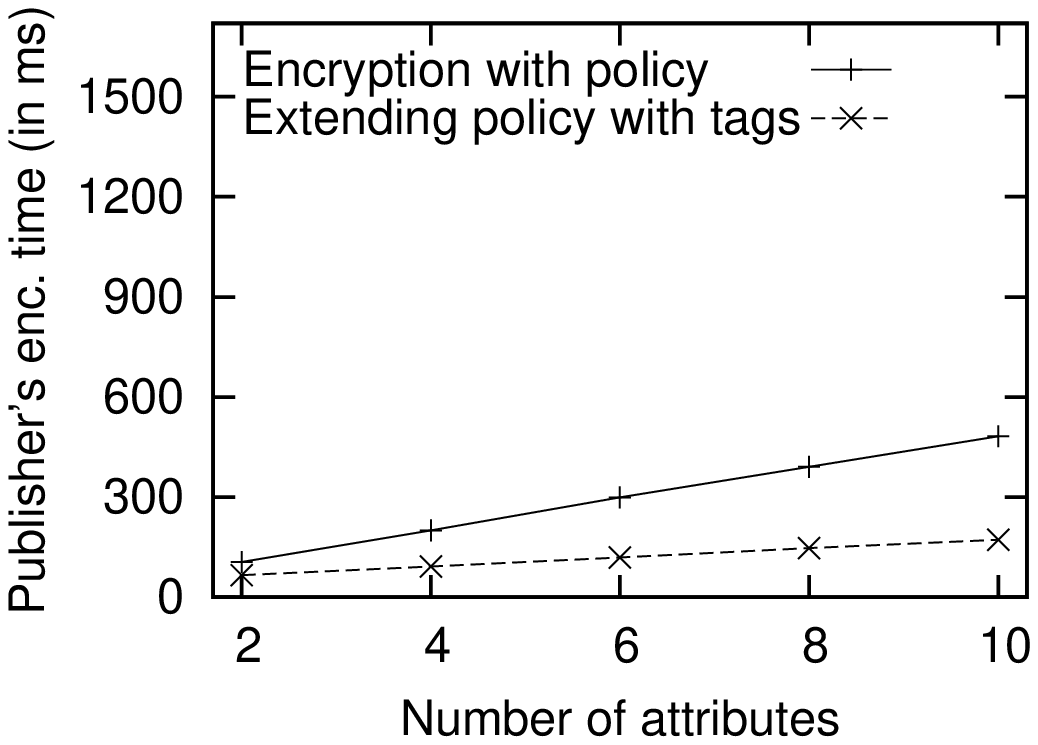} % .31
\label{gra:enc-attributes}
}
\subfigure[]{
\includegraphics[angle=-90,width=.48\textwidth]{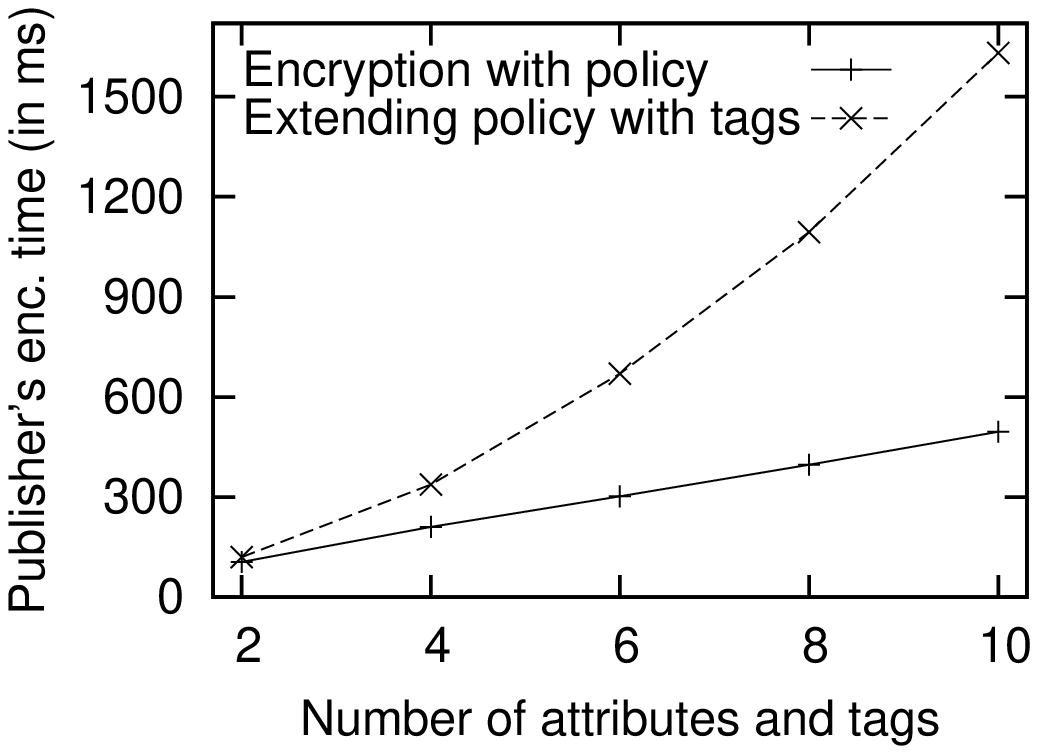} % .31
\label{gra:enc-attributes-tags}
}

\caption[Effect of tags and attributes on publisher's encryption time]{Effect of \subref{gra:enc-tags} tags, \subref{gra:enc-attributes} attributes and \subref{gra:enc-attributes-tags} both tags and attributes on publisher's encryption time}
\label{fig:pidgin-pub-enc}
\end{figure}

\iffalse
Tag = 51.78
Trapdoor = 28.47
Match = 13.28
\fi

\subsection{The Publisher's Encryption Phase}
In this phase, a publisher encrypts content with a randomly generated symmetric key. In our prototype we use \gls{AES} keys. 
The symmetric key is encrypted with the \gls{CPABE} policy. The \gls{CPABE} policy is extended with tags that are also encrypted. 
Figure \ref{gra:contentsize} shows the symmetric encryption time, which grows linearly with the increase in size of content ($C$). Encryption of a piece of content of size 40 \gls{KB} takes 0.105 \gls{ms} (i.e., an average of 0.026 \gls{ms} per \gls{KB}). To measure the performance overhead for the encryption time, we varied the numbers of tags and/or attributes ($A_P^*$), as shown in Figure \ref{fig:pidgin-pub-enc}. In Figure \ref{gra:enc-tags} and Figure \ref{gra:enc-attributes}, we observe the effect of tags and attributes on publisher's encryption time, respectively. In Figure \ref{gra:enc-tags}, we observe effect of tags (ranging from 2 to 10) while keeping the number of attributes constant (i.e., 2 attributes - the minimum attributes required to make AND/OR policy). As we can expect, the time to extend a policy with tags grows linearly with increase in number of tags. In Figure \ref{gra:enc-attributes}, we observe the effect of attributes (ranging from 2 to 10) in a policy while considering a single tag. The time for encryption of the symmetric key with the policy grows linearly with increase in number of attributes. Since the number of attributes increases, it also linearly increases the time to extend the policy with tags. In Figure \ref{gra:enc-attributes-tags}, we show the most complex case in which we increase both attributes and tags simultaneously. The growth of the time needed to extend a policy with tags is quadratic, depending on the number of attributes and the number of tags. In our experimentation, we considered the number of tags as equal to the number of attributes. In a policy with 2 attributes each with 2 tags, it takes approximately 120 \gls{ms} to extend the policy tags, while in a policy with 10 attributes with 10 tags each, it takes approximately 1632 \gls{ms}. Generally, the asymptotic complexity of publisher's encryption is ${\Theta}(|A_P^*| \cdot |T| + |C|)$. %JS
% TODO: check average case for 2 and 10.

\begin{figure}
\centering

\subfigure[]{
\includegraphics[angle=-90,width=.48\textwidth]{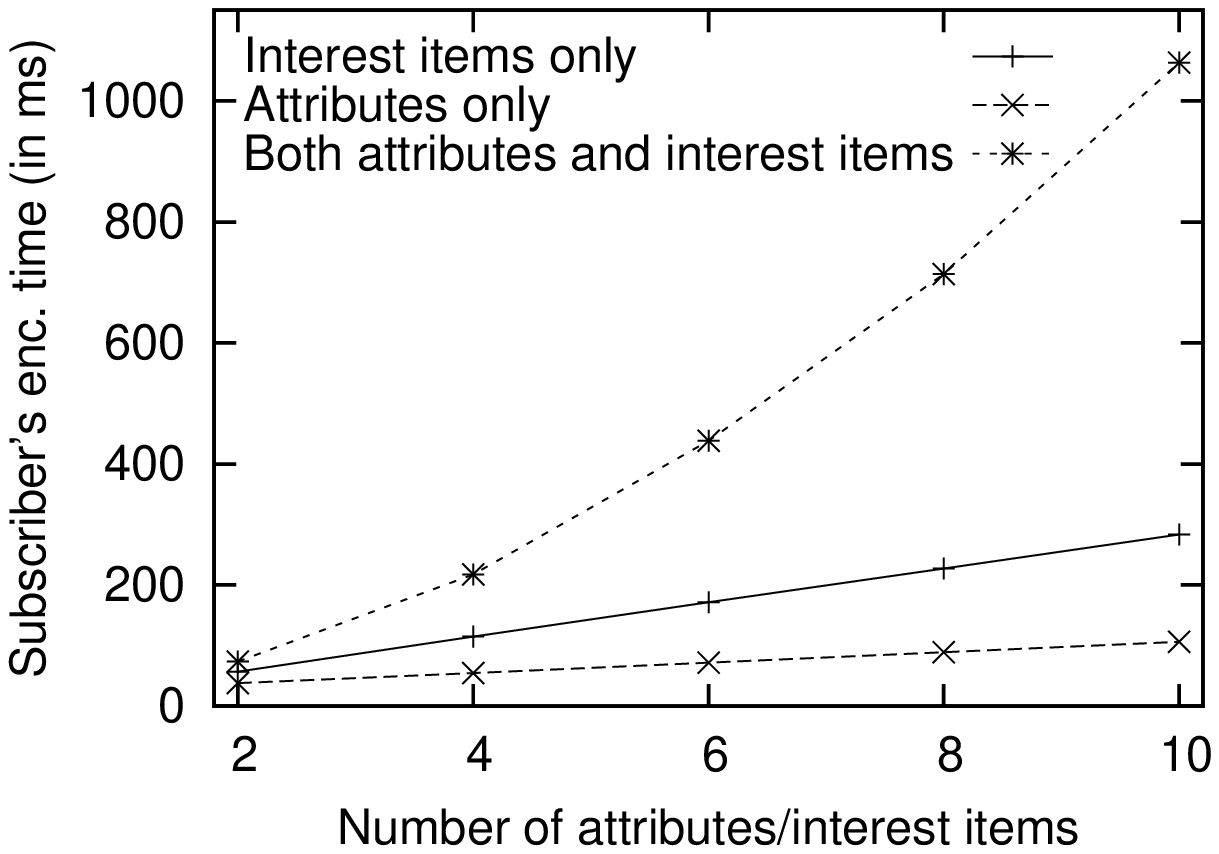} % .31
\label{gra:sub}
}
\subfigure[]{
\includegraphics[angle=-90,width=.48\textwidth]{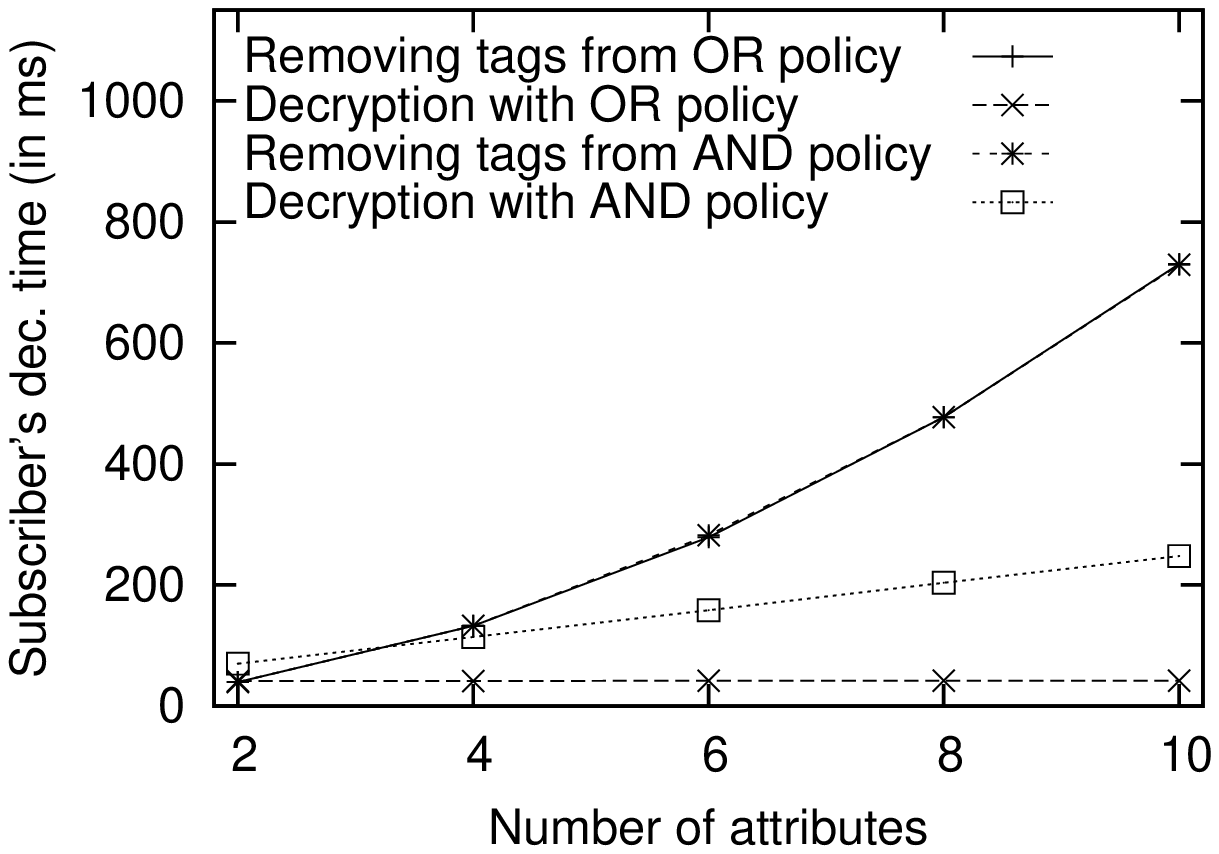} % .31
\label{gra:dec-attributes}
}
\subfigure[]{
\includegraphics[angle=-90,width=.48\textwidth]{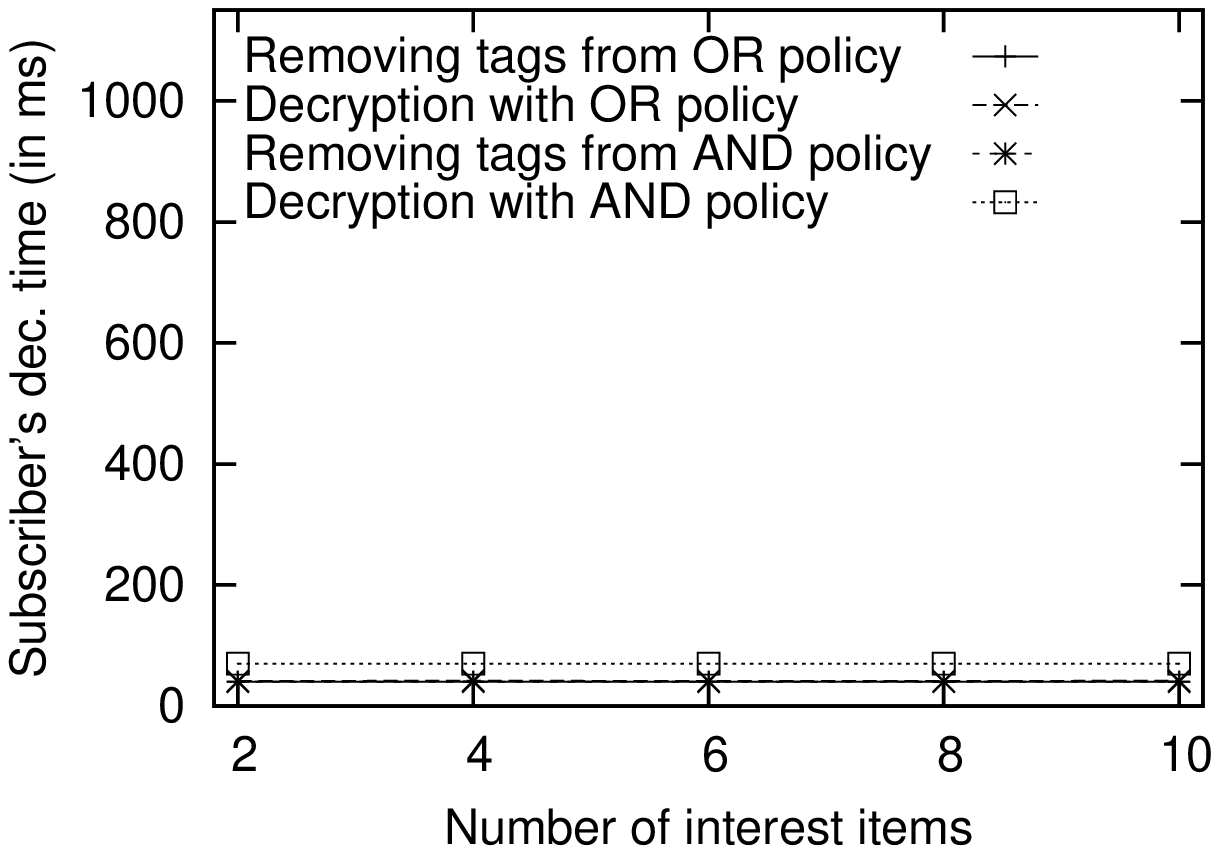} % .31
\label{gra:dec-tags}
}

\caption[Effect of attributes/items on the subscriber's encryption/decryption time]{Effect of \subref{gra:sub} attributes/interest items on the subscriber's encryption time and effect of \subref{gra:dec-attributes} attributes and \subref{gra:dec-tags} tags on the subscriber's decryption time}
\label{fig:pidgin-sub-enc-dec}
\end{figure}

\subsection{The Subscriber's Encryption Phase}
Figure \ref{fig:pidgin-sub-enc-dec} shows the performance overhead incurred during the encryption (see Figure \ref{gra:sub}) and decryption phases (see Figure \ref{gra:dec-attributes} and Figure \ref{gra:dec-tags}). In the subscriber's encryption phase, a subscriber encrypts the subscription, which is based on the number of interest items ($I$) and attributes ($A_S^*$). In our experimentations, we observed the effect of how different values for the number of attributes and interest separately and together affect the subscription's encryption time. To observe the effect of the number of attributes, we increased the attributes from 2 to 10 while keeping interest items constant (i.e., 1 interest item). Generation of trapdoors for 10 attributes with a single interest item each took approximately 106 \gls{ms}. Second, we observed the effect of number of interest items on the subscription's time by increasing interest items from 2 to 10 while keeping attributes constant (i.e., 2 attributes conjuncted with either AND or OR). The subscriber took approximately 284 \gls{ms} to encrypt an interest containing 10 items. As illustrated in Figure \ref{gra:sub}, attributes alone or interest items alone linearly affect the subscriber's encryption time. However, we also consider the case when we see effects of both attributes and interest items together. For this purpose, we assumed that number of attributes is equal to that of interest items; that is, if there are two attributes, it means there are two interest items per attribute. Similarly, we assumed 10 attributes with 10 interest items each, which took 1063 \gls{ms}. The combined effect of attributes and interest items indicates that its growth has quadratic effect on the subscriber's encryption time as shown in Figure \ref{gra:sub}. The asymptotic complexity of the subscriber's encryption is: ${\Theta}(|A_S^*| \cdot |I|)$.

\begin{figure}
\centering

\subfigure[]{
\includegraphics[angle=-90,width=.48\textwidth]{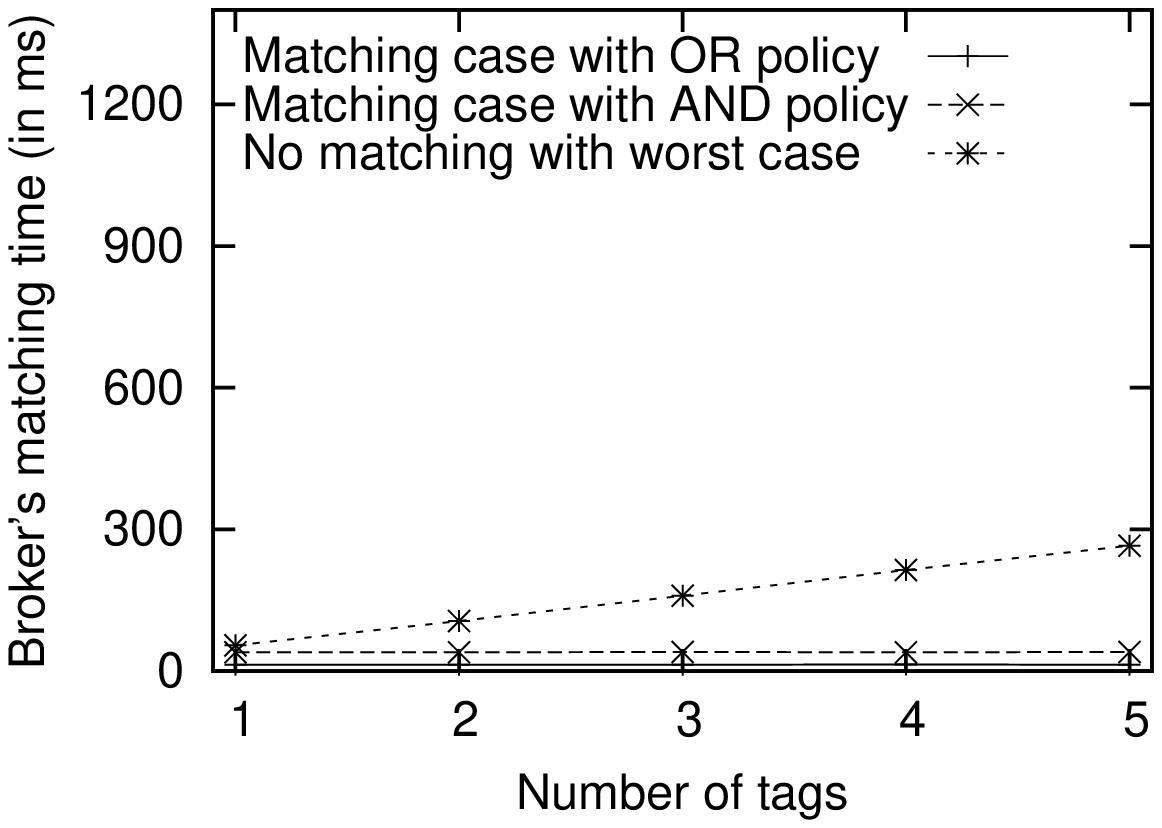} % .31
\label{gra:bro-tags}
}
\subfigure[]{
\includegraphics[angle=-90,width=.48\textwidth]{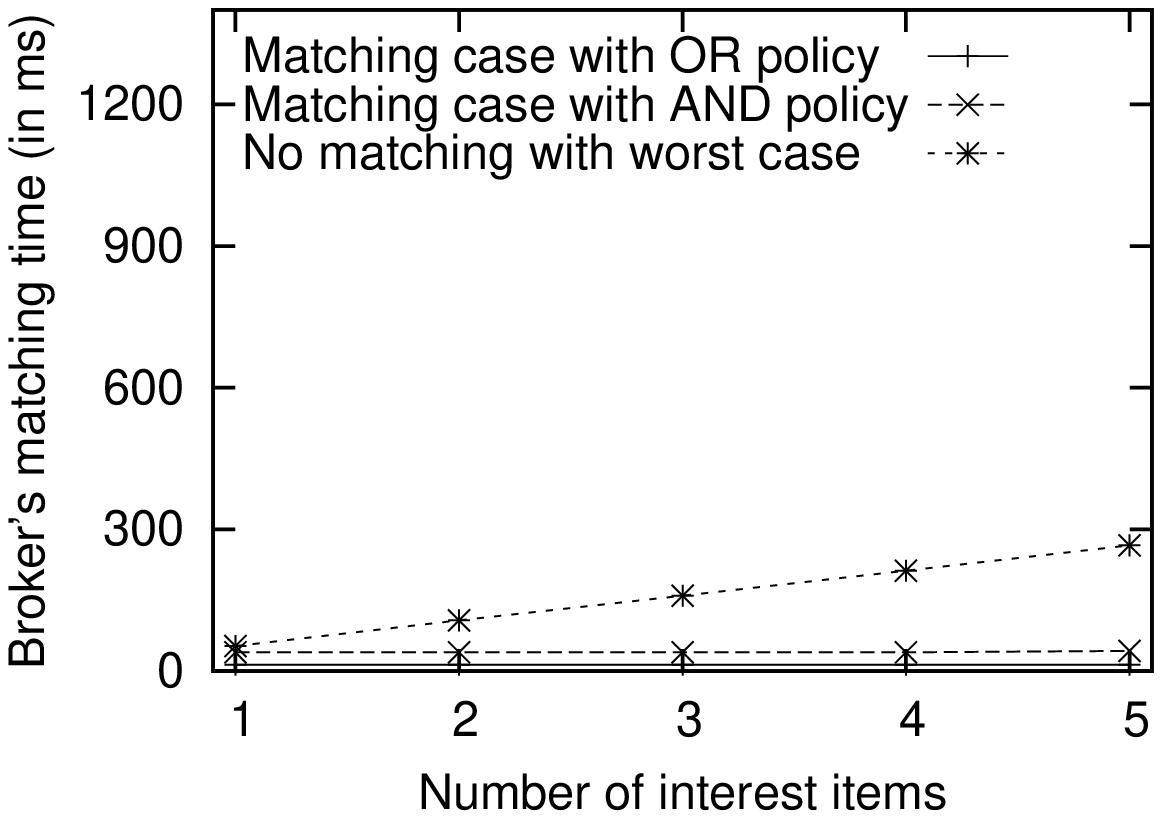} % .31
\label{gra:bro-interest}
}
\subfigure[]{
\includegraphics[angle=-90,width=.48\textwidth]{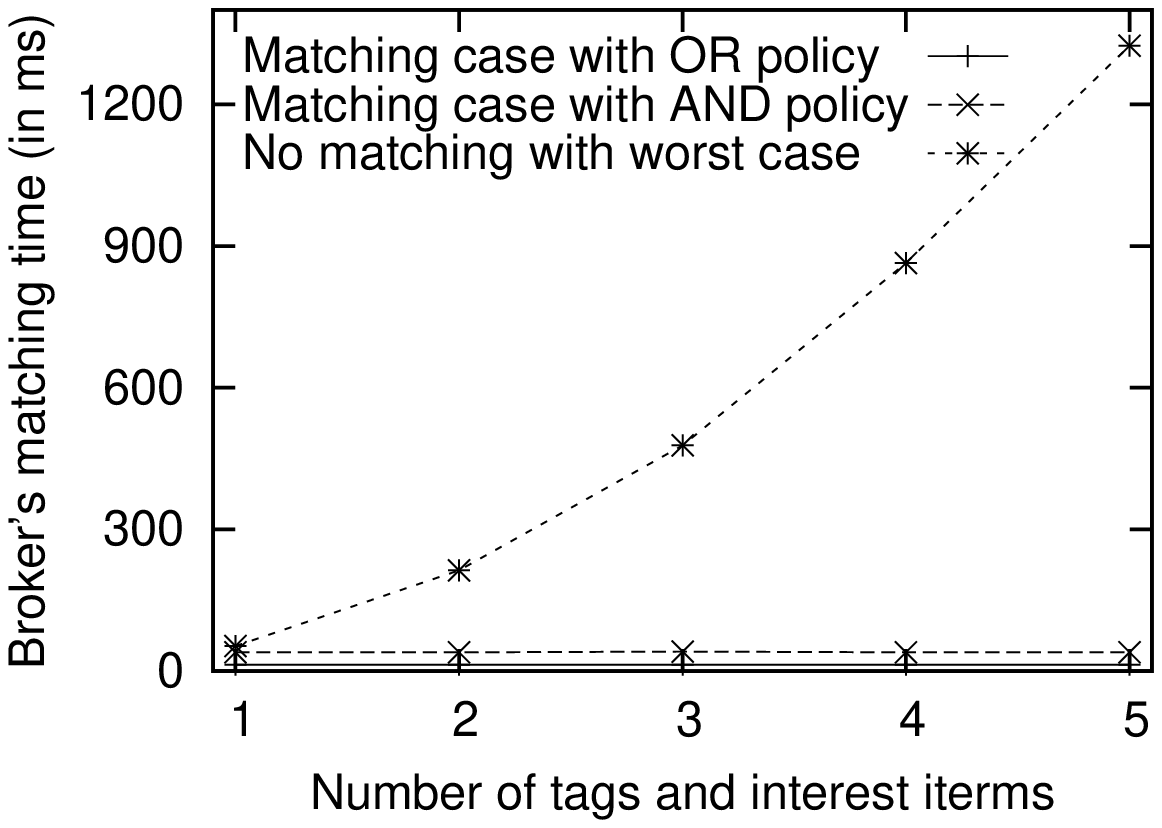} % .31
\label{gra:bro-tag-interest}
}

\caption[Effect of tags and interest items on the broker's encrypted matching time]{Effect of \subref{gra:bro-tags} tags, \subref{gra:bro-interest} interest items and \subref{gra:bro-tag-interest} both tags and interest items on the broker's encrypted matching time}
\label{fig:pidgin-bro}
\end{figure}

\subsection{The Broker's Matching Phase}
This is the key phase in the lifecycle of \gls{PIDGIN}. During this phase, a broker matches the encrypted subscription against the encrypted policy associated with the encrypted content. In our analysis, we observe the effect of the numbers of tags and interest items separately and together while keeping the number of attributes constant (i.e., 2 attributes, necessary to have OR or AND policy). Furthermore, we consider the matching case with both OR and AND policies, as well as a zero match case which is the worst case situation. Figure \ref{fig:pidgin-bro} shows the performance analysis of this phase. In Figure \ref{gra:bro-tags}, we observe the effect of number of tags on the matching time while keeping the number of interest items as constant, i.e., 1. As the graph shows, the matching time increases linearly with the increase in number of tags. Similarly, we measure the effect of the number of interest items on the matching time while keeping the number of tags constant, i.e., 1. As Figure \ref{gra:bro-interest} indicates, the matching time grows linearly with the increase in the number of interest items. In both Figure \ref{gra:bro-tags} and Figure \ref{gra:bro-interest}, the OR policy takes less time as compared to that of the AND policy when we consider the matching case because we use a short circuit evaluation (explained in Section \ref{sec:pidgin-optimisation}) to evaluate both OR and AND gates. Finally, we consider the most complex case in which we increase the number of tags and the number of interest items together (equally) with 2 attributes. Similar to Figure \ref{gra:bro-tags} and Figure \ref{gra:bro-interest}, it takes less time to evaluate the OR policy as compared to that of the AND policy. Next, we consider the worst case in which there are 5 tags and 5 interest items with a 2-attribute policy conjuncted using OR. Since there are 2 attributes in the policy tree with 5 tags each, there will be 10 leaf nodes in the encrypted policy. Furthermore, 2 attributes with 5 interest items each will make 10 trapdoors in the subscription list. The broker checks whether any encrypted leaf node in the policy matches with any trapdoor in the subscription list. In this worst case, the broker runs the \textbf{Test} function 100 times, thus taking approximately 1324 \gls{ms}. In addition to this experiment, we measured the overhead for running the \textbf{Test} function and discovered that it takes 13.28 \gls{ms}. This implies that the real overhead comes from the \textbf{Test} function, that is, in fact, a bilinear pairing operation. Hence, the matching operation is dependent on how efficient the bilinear pairing is. The best and worst case complexities of this phase are ${\Omega}(1)$ and $O(|A_P^*| \cdot |T| \cdot |A_S^*| \cdot |I|)$, respectively.

\subsection{The Subscriber's Decryption Phase}
% TODO: mention story of list ???
A subscriber receives the encrypted content (along with the encrypted policy) from the broker if the encrypted interest satisfies the encrypted policy associated with the encrypted content. During the decryption phase, first a subscriber strips off the tags from the policy and then performs decryption with the policy to recover the symmetric key, which is finally used to decrypt the contents. Figure \ref{gra:dec-attributes} and Figure \ref{gra:dec-tags} show the effect of the number of attributes and interest items, respectively, on the subscriber's decryption time. In Figure \ref{gra:dec-attributes}, where we increase attributes from 2 to 10 while keeping the number of interest items constant i.e., 1, we consider both OR and AND policies to see the effect of attributes on the stripping of tags from the policy. Also, we show the performance overhead for the decryption that recovers the symmetric key. In Figure \ref{gra:dec-tags}, we describe the case in which the number of interest items are increased from 2 to 10 (but attributes are kept constant i.e., 2), assuming the matching case, i.e., both the publisher and the subscriber are using the same tags and interest items, respectively. Here, the overhead does not increase with the increase in the number of interest items because we have implemented the short circuit evaluation to evaluate AND, OR and threshold gates. In fact, the trapdoor in the subscription matches interest items against tags in the policy, thus making policy evaluation successful without requiring further matches. Finally, the encrypted contents are decrypted using the symmetric key, which is recovered after we perform the \gls{CPABE} decryption. Figure \ref{gra:contentsize} shows the time required for decryption of the content using the \gls{AES} key. Decryption of a piece of content of size 40 \gls{KB} takes 0.87 \gls{ms} (i.e., an average of 0.22 \gls{ms} per \gls{KB}). Overall, the complexity of subscriber's decryption is: $O(|A_P^*| \cdot |T| \cdot |A_S^*| \cdot |I| + |C|)$ in the worst case and ${\Omega}(|C|)$ in the best case.

\begin{table} [htp]
\centering
\caption[Time complexity of each phase in the lifecycle of PIDGIN]{Summary of time complexity of each phase in the lifecycle of \gls{PIDGIN}}
\label{tab:pidgin-complexity-summary}
\begin{tabular}{ |l|c|c| } 

\hline

\textbf{Phase Name} & \textbf{Best Case} & \textbf{Worst Case} \\ \hline

Key generation & \multicolumn{2}{c|}{${\Theta}(| A |)$} \\ \hline

Publisher encryption & \multicolumn{2}{c|}{${\Theta}(|A_P^*| \cdot |T| + |C|)$} \\ \hline

Subscriber encryption & \multicolumn{2}{c|}{${\Theta}(|A_S^*| \cdot |I|)$} \\ \hline

Broker matching & ${\Omega}(1)$ & $O(|A_P^*| \cdot |T| \cdot |A_S^*| \cdot |I|)$ \\ \hline

Subscriber decryption & ${\Omega}(|C|)$ & $O(|A_P^*| \cdot |T| \cdot |A_S^*| \cdot |I| + |C|)$ \\ \hline

\end{tabular}
\end{table}

Table \ref{tab:pidgin-complexity-summary} summarises time complexity of each phase in the lifecycle of \gls{PIDGIN}.

\begin{table} [htp]
\centering
\caption{Space overhead of generating encrypted tags and trapdoors}
\label{tab:pidgin-storage}
\begin{tabular}{ |l|c| }
\hline
\textbf{Function} & \textbf{Size (in Bytes)} \\ \hline
Encrypted Tag (by a publisher) & 256 \\ \hline
Trapdoor (by a subscriber) & 128 \\ \hline
\end{tabular}
\end{table}

\section{Discussion}
\label{sec:pidgin-discussion}

%This section discusses various aspects of \gls{PIDGIN}.

\subsection[Storage Analysis of PIDGIN]{Storage Analysis of \gls{PIDGIN}}

As we explained in Section \ref{sec:pidgin-analysis}, the curve we use in our experimentation is of type A. Using this curve, the space complexity of an encrypted tag a trapdoor are 256 and 128 Bytes, respectively. Table \ref{tab:pidgin-storage} shows space overhead of generating encrypted tags and trapdoors.

\subsection{Optimisation and Scalability}
\label{sec:pidgin-optimisation}

\noindent \textbf{Optimisation using Short Circuit Evaluation.} The real bottleneck is matching at brokers a set of encrypted policies against encrypted subscriptions. The large scale matching requires efficiency and some optimisations. One of the optimisations at brokers is implementation of short circuit evaluation for evaluating internal (i.e., non-leaf) nodes of the encrypted policy tree including OR and AND gates. That is, if the node is an OR gate then a broker can stop its evaluation and mark it satisfied once a single child node is satisfied, without performing further matches. Similarly, a broker can mark an AND gate unsatisfied when a single child node is marked unsatisfied. The short circuit evaluation can significantly reduce number of encrypted matches at brokers. This might be useful for the large policies involving a number of children in the policy tree. However, this might not speed up the performance when the set of policies or the number of subscriptions is very large. \\ \\
\noindent \textbf{Scalability.} For matching a large set of encrypted policies against a large number of encrypted subscriptions, \gls{PIDGIN} can take into account additional information that can drastically improve the overall performance. That is, a publisher can specify the content creation date while a broker can log time when the content was received. A subscriber can take advantage of this extra information by expressing additional constraints in subscription. For instance, a subscriber can express her subscription as: \emph{all pieces of content matching with my interest, where the content is created or received in last two hours}. The content creation date and the content received time may help brokers to check whether subscriptions satisfy the published contents, without requiring encrypted matching. Furthermore, a publisher can publish content with \gls{TTL}, meaning brokers should remove that particular piece of content after expiration of \gls{TTL}. Similarly, subscribers also can include \gls{TTL} with subscriptions to indicate that brokers can remove subscriptions from the network after expiration of \gls{TTL}. The inclusion of \gls{TTL}, in both the content and the subscription, will reduce both computation and storage needs.

\subsection{Key Management}

\noindent \textbf{Deployment in Practical Scenarios.}
There are various options to setup the \gls{TKMA}, an offline trusted entity that distributes keying material. It mainly depends on the scenario for which \gls{PIDGIN} is deployed. For instance, for the military scenarios, it can be administrated by a military headquarter; similarly, in organisations, the admin department can manage it. However, it is challenging to setup the \gls{TKMA} for various civilian applications. For those kinds of applications, the town or city administration could be one option. For emerging scenarios, such as social events, the organising authorities (such as event organisers) might own the \gls{TKMA}. \\ \\
\noindent \textbf{Distributed \gls{TKMA}.} 
Without loss of generality, we can make the \gls{TKMA} distributed. There are two main types of keys that are generated by the \gls{TKMA}, the \gls{CPABE} and the search keys. There are alraedy solutions for setting up multi-authority \gls{ABE} \cite{Chase:2009, Chase:2007}, where the \gls{CPABE} key authorities can be distributed. Whereas, the key authority for generating the search keys is inherently distributed.

\section{Related Work}
\label{sec:pidgin-related-work}

The problem of encrypted matching in opportunistic networks is an instance of the wider problem of a search over encrypted data. Song \emph{et al.} \cite{Song:2000} propose a search scheme over encrypted data based on symmetric keys. The symmetric nature of the scheme rules out its applicability where mobile nodes communicate with each other without any prior contacts. The \gls{PEKS} scheme \cite{Boneh:2004} supports a search on encrypted data in the public key setting. In \gls{PIDGIN}, we use the \gls{PEKS} scheme as a building block; moreover, its usage in isolation does not solve privacy and confidentiality issues in opportunistic networks because it lacks the ability to regulate access on content while providing collusion-resistant decryption keys.

The \gls{ABE} schemes can regulate access to content while guaranteeing collusion resistance. However, both variants of \gls{ABE} including \gls{CPABE} \cite{Bethencourt:2007} and \gls{KPABE} \cite{Goyal:2006} do not protect the policies and attributes associated with content, respectively. In \gls{PIDGIN}, we use \gls{CPABE} \cite{Bethencourt:2007} as a building block but only after we protect the policies because the original \gls{CPABE} scheme does not specifically protect them. The complimentary \gls{KPABE} \cite{Goyal:2006} scheme does not protect attributes. While, Goyal \emph{et al.} leave the problem of encrypted attributes as open \cite{Goyal:2006}, we address this challenging issue in this chapter.

\gls{ESPOON} \cite{Asghar2011-ARES} can protect security policies in outsourced environments. In \cite{Asghar2013-COSE}, we propose \gls{ESPOONERBAC} that extends \gls{ESPOON} with \gls{ERBAC} that is deployable in outsourced environments. However, these solutions \cite{Asghar2013-COSE, Asghar2011-ARES, Kapadia:2007} assume no collusion between a user and a server. Thus, none of these solutions \cite{Asghar2013-COSE, Asghar2011-ARES, Kapadia:2007} are applicable to opportunistic networks in which each node can serve as all three roles including publisher, broker and subscriber.

There are schemes that protect policies \cite{Shen:2009, Katz:2013, Nishide:2008, Lai:2011} and assume that the policy is evaluated at the receiver's end. Furthermore, schemes offering hidden credentials \cite{Holt:2003} and hidden policies \cite{Frikken:2006} assume direct interaction between the sender and the receiving parties. Unfortunately, all such schemes cannot work in opportunistic networks where policy enforcement is delegated to untrusted brokers.

Shikfa \emph{et al.} \cite{Shikfa:2010} propose a method that provides privacy and confidentiality in context-based forwarding. However, their method is a different dimension of work than ours. In fact, their proposed scheme disseminates information in one direction, i.e., from publishers, without taking into account whether a subscriber is interested or not. In other words, it does not provide opportunity for a subscriber to subscribe. Moreover, our proposed scheme regulates access to content while offering more expressive and fine-grained policies as compared to the one proposed in \cite{Shikfa:2010}.

%\cite{Shikfa:2009}, also see

%CP-A$^3$BE \cite{Li:2009}, optional

Nabeel \emph{et al.} \cite{Nabeel:2012} provide a solution for preserving privacy in content based publish-subscribe systems. In their approach, brokers in outsourced environments make routing decisions without knowing the content. However, they assume that subscribers get registered with publishers prior to any communication and publishers share the symmetric key with subscribers. This solution cannot work in opportunistic network settings where loosely-coupled publishers and subscribers do not require any registration or key sharing with each other. 

In the context of publish-subscribe systems, there are many solutions that address privacy and security issues \cite{Choi:2010, Shang:2010, Srivatsa:2007}. However, state-of-the-art techniques are mainly based on centralised solutions that cannot be applied to opportunistic networks, where each node may serve as a publisher, a broker and a subscriber.

\section{Chapter Summary}
\label{sec:pidgin-summary}

This chapter presented \gls{PIDGIN}, a privacy preserving interest and content sharing scheme for opportunistic networks. In \gls{PIDGIN}, access policies are enforced by brokers such that they neither learn content and associated policies nor compromise privacy of subscribers. To show the feasibility of our approach, we implemented \gls{PIDGIN} and evaluated its performance by measuring the overhead incurred by cryptographic operations when run on a smartphone.

In Chapter \ref{cha:espoon}-\ref{cha:egrant}, we have investigated how to enforce sensitive security policies in outsourced environments while this chapter has addressed how sensitive policies can be enforced in distributed environments. Hence, we covered enforcement of sensitive security policies in both outsourced and distributed environments. In the next chapter, we summarise our contributions and highlight some directions for future work.
%

%%%%%%%%%%%%%%%%%%%%%%%%% CHAPTER CONCLUSIONS %%%%%%%%%%%%%%%%%%%%%%%%%

\chapter{Conclusions and Future Work}
\label{cha:conclusion}

%Conclusions should summarize the problem, the solution and its main innovative features, outlining future work on the topic or application scenarios of the proposed scheme.

% NOTE!!!!! After reading thousands of words, students must be able to present effective, convincing conclusions, restating the original contribution to knowledge, the significance of the research, the problems and flaws and further areas of scholarship. Short conclusions are created by tired doctoral students. They run out of words.

In this dissertation, we have addressed a fundamental issue of establishing trust in untrusted environments by protecting access policies and data. In particular, we have investigated how to enforce sensitive policies in outsourced and distributed environments. In our approach, the data is encrypted under expressive access control policies that are attached with the encrypted data. Our proposed mechanisms enforce those policies such that private information is not revealed during the policy deployment and evaluation phases. Furthermore, we offer the full-fledged \gls{RBAC} mechanism (including role hierarchies and dynamic security constraints) for large enterprises with complex user management.

In our work, we have presented some motivational scenarios. Besides what we have considered, there could be other application scenarios as well. For data outsourcing, we can imagine investigation and security agencies that might require data protection, as well as secure enforcement of sensitive policies. Similarly, we can apply our policy enforcement mechanism in opportunistic networks to report and control crimes in developing countries, where the Internet connectivity is poor or unaffordable. In developed countries, we can think of more sophisticated use cases, such as partially offloading the central \gls{CDN} by employing our proposed mechanism so that subscribers can download content from neighbourhood, thus reducing the burden on the centralised server.

In this chapter, we briefly summarise the research contributions of the dissertation and outline some future directions emerging from this work.

\section{Summary of the Contributions}
The core contributions of this dissertation are stated as follows: \\ \\
\noindent \textbf{\gls{ESPOON}: Enforcement of Sensitive Policies in Outsourced Environments.}
In Chapter \ref{cha:espoon}, we have addressed the challenging issue of enforcing sensitive policies in outsourced environments while protecting confidentiality of access policies. Our proposed solution, \gls{ESPOON}, provides a clear separation between security policies and the enforcement mechanism. \gls{ESPOON} does not reveal private information about access policies or the access request. In fact, we implement \gls{ESPOON} as an outsourced service without compromising the confidentiality of policies under the assumption that the service provider is honest-but-curious. Furthermore, \gls{ESPOON} supports contextual conditions and incorporates contextual information during the policy evaluation phase. Contextual conditions are expressive because they include non-monotonic boolean expressions and range queries. The system entities do not share any keys; therefore, if a user is deleted or revoked, the system is still able to perform its operations without requiring any re-encryption of policies. Last but not least, we have implemented a prototype of \gls{ESPOON} to measure overheads incurred by cryptographic operations during the policy deployment and evaluation phases. \\ \\
\noindent \textbf{\gls{ESPOONERBAC}: Supporting \gls{RBAC} Policies and Role Hierarchies.}
In Chapter \ref{cha:erbac}, we have extended \gls{ESPOON} with \gls{RBAC} policies and proposed \gls{ESPOONERBAC}. In \gls{ESPOONERBAC}, users are assigned roles and permissions are assigned to roles. A user can execute the permission if she is active in a role managed by the session maintained in outsourced environments. Besides the basic \gls{RBAC} policies, \gls{ESPOONERBAC} incorporates roles hierarchies, where roles can be inherited. For developing prototype of \gls{ESPOONERBAC}, we have extended prototype of \gls{ESPOON}. Finally, we have measured the computational overheads incurred by \gls{ESPOONERBAC} operations. \\ \\
\noindent \textbf{\gls{EGRANT}: Facilitating \gls{RBAC} with Dynamic Constraints.}
In Chapter \ref{cha:egrant}, we have focused on the enforcement of dynamic security constraints without revealing sensitive information to the untrusted infrastructure. The dynamic constraints include \gls{DSoD} and \gls{CW}. For enforcement of dynamic constraints, we have developed \gls{EGRANT}. \gls{EGRANT} can seamlessly be integrated with \gls{ESPOONERBAC}. Finally, we have developed the prototype and reported performance overhead of \gls{EGRANT}. \\ \\
We would like to mention that \gls{ESPOON}, \gls{ESPOONERBAC} and \gls{EGRANT} can be deployed as \gls{SaaS}. \\ \\
\noindent \textbf{\gls{PIDGIN}: Protecting Privacy and Confidentiality in Opportunistic Networks.}
In Chapter \ref{cha:pidgin}, we have investigated how to exchange content and interest without (i) providing any access to unauthorised brokers and compromising privacy of subscribers. The solution we propose is \gls{PIDGIN} that aims at regulating access by encrypting content using \gls{CPABE} policies. The \gls{CPABE} policies are very expressive and specify who can gain access to content. In \gls{PIDGIN}, \gls{CPABE} policies and tag associated with content are further encrypted using the \gls{PEKS} scheme. Therefore, brokers match subscriber's interest against content polices without compromising privacy of subscribers. Furthermore, unauthorised brokers do not gain access to content and nodes gain access to content if they satisfy fine-grained policies specified by the publishers. Moreover, the system provides a scalable key management, where loosely-coupled publishers and subscribers do not share any keys. Finally, we have developed a prototype of \gls{PIDGIN} and analysed the performance of involved cryptographic algorithms by running \gls{PIDGIN} on smartphones.

\section{Future Directions}

The research work described in this dissertation can be extended along several directions. \\ \\
\noindent \textbf{Accountable Access Control Mechanisms.}
In this dissertation, we have proposed how sensitive policies can be enforced. One possible direction of future research is to explore ways of making the enforcement architecture accountable in untrusted environments, thus preventing service providers (or brokers) to repudiate the operations that have been performed. The mechanism should allow service providers to generate genuine audit logs without revealing private information about both data and access policies. However, an auditing authority must be able to retrieve information about who accessed the data and what policy was enforced against any access request. \\ \\
\noindent \textbf{Negative Authorisation Policies and Conflict Resolution.}
In our proposed solutions, we have considered positive authorisation policies in untrusted environments. It would be interesting to investigate how to support negative authorisation policies. Since negative and positive authorisation policies might raise conflicts, conflict resolution of policies in untrusted environments might be another interesting topic of research. \\ \\
\noindent \textbf{Making Policy Outsourcing Distributed.}
Another substantial part of our future research aims at re-engineering the architecture in a distributed manner in order to run several instances of the proposed system on multiple nodes of the service provider. One of the key aspects here is to adapt the number of instances to the actual request load for offering a reasonable \gls{QoS} without over-provisioning the resources. \\ \\
\noindent \textbf{Scalable and Collusion-Resistant Access Control Models.}
Generally, access control models in the literature are only either scalable or collusion-resistant. In our view, proposing a scalable and collusion-resistant access model for outsourced environments is still an open challenge. Besides that, developing an efficient cryptographic construction and implementing it efficiently are also among open research challenges. \\ \\
\noindent \textbf{Protection of Policy Structure.}
In our proposed mechanisms, we express an access control policy as a tree, where leaf nodes of the tree are encrypted while internal nodes (including AND, OR and threshold gates) are in cleartext. Protection of this policy structure is also an open challenge. More specifically, it is a challenging issue to support expressive access control policies such that service providers do not learn any information about structure of policies being enforced. \\ \\
\noindent \textbf{Key Revocation in Distributed Settings.}
In distributed settings (including opportunistic networks), revoking a key is quite problematic. The issue is that one cannot inform all nodes about keys that have been revoked because there is no centralised authority for management of key revocation. That is, the key revocation information could epidemically be disseminated only through nodes, say from a group of nodes to other nodes in the network. We believe that investigating an approach to efficiently address the key revocation problem would make distributed networks more practical. \\ \\
\noindent \textbf{Efficient Pairing Implementation.}
As evident from the performance evaluation, the real bottleneck is the overhead incurred by pairing operations at brokers in opportunistic networks. Basically, an efficient pairing implementation would drastically improve the performance of the system. As future work, we would investigate possible optimisations and the use of an efficient pairing implementation, such as the one proposed in \cite{Grewal:2013}. Alternatively, we can consider implementation of cryptographic constructs at processor level, i.e., support of pairing operations in a cryptographic processor.
%

%\noindent \textbf{Write Authorisation.}
%Cite %http://spdp.di.unimi.it/papers/Foresti_phd_thesis.pdf

\section{Closing Remarks}
This work has appeared in international journals, conferences and workshops (See Appendix \ref{app:publications}). In particular, the basic architecture for enforcing sensitive security policies in outsourced environments has been presented in \cite{Asghar2011-ARES}. The proposed architecture has been extended to support \gls{RBAC} style of access policies in outsourced environments and is described in \cite{Asghar2013-COSE, Asghar2011-CCS}. The work on enforcing \gls{RBAC} in outsourced environments has further been extended by incorporating security constraints in \gls{RBAC}, which is presented in \cite{Asghar2013:IJIS:EGRANT}. The data protection issues have been tackled in \cite{Asghar2013:CCSW, Asghar2011-iNetSec-Provenance}. The scenario based security and privacy issues have been listed in \cite{Asghar2012-SmartGridSec}. For brevity reasons, we have included only the research work that fall within the core topic of this dissertation and excluded some published work \cite{Asghar2013:CCSW, Asghar2011-iNetSec-Provenance, Asghar2012-SmartGridSec}. Finally, the issue of policies and data protection in distributed environments has been analysed and addressed in \cite{Asghar2013:NDSS:PIDGIN}.

\clearemptydoublepage

%----

\thispagestyle{empty}
%----
\makeatletter
%\addcontentsline{toc}{chapter}{Bibliography}
\bibliographystyle{ieeetr}
\bibliography{Asghar-PhD-Dissertation}

\clearemptydoublepage

\appendix

\iffalse

\chapter{Basic Algorithms}
Proxy re-encryption and
Keywork encryption

\chapter{Security Proof of Basic Algorithms}
Proofs of both 
Proxy re-encryption and 
Keywork encryption

\fi

\chapter{Research Publications}
\label{app:publications}

%\nobibliography*
\section{Related Publications}
\label{app:selected-publications}
 
\noindent {\bf In International Journals}
 
\begin{enumerate}%[labelindent=.5em,labelsep=-.5cm,leftmargin=*]

  \item %\bibverse{Asghar2013-COSE}
\textbf{Muhammad Rizwan Asghar}, Mihaela Ion, Giovanni Russello, Bruno Crispo, \textit{\textbf{\gls{ESPOONERBAC}: Enforcing Security Policies in Outsourced Environments}, Elsevier Computers \& Security (COSE), volume 35, pages 2-24, 2013. \textbf{One of three papers from ARES 2011 invited to this journal.}} \vspace{3mm} \\
\textbf{Abstract:} Data outsourcing is a growing business model offering services to individuals and enterprises for processing and storing a huge amount of data. It is not only economical but also promises higher availability, scalability, and more effective quality of service than in-house solutions. Despite all its benefits, data outsourcing raises serious security concerns for preserving data confidentiality. There are solutions for preserving confidentiality of data while supporting search on the data stored in outsourced environments. However, such solutions do not support access policies to regulate access to a particular subset of the stored data. 

	For complex user management, large enterprises employ \acrfull{RBAC} models for making access decisions based on the role in which a user is active in. However, \gls{RBAC} models cannot be deployed in outsourced environments as they rely on trusted infrastructure in order to regulate access to the data. The deployment of \gls{RBAC} models may reveal private information about sensitive data they aim to protect. In this chapter, we aim at filling this gap by proposing \textbf{\gls{ESPOONERBAC}} for enforcing \gls{RBAC} policies in outsourced environments. \gls{ESPOONERBAC} enforces \gls{RBAC} policies in an encrypted manner where a curious service provider may learn a very limited information about \gls{RBAC} policies. We have implemented \gls{ESPOONERBAC} and provided its performance evaluation showing a limited overhead, thus confirming viability of our approach. \vspace{3mm} \\
\textbf{Keywords:} Encrypted \gls{RBAC}, Policy Protection, Sensitive Policy Evaluation, Secure Cloud Storage, Confidentiality
  
  \item %\bibverse{Asghar2013:IJIS:EGRANT}
\textbf{Muhammad Rizwan Asghar}, Mihaela Ion, Giovanni Russello, Bruno Crispo, \textit{\textbf{\gls{EGRANT}: Enforcing Encrypted Dynamic Security Constraints in the Cloud}, 2013. (In submission).} \vspace{3mm} \\
\textbf{Abstract:} Cloud computing is an emerging paradigm offering outsourced services to enterprises for storing and processing huge amount of data at very competitive costs. For leveraging the cloud to its fullest potential, organisations require security mechanisms to regulate access on data, particularly at runtime. One of the strong obstacles in widespread adoption of the cloud is to preserve confidentiality of the data. In fact, confidentiality of the data can be guaranteed by employing existing encryption schemes; however, access control mechanisms might leak information about the data they aim to protect. State of the art access control mechanisms can statically enforce constraints such as static separation of duties. The major research challenge is to enforce constraints at runtime, i.e., enforcement of dynamic security constraint (including \acrlong{DSoD} and \acrlong{CW}) in the cloud. The main challenge lies in the fact that dynamic security constraints require notion of sessions for managing access histories that might leak information about the sensitive data if they are available as cleartext in the cloud. In this chapter, we present \gls{EGRANT}: an architecture able to enforce dynamic security constraints without relying on a trusted infrastructure, which can be deployed as \gls{SaaS}. In \gls{EGRANT}, sessions' access histories are encrypted in such a way that enforcement of constraints is still possible. As a proof-of-concept, we have implemented a prototype and provide a preliminary performance analysis showing a limited overhead, thus confirming the feasibility of our approach. \vspace{3mm} \\
\textbf{Keywords:} Secure Cloud Services, Sensitive Dynamic Constraints, Encrypted \gls{DSoD},  Encrypted \acrlong{CW}, \acrshort{SaaS} Enforcement Mechanism
  
\end{enumerate}

  \noindent {\bf In International Conferences and Workshops}
  
\begin{enumerate}%[labelindent=.5em,labelsep=-.5cm,leftmargin=*]
  \setcounter{enumi}{2}
  
  \item %\bibverse{Asghar2013:NDSS:PIDGIN}
\textbf{Muhammad Rizwan Asghar}, Ashish Gehani, Giovanni Russello, Bruno Crispo, \textit{\textbf{\gls{PIDGIN}: Privacy-preserving Interest and Content Sharing in Opportunistic Networks}, 2013. (In submission).} \vspace{3mm} \\
\textbf{Abstract:} Opportunistic networks have recently received considerable attention from both industry and researchers. These networks can be used for many applications without the need for a dedicated \gls{IT} infrastructure. In the context of opportunistic networks, the application to content sharing in particular has attracted specific attention. To support content sharing, opportunistic networks may implement a publish-subscribe system in which users may publish their own content and indicate interest in other content through subscription. Using a smartphone, any user can act as a broker by opportunistically forwarding both published content and interest within the network. Unfortunately, despite their provision of this great flexibility, opportunistic networks raise serious privacy and security issues. Untrusted brokers can not only compromise the privacy of subscribers by learning their interest but also can gain unauthorised access to the disseminated content. This chapter addresses the research challenges inherent to the exchange of content and interest without: (i) compromising the privacy of subscribers and (ii) providing unauthorised access to untrusted brokers. Specifically, this chapter presents an interest and content sharing solution that addresses these security challenges and preserves privacy in opportunistic networks. We demonstrated the feasibility and efficiency of this solution by implementing a prototype and analysing its performance on real smart phones. \vspace{3mm} \\
\textbf{Keywords:} Secure Opportunistic Networks, Privacy-preserving Content Sharing, Sensitive Policy Enforcement, Encrypted \acrshort{CPABE} Policies, Secure Haggle
  
  \item %\bibverse{Asghar2013:CCSW}
\textbf{Muhammad Rizwan Asghar}, Giovanni Russello, Bruno Crispo, Mihaela Ion, \textit{\textbf{Supporting Complex Queries and Access Policies for Multi-user Encrypted Databases}, In Proceedings of The 5th ACM Workshop on Cloud Computing Security Workshop (CCSW) in conjunction with the 20th ACM Conference on Computer and Communications Security (CCS), Berlin, Germany, November 2013.} \vspace{3mm} \\
\textbf{Abstract:} Cloud computing is an emerging paradigm offering companies (virtually) unlimited data storage and computation at attractive costs. It is a cost-effective model because it does not require deployment and maintenance of any dedicated \gls{IT} infrastructure. Despite its benefits, it introduces new challenges for protecting the confidentiality of the data. Sensitive data like medical records, business or governmental data cannot be stored unencrypted on the cloud. Companies need new mechanisms to control access to the outsourced data and allow users to query the encrypted data without revealing sensitive information to the cloud provider. State-of-the-art schemes do not allow complex encrypted queries over encrypted data in a multi-user setting. Instead, those are limited to keyword searches or conjunctions of keywords. This chapter extends work on multi-user encrypted search schemes by supporting SQL-like encrypted queries on encrypted databases. Furthermore, we introduce access control on the data stored in the cloud, where any administrative actions (such as updating access rights or adding/deleting users) do not require re-distributing keys or re-encryption of data. Finally, we implemented our scheme and presented its performance, thus showing feasibility of our approach. \vspace{3mm} \\
\textbf{Keywords:} Encrypted Databases, Complex Encrypted Queries, Access Control, Data Outsourcing
  
  \item %\bibverse{Asghar2012-SmartGridSec}
\textbf{Muhammad Rizwan Asghar}, Daniele Miorandi, \textit{\textbf{A Holistic View of Security and Privacy Issues in Smart Grids}, In Jorge Cuellar, editor, Smart Grid Security, volume 7823 of Lecture Notes in Computer Science, pages 58-71, Springer Berlin Heidelberg, 2013.} \vspace{3mm} \\
\textbf{Abstract:} The energy system is undergoing a radical transformation. The coupling of the energy system with advanced information and communication technologies is making it possible to monitor and control in real-time generation, transport, distribution and consumption of energy. In this context, a key enabler is represented by smart meters, devices able to monitor in near real-time the consumption of energy by consumers. 

	If, on one hand, smart meters automate the process of information flow from endpoints to energy suppliers, on the other hand, they may leak sensitive information about consumers. In this chapter, we review the issues at stake and the research challenges that characterise smart grids from a privacy and security standpoint. \vspace{3mm} \\
\textbf{Keywords:} Privacy, Data Security, Smart Meters, Smart Grids, Prosumers
  
	\item %\bibverse{Asghar2011-iNetSec-Provenance}
\textbf{Muhammad Rizwan Asghar}, Mihaela Ion, Giovanni Russello, Bruno Crispo, \textit{\textbf{Securing Data Provenance in the Cloud}, In Jan Camenisch and Dogan Kesdogan, editors, Open Problems in Network Security, volume 7039 of Lecture Notes in Computer Science, pages 145-160, Springer Berlin Heidelberg, 2012.} \vspace{3mm} \\
\textbf{Abstract:} Cloud storage offers the flexibility of accessing data from anywhere at any time while providing economical benefits and scalability. However, cloud stores lack the ability to manage data provenance. Data provenance describes how a particular piece of data has been produced. It is vital for a post-incident investigation, widely used in healthcare, scientific collaboration, forensic analysis and legal proceedings. Data provenance needs to be secured since it may reveal private information about the sensitive data while the cloud service provider does not guarantee confidentiality of the data stored in dispersed geographical locations. This chapter proposes a scheme to secure data provenance in the cloud while offering the encrypted search. \vspace{3mm} \\
\textbf{Keywords:} Secure Data Provenance, Encrypted Cloud Storage, Security, Privacy
	
\item %\bibverse{Asghar2011-CCS}
%
% TODO: on update remove \\ that is added to fix format issue
%
\textbf{Muhammad Rizwan Asghar}, Giovanni Russello, Bruno Crispo, \textit{\textbf{Poster: \\ \gls{ESPOONERBAC}: Enforcing Security Policies in Outsourced Environments with Encrypted \gls{RBAC}}, In Proceedings of the 18th ACM Conference on Computer and Communications Security, CCS'11, pages 841-844. ACM, 2011.} \vspace{3mm} \\
\textbf{Abstract:} The enforcement of security policies is an open challenge in environments where the \gls{IT} infrastructure has been outsourced to a third party. Although the outsourcing allows companies to gain economical benefits and scalability, it imposes the threat of leaking the private information about the sensitive data managed and processed by untrusted parties. In this work, we propose an architecture to enforce \acrfull{RBAC} style of authorisation policies in outsourced environments. As a proof of concept, we have implemented a demo and measured the performance overhead incurred by the proposed architecture. \vspace{3mm} \\
\textbf{Keywords:} Encrypted \gls{RBAC}, Encrypted Policy Enforcement, Data Outsourcing, Security, Privacy
	
	\item %\bibverse{Asghar2011-ARES}
\textbf{Muhammad Rizwan Asghar}, Mihaela Ion, Giovanni Russello, Bruno Crispo, \textit{\textbf{\gls{ESPOON}: Enforcing Encrypted Security Policies in Outsourced Environments}, In The Sixth IEEE International Conference on Availability, Reliability and Security, ARES'11, pages 99-108. IEEE Computer Society, August 2011 (Full paper \textbf{acceptance rate: 20\%}).} \vspace{3mm} \\
\textbf{Abstract:} The enforcement of security policies in outsourced environments is still an open challenge for policy-based systems. On the one hand, taking the appropriate security decision requires access to the policies. However, if such access is allowed in an untrusted environment then confidential information might be leaked by the policies. Current solutions are based on cryptographic operations that embed security policies with the security mechanism. Therefore, the enforcement of such policies is performed by allowing the authorised parties to access the appropriate keys. We believe that such solutions are far too rigid because they strictly intertwine authorisation policies with the enforcing mechanism.

	In this paper, we want to address the issue of enforcing security policies in an untrusted environment while protecting the policy confidentiality. Our solution \gls{ESPOON} is aiming at providing a clear separation between security policies and the enforcement mechanism. However, the enforcement mechanism should learn as less as possible about both the policies and the requester attributes. \vspace{3mm} \\
\textbf{Keywords:} Encrypted Policies, Policy Protection, Sensitive Policy Evaluation, Data Outsourcing, Cloud Computing, Privacy, Security
	
\end{enumerate}

\section{Other Publications}

\noindent {\bf In International Conferences and Workshops}
  
\begin{enumerate}%[labelindent=.5em,labelsep=-.5cm,leftmargin=*]
  \setcounter{enumi}{8}
  
  \item
Soudip Roy Chowdhury, Muhammad Imran, \textbf{Muhammad Rizwan Asghar}, Sihem Amer-Yahia, Carlos Castillo, \textit{\textbf{Tweet4act: Using Incident-Specific Profiles for Classifying Crisis-Related Messages}, In Proceedings of The 10th International Conference on Information Systems for Crisis Response and Management (ISCRAM), Baden-Baden, Germany, May 2013.}

  \item
\textbf{Muhammad Rizwan Asghar}, Giovanni Russello, \textit{\textbf{ACTORS: A Goal-Driven Approach for Capturing and Managing Consent in e-Health Systems}, In 2012 IEEE International Symposium on Policies for Distributed Systems and Networks (POLICY), pages 61-69, 2012.}

	\item
%
% TODO: on update remove \\ that is added to fix format issue
%
\textbf{Muhammad Rizwan Asghar}, Giovanni Russello, \textit{\textbf{Flexible and Dynamic \\ Consent-Capturing}, In Jan Camenisch and Dogan Kesdogan, editors, Open Problems in Network Security, volume 7039 of Lecture Notes in Computer Science, pages 119-131, Springer Berlin Heidelberg, 2012.}

\end{enumerate}

\clearemptydoublepage

\chapter{Vitae}
\label{app:biography}

% left bottom right top
%\parpic{\includegraphics[width=2in,height=3in,trim=10mm 10mm 10mm 10mm,clip]{images/asghar}} % ,keepaspectratio
\parpic{\includegraphics[width=1.9in,keepaspectratio,trim=20mm 30mm 20mm 15mm,clip]{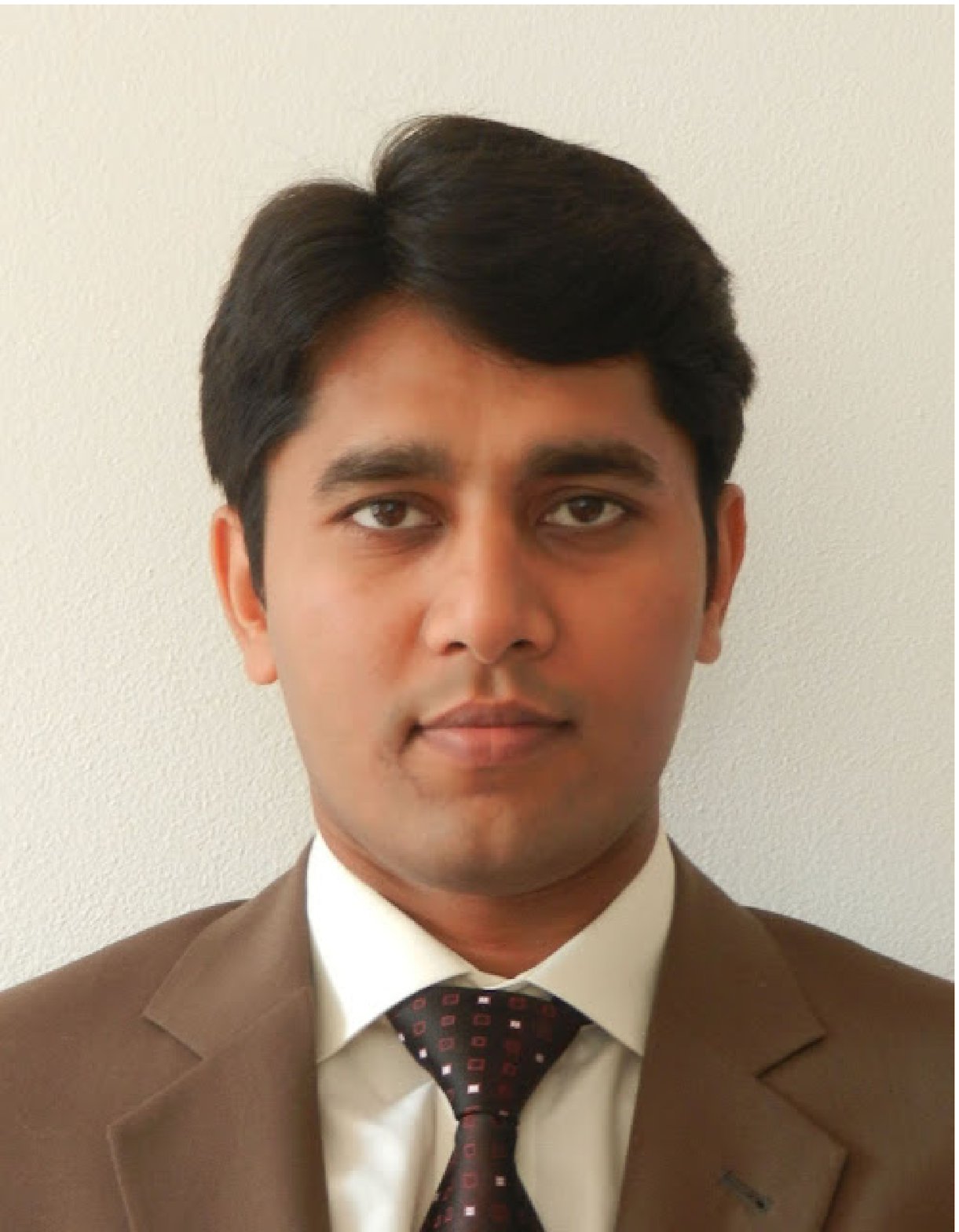}}
\noindent {\bf Muhammad Rizwan Asghar} was born in Faisalabad, Pakistan on July 26, 1983. He is a Researcher at CREATE-NET (an International Research Centre based in Trento, Italy) since September 2010. For pursuing his Ph.D., he joined the Security Group at the Department of Information Engineering and Computer Science (DISI), University of Trento, Italy in November 2010. In his Ph.D. research, he investigated privacy preserving enforcement of sensitive policies in outsourced and distributed environments (presented in this dissertation), under the supervision of Associate Prof. Dr. Bruno Crispo and Dr. Giovanni Russello. \\ \\
He was a Visiting Fellow in the Computer Science Laboratory at the Stanford Research Institute (SRI), California, USA from July to December 2012. Prior to joining CREATE-NET, he was a Research Assistant at the University of Trento from September 2009 to August 2010. He received his M.Sc. degree in Information Security Technology from the Department of Mathematics and Computer Science, Eindhoven University of Technology (TU/e), The Netherlands in 2009 and carried out his research on ``DRM Convergence: Interoperability between DRM Systems'' as a Master Thesis Student at Ericsson Research Eurolab, Germany. He obtained his B.Sc. (Honours) degree in Computer Science from Punjab University College of Information Technology (PUCIT), University of the Punjab, Lahore, Pakistan in 2006. During his career, he served also as a Software Engineer at international software companies. \\ \\
%
% on update remove \\ that is added to fix format issue
His research interests include access control, applied cryptography, cloud computing, \\ security and privacy. \\ \\
\textbf{Homepage:} \href{http://disi.unitn.it/~asghar/}{http://disi.unitn.it/\texttt{\char`\~}asghar}

\end{document}